# <u>Post-2000 Nonlinear Optical Materials and Measurements:</u>
# <u>Data Tables and Best Practices</u>

Nathalie Vermeulen,[1,*,◊] Daniel Espinosa,[2] Adam Ball,[3] John Ballato,[4,*] Philippe Boucaud,[5] Georges Boudebs,[6] Cecília L. A. V. Campos,[7] Peter Dragic,[8] Anderson S. L. Gomes,[7,*] Mikko J. Huttunen,[9,*] Nathaniel Kinsey,[3,*] Rich Mildren,[10] Dragomir Neshev,[11] Lázaro A. Padilha,[12] Minhao Pu,[13,*] Ray Secondo,[3] Eiji Tokunaga,[14] Dmitry Turchinovich,[15,*] Jingshi Yan,[11] Kresten Yvind,[13] Ksenia Dolgaleva,[2] and Eric W. Van Stryland[16,*]

(1) Brussels Photonics (B-PHOT), Department of Applied Physics and Photonics (IR-TONA), Vrije Universiteit Brussel, Pleinlaan 2, 1050 Brussel, Belgium

(2) School of Electrical Engineering and Computer Science, University of Ottawa, Ottawa, ON K1N 6N5, Canada

(3) Dept. Of Electrical & Computer Engineering, Virginia Commonwealth University, Richmond, VA, 23284, USA

(4) Dept of Materials Science and Engineering, Clemson University, Clemson, SC 29634, USA

(5) Université Côte d'Azur, CNRS, CRHEA, Rue Bernard Grégory, 06905 Sophia-Antipolis, France

(6) Univ Angers, LPHIA, SFR MATRIX, F-49000 Angers, France

(7) Departamento de Física, Universidade Federal de Pernambuco, 50670-901 Recife-PE, Brazil

(8) Dept. Of Electrical & Computer Engineering, University of Illinois at Urbana-Champaign, Urbana, IL 61801, USA

(9) Tampere University, Photonics Laboratory, Physics Unit, Tampere FI-33014, Finland

(10) MQ Photonics Research Centre, Macquarie University 2109 NSW Australia

(11) ARC Centre of Excellence for Transformative Meta-Optical Systems (TMOS), Department of Electronic Materials Engineering, Research School of Physics, Australian National University, Canberra, ACT, 2601, Australia

(12) Instituto de Fisica "Gleb Wataghin", Universidade Estadual de Campinas, Campinas, Sao Paulo, Brazil 13083-970

(13) Department of Electrical and Photonics engineering, Technical University of Denmark, DK-2800 Kgs. Lyngby, Denmark

(14) Department of Physics, Faculty of Science, Tokyo University of Science, 1-3 Kagurazaka, Shinjuku-ku, Tokyo 162-8601, Japan

(15) Fakultät für Physik, Universität Bielefeld, Universitätsstr. 25, 33615 Bielefeld, Germany

(16) CREOL, The College of Optics and Photonics, University of Central Florida, Orlando, FL, USA

* Team leaders for the material categories considered in the article

◊ Corresponding author: Nathalie.Vermeulen@vub.be

**Abstract**

In its 60 years of existence, the field of nonlinear optics has gained momentum especially over the past two decades thanks to major breakthroughs in material science and technology. In this article, we present a new set of data tables listing nonlinear-optical properties for different material categories as reported in the literature since 2000. The papers included in the data tables are representative experimental works on bulk materials, solvents, 0D-1D-2D materials, metamaterials, fiber waveguiding materials, on-chip waveguiding materials, hybrid waveguiding systems, and materials suitable for nonlinear optics at THz frequencies. In addition to the data tables, we also provide best practices for performing and reporting nonlinear-optical experiments. These best practices underpin the selection process that was used for including papers in the tables. While the tables indeed show strong advancements in the field over the past two decades, we encourage the nonlinear-optics community to implement the identified best practices in future works. This will allow a more adequate comparison, interpretation and use of the published parameters, and as such further stimulate the overall progress in nonlinear-optical science and applications.

## List of symbols

$d_{(eff)}$: effective second-order nonlinear coefficient
$dk/dT$: change in wave-number with temperature
$dn/dT$: change in linear refractive index with temperature
$g_{Brillouin}$: Brillouin gain coefficient
$g_{Raman}$: Raman gain coefficient
$I_{sat(eff)}$: (effective) saturation irradiance
$n_0$: linear refractive index
$n_{2(eff)}$: (effective) nonlinear index
$\alpha_0$: linear loss coefficient
$\alpha_2$: two-photon absorption coefficient
$\alpha_3$: three-photon absorption coefficient
$\gamma_{(eff)}$: effective nonlinear coefficient
$\gamma_{Brillouin}$: Brillouin gain factor
$\gamma_{Raman}$: Raman gain factor
$\Delta\alpha$: change in absorption
$\Delta n$: change in refractive index
$\Delta n_g$: change in group index
$\Delta OD$: change in optical density
$\Delta T$: change in pulse duration
$\varepsilon$: dielectric permittivity
$\eta$: conversion efficiency
$\lambda$: wavelength
$\lambda_{pump}$: pump/excitation wavelength
$\lambda_{probe}$: probe/signal wavelength
$\sigma$: nonlinear conductivity
$\tau_{pump}$: pump/excitation pulse duration
$\tau_{probe}$: probe/signal pulse duration
$\chi^{(2)}$: second-order susceptibility
$\chi^{(3)}$: third-order susceptibility
$\omega$: frequency

## List of abbreviations

0D: zero-dimensional
1D: one-dimensional
2D: two-dimensional
2FBS: two-frequency beat signal
2PA: two-photon absorption
3PA: three-photon absorption
3WM: three-wave mixing
BD: beam deflection
BIC: bound states in the continuum
CAIBE: chemically assisted ion-beam etching
CDQW: coupled double quantum well
ChG: chalcogenide glass
CVD: chemical vapor deposition
CVT: chemical vapor transport
CW: continuous-wave
dB: decibels
DB: diffusion bonded
DFG: difference-frequency generation

DUV: deep ultraviolet
EFISH: electric-field-induced second-harmonic generation
ENZ: epsilon near zero
EO: electro-optic
ER: ellipse rotation
FCA: free-carrier absorption
FEL: free-electron laser
FEOS: free-space electrooptic sampling
FIB: focused ion beam
FSR: free spectral range
FWHM: full width at half maximum
FWM: four-wave mixing
GO: graphene oxide
GVD: group velocity dispersion
GVM: group velocity mismatch
HHG: high-harmonics generation
HPHT: high pressure high temperature
HRS: hyper-Rayleigh scattering
HVPE: hydride vapor phase epitaxy
HW1/e$^2$M: half width at 1/e$^2$ maximum
IGA: induced-grating autocorrelation
IR: infrared
IRS: inverse Raman scattering
ISBT: intersubband transitions
ITO: indium tin oxide
LED: light emitting diode
LIDAR: light detection and ranging
LN: lithium niobate
LO: longitudinal optical mode
LPCVD: low-pressure chemical vapor deposition
LSPR: localized surface plasmon resonance
MBE: molecular-beam epitaxy
MFD: mode field diameter
MOCVD: metal-organic chemical vapor deposition
MOVPE: metal-organic vapor-phase epitaxy
MPAPS: multiphoton absorption photoluminescence saturation
MQWs: multiple quantum wells
NA: numerical aperture
NLA: nonlinear absorption
NLO: nonlinear optics / nonlinear-optical
NLR: nonlinear refraction
NSM: nanostructured material
NZI: near-zero index
OC: optically contacted
OP: orientation patterned
OPA: optical parametric amplification
OPG: optical parametric generation
OPO: optical parametric oscillation
PAMBE: plasma-assisted molecular beam epitaxy
PCF: photonic-crystal fiber
PDI: periodic domain inversion
PECVD: plasma-enhanced chemical vapor deposition
PhC: photonic crystal
PI: periodically inverted
PIC: photonic integrated circuit

PVT: physical vapor transport
PLD: pulse laser deposition
QCSE: quantum-confined Stark effect
QD: quantum dot
QPM: quasi-phase-matching
QW: quantum well
rGO: reduced graphene oxide
RIE: reactive ion etching
SA: saturable absorption
SBS: stimulated Brillouin scattering
SCG: supercontinuum generation
SEED: self-electro-optic effect device
SEM: scanning electron microscope
SESAM: semiconductor saturable absorber mirror
SFG: sum-frequency generation
SHG: second-harmonic generation
SLR: surface lattice resonance
SOI: silicon-on-insulator
SpBS: spontaneous Brillouin scattering
SPDC: spontaneous parametric down-conversion
SPM: self-phase modulation
SpRS: spontaneous Raman scattering
SRR: split-ring resonator
SRS: stimulated Raman scattering
SRTBC: spectrally resolved two-beam coupling
SWCNT: single-wall carbon nanotube
TDS: time-domain spectroscopy
TE: transverse-electric
THG: third-harmonic generation
TM: transverse-magnetic
TMD: transition metal dichalcogenide
TO: transverse optical mode
TPFP: tilted pulse front pumping
TRI: time-resolved interferometry
WZ: wurtzite
XPM: cross-phase modulation
ZB: zinc blende

## Table of contents

# 1 General introduction

The idea of composing a new set of data tables for nonlinear-optical (NLO) materials emerged in 2020 on the occasion of 60 years of NLO research [Franken1961, Kaiser1961]. In those 60 years, the field has witnessed tremendous growth, and several NLO data tables were published before the turn of the century [Robinson1967, Chase1994, Van Stryland1994, Sutherland1996, Dmitriev1999]. After the year 2000, additional data tables were introduced for specific material types (see, for example, [Nikogosyan2005, Smith2018]), but a data table that focuses on the post-2000 NLO research developments for a wide range of materials has not yet been presented. Nevertheless, there have been major advances in materials science and technology since 2000, and these have also accelerated the overall progress in NLO research. Whereas the idea of creating a new set of NLO data tables was originally introduced by John Dudley, we further elaborated on it along the following approach: the data tables presented here have been composed based on a representative set of experimental works published since 2000. In other words, the list of publications included in the table is not comprehensive. We mostly focused on experimental papers that not only provided NLO coefficients, but also reported experimental parameters that give the context and limits of validity for using the quoted coefficient values. In this regard, we also listed best practices for performing and reporting NLO experiments. Some of these best practices are appropriate for any NLO measurement, while others are specific for the chosen NLO characterization technique, e.g., second-harmonic generation (SHG), Z-scan, four-wave mixing (FWM), etc. In turn, many of these NLO techniques are appropriate only for specific categories of NLO materials: bulk materials, 0D-1D-2D materials, metamaterials, fiber waveguiding materials, on-chip waveguiding materials, and/or hybrid waveguiding systems. Both the NLO techniques and the material categories are defined in more detail in the specific separate sections. With this work, besides providing a set of NLO data tables focused on the progress since 2000, we also aim at stimulating the use of the listed best practices in future NLO publications to allow a better comparison, interpretation and use of the published parameters. The long-term goal of this article is to help advance the development of innovative NLO materials, devices and systems for real-life applications in optical data communication, signal processing, metrology, medical imaging, sensing, laser and quantum light generation, and many other areas.

Most of the NLO materials listed in the data tables are solids, but we also included some solvents as they are often used as reference materials in NLO measurements or for preparing solutions or dispersions. However, we did not consider organic/polymeric NLO materials since these are so numerous that tabulating them is outside the scope of this work. Gases are not included either, except for a few examples in the hybrid waveguiding systems category. More specifics of what is and is not included are given in the respective material category sections. We point out that these will be 'living' data tables that can be updated, so the materials that are currently absent might be added in the future.

To build the data tables presented here, we started by identifying the different material categories while also listing the different NLO techniques and their associated best practices. We then performed a literature search for experimental NLO papers published since 2000 and made a selection based on the listed best practices. Note that we did not limit our search to optical-wavelength-based experiments only but also included works on THz NLO. Finally, we filled out the data in dedicated table templates per material category. To minimize errors, the data filled out by each co-author were also cross-checked by another co-author. Hence, the parameter values listed in the tables should match with those provided in the selected papers.

The main NLO coefficients that we considered for composing the data tables are: the second-order nonlinear susceptibility $\chi^{(2)}$, the effective second-order nonlinear coefficient $d_{eff}$, the third-order nonlinear susceptibility $\chi^{(3)}$, the effective nonlinear index $n_{2,eff}$, the effective third-order nonlinear coefficient $\gamma_{eff}$, the two- and three-photon absorption coefficients $\alpha_2$, $\alpha_3$, the saturation irradiance $I_{sat}$ specified for saturation of the linear absorption, and the Raman and Brillouin gain coefficients $g_{Raman}$, $g_{Brillouin}$. Further information about the underlying physics for each of these coefficients can be found, for example, in [Sutherland1996, Bloembergen1996, Shen2002, Stegeman2012, Boyd2020]. We have not tabulated hyperpolarizabilities, photorefractive effects, electro-optic effects, stimulated polariton scattering, nor cascaded NLO processes. Note that the meaning of 'effective' is different for $d_{eff}$ than for $n_{2,eff}$ and $Re(\gamma_{eff})$. For $d_{eff}$ the subscript 'effective' implies that the coefficient comprises all contributions from the different tensor components being studied during the experiment. In contrast, $n_{2,eff}$ and $Re(\gamma_{eff})$ are effective coefficients in the sense that they might not be solely due to bound-electronic nonlinear transitions as one would expect for a 'pure' $n_2$ and $Re(\gamma)$, but instead could also contain contributions from, e.g., nuclear and/or thermal effects (see also Section 2.1). Finally, we point out that some works in the data tables do not only report a NLO coefficient but also provide NLO conversion efficiencies η and bandwidths in case techniques such as SHG, THG, Raman, Brillouin and FWM are used. These parameters are then also tabulated for those works since they are important to assess the practical use of the material in wavelength conversion applications.

The outline of the manuscript is as follows: in Section 2 we list various NLO measurement techniques and describe some best practices for performing and reporting NLO experiments. Here we make a distinction between best practices that can be applied in general and those that are technique-specific. In Section 3 we present the actual data tables per material category, together with background information of the status prior to 2000, a brief discussion of the advancements since 2000 and of the remaining challenges, and some recommendations for future works. Finally, we summarize and conclude in Section 4.

As an intermezzo before the main body of the manuscript, we want to visit what is arguably the single most studied material in NLO, namely fused silica. We think this digression is instructive on the difficulties inherent in making NLO measurements as well as in obtaining a complete theoretical understanding. It is also of significant importance because fused silica has been used as a reference in many studies of other materials, i.e., if the reference value is in error, so is the reported value of the measured material. This history of NLO measurements in fused silica is also illustrative of how NLO materials properties are not as well understood as one often assumes.

## 1.1 Fused silica nonlinearity

To illustrate the difficulties in reporting accurate values of nonlinear parameters, let us look at $n_2$ of fused silica (see Table 1 below). Fused silica is often used as a reference for determining the $n_2$, or $n_{2,eff}$ of other materials. We use $n_{2,eff}$ since while the bound-electronic nonlinearity is essentially instantaneous, other nonlinearities, including those involving nuclei, depend on the pulsewidth used. The values found in the literature for fused silica do not always agree, and, of course, there is dispersion. [Milam1998] attempted to determine the best experimental values at various wavelengths by taking a weighted average, determined by experimental error bars, of the published values up to 1998. This approach gave a value of ~2.7 x $10^{-20}$ m$^2$/W in the near-infrared (IR) with little dispersion from 1–1.6μm, with values increasing toward the UV. As summarized in [Agrawal2013] the nuclear (Raman) contribution to the nonlinear refractive index of fused silica may be of the order of ~20% as first discussed by the seminal work of [Hellwarth1975, Hellwarth1977]. Under this assumption, the bound-electronic nonlinearity is $n_2 \cong 2.2 \times 10^{-20}$ m$^2$/W. However, the nuclear contribution estimated in other publications varies from 13% [Smolorz1999] to 26% [Heiman1979]. Following [Agrawal2013], for pulses much longer than ~1 ps, the nuclear contribution should be fully established. However, [Santran2004] calculates a response function, from which they determine a pulsewidth dependence curve indicating that the nuclear contribution would not be fully developed until >10 ps. It appears that [Stolen1992] was the first to predict the pulsewidth dependence of $n_{2,eff}$ for fused silica, showing the effective Raman contribution decreasing for pulsewidths ≤100 fs, going to zero for a pulsewidth of ~30 fs, and then predicting that it actually turns negative for even shorter pulses. These two publications are the only ones we find that present the projected pulsewidth dependence for $n_{2,eff}$. Note that when using femtosecond experiments, the finite duration/finite bandwidth of the pulses results in spectral-filtering effects on the intrinsic material response that can result in difficulties in interpreting the experimental data [McMorrow1990]. For much longer pulses (>1ns) electrostriction can become important and can further significantly increase the measured $n_{2,eff}$. This is particularly important in fibers [Buckland1996]. Additionally, more recent measurements tend to use shorter pulses and yield smaller values of $n_{2,eff}$ (see Table 1). [Buckland1996] also calculates that $n_{2,eff}$ is reduced by a factor of $^8/_9$ due to the polarization randomization in fibers as indicated in Table 1. We also note that measurements of $n_{2,eff}$ reported for fibers have rarely been corrected for any modal overlap with the cladding. In principle, there could be a small systematic difference between fiber measurements and bulk measurements, but given the spread of data it is challenging to discern.

Among the several publications where attempts were made to measure the temporal response function, e.g., [Kang1996, Aber2000, Santran2004, Patwardhan2021], the only one to see a temporal dependence is [Smolorz1999], which used "spectrally-resolved two-beam coupling" with 18 fs, 800 nm pulses. They observed a small oscillatory signal in fused silica lasting for 100s of femtoseconds after excitation indicating a ~13% nuclear contribution to $n_{2,eff}$.

In all the papers of which we are aware where fused silica is used as a reference, its $n_{2,eff}$ is assumed constant. For all the materials reported in the tables in this publication that have been referenced to fused silica, this adds to the uncertainly of the reported $n_{2,eff}$ values. For very short pulses, ~10 fs, the nonlinear response could be reduced by as much as the above quoted fractions of the Raman contribution, the highest estimate being 26%.

We are not aware of any direct measurement of the variation of $n_{2,eff}$ with pulsewidth in the picosecond to femtosecond regime to determine its time dependence. Future measurements of the temporal dependence of the nonlinear response of fused silica and other materials would be quite useful, as has been done for solvents (see Section 3.2).

**Table 1**: Nonlinear refractive index of fused silica giving some historical references to early works as well as a representative group of more recent publications (not inclusive) – listed in order of year published. Note: fused silica is high purity synthetic glass (amorphous $SiO_2$) and different from naturally occurring or synthetic quartz (crystalline $\alpha$-$SiO_2$). All measurements listed use linearly polarized inputs except where noted. Legend for superscripts: see below the table. The following abbreviations have been used: de-pol=de-polarized; pol-maint=polarization maintaining; ER=ellipse rotation; SF-Pcr=self-focusing using the critical power for self-focusing, Pcr; Freq Mix=frequency mixing; TRI=time-resolved interferometry; SPM=self-phase modulation; 3WM=three-wave mixing; XPM=cross-phase modulation; 2FBS=two frequency beat signal; Multiple=multiple techniques were used in this compilation of publication data; SRTBC=spectrally resolved two-beam coupling; IGA=induced-grating autocorrelation-based upon time-delayed four-beam coupling in a photorefractive crystal

| Method | $n_{2,eff}$ (x$10^{-20}$ m$^2$/W) | $\lambda$ (nm) | Fiber/Bulk | Pulse width | Notes | Ref. |
|---|---|---|---|---|---|---|
| ER | 3.2 | 694 | bulk | 13 ns | fused quartz | [Owyoung1972, 1973] |
| SF Pcr | 3.94 | 1064 | bulk | 30 ps | | [Smith1975] |
| Freq Mix | 5.2 | ~525 | bulk | 3 ns | | [Levenson1974] |
| TRI | 2.73 | 1064 | bulk | 125 ps | | [Milam1976] |
| TRI | 2.73 | 1064 | bulk | 100-150 ps | | [Weber1978] |
| SPM | 2.7; 3.3 | 515 | 100 m fiber | 90 ps | 2nd entry x9/8 | [Stolen1978] |
| 3WM | 2.4 | 1064 | bulk | 3 ns | CS$_2$ used as ref | [Adair1992] |
| SPM | 2.36; 2.66 | 1319 | 250 m fiber | 110 ps | 2nd entry x9/8 | [Kim1994] |
| XPM | 2.48; 2.79 | 1550 | fiber | CW | 1% F doping | [Kato1995] |
| 2FBS | 2.2; 2.5 [#] | 1550 | fiber | CW | 2nd entry x9/8 | [Boskovic1996] |
| XPM | 2.47 | 1550 | fiber | CW | de-pol input not x9/8 | [Wada1996] |
| Z-scan | 2.14 | 1064 | bulk | 30 ps | | [DeSalvo1996] |
| Z-scan | 2.24 | 532 | bulk | 22 ps | | [DeSalvo1996] |
| Z-scan | 2.41 | 355 | bulk | 17 ps | | [DeSalvo1996] |
| Z-scan | 7.8 | 266 | bulk | 15 ps | | [DeSalvo1996] |
| Multiple | 2.56 | 800-1550 | bulk | ps to ns | compilation of ref. data | [Milam1998] |
| SRTBC | 2.3 | ~800 | bulk | 18 fs | | [Smolorz1999, 2000] |
| IGA | 2.44 | 1064 | 24 m fiber pol-maint | 53 ps | not x9/8 | [Garcia2003] |
| IGA | 2.2; 2.5 | 1064 | 23 m fiber | 50-70 ps | 2nd entry x9/8 | [Oguama2005a] |
| IGA | 1.81; 2.04 | 1064 | ~100 m fiber | 56 ps | 2nd entry x9/8 | [Oguama2005b] |
| IGA | 2.22 | 1064 | short fiber pol-maint | 50-70 ps | not x9/8 | [Oguama2005c] |
| ER | 2.5 | 775 | bulk | 150 fs | | [Miguez2015] |
| Z-scan | 2.23 | 1030 | bulk | 140 fs | | [Flom2015] |
| SPM | 2.1;2.4 | 1550 | fiber | telecom | Vascade fiber | [Makovejs2016] |
| SRTBC | 1.94 | 2300 | bulk | >65 fs | | [Patwardhan2021] |
| SRTBC | 2.0 | 3500 | bulk | >65 fs | | [Patwardhan2021] |

[#] Values should be reduced by 16% [Smolorz1999] due to electrostrictive contribution estimated by [Buckland1996].

# 2 NLO characterization techniques and their best practices

## 2.1 General best practices to obtain and report high-quality NLO measurement data

Before addressing the existing NLO characterization techniques, we provide some general 'best practices' that apply to most of them, i.e., some general rules for obtaining and reporting high-quality, useful NLO measurement data regardless of the technique employed:

- First, the material's composition, dimensions, method of fabrication and linear optical properties as well as temperature need to be known. Particularly, the linear loss/absorption at the wavelengths used in the NLO measurements is an essential parameter, but also other linear optical characteristics such as dispersion coefficients and the properties of optically excited free carriers can be relevant.

- Second, a careful quantification of the NLO pump/excitation parameters *inside* the material and, if relevant for the technique used, also of the signal/probe parameters is required. These parameters, relevant for both pulsed and continuous-wave (CW) operation, include (but are not limited to):
  - ➢ Wavelength
  - ➢ Peak power, irradiance (or in some cases electric field amplitude), and/or pulse energy
  - ➢ Beam size (specified as, e.g., FWHM, $HW1/e^2M$) and beam shape
  - ➢ Pulse width (specified as, e.g., FWHM, $HW1/e^2M$) and pulse shape
  - ➢ Pulse repetition rate (for pulsed operation)
  - ➢ Beam polarization and orientation with respect to crystal axes if appropriate and relative polarization for 2-beam experiments

- Last, the generated NLO response needs to be carefully measured, and the measurement results should be appropriately analyzed to extract the NLO coefficient(s). This can be done by means of a model for the technique used, by means of a benchmark measurement on a sample with well-known characteristics, etc. Models should also account for other effects that may interfere with the NLO measurements, and additional supporting information for the conclusions drawn (e.g., investigation of the dependence on pump power/irradiance/energy, the wavelength dependence, or the temporal dependence of the NLO response) is often required (see previous Section 1.1). In addition, special attention should be paid to the influence of the sample substrate (if any) and to the possible occurrence of sample damage when performing high-irradiance NLO experiments. When specifying conversion efficiencies and bandwidths, it should be clear which formulas were used for determining them. Finally, the inclusion of error bars on the data sets is very valuable to understand the limitations of the measurements and the extracted parameters.

We consider these general best practices to be an appropriate set of rules for carrying out, analyzing and reporting NLO measurements, independently from the type of technique used. It is also important to keep in mind the following general insights valid for any kind of NLO measurement: whenever long pulse durations and/or high pulse repetition rates are employed in NLO experiments, thermal effects can occur. They can build up collectively over many pulses or arise within the pulse duration, depending upon the material absorption and pulse width. These irradiance-dependent thermal effects can mask the NLO effect one aims to study. More generally speaking, it is unusual for a single nonlinearity to fully determine the overall NLO response. Hence, one should try to unravel the various contributions, e.g., by studying the power/irradiance/energy dependence, the wavelength dependence, the temporal dependence, etc., or at least consider the measured nonlinearity as an 'effective' coefficient for the input parameters quoted. We note that long and short pulses can trigger very different NLO processes (see, e.g., Section 3.2 on the NLO response of solvents) and could also exhibit different damage thresholds for the material under study. Also, measuring the wavelength dependence of the NLO response provides crucial information, since knowing the spectrum of the nonlinear absorption (NLA) and the dispersion curve of the nonlinear refraction (NLR) of a material is extremely helpful for understanding the fundamental physical interactions leading to the observed NLO effects. In some cases, the nonlinear absorption spectra/dispersion curves are necessary information for a proper interpretation of the experimental results. A more detailed discussion of the underlying physics can be found in [Christodoulides2010, Van Stryland2009a, Van Stryland2009b, Boyd2020, Stegeman2012, Sutherland2003, Shen2002, Bloembergen1996].

## 2.2 Technique-specific best practices

Besides the general best practices identified above, one should also consider additional best practices that are specific for the NLO technique used. In what follows, we list the most common technique families[1], the NLO coefficients they yield and the technique-specific best practices we have identified for each of them.

### 2.2.1 Best practices for second-order NLO techniques

#### 2.2.1.1 Second-order wave mixing

Second-order wave mixing processes, mainly second-harmonic generation (SHG), sum-frequency generation (SFG) and difference-frequency generation (DFG) [Shen1989, Shen2002, Sutherland2003, Wang2009, Liang2017, Prylepa2018], are often utilized in characterization of bulk materials, 0D-1D-2D materials, metamaterials, fiber waveguiding materials, on-chip waveguiding materials, and hybrid waveguiding systems. The developed techniques allow one to extract the material's $\chi^{(2)}$ and $d_{eff}$ coefficients, and the conversion efficiency of the associated process. In some cases, the techniques also allow characterization of the relevant components of the $\chi^{(2)}$ tensor, or the conversion bandwidth of the process. Second-order wave mixing is also at the heart of optical parametric oscillation (OPO), generation (OPG), and amplification (OPA) [Shen2002, Sutherland2003]. We note that electric-field-induced second-harmonic generation (EFISH) is in fact a third-order wave mixing process that in the presence of an external static electric field gives rise to second-harmonic generation.

Besides the general best practices in Section 2.1, it is important to verify successful phase matching of the NLO interaction in the material and to determine the parameters on which the considered phase matching approach relies (e.g., dispersion, poling parameters in some cases of quasi-phase-matching (QPM), etc.). Furthermore, when reporting the conversion efficiency and conversion bandwidth of the material, the used parameter definitions should be clearly stated. Finally, when performing SHG experiments, it is important to verify that the SHG response is not masked by two-photon-excited fluorescence effects.

#### 2.2.1.2 Second-order nonlinear imaging

Second-order nonlinear imaging techniques, most notably SHG imaging [Gannaway1978, Kumar2013, Yin2014, Woodward2016], can be meaningfully applied to material categories exhibiting spatially varying second-order nonlinear parameters (0D-1D-2D materials, metamaterials, on-chip waveguiding materials, and hybrid waveguiding systems). Second-order nonlinear imaging allows one to characterize the magnitude of $\chi^{(2)}$ along with the conversion efficiency associated with the material and the performed experiments.

---

[1] There exist very specialized techniques that are not addressed in this section as they are only employed in very specific or rather uncommon experiments. Some of these specialized techniques are used in a few papers included in the tables of this work. The reader is directed to those papers for further information on these techniques.

While the general best practices listed in Section 2.1 also apply for second-order nonlinear imaging, we further stress the importance of carefully characterizing the NLO excitation parameters (the peak power/irradiance/energy, beam size, pulse width and polarization) inside the material. This is particularly important, because optical components such as microscope objective lenses may considerably modify the properties of incident pulses. Furthermore, when using SHG imaging, the occurrence of two-photon-excited fluorescence effects should be avoided.

## 2.2.2 Best practices for third-order NLO techniques

### 2.2.2.1 Third-order parametric wave mixing

Third-order parametric wave mixing refers to techniques such as four-wave mixing (FWM) [Jain1979, Friberg1987, Agrawal2001], third-harmonic generation (THG) [Ward1969, Soon2005], and self-/cross-phase modulation (SPM/XPM) [Alfano1986, Agrawal2001], and can be applied to all material categories considered in this work. Third-order parametric wave mixing allows measuring the magnitude of Re($\chi^{(3)}$), n$_{2,\text{eff}}$ (which is usually assumed to be real), and Re($\gamma_{\text{eff}}$). Specifically in the case of SPM/XPM one can also extract the sign of these nonlinearities [Vermeulen2016a]. When using THG one most often quantifies a Re($\chi^{(3)}$) value that is specifically associated with bound-electronic nonlinear transitions. Note also that FWM may be more complex than THG and SPM/XPM as often the imaginary part of the third-order nonlinearity cannot be neglected. To separate the real and imaginary parts of the nonlinearity in a FWM signal, special experimental modifications and careful analysis involving also two-photon absorption (2PA) and Raman contributions, which will be described in the following sections, are generally required [Burris1985].

In FWM and THG experiments one most often measures the power/energy of the new signal generated by these wave mixing interactions, whereas SPM/XPM experiments are focused on measuring the phase modulation effects (e.g., spectral broadening, supercontinuum generation (SCG) [Dudley2010], …) that these processes induce. Besides the general best practices listed in Section 2.1, FWM and THG also require that their phase matching conditions are addressed. When the conversion efficiency of FWM or THG is measured, attention needs to be paid as to how the conversion efficiency and bandwidth are defined. For SPM experiments the spectral shape and phase information, i.e., chirp, of the input pulses also need to be known to allow for a correct measurement of the NLO coefficients [Vermeulen2016a]. However, in the case of wideband SCG, the input chirp typically has a negligible impact on the output spectrum and hence does not necessarily have to be known to allow for a correct extraction of the NLO coefficients. Finally, it is important to keep in mind that, besides the material's bound-electronic (Kerr) nonlinearity, other effects can also contribute to its third-order parametric wave mixing response (e.g., free-carrier effects contributing to SPM- and XPM-induced spectral broadening [Vermeulen2018, Zhang2016]).

## 2.2.2.2 Raman/Brillouin gain measurement techniques

The material intrinsic NLO parameter that most often characterizes stimulated scattering for the Raman/Brillouin processes is the gain coefficient $g_{Raman/Brillouin}$ (m/W). In the context of waveguides, the parameter can be expressed as a gain factor $\gamma_{Raman/Brillouin}$ (m$^{-1}$W$^{-1}$). Two classes of measurement methods are distinguishable: (1) direct methods – which rely on the amplification of a Stokes field, and (2) indirect methods – which rely on the determination of the spontaneous scattering cross-section and the dephasing time. Both methods can in principle be applied to all material categories considered in this work. Although most of the literature has focused on the Raman gain coefficient, the considerations often apply analogously to Brillouin scattering. Besides the general best practices listed in Section 2.1, we have identified the following best practices specifically for Raman and Brillouin gain measurements:

- Direct methods are based on measuring the threshold for stimulated Raman/Brillouin scattering (SRS/SBS) [Ippen1972, Zverev1999], the threshold of laser action [McQuillan 1970] and probing gain in an amplifier [Stappaerts1980, Niklès1997]. In the tables presented further on, these are indicated as "SRS/SBS threshold," "Raman/Brillouin laser threshold," and "SRS/SBS pump-probe," respectively. Major contributors to the measurement uncertainty include the precision of beam sizes and shapes, and the specification of the relative polarization of pump and Stokes (in relation to crystal axes in the case of crystals) and the pulse shape. In cases where the Rayleigh range of the beams are comparable to or shorter than the material length, the variation in beam waist upon propagation through the medium should be factored-in [Boyd1969]. The gain is a function of the pump and Stokes linewidths, decreasing markedly as the widths approach that of the phonon resonance [Georges1991, Agrawal2001, Bonner2014] or, equivalently, as pulse durations approach the $T_2$ phonon relaxation time [Basiev1999b]. For methods involving co-propagating multi-longitudinal mode beams, correlations between the pump and Stokes irradiances also need to be considered [Sabella2015] [Stappaerts1980]. Benchmarking against better known values of other materials [Basiev1999b], or the Kerr nonlinearity [Sabella2015], are valuable ways to increase confidence in the obtained values. For measurements at frequencies approaching the bandgap, a model is needed that includes a description of other nonlinearities occurring in the material (2PA, multiphoton absorption, free-carrier absorption (FCA), …). It should be noted that measurements of the stimulated scattering threshold often use a chosen exponential gain value $G$ in the vicinity of $G = 20$, i.e., it is assumed that for typical spontaneous seeding, an irradiance growth of $e^G$ with $G \sim 20$ leads to a readily observable stimulated scattering signal [Grasyuk1998, Ippen1972]. However, there is no universally accepted value for $G$ which may lead to some variability between reported values. In waveguides, two-color pump-probe techniques have been used to better separate the gain signal from background noise and other complications [Yang2020, Renninger2016].

- The gain coefficient may be determined indirectly by measuring the total cross-section for spontaneous scattering and its peak linewidth ($T_2$ dephasing time) [Basiev1999a], which are indicated in the tables, by the terms "SpRS/SpBS cross-section" and "SpRS, SpBS linewidth". Absolute measurements of the cross section are rare due to the increased practical difficulties associated with precise photometric measurements; a problem often mitigated by benchmarking against better known materials. It has also been shown that inverse Raman scattering (IRS) of a probe beam at the anti-Stokes frequency also gives access to absolute values without benchmarking [Schneebeli2013]. For Brillouin scattering, the cross-section may be determined from the photoelastic tensor [Ippen1972]. Determination of the linewidth may be readily achieved using a spectrometer provided that the instrument width is sufficiently small or is adequately deconvolved. The linewidth may also be determined from direct measurements of $T_2$ by using fast pump – probe optical pulses [Waldermann2008].

### 2.2.2.3 Third-order polarization rotation

Third-order polarization rotation refers to techniques where the change of polarization of an intense pump beam, or of a probe beam, is monitored to determine the real (or imaginary) part of the nonlinear susceptibility $\chi^{(3)}$, which may also yield the nonlinear refractive index $n_{2,\text{eff}}$ or a nonlinear absorption coefficient, e.g., $\alpha_2$. Examples are ellipse rotation (ER) measurements [Maker1964, Owyoung1972, Miguez2014], specifically for linearly isotropic materials such as liquids (see Section 3.2 on solvents), or Optical Kerr Gating [Duguay1969, McMorrow1988]. If the polarization is changed solely by refractive index changes, this refers to the quadratic Kerr effect. These techniques are most often used for bulk materials, such as isotropic liquids or linearly isotropic or cubic symmetry solids, but can also be applied to other material categories such as fiber materials and on-chip waveguiding materials.

Besides the general best practices listed in Section 2.1, third-order polarization rotation measurements also require special attention to precisely knowing the polarization of the source, or sources in the case of pump-probe experiments where knowledge of the relative polarization is also needed. The example of ER is illustrative. In that method for isotropic materials, due to symmetry, there are three unknown susceptibility elements. The key in an ER experiment is to create an elliptically polarized beam of well-known orientation. This can be done, e.g., by starting with a linearly polarized beam and introducing a quarter-wave plate, and then carefully measuring the transmission of the beam through a polarizer as a function of orientation, both before and after propagation through the NLO medium.

For pump-probe experiments, the optical Kerr gate is a good example. Starting with linearly polarized pump and probe beams with a relative polarization at 45 degrees, the pump induces a birefringence in the nonlinear sample that rotates the transmitted polarization. Monitoring the transmittance of the probe through a polarizer crossed to give zero transmittance without the pump gives a sensitive method for measuring the induced refractive index or absorption changes [Stegeman2012, Duguay1969, McMorrow1998]. Again, careful attention to the polarization is needed in these experiments.

### 2.2.2.4 Beam distortion/absorption

In beam distortion measurements, one quantifies the spatial variation of the beam due to a NLO process by measuring the power that passes through a fixed reference (e.g., pin-hole) or using a spatially resolved detector (e.g., quad-cell). For beam absorption measurements, one seeks to measure the nonlinearity-dependent power reflected/transmitted (also referred to as "nonlinear R/T") from the material, with the transmission dependent on both nonlinear absorption and scattering. In either case, it is possible to utilize single-beam or pump-probe style measurements. While more complex, the latter case provides more versatility to explore the temporal, polarization, and angular dependences of the nonlinearity as well as degenerate and non-degenerate spectral dependences.

Beam distortion and absorption refer to techniques such as Z-scan [Sheik-Bahae 1989, Sheik-Bahae 1990, Balu 2004, de Araújo 2016], I-scan [Taheri 1996], nonlinear reflection/transmission (loss) [Bloembergen 1962, Clays 1991, Sutherland 2003, Cheng 2018, Dong 2018, da Silva-Neto 2020], beam deflection [Ferdinandus 2013, Ferdinandus 2017,], and power limiting [Siegman 1962, Smirl 1975,

Soileau 1983, Gu 2006, Maas 2008, Liaros 2013, Riggs 2000]. Note that the I-scan technique had previously been used before being named as such (see, for example, [Bechtel 1976, Van Stryland 1985]). As these techniques involve spatial distortion/pointing and/or loss of the beam, they are generally used for the extraction of nonlinear parameters in free space. As a result, bulk materials, 0D-1D-2D materials, and metamaterials are commonly evaluated using these techniques. However, nonlinear loss measurements can also be used in waveguides and fibers, and is common for characterizing, e.g., 2PA and saturable absorption [Sorin 1984, Colin 1996, Set 2004, Jiang 2018].

In addition to the general best practices listed in Section 2.1, beam distortion techniques are the simplest to analyze using clean Gaussian beam profiles as this is currently an assumption of many subsequent analysis techniques [Sheik-Bahae 1990, Gu 2005] to extract the NLO coefficients; however, by using reference samples of known nonlinearity, such techniques can be calibrated for non-Gaussian beams [Bridges1995]. Furthermore, for thick samples and/or short pulses, it is useful to characterize the pulse chirp to correct for modifications in the temporal response of the nonlinearity due to varying group velocities. We also note that nonlinear beam distortion and loss, when used together, provide a complete methodology for characterizing lossy and/or lossless materials, enabling accurate evaluation of the magnitude, sign, and dispersion of the nonlinear refractive index $n_2$, $n_{2,eff}$, $\alpha_2$, $\alpha_{2,eff}$, thermal index ($dn/dT$ and $dk/dT$), etc. through the fitting of experimental quantities. With subsequent analysis [del Coso 2004, Christodoulides 2010], the complex value and sign of, e.g., a pure $\chi^{(3)}$ can be determined as well. When only one of the techniques is used, the information is often restricted. Beam distortion alone is sometimes sufficient to discern the sign and magnitude of the complex nonlinear index and susceptibility in the case of lossless (or low-loss) materials. However, in the case of very lossy materials ($Im(\varepsilon) \sim Re(\varepsilon)$), the terms of the complex susceptibility are inextricably coupled in measurements of both nonlinear losses and nonlinear refraction. The use of beam distortion (Z-scan or other methods) alone results in an effective coefficient which combines contributions from both nonlinear absorption and refraction [Del Coso 2004].

In general, it can be difficult to separate the different contributions using a single methodology (e.g., Z-scan requires both open and closed aperture experiments), although information on the sign of the respective nonlinearities can often be obtained. Beam absorption alone is typically sufficient to discern the sign and magnitude of nonlinear absorption but does not give information on the nonlinear refraction, although in some cases the sign of the nonlinear refraction can be predicted [Smith 1999] (here negative absorption would correspond to either absorption saturation or gain depending on circumstances). Since both components cannot be individually interpreted, absorption alone also cannot provide complete information on the $\chi^{(3)}$ although it may be possible to discern the sign of the terms. Pump-probe experiments such as beam deflection [Ferdinandus 2013] again require simultaneous transmittance and deflection signals to determine both absorptive and refractive components [Ferdinandus 2017]. These restrictions can be easily understood by comparing the number of unknowns (nonlinear refraction and absorption coefficients) for a given material with the constraints (distortion and absorption measurements) imposed by the experimental information. In the case of an under-constrained problem such as when attempting to extract two coefficients with a single measurement, the sign can be argued based on Kramers-Kronig causality if one knows the underlying mechanism(s) and/or the spectral response of one of the nonlinear terms. Lastly, we reiterate that although in some cases a singular technique is sufficient to obtain information on the nonlinearity of a material, if one assumes a pure $\chi^{(3)}$ process, it is rare for this assumption to be absolutely accurate (e.g., higher-order processes may provide non-negligible contributions). In this scenario, a single measurement will return only an effective nonlinear response, comprising multiple underlying physical mechanisms. Thus, the use of several analysis techniques is highly encouraged. This will provide more data to constrain analysis and elucidate the response of various contributing effects.

### 2.2.2.5  Nonlinear interferometry

The purpose of NLO interferometry is usually to measure the real part of the nonlinear susceptibility ($\chi^{(2,3)}$), but this technique can also be used to measure both the real and imaginary parts by suitable experimental modifications [Chang 1965, Sutherland 2003]. When using a wavelength of operation at which the NLO material is transparent, the most important part is the non-resonant $\mathrm{Re}(\chi^{(n)})$. In contrast, if we are interested in the enhancement of $\chi^{(n)}$ by resonance effects, information on both $\mathrm{Re}(\chi^{(n)})$ and $\mathrm{Im}(\chi^{(n)})$ spectra is needed. Although dispersion relations similar to the Kramers-Kronig relations hold for NLO coefficients [Sheik-Bahae 1990, Bassani 1991, Sheik-Bahae 1991, Hutchings 1992], it is practically impossible to strictly apply the Kramers-Kronig relations and there are restrictive conditions under which the Kramers-Kronig relations hold in time-resolved spectroscopy [Tokunaga 1995]. Thus, direct measurement of the real and imaginary parts is often required. NLO interferometry is most often used for bulk materials, but can be applied to other material categories as well. Although interferometric methods are also modified as heterodyne detection to be employed in SFG [Ostroverkhov 2005, Nihonyanagi 2013, Wang 2017] (otherwise only $|\chi^{(2)}|$ is obtained), most of the targeted nonlinearities measured by NLO interferometry are third-order nonlinearities in a pump-probe setup. In this case, a NLO interferometer is often based on pump, probe (signal), and reference beams. There are multiple interferometric configurations possible. They can be categorized in two-arm types (Michelson, Mach-Zehnder, Sagnac, etc., or two-beam interference) and resonator types (Fabry-Perot, or multiple-beam interference), and hybrids of them, e.g., Laser Interferometer Gravitational-Wave Observatory (LIGO) combining Michelson and Fabry-Perot [Abbott 2016]. There are also other classifications such as common path (Sagnac) and non-common path (Michelson) for the two arms, and also beam division methods are specified (space, time, direction, polarization, etc.).

In addition to the general best practices outlined in Section 2.1, it is important to evaluate the measurement sensitivity of the NLO interferometric method used and to account for a possible surface reflectance change due to changes in both the $\mathrm{Re}(\chi^{(3)})$ and $\mathrm{Im}(\chi^{(3)})$ parts. The bulk nonlinear refraction can also change the propagation time of pulses. With pulsed lasers, it is convenient to use time division (i.e., in the same sample path, reference, pump, and probe pulses arrive at the sample in that order) and spectral interference [Tokunaga 1995, Chen 2007]. In addition to time division, common path interferometry such as division by direction [Misawa 1995] or by polarization [van Dijk 2007] (where the reference and probe share a common optical path) is advantageous for obtaining stable interference, but otherwise active control to stabilize the optical path length is often required [Cotter 1989]. In order to cancel instabilities of the laser irradiance, balanced detection can also be employed by dividing the probe beam in two to be detected with an identical detector for subtractive detection [Waclawek 2019]. Resonance cavity effects with Fabry-Perot interferometers are also useful for enhancing the NLO effects by increasing the light-matter interaction [Birnbaum 2005, Fryett 2018, Wang 2021].

### 2.2.2.6  Third-order nonlinear imaging

Third-order nonlinear imaging techniques, most notably THG imaging [Squier1999, Woodward2016, Karvonen2017], can be meaningfully applied to material categories exhibiting spatially varying third-order nonlinear parameters (0D-1D-2D materials, metamaterials, on-chip waveguiding materials, and hybrid waveguiding systems). Third-order nonlinear imaging allows one to characterize the magnitude of $\chi^{(3)}$ along with the conversion efficiency associated with the material and the performed experiments. While the general best practices listed in Section 2.1 also apply for third-order nonlinear imaging, we further stress the importance of carefully characterizing the NLO excitation parameters (the peak power/irradiance/energy, beam size, pulse width and polarization) inside the material. This is particularly important, because optical components such as microscope objective lenses may considerably modify the properties of incident pulses.

## 2.2.3  Remarks on choice of NLO technique

The technique-specific best practices listed above combined with the general best practices in Section 2.1 should form a solid basis to obtain high-quality NLO measurement data and appropriate analysis results. When selecting a technique for characterizing a given NLO coefficient, it should be noted that for single-wavelength experiments only a specific frequency component of this nonlinearity will be revealed. For example, when applying both THG and SPM with pump frequency $\omega$ to characterize the $\chi^{(3)}$ of a given material, THG will yield $\chi^{(3)}$ ($3\omega$: $\omega$, $\omega$, $\omega$) whereas SPM will provide information about $\chi^{(3)}$ ($\omega$: -$\omega$, $\omega$, $\omega$) plus permutations [Bloembergen1996, Shen2002, Sutherland2003, Stegeman2012, Boyd2020]. This underlines the importance of including the technique used in each entry of the data tables presented below. And, as we have previously stated, it is highly desirable to use multiple techniques and/or multiple wavelengths and/or multiple pulse widths etc., as this both helps to distinguish different nonlinearities (with perhaps multiple being present simultaneously) and yields more physical insight into the NLO response(s) of the material under study.

# 3 Data tables and discussion

After having listed the most common NLO technique families and their best practices, we present here the actual data tables for bulk materials, 0D-1D-2D materials, metamaterials, fiber waveguiding materials, on-chip waveguiding materials, and hybrid waveguiding systems. The tables contain NLO data of representative experimental papers published since 2000 that we selected after verifying both the general best practices and the best practices specifically for the NLO techniques used in these works. In some cases, publications missing just one important parameter as specified in the best practices could not be included, although they seemed technically sound. The NLO coefficients that we searched for in the literature since 2000 are: $\chi^{(2)}$, $d_{eff}$, $\chi^{(3)}$, $n_{2,eff}$, $\gamma_{eff}$, $\alpha_2$, $\alpha_3$, $I_{sat}$, $g_{Raman}$, and $g_{Brillouin}$. We particularly looked for works that provide the most extensive NLO information, such as the dependence of the NLO coefficients on wavelength, on pulse duration, on doping level, on material composition, etc. For some recently developed materials in the data tables such extensive studies are yet to be performed; in those cases we selected papers that provide at least some quantitative NLO information, e.g., a single NLO coefficient and/or conversion efficiency, while complying with all (or almost all) the best practices. Note that no hyperpolarizabilities have been listed in the tables as these are typically only characterized at the molecular level, and that all the considered nonlinearities are of the second or third order, except for the three-photon absorption (3PA) coefficient $\alpha_3$. However, nonlinearities from, e.g., 2PA-induced free carriers, appear as effective fifth-order nonlinearities and may possibly contribute to some values reported in the literature. Note also that we have not included photorefractive effects, electro-optic effects, stimulated polariton scattering, nor cascaded second-order nonlinearities. The latter can very closely mimic bound-electronic nonlinearities but require propagation [DeSalvo1992, Torruellas1995, Stegeman1996]. These can sometimes be difficult to separate from other NLO responses. It is also important to keep in mind that some nonlinearities cannot properly be classified as either second or third order. As an example, we cite absorption saturation which in principle contains all orders of nonlinearities; however, it is often discussed as third order since at the lowest inputs it behaves that way. Furthermore, for several NLO coefficients the 'effective' values are often reported in the literature. When taking the example of the nonlinear index $n_2$, strictly speaking it originates only from the Kerr-nonlinear response of bound electrons in the material, but in practice the measured value can also comprise the nonlinear response from free electrons, electrostriction, thermal effects, etc. so that the terminology of an 'effective' nonlinear index $n_{2,eff}$ is more appropriate [Christodoulides2010, Buckland1996]. Also, some $\chi^{(3)}$ values in the literature are in fact 'effective' $\chi^{(3)}$ values containing multiple nonlinearity contributions rather than just the bound-electronic contribution. Finally, we point out that, besides tabulating NLO coefficients, the tables also contain NLO conversion efficiencies and bandwidths specified in works based on, e.g., SHG, THG or FWM. This way we also pay attention to those materials for which the NLO coefficients were reported before 2000 but that witnessed strong progress after 2000 in terms of attainable conversion efficiencies and bandwidths.

The general approach used for filling out the data tables has been to have a single entry per paper. For example, for a paper where the wavelength dependence of a given material nonlinearity is reported, there is typically only a single table entry, specifying a prominent nonlinearity value (in many cases the highest value) and the corresponding wavelength, but with a special annotation in this entry that shows the paper also contains information on the wavelength dependence. Note that a paper reporting on multiple materials is documented in the table with multiple entries, i.e., one entry per material. Another general policy has been to include only those parameter values that are explicitly specified in the paper. In other words, conversions from one physical quantity to another have been avoided to rule out calculation errors from our side. As such, when for a given entry in the tables there is no value for, e.g., pump peak power or irradiance, the paper under consideration might have specified the pump excitation in terms of fluence or average power rather than peak power or irradiance.

The data tables in the following subsections (Sections 3.1 – 3.8) are presented per material category (bulk materials, solvents, 0D-1D-2D materials, metamaterials, fiber waveguiding materials, on-chip waveguiding materials, and hybrid waveguiding systems) and include those parameters that are most important for the material category under consideration. The entries in these tables present NLO data at optical excitation wavelengths, and some of the tabulated works report on THz generation through, e.g., DFG of two optical excitation beams. However, for those papers where THz radiation was used as the excitation source, we added a dedicated THz table[2] at the end of the manuscript (see Section 3.8). The different data tables are accompanied by an introductory text addressing relevant background information prior to 2000, followed by a discussion of the general trends seen in the data tables (e.g., how the new post-2000 data represent an advancement) and some recommendations for future NLO research. Each of the subsections (Section $3.1 - 3.8$) starts with the names of the contributing co-authors in alphabetical order ("team") and the team leader.

We would like to emphasize again that it has not been our goal to include a comprehensive overview of papers but rather a representative set of works in these data tables. The current version of the tables contains papers with a publication date up to May 2021. Also, we point out that the existing literature already provides very valuable review papers that give an overview of NLO works for a specific material or material category, with a detailed discussion of the underlying physics. We refer to several of these recent review papers in the introductory texts and discussions below.

---

[2] The focus of the THz table is on the intrinsic THz NLO properties of various materials. It is useful to note that active electronic devices, such as Schottky diodes, high electron mobility transistors, resonant tunneling diodes, etc. can also generate (sub-)THz radiation when paired with appropriate in- and out-coupling structures. However, due to the fundamentally different nonlinear mechanisms of these devices, they are not included in this manuscript.

## 3.1 Bulk materials: data table and discussion

***Team:*** *Adam Ball, Philippe Boucaud, Georges Boudebs, Ksenia Dolgaleva, Daniel **Espinosa**, Anderson Gomes, **Nathaniel Kinsey (team leader)**, Rich Mildren, Ray Secondo, Eiji Tokunaga, Eric Van Stryland*

### 3.1.1 Introduction

#### 3.1.1.1 Bulk materials and their NLO applications

The bulk materials category is a highly diverse segment that presents NLO data on macroscopic 3D-material samples (ranging from thin films of ~100 nm thickness to crystalline material samples on the scale of cm's), where material structuring (see metamaterials category in Section 3.4) and quantum size effects (see 0D-1D-2D materials category in Section 3.3) are not important. The properties of bulk materials are typically measured using free-space diffracting beams (usually 1 or 2-beam experiments). Most experiments are performed with sample lengths less than the diffraction length (Rayleigh range) to allow for simple analysis.

Many bulk materials themselves are key for applications including frequency synthesis, stimulated scattering, ultrafast pulse characterization, and lasers. In addition, bulk materials generally serve as a starting point for the investigation of new materials before they are integrated into devices for applications across fields such as optoelectronics and fiber optics including uses in sensing, optical communications, and fiber lasers, as well as other types of active fibers and integrated photonics technology.

The listed materials, which include high-purity as well as impurity-doped materials, have been grouped into three sub-categories of insulators, semiconductors, and conductors. Within these groups, we focus our efforts on the development of inorganic materials only. Solvents are included in a separate section of this article (see Section 3.2), and we also draw attention to Section 3.8 on the characterization of material nonlinearities in the THz regime, many of which are bulk materials. These works have been separated to account for their unique considerations and in view of the rising area of THz NLO. However, works which utilize optical beams to generate THz waves remain in this section.

In keeping with this article's focus on post-2000 developments, the listed NLO materials emphasize those featured in research since 2000 but is not all-inclusive, giving priority to those publications that contain the most experimental information (see best practices in Section 2).

#### 3.1.1.2 Background prior to 2000

##### 3.1.1.2.1 Background for insulators/dielectrics

When the laser was invented in 1960 electro-optic (EO) crystals were already in use. Nevertheless, the terminology of "nonlinear optics" was not typically employed in this context. The first use of dielectrics for NLO is considered to be the wavelength conversion experiment in the seminal work of [Franken1961]. Some of the most important NLO materials are insulating crystals such as lithium niobate [Midwinter1968; Schaufele1966; Smith1965], beta-barium borate [Chen1987], and potassium titanyl phosphate [Bierlein1989], to name a few. Works from the earliest days of NLO have demonstrated efficient nonlinear actions accompanied by well-developed theories of nonlinear light-matter interaction and crystal optics [Boyd2020; Stegeman2012; Sutherland2003]. Thus, these nonlinear crystals have been key enablers for many past and current technologies including harmonic generators, OPO/OPG/OPA, optical modulators and switches making use of numerous second- and third-order nonlinear phenomena such as the Pockels effect, parametric mixing, and Raman scattering. In addition, efforts worked to improve the efficiency limits of phase matching by introducing periodically poled crystals with QPM [Lim1989; Myers1997; Wang1999]. In this regard lithium niobate is by far the most developed. Of course, much work went into the study of glasses as well. For example, the laser fusion program developed low nonlinear index glasses to prevent optical damage from catastrophic self-focusing in laser host glasses as well as other glass optical elements [Agrawal2013;

Tollefson2021]. Another interesting development was the measurement of third-order effects in popular second-order NLO crystals [Sheik-Bahae1997a] using Kerr lens autocorrelation [Sheik-Bahae1997b]. While third-order nonlinearities are generally masked by second-order processes, and thus might be neglected, certain applications can be limited by third-order effects such as self-focusing within the nonlinear crystal. Thus, understanding and correcting for these higher-order nonlinearities is a key feature for high-power applications or scenarios requiring especially large crystals.

Another key advance occurred during the late 1960s to 1970s with the invention of the laser diode, and the development of optical fibers. Together these opened new avenues to pursue all-optical devices aimed at applications in communications [Willner2019], and motivated studies of the nonlinearities in fused silica and other optical glasses [Agrawal2013]. Despite the small nonlinearity of silica-based glasses, the low propagation loss in fused silica fibers provided a platform to explore various nonlinear phenomena with appreciable efficiency including pulse compression [Dietel1983; Palfrey1985], temporal soliton propagation [Agrawal2013; Boyd2020; Stegeman2012], and all-optical switching [Agrawal2013; Boyd2020; Mollenauer1980; Stegeman2012] that have led to applications in fiber lasers [Ter-Mikirtychev2014] and communication systems [Willner2019] (see fiber waveguiding materials category in Section 3.5 for more information). Similarly, efforts also focused on increasing the typically weak nonlinearity of glasses by employing doped glasses [Alekseev1980; Ekimov1988; Kityk2002], photorefractive glasses [Hall1985] (not included in the table), and more recently chalcogenide glasses (ChGs) [Asobe1997]. Models to describe the nonlinearities in these dielectrics also evolved alongside experimental studies [Agrawal2013; Boyd2020; Stegeman2012]. The linear dispersion can be used in a semi-empirical way to predict nonlinear index changes in low-loss optical glasses. This so-called "BGO" model, named after Boling, Glass and Owyoung [Boling1978], was found to be valid to estimate $n_2$ of many glasses. This includes chalcogenide glasses [Fedus2010] despite the presence of NLA. It was also noted that for dielectrics, the bandgap determines the NLA spectrum as well as the dispersion of the NLR [DeSalvo1996; Sheik-Bahae1990; Sheik-Bahae1991]. Thus, the BGO model and bandgap scaling are complementary.

### 3.1.1.2.2   Background for semiconductors

Bulk semiconductors from groups II-VI, III-V, or IV, like ZnSe, GaAs, or Si, respectively, have been a mainstay in NLO since the invention of the laser. While the study of such materials in their bulk form has been a key starting point for the study of new semiconductors, structuring and combining semiconductors in novel ways has continually ignited many unique and important technologies such as detectors, lasers, and amplifiers. Much of this research excitement in semiconductors originated from the near-gap resonant nonlinearities which produced large NLO responses, with potential application in all-optical switching, all-optical computing, etc., albeit with limited bandwidth and recovery times [Hardy2007; Miller2010]. Applications involving resonant nonlinearities have included semiconductor optical amplifiers [Olsson1989; Urquhart2011], laser diodes [Bhattacharya1997], and more recently the generation of broadband infrared and THz waves [Krotkus2010; Shan2004]. Additionally, the introduction of quantum well structures provided large second-order nonlinearities, able to be engineered by altering the layer stack [Schmitt-Rink1989] (see the 0D-1D-2D materials category in Section 3.3 for more information). Such approaches have led to several useful technologies including semiconductor saturable absorption mirrors (SESAMs) [Jung1997; Keller1996; Kim1989], which have directly enabled ultrashort mode locked laser technologies, and self-electro-optic effect devices (SEEDs) [Miller1989]. Theoretical descriptions for the response of resonant nonlinearities in semiconductors evolved similarly, encompassing effects such as band-filling, saturation, bandgap renormalization, and exciton resonance [Garmire1998].

Similarly, below-gap non-resonant nonlinearities including harmonic generation, and wave mixing were also studied [Boyd2020; Stegeman2012], and while they provide nearly an instantaneous response, these were noted as being weaker than their resonant counterparts and challenging to phase match due to the limited anisotropy of many semiconductors. Through fabrication advances, these challenges were addressed via the use of QPM approaches [Armstrong1962; Gordon1993;

Thompson1976; Hum2007]. Another key advance in semiconductor nonlinearities came from a unified model of the nonlinear refractive index. In the case of direct bandgap semiconductors in their transparency range, they were understood in terms of nonlinear Kramers-Kronig relations [Hutchings1992; Sheik-Bahae1991] as a consequence of the NLA mechanisms of 2PA, Raman, and AC-Stark effects. Through this, it was noted that as the incident photon energy moves from resonant linear absorption saturation near the bottom of the conduction band to a non-resonant transition, the nonlinearity smoothly transitions from real saturation to "virtual" saturation otherwise known as the AC-Stark effect. Scaling rules and nonlinear dispersion relations based on the material bandgap were subsequently developed which supplied theoretical models to predict NLO coefficients, and to analyze the variety of reported NLO data [Hutchings1992; Sheik-Bahae1990; Sheik-Bahae1991; Wherrett1984]. Since 2000, Dinu's [Dinu2003a] and Garcia's [Garcia2012] models have been widely used for similar descriptions of indirect bandgap semiconductors.

### 3.1.1.2.3    Background for conductors

Nonlinearities in metals and conducting materials have been studied for nearly as long as NLO with early works examining harmonic generation in thin films [Bloembergen1968, Bloembergen1969; Sipe1980] as well as in percolated or random films [Shalaev1998]. The majority of research focused on elemental metals such as Au [C. K. Sun1994], Ag [Bloembergen1968], Cu [Elsayed-Ali1987], etc., and was closely tied with the fields of transient thermoreflectance to measure the thermal properties of metals [Brorson1990; Hohlfeld2000]. While bulk metals have exhibited large nonlinearities, they have not led to many applications in large part due to their high loss and propensity to be damaged under the high irradiances needed for NLO. One key exception to this has been surface enhanced nonlinearities, useful for achieving large improvement in processes such as Raman scattering [Wang2020]. Yet due to their limitations, the study of bulk metals was largely relegated to fundamental measurements and understanding of various light-induced effects such as so-called "Fermi-smearing," interband transitions, and free-electron photoionization [Allen1987; Anisimov1974]. More recently, metals are finding great interest in areas including plasmonics [Maier2007; Maradudin2014], metamaterials and nanoparticles embedded in dielectrics [Cai2009] (see metamaterials category in Section 3.4), as well as in the closely related area of epsilon-near-zero (ENZ) effects [Liberal2017; Reshef2019] (see Section 3.1.2.1.2).

## 3.1.1.3   Considerations for bulk materials when performing NLO measurements

For solid materials, attention should be paid to thermal effects when using high-repetition-rate lasers or with materials showing linear absorption. Thermal effects may build up collectively over many pulses or arise within the pulse width depending upon the loss of the material. As a result, one must clearly identify the different roles of thermal and non-thermal nonlinearities [Bautista2021; Falconieri1999]. Also, the role of the pulse duration is very important in determining the origin of the nonlinearity being exploited [Christodoulides2010; De Araújo2016] and should be considered accordingly. Additionally, in many cases, bulk materials are explored as thin films grown on substrates. In this scenario, researchers should carefully characterize and remove the response of the substrate. Works should contain a detailed description of how the substrate contribution was removed as well as a report of the substrate-only response for comparison. In addition to these specific considerations for bulk materials, also the best practices described in Section 2 should be taken into account when performing NLO measurements.

## 3.1.1.4   Description of general table outline

Tables 2A and 2B show a representative list of, respectively, the second- and third-order NLO properties reported since 2000 for our range of bulk materials, across our 3 subcategories. The works included in Tables 2A and 2B were selected based upon the best practices in Section 2 and the considerations outlined above. Table 2B lists, besides third-order nonlinearities, also 3PA coefficients. Each entry in the Tables contains information about the material and its properties, linear optical properties, NLO technique used, the excitation parameters, and the resulting NLO parameter(s).

The entries are arranged in alphabetical order, and within each material, entries are ordered by NLO technique (referred to as "Method" in the tables). The works that report dispersive datasets or dependences of the NLO parameter on multiple parameters (e.g., thickness, doping, composition, etc.) are denoted by "ᵃ ᵇ" respectively, in the Tables. The papers included in the Tables nominally report data obtained at room temperature, unless denoted otherwise by "ᶜ". The works reporting different compositions or material variations are split into different entries where applicable. For papers with multiple thicknesses of the same material, the thickest material is reported. The data is sub-divided into "Material properties," "Measurement details," and "Nonlinear properties." Measurements using two beams have the properties of the pump and probe beam listed separately, where provided, while single beam measurements are contained within the pump column. Units are contained within the table unless noted in the header. Within each column the information is given in the order of the header description, and powers of 10 (e.g., $10^{\pm\alpha}$) are written as E±α for compactness. If dispersive values for the NLO parameter were provided, the cited value represents the peak value for the material within the stated measurement range. For works that report tensor components, the component is shown as a leading number followed by a colon and the value (e.g., $\chi^{(2)}_{15} = $ 1E-12 [m/V] → 15: 1E-12 [m/V]). Indices are shown as listed in the original paper. Lastly, some papers relied on non-standard NLO measurement techniques (e.g., method listed as "Other") or have notes associated with their measurement/analysis (e.g., identifying the definition used for the conversion efficiency η). These values and information are listed within the "Comments" column with the value and unit identified.

## 3.1.2  Discussion

### 3.1.2.1  Advancement since 2000 and remaining challenges

#### 3.1.2.1.1  Advancement and challenges for insulators/dielectrics

Dielectrics are one of the most well-developed and well-understood NLO material platforms. While several works have continued to refine the understanding of nonlinearities in dielectrics such as fused silica (see Section 1.1) and sapphire [Nikolakakos2004] due to their use as reference materials, after 2000 most of the advancement has been focused on applications, providing new devices and enhancing performance of existing methods, including pushing the efficiency of cascaded NLO processes (not included in tables). While new material research is still ongoing, it has been more focused. In particular, advance in IR NLO materials such as chalcogenide glasses (ChGs), pnictide, and oxides has been realized, driven by emerging lasers, spectroscopy, sensing, and imaging applications in the mid- to far-IR. Among them, ChGs have received attention due to their ability to be integrated with optical fiber [Dudley2009; Knight2003]. When doped with rare earth elements such as Er, Nd, Pr, etc. or undoped, many applications of active and passive optical devices have been proposed [Mairaj2002; Sanghera2009; Ta'eed2005; Yan2016]. The high optical nonlinearities of these glasses have been studied over a wide range of wavelengths to optimize their compositions for applications related to all-optical switching or optical limiting [Chen2020; Cherukulappurath2004; Ensley2019; Fedus2010; Petit2009]. Furthermore, experiments have been made on mid-IR supercontinuum generation (SCG) by pumping bulk or fiber chalcogenide with fs pulses [Gattass2012; Granzow2011; Marandi2012; Møller2015; Petersen2018; Yu2013, Yu2014]. Similar advances in bulk form have been realized for cm-scale materials such as CdSiP$_2$ and lead oxyhalides, wherein lead oxides demonstrate both high transparency and strong SHG [Abudurusuli2021].

We also note the continued expansion of organic and organic-crystal materials for NLO applications from the mid-1990s and well into the 2000s [Bosshard2020; Nalwa1997]. While our effort here focuses on the development of inorganic materials, advances in organic nonlinearities have nonetheless constituted a major effort since 2000. For more information we direct the reader to reviews on organic NLO [Wang2012; Yesodha2004].

### 3.1.2.1.2    Advancement and challenges for semiconductors

Major advancements in semiconductor nonlinearities have been made largely due to improved fabrication and characterization methods. New measurements were completed on many semiconductors, such as Si, diamond, GaN, AlN, GaAs, GaP, InN, and InSb over a wide range of wavelengths, and compared to existing models of the NLR and NLA coefficients, leading to better understanding of the 3rd-order NLO coefficient's spectral behavior [Almeida2019; Bristow2007a; Chen2017; Fishman2011; Hurlbut2007; Lin2007; Oishi2018a, Oishi2018b; Olszak2010; Wang2013]. A unifying theory of 3PA has also been reported and compared to Wherrett's scaling model for most binary II-VI and III-V semiconductors [Wherrett1984; Benis2020].

Due to these advances in fabrication and characterization, new potential applications of semiconductors have begun to emerge. For example, QPM techniques in orientation-patterned (OP) GaAs and GaP have surpassed the strict constraint of birefringent phase-matching to produce efficient SHG, DFG, OPO, and OPG. As a result, a wide range of IR (wavelengths from 2500 nm to 14200 nm) [Feaver2013; Kuo2006; O'Donnell2019; Smolski2018; Vodopyanov2014] and THz (from 0.4 THz to 4.5 THz) frequencies [Kiessling2013; Schaar2008] can be generated.

Spurred by parallel advancements in plasmonics and metamaterials, in the last 5 years, another key area of advancement has been in the study of semiconductors that exhibit epsilon-near-zero (ENZ) or near-zero-index (NZI) properties [Kinsey2019; Liberal2017; Reshef2019] occurring from either free-electrons or phononic resonances (see also the metamaterials category in Section 3.4). Although optically lossy, which must be carefully considered for its impact upon applications, homogeneous ENZ materials have demonstrated extraordinarily large index modification ($\Delta n \sim 0.5 - 1$) [Alam2016; Benis2019; Caspani2016; Clerici2017; Kinsey2015a], enhanced harmonic generation [Capretti2015a, Capretti2015b], negative refraction [Bruno2020], optically defined surfaces [Saha2020], and recently temporal interfaces and interactions such as frequency shifting [Khurgin2020; Shaltout2016; Zhou2020]. While initial works demonstrated effects through experimental efforts, more recently theory has followed to provide a deterministic and predictive explanation of nonlinearities in popular ENZ materials [Khurgin2021; Secondo2020; Solís2021]. Currently, efforts have focused on the exploration of doped oxides such as Dy:CdO, Al:ZnO, and Sn:InO in the near-IR (~1300 – 1600 nm) but undoubtedly other materials will be explored in the future expanding into the mid- and far-IR spectral ranges.

### 3.1.2.1.3    Advancement and challenges for conductors

Since 2000, the growth of NLO research in plasmonics, metamaterials, metal nanoparticles, and nanophotonics [Maier2007; Maradudin2014] has brought a renewed interest in the study of nonlinearities in bulk metals and conducting materials [Kauranen2012]. The ability to nanostructure metals with ultra-tight light confinement has led to studies that make use of the inherent metal nonlinearity [MacDonald2009; Mayer2011; Mesch2016] as well as for enhancing the NLO response of an external material [Lee2014]. One key application has been in the enhancement of Raman interactions for sensing and material study [Jiang2010]. This has been the primary advance in the study, application, and understanding of metals in NLO since 2000 (see metamaterials category in Section 3.4 for more information). Ignited by materials exploration within the field of plasmonics, another advancement in conductors was the expansion of studied materials for optical applications, with many heavily doped semiconductors, conducting ceramics, and alloys being investigated [Saha2020]. These compound metallic materials provided additional benefits such as robustness, tunability of the NLO response, and new potential applications in sensing [Golubev2018; Robert2016] while exhibiting similar NLO responses to previously explored materials.

Similarly, the rise of topological materials and topological photonics has also affected NLO in the 21$^{st}$ century. While the area is closely linked to low dimensional materials (see 0D-1D-2D materials category in Section 3.3), some multilayer or bulk variants have also been studied. In particular, large second-order NLO effects have been reported in Weyl semimetals such as TaAs, TaP, and NbAs [Osterhoudt2019; Patankar2018; Wu2017; Yan2017], as well as enhanced surface nonlinearities in Dirac nodal-line ZrSiS semi-metals [Chi2020].

While the losses that occur in conducting materials and their propensity for damage remain a primary concern for the broad adoption of bulk conductors in NLO applications, the expansion of available materials combined with new physical understanding and physics may provide the ingredients for metals to make a contribution to the NLO field. Regardless, they remain key supporting materials for various fields of photonics, and understanding their inherent nonlinearities remains an important task for the field of NLO.

### 3.1.2.2  Recommendations for future works on bulk materials

As lasers and various methods of measurement continue to advance, our ability to characterize materials evolves in parallel. Systems with several tunable excitation beams, supercontinuum probes, variable repetition rates, etc. are expanding characterization capabilities and allowing more flexibility, which before 2000 was difficult to put into practice. We encourage future NLO works to embrace these new capabilities, exploring characterization and datasets which include more than just single wavelengths and pulse widths. Among various parameters, temporal and spectral information are perhaps the most useful information to be added for bulk materials, although this depends upon the application area of the material. In addition, the more thorough characterization will support an improved understanding of the NLO interactions at play, helping to delineate the roles of unwanted effects such as thermal nonlinearities and nonlinear scattering, and improving the reliability of results.

Another area of recommendation for the community is in the reporting of key experimental metrics. In many cases, papers are missing certain parameters of the experiment (peak/average power, beam waist, repetition rate, pulse width, etc., see tables) or do not report the method for calculation (flat top, Gaussian, full-width-half-max, etc.). This can lead to difficulty when interpreting the results as well as invoke questions as to the experimental rigor. By meticulously reporting the metrics of experiments, we can improve the trust in the results, aid in reproducibility, and help to avoid errors.

Lastly, the new millennium has seen an explosion in new materials available for study. In many cases, it can be observed that works jump directly to the demonstration of example applications without performing rigorous characterization of the material's NLO response. While such studies are useful, we remind the community that carefully conducted experiments and reporting of inherent NLO coefficients ($\chi^{(2,3)}$, $n_2$, $g_{Raman}$, etc.) is of key importance to the field as new materials and applications continue to be explored.

### 3.1.3 Data table for bulk materials

**Table 2A**. Second-order NLO properties of bulk materials from representative works since 2000. Legend for superscripts: see below the table.

| Material | Method | Material Properties | | | Measurement Details | | Nonlinear Properties | | | |
|---|---|---|---|---|---|---|---|---|---|---|
| | | Thickness Fabrication Substrate | Index Abs. Coeff. Wavelength | Crystallinity Bandgap Doping Level | **Pump** *Wavelength Peak Power or Irradiance Pulse Width Rep. Rate* | **Probe** *Wavelength Peak Power or Irradiance Pulse Width* | $\chi^{(2)}$ [m/V]* | d [m/V]* | $\eta$ [%]*,† | Comments<br>Reference |
| **Conductors** | | | | | | | | | | |
| NbAs [a] | SHG | -<br>Vapor Transport<br>- | -<br>5 [1/μm]<br>800 [nm] | Monocrys.<br>-<br>- | 800 [nm]<br>-<br>100 [fs]<br>- | -<br>-<br>- | - | 33: 2.7E-9 | - | -<br><br>[Wu2017] |
| TaAs [a] | SHG | -<br>Vapor Transport<br>- | -<br>5 [1/μm] | Monocrys.<br>-<br>- | 800 - 2500 [nm]<br>-<br>50 [fs]<br>- | -<br>-<br>- | - | - | - | Nonlinear conductivity σ = 5E-3 [1/Ω-V]<br><br>[Patankar2018] |
| TaAs [a,c] | SHG | -<br>Vapor Transport<br>- | -<br>5 [1/μm]<br>800 [nm] | Monocrys.<br>-<br>- | 800 [nm]<br>-<br>100 [fs]<br>- | -<br>-<br>- | - | 33: 3.6E-9 | - | -<br><br>[Wu2017] |
| TaAs | Other (see comments) | -<br>Chemical Vapor Transport<br>- | -<br>-<br>- | Monocrys.<br>-<br>$N_e$ = 8.08E+23<br>$N_h$ = 9.35E+23 [1/m³] | 10600 [nm]<br>CW<br>- | -<br>-<br>-<br>- | - | - | - | Method: Bulk Photovoltaic Effect<br><br>Nonlinear conductivity $\sigma_{aac}$ = 154 ± 17 [μA/V²]<br><br>[Osterhoudt2019] |
| TaP [a] | SHG | -<br>Vapor Transport<br>- | -<br>5 [1/μm]<br>800 [nm] | Monocrys.<br>-<br>- | 800 [nm]<br>-<br>100 [fs]<br>- | -<br>-<br>- | | 33: 3.2E-9 | - | -<br><br>[Wu2017] |

|  |  |  | |  | |  |  | |  |  |  |  |
|---|---|---|---|---|---|---|---|---|---|---|---|---|
| ZrSiS | SHG | -<br>Chemical Vapor Transport<br>- | -<br>-<br>- | Monocrys.<br>-<br>- | -<br>-<br>- | 1300 [nm]<br>7.6 [MW/m$^2$]<br>120 [fs]<br>1 [kHz] | -<br>-<br>- | 4780E-12 | - | 0.011 | -<br><br>[Chi2020] |
| **Semiconductors** | | | | | | | | | | | |
| Al$_{0.08}$Ga$_{0.92}$N [b] | SHG | 980 [nm]<br>MOCVD<br>(0001) c-Sapphire | -<br>-<br>- | Monocrys.<br>-<br>- | -<br>-<br>- | 1064 [nm]<br>5 [ns]<br>14 [Hz] | -<br>-<br>- | - | 31: 1.45E-12<br>15: 1.45E-12<br>33: -2.9E-12 | - | -<br><br>[Passeri2004] |
| Al$_{0.5}$Ga$_{0.5}$N | SHG | 63 [nm]<br>MOCVD<br>(0001) c-Sapphire | -<br>-<br>- | Monocrys.<br>-<br>- | -<br>-<br>- | 1064 [nm]<br>5 [ns]<br>13 [Hz] | -<br>-<br>- | - | 33: 1.20e-12 | - | -<br><br>[Larciprete2006] |
| Al$_x$Ga$_{1-x}$N [b] | SHG | 2.13 [μm]<br>MOCVD and HVPE<br>(0001) c-Sapphire | -<br>-<br>- | Monocrys.<br>-<br>- | -<br>-<br>- | 1064 [nm]<br>-<br>82 [MHz] | -<br>-<br>- | 31: 5.3E-12<br>33: -7.4E-12 | - | - | -<br><br>[Sanford2005] |
| AlN | SHG | 340 [nm]<br>MBE<br>Sapphire | -<br>-<br>- | Monocrys.<br>-<br>- | -<br>-<br>- | 800 [nm]<br>360 [kW]<br>7 [fs]<br>80 [MHz] | -<br>-<br>- | - | 3.22E-12 | - | -<br><br>[Kobayashi2007] |
| AlN [b] | SHG | 620 [μm]<br>PVT<br>- | -<br>60 [1/m]<br>1030 [nm] | Monocrys.<br>-<br>- | -<br>-<br>- | 1030 [nm]<br>2.2 [TW/m$^2$]<br>10 [ns]<br>1 [kHz] | -<br>-<br>- | - | 31: 9.55E-14<br>33: 4.3E-12 | - | -<br><br>[Majkić2017] |
| DB-GaAs [a] | DFG | 5.05 [mm]<br>-<br>- | -<br>-<br>- | -<br>-<br>- | -<br>-<br>- | 2100-2130 [nm]<br>-<br>6 [ps]<br>50 [MHz] | 2130-2160 [nm]<br>-<br>- | - | - | 0.012<br>1.2 | Optical-to-THz Conversion Eff.<br>Quantum Eff.<br><br>DFG was also performed in OC-GaAs and OP-GaAs samples.<br><br>[Schaar2008] |
| PI-GaAs | DFG | 653 [μm]<br>-<br>- | -<br>-<br>- | -<br>-<br>- | -<br>-<br>- | 2119.6 [nm]<br>4.6 [ns]<br>24 [kHz] | 2138.4 [nm]<br>-<br>4.6 [ns] | - | - | 4.31E-5<br>0.0103 | DFG Conversion Eff.<br>Quantum Eff.<br><br>[Mei2016] |

| Material | Method | Length / Growth | | | Source | | Value 1 | Value 2 | Value 3 | Comments |
|---|---|---|---|---|---|---|---|---|---|---|
| OP-GaAs | DFG | 1.8 [mm]<br>-<br>- | -<br>-<br>- | -<br>-<br>- | 1053 [nm]<br>-<br>17 [ns]<br>2 [kHz] | -<br>-<br>- | - | - | 0.6<br>4.2 | DFG Energy / Pump Energy Quantum Conversion Eff.<br><br>[Boyko2018] |
| GaAs [b] | SHG | -<br>-<br>- | -<br>-<br>- | Monocrys.<br>1.42 [eV]<br>- | 852 [nm]<br>-<br>14 [ps]<br>- | -<br>-<br>- | 7.5E-10 | - | - | -<br><br>[Bergfeld2003] |
| OP-GaAs | SHG | 500 [μm]<br>MBE and HVPE<br>4135 [nm] | -<br>8 [1/m]<br>4135 [nm] | -<br>-<br>- | 4135 [nm]<br>-<br>63 [ns]<br>25 [Hz] | -<br>-<br>- | - | 1.0857E-10 | 33 | -<br><br>[Skauli2002] |
| OP-GaAs [b] | Other (see comments) | 1 [mm]<br>-<br>- | -<br>-<br>- | -<br>-<br>- | 1952 [nm]<br>-<br>46 [ps]<br>1 [MHz] | -<br>-<br>- | - | - | 22.5 | Method: Optical parametric generation<br>OPA Power Conversion Eff.<br><br>[Fu2018] |
| OP-GaAs [b] | Other (see comments) | 1 [mm]<br>-<br>- | -<br>-<br>- | -<br>-<br>- | 1992 [nm]<br>150 [GW/m$^2$]<br>95 [ps]<br>100 [MHz] | -<br>-<br>- | - | - | 32.6 | Method: Optical Parametric Oscillation<br><br>[Fu2020] |
| DB-GaAs [b] | Other (see comments) | -<br>MBE and HVPE | 3.33<br>-<br>4400 [nm] | -<br>-<br>- | 4400 [nm]<br>100 [TW/m$^2$]<br>100 [fs]<br>1 [kHz] | -<br>-<br>- | - | - | 0.087<br>3.3 | Method: Optical Rectification<br>Internal Optical-to-THz Conversion Eff.<br>Internal Photon Conversion Eff.<br><br>[Vodopyanov2006] |
| OP-GaAs | Other (see comments) | 500 [μm]<br>MBE and HVPE | -<br>-<br>- | -<br>-<br>- | 3250 [nm]<br>-<br>1 [ps]<br>- | -<br>-<br>- | - | - | 51<br>15 | Method: Optical parametric generation<br>Slope Eff.<br>External Conversion Eff.<br><br>[Kuo2006] |
| OP-GaAs | Other (see comments) | 400 [μm]<br>MBE and HVPE<br>- | -<br>-<br>- | -<br>-<br>- | 1980 [nm]<br>-<br>120 [fs]<br>100 [MHz] | -<br>-<br>- | - | - | 1.6E-4<br>4.7E-4 | Method: Parametric down conversion<br>Optical-to-THz Conversion Eff.<br>Internal Optical-to-THz Conversion Eff. |

| Material | Type | Sample | n / λ | Bandgap | Pump (λ / power / pulse / rep) | d-coeff (1) | d-coeff (2) | d-coeff (3) | Reference / Comments |
|---|---|---|---|---|---|---|---|---|---|
| | | | | | | | | | [Imeshev2006] |
| OP-GaAs | Other (see comments) | 250 [μm] / - / - | 3.01 / - / 4200 [nm] | - | 1000 [nm] / - / 70 [fs] / 250 [MHz] | - | - | 25 | Method: Half-Harmonic Generation; Slope Conversion Eff. [Sorokin2018] |
| GaN [b] | SHG | 229.67 [μm] MOCVD and HVPE Free-standing | - / - / - | Monocrys. | 1064 [nm] / - / 82 [MHz] | 31: 5.7E-12 33: -9.2E-12 | - | - | - [Sanford2005] |
| GaN [b] | SHG | 4.2 [μm] Sapphire | 2.2978 / 1064 [nm] | Monocrys. | 1064 [nm] / - / 10 [ns] / 10 [Hz] | zxx: 15.3E-12 xzx: 14.8E-12 zzz: 30.3E-12 | - | - | [Fujita2000] |
| GaN | SHG | 3.5 [μm] MBE (0001) c-Sapphire | - / - / - | - | 1980 [nm] / - / 6 [ns] / 30 [Hz] | 2.8E-11 | - | - | - [Nevou2006] |
| GaN | SHG | 2 [μm] MOCVD (0001) c-Sapphire | - | Monocrys. 3.42 [eV] | 1064 [nm] / - / 5 [ns] / 14 [Hz] | - | 31: 2.4E-12 15: 1.8E-12 33: -3.7E-12 | - | - [Passeri2004] |
| GaN | SHG | 302 [nm] MOCVD (0001) c-Sapphire | - | Monocrys. 3.42 [eV] | 1064 [nm] / - / 5 [ns] / 13 [Hz] | - | 33: 4.82e-12 | - | - [Larciprete2006] |
| GaN | SHG | 340 [nm] MBE Sapphire | - | Monocrys. | 800 [nm] / 360 [kW] / 7 [fs] / 80 [MHz] | - | 1.59E-11 | - | - [Kobayashi2007] |
| GaN | SHG | 1 [μm] MOCVD (0001) c-Sapphire | 2.29 / 1064 [nm] | Monocrys. 3.4 [eV] | 1064 [nm] / - / 7 [ns] / 10 [Hz] | - | 31: 5.45E-12 15: 5.48E-12 33: -1.107E-11 | - | - [Kravetsky2000] |

| Material | Method | | | | | | | | | Comments / Reference |
|---|---|---|---|---|---|---|---|---|---|---|
| GaN | SHG Imaging | 2.5 [µm] <br> - <br> - | - <br> - <br> - | - <br> - <br> - | 1230 [nm] <br> 150 [fs] <br> 125 [MHz] | - <br> - <br> - | 3E-12 | - | - | - <br><br> [Sun2000] |
| ITO [a] | SHG | 37 [nm] <br> Sputtering <br> Silicon | - <br> - <br> - | - <br> - <br> - | 1100 [nm] <br> 420 [GW/m²] <br> 150 [fs] <br> 81 [MHz] | - <br> - <br> - | xzx: 5E-14 <br> zzz: 1.8E-13 | - | 3.00E-13 | Photon Conversion Eff. <br><br> [Capretti2015a] |
| (110)-GaP [b] | DFG | 663 [µm] <br> - <br> - | - <br> - <br> - | - <br> - <br> - | 1074 [nm] <br> 1.3 [TW/m²] <br> 5 [ns] <br> 10 [Hz] | 1064 [nm] <br> 1.5 [MW] <br> 10 [ns] | - | - | 0.182 <br> 39.6 | THz/Input <br> Photon Conversion Eff. <br><br> [Jiang2011] |
| (110)-GaP | DFG | 663 [µm] <br> - <br> - | 3.217 <br> 220 [1/m] <br> 120000 [nm] | - <br> - <br> - | 1064 [nm] <br> - <br> 10 [ns] | - <br> 1.2 [TW/m²] <br> 5 [ns] | - | - | 0.22 <br> 25 | Internal Eff. <br> Photon Conversion Eff. <br><br> [Jiang2010] |
| OP-GaP [a] | DFG | 1 [mm] <br> - <br> - | - <br> - <br> - | - <br> - <br> - | 1550 [nm] <br> 110 [fs] <br> 93.4 [MHz] | 1800-1960 [nm] <br> 60 [fs] | - | - | 19 | Photon Conversion Eff. <br><br> [Lee2017] |
| OP-GaP [b] | DFG | 1.7 [mm] <br> - <br> - | - <br> 32 [1/m] <br> 1064 [nm] | - <br> - <br> - | 1064 [nm] <br> 23 [ns] <br> 80 [kHz] | 1748 [nm] | - | - | 1 <br> 2.5 | Pump-to-DFG Eff. <br> Photon Conversion Eff. <br><br> [Wei2018] |
| OP-GaP | DFG | 1.7 [mm] <br> - <br> - | - <br> - <br> - | - <br> - <br> - | 1064 [nm] <br> 23 [ns] <br> 80 [kHz] | 1748 [nm] <br> 16 [ns] | - | 1.3E-11 | 1.2 | - <br><br> [Wei2017] |
| OP-GaP | Other (see comments) | 1 [mm] <br> - <br> - | - <br> - <br> - | - <br> 2.27 [eV] <br> - | 1048 [nm] <br> 140 [fs] <br> 80 [MHz] | - <br> - <br> - | - | - | 8.6 <br> 28.9 | Method: Optical Parametric Oscillation <br><br> Idler Slope Conversion Eff. <br> Quantum Eff. <br><br> [O'Donnell2019] |

| | | | | | | | | | |
|---|---|---|---|---|---|---|---|---|---|
| OP-GaP | Other (see comments) | 500 [µm]<br>-<br>- | 3.3<br>-<br>4200 [nm] | -<br>2.26 [eV]<br>- | 1000 [nm]<br>-<br>70 [fs]<br>250 [MHz] | -<br>-<br>- | - | - | 59 | Method: Half-Harmonic Generation<br><br>Slope Conversion Eff.<br><br>[Sorokin2018] |
| 4H SiC | SHG | -<br>SiC | 2.54<br>1064 [nm] | Monocrys.<br>- | 1064 [nm]<br>3.71 [GW/m²]<br>100 [ns]<br>5 [kHz] | -<br>- | - | 31: 6.5E-12<br>15: 6.7E-12<br>33: -11.7E-12 | - | -<br>[Sato2009] |
| 6H SiC | SHG | -<br>SiC | 2.54<br>1064 [nm] | Monocrys.<br>- | 1064 [nm]<br>3.71 [GW/m²]<br>100 [ns]<br>5 [kHz] | -<br>- | - | 31: 6.7E-12<br>15: 6.5E-12<br>33: -12.5E-12 | - | -<br>[Sato2009] |

[a] Multiple wavelengths reported, [b] Multiple parameters (e.g., thickness, crystal orientation) reported, [c] Measurement taken at a temperature other than room temperature. *Units as illustrated unless otherwise indicated in table, † See Comments for a description of conversion efficiencies

**Table 2B.** Third-order NLO properties of bulk materials from representative works since 2000. In addition, 3PA coefficients are provided at the end of the table. Legend for superscripts: see below the table.

| | | Material Properties | | | Measurement Details | | Nonlinear Properties | | |
|---|---|---|---|---|---|---|---|---|---|
| Material | Method | Thickness Fabrication Substrate | Index Abs. Coeff. Wavelength | Crystallinity Bandgap Doping Level | Pump *Wavelength Peak Power or Irradiance Pulse Width Rep. Rate* | Probe *Wavelength Peak Power or Irradiance Pulse Width* | $\chi^{(3)}$ [m²/V²]* | $n_2$ [m²/W] $\alpha_2$ [m/W]† $\alpha_3$ [m³/W²] | Comments Reference |
| **Conductors** | | | | | | | | | |
| TiN | Z-Scan | 52 [nm] Sputtering Fused Silica | 2.66 35 [1/µm] 1550 [nm] | Polycrys. - - | 1550 [nm] 14.1 [TW/m²] 150 [fs] 1 [kHz] | - - - | -5.9E-17-i1.7E-16 | -3.70E-15 -6.60E-09† | - [Kinsey2015b] |
| TiN | Beam Defl. | 30 [nm] Sputtering Fused Silica | - 54.8 [1/µm] 650 [nm] | Polycrys. 5 [eV] - | 800 [nm] - 55 [fs] 1 [kHz] | 650 [nm] - 110 [fs] | - | 4.70E-13 2.15E-07 | - [George2019] |
| **Insulators** | | | | | | | | | |
| Al₂O₃ [a] | Z-Scan | 1 [mm] - - | - - - | Monocrys. 7.3 [eV] - | 550 [nm] - 1[ps] 1 [kHz] | - - - | - | 3.30E-20 - - | - [Major2004] |
| As₂S₃ | Z-Scan | - Amorphous Mat. Inc. - | - - - | - - - | 2000 [nm] 20 [TW/m²] 90 [fs] 26 [Hz] | - - - | - | 2.50E-18 - - | - [Ensley2019] |
| AsSe | Z-Scan | - Amorphous Mat. Inc. - | - - - | - - - | 2000 [nm] 20 [TW/m²] 90 [fs] 26 [Hz] | - - - | - | 6.20E-18 - - | - [Ensley2019] |
| (Bi₂O₃)$_{0.25}$(ZnO)$_{0.375}$ (B₂O₃)$_{0.375}$ | Z-Scan | - Melt Quench - | - - - | - - - | 532 [nm] - 80 [ps] 10 [Hz] | - - - | - | 3.00E-18 5.50E-11 - | - [Gomes2007] |

| Ga$_{10}$Sn$_{20}$Se$_{70}$ | Z-Scan | -<br>Melt Quench<br>-<br>- | 2.65<br>1000 [1/m]<br>1064 [nm] | 1.58 [eV]<br>- | 1064 [nm]<br>10 [TW/m$^2$]<br>17 [ps]<br>10 [Hz] | -<br>-<br>- | - | 6.50E-17<br>1.14E-10<br>- | -<br><br>[Chen2020] |
|---|---|---|---|---|---|---|---|---|---|
| Ge$_{0.18}$Ga$_{0.05}$Sb$_{0.07}$S$_{0.3}$Se$_{0.4}$ | Z-Scan | -<br>Melt Quench<br>-<br>- | -<br>-<br>- | 1.66 [eV]<br>- | 1064 [nm]<br>20 [TW/m$^2$]<br>15 [ps]<br>10 [Hz] | -<br>-<br>- | - | 4.60E-18<br>9.00E-12<br>- | -<br><br>[Petit2006] |
| Ge$_{0.1}$As$_{0.1}$Se$_{0.6}$Te$_{0.2}$ | Z-Scan | -<br>Melt Quench<br>-<br>- | 2.9<br>317 [1/m]<br>1064 [nm] | -<br>- | 1064 [nm]<br>14 [TW/m$^2$]<br>15 [ps]<br>10 [Hz] | -<br>-<br>- | - | 2.00E-17<br>8.00E-11<br>- | -<br><br>[Cherukulappurath2004] |
| Ge$_{0.115}$As$_{0.24}$Se$_{0.645}$ | Z-Scan | -<br>Melt Quench<br>-<br>- | 2.265<br>-<br>1550 [nm] | 1.75 [eV]<br>- | 1550 [nm]<br>10 [TW/m$^2$]<br>260 [fs]<br>1 [kHz] | -<br>-<br>- | - | 7.90E-18<br><1.00E-13<br>- | -<br><br>[Wang2014] |
| Ge$_{0.15}$Sb$_{0.1}$Se$_{0.75}$ | Z-Scan | -<br>Melt Quench<br>-<br>- | 2.598<br>-<br>1550 [nm] | -<br>1.72 [eV]<br>- | 1550 [nm]<br>10 [TW/m$^2$]<br>260 [fs]<br>1 [kHz] | -<br>-<br>- | - | 7.50E-18<br><1.00E-13<br>- | -<br><br>[Wang2014] |
| Ge$_{0.16}$Sb$_{0.14}$S$_{0.7}$ | Z-Scan | -<br>Melt Quench<br>-<br>- | 2.31<br>-<br>1064 [nm] | -<br>2.1 [eV]<br>- | 1064 [nm]<br>20 [TW/m$^2$]<br>15 [ps]<br>10 [Hz] | -<br>-<br>- | - | 2.10E-18<br><1.00E-12<br>- | -<br><br>[Petit2009] |
| Ge$_{0.23}$Sb$_{0.07}$Se$_{0.7}$ | Z-Scan | -<br>Melt Quench<br>-<br>- | 2.58<br>-<br>1064 [nm] | -<br>1.66 [eV]<br>- | 1064 [nm]<br>5 [TW/m$^2$]<br>15 [ps]<br>10 [Hz] | -<br>-<br>- | - | 1.03E-17<br>2.40E-11<br>- | -<br><br>[Petit2007] |
| Ge$_{0.33}$As$_{0.12}$Se$_{0.55}$ | Z-Scan | -<br>Amorphous Mat. Inc.<br>-<br>- | -<br>-<br>- | -<br>- | 2000 [nm]<br>20 [TW/m$^2$]<br>90 [fs]<br>26 [Hz] | -<br>-<br>- | - | 3.80E-18<br>-<br>- | -<br><br>[Ensley2019] |
| Ge$_{0.33}$As$_{0.12}$Se$_{0.55}$ | Z-Scan | -<br>Amorphous Mat. Inc.<br>-<br>- | -<br>-<br>- | -<br>- | 3900 [nm]<br>20 [TW/m$^2$]<br>240 [fs]<br>10 [Hz] | -<br>-<br>- | - | 7.50E-18<br>-<br>- | -<br><br>[Ensley2019] |
| (GeS$_2$)$_{0.1}$(Sb$_2$S$_3$)$_{0.75}$(CsI)$_{0.15}$ | Z-Scan | -<br>Melt Quench<br>-<br>- | 2.8<br>-<br>1064 [nm] | -<br>- | 1064 [nm]<br>20 [TW/m$^2$]<br>17 [ps]<br>0.1 [Hz] | -<br>-<br>- | - | 8.10E-18<br>1.20E-11<br>- | -<br><br>[Fedus2010] |

| Material | Method | | | | | | | | Reference |
|---|---|---|---|---|---|---|---|---|---|
| $(NaPO_3)_{0.4}(BaF_2)_{0.1}(WO_3)_{0.5}$ | Z-Scan | -<br>Melt Quench<br>- | -<br>-<br>- | -<br>-<br>- | 532 [nm]<br>80 [ps]<br>10 [Hz] | -<br>-<br>- | - | 6.00E-19<br>5.00E-12<br>- | -<br><br>[Falcão-Filho2004] |
| $(Pb(PO_3)_2)_{0.4}(WO_3)_{0.6}$ | Z-Scan | -<br>Melt Quench<br>- | 1.93<br>116 [1/m]<br>1064 [nm] | -<br>-<br>- | 1064 [nm]<br>50 [TW/m²]<br>17 [ps]<br>10 [Hz] | -<br>-<br>- | - | 4.50E-19<br><2.00E-13<br>- | -<br><br>[Oliveira2010] |
| $(PbO)_{0.46}(Ga_2O_3)_{0.1}$<br>$(Bi_2O_3)_{0.426}$<br>$(BaO)_{0.014}$ | Z-Scan | -<br>Melt Quench<br>- | 2.3<br>-<br>1064 [nm] | -<br>-<br>- | 1064 [nm]<br>30 [TW/m²]<br>15 [ps]<br>10 [Hz] | -<br>-<br>- | - | 1.60E-18<br><1.00E-12<br>- | -<br><br>[De Araújo2005] |
| $(TeO_2)_{0.7}(GeO_2)_{0.15}$<br>$(K_2O)_{0.05}(Bi_2O_3)_{0.1}$ | Z-Scan | -<br>Melt Quench<br>- | 2.1<br>11 [1/m]<br>1064 [nm] | 3.08 [eV]<br>-<br>- | 1064 [nm]<br>50 [TW/m²]<br>17 [ps]<br>10 [Hz] | -<br>-<br>- | - | 7.50E-20<br><2.00E-13<br>- | -<br><br>[Oliveira2014] |
| **Semiconductors** | | | | | | | | | |
| Al:ZnO | Nonlin. R/T | 900 [nm]<br>PLD<br>- | -<br>-<br>- | -<br>-<br>- | 785 [nm]<br>13000 [TW/m²]<br>100 [fs]<br>100 [Hz] | 1258 [nm]<br><br>100 [fs] | 8E-20+i2E-20 | 3.50E-17<br>-2.50E-10†<br>- | -<br><br>[Caspani2016] |
| Al:ZnO [b] | Nonlin. R/T | 900 [nm]<br>PLD<br>- | -<br>-<br>- | -<br>-<br>- | 787 [nm]<br>900 [TW/m²]<br>100 [fs]<br>100 [Hz] | 1120-1550 [nm]<br>- | 3.5E-19-i2E-19 | 5.20E-16<br>-7.10E-09†<br>- | -<br><br>[Carnemolla2018] |
| Diamond | Brillouin Laser Threshold | 5 [mm]<br>CVD<br>- | -<br>-<br>- | Monocrys.<br>-<br>40 [ppb] | 532 [nm]<br>-<br>CW<br>- | -<br>-<br>- | - | -<br>-<br>- | $g_{Brillouin}$ = 6.0E-10 [m/W]<br>Brillouin shift = 167 [GHz]<br><br>[Bai2020] |
| Diamond | I-scan | 100 [μm]<br>Natural<br>- | -<br>-<br>- | Monocrys.<br>-<br>- | 250 [nm]<br>500 [TW/m²]<br>100 [fs] | -<br>-<br>- | - | -<br>2.20E-11<br>- | -<br><br>[Gagarskii2008] |
| Diamond | Nonlin. R/T | 100 [μm]<br>Type IIa<br>- | -<br>-<br>- | Monocrys.<br>-<br>- | 273 [nm]<br>90 [TW/m²]<br>150 [fs]<br>1 [kHz] | 410 [nm]<br>-<br>- | - | -<br>2.40E-11<br>- | -<br><br>[Roth2001] |

| Diamond | Nonlin. R/T | 300 [µm]<br>CVD<br>- | -<br>-<br>- | Monocrys.<br>-<br>- | 800 [nm]<br>-<br>40 [fs]<br>100 [kHz] | -<br>-<br>- | - | 7.30E-21<br>9.00E-13<br>- | -<br><br>[Motojima2019] |
|---|---|---|---|---|---|---|---|---|---|
| Diamond | Nonlin. R/T | 300 [µm]<br>CVD<br>- | -<br>-<br>- | Monocrys.<br>-<br>$N_{ion\ flux}$ = 2E+11 [1/cm$^2$] | 800 [nm]<br>-<br>40 [fs]<br>100 [kHz] | -<br>-<br>- | - | 1.20E-20<br>1.75E-12<br>- | -<br><br>[Motojima2019] |
| Diamond | Nonlin. R/T | 300 [µm]<br>CVD<br>- | -<br>-<br>- | Monocrys.<br>-<br>$N_{ion\ flux}$ = 1E+12 [1/cm$^2$] | 800 [nm]<br>-<br>40 [fs]<br>100 [kHz] | -<br>-<br>- | - | 2.42E-19<br>1.01E-12<br>- | -<br><br>[Motojima2019] |
| Diamond | SRS Pump-Probe | 8 [mm]<br>CVD<br>- | -<br>-<br>- | Monocrys.<br>-<br>- | 1864 [nm]<br>100 [kW]<br>4 [ns]<br>- | 2480 [nm]<br>3.75 [kW]<br>4 [ns] | - | -<br>-<br>- | $g_{Raman}$ = 3.8E-11 [m/W]<br>Raman shift = 1.33E+5 [1/m]<br><br>[Sabella2015] |
| Diamond | THG | 1 [µm]<br>CVD<br>Si | -<br>-<br>- | Nanocrystalline<br>-<br>- | 1055 [nm]<br>150 [GW/m$^2$]<br>90 [fs]<br>1 [kHz] | -<br>-<br>- | 5E-22 | -<br>-<br>- | Conversion Eff. = 5.6E-6 [%]<br><br>[Trojánek2010] |
| Diamond | Z-Scan | 300 [µm]<br>CVD<br>- | -<br>-<br>- | Monocrys.<br>-<br>- | 800 [nm]<br>-<br>40 [fs]<br>100 [kHz] | -<br>-<br>- | - | 4.16E-20<br>9.93E-14<br>- | -<br><br>[Motojima2019] |
| Diamond | Z-Scan | 300 [µm]<br>CVD<br>- | -<br>-<br>- | Monocrys.<br>-<br>$N_{ion\ flux}$ = 1E+12 [1/cm$^2$] | 800 [nm]<br>-<br>40 [fs]<br>100 [kHz] | -<br>-<br>- | - | 5.50E-20<br>1.61E-13<br>- | -<br><br>[Motojima2019] |
| Diamond | Z-Scan | 1 [µm]<br>CVD<br>Si | -<br>-<br>- | Nanocrystalline<br>-<br>- | 580 [nm]<br>2500 [TW/m$^2$]<br>90 [fs]<br>1 [kHz] | -<br>-<br>- | -4E-19 | -2.00E-17<br>-<br>- | -<br><br>[Trojánek2010] |
| Diamond | Z-Scan | -<br>CVD<br>- | -<br>-<br>- | Monocrys.<br>-<br>- | 310 [nm]<br>700 [TW/m$^2$]<br>100 [fs]<br>1 [kHz] | -<br>-<br>- | - | -<br>9.00E-12<br>- | -<br><br>[Kozák2012] |

| Material | Method | | | | | | | | Comments |
|---|---|---|---|---|---|---|---|---|---|
| Diamond | Z-Scan | -<br>CVD<br>-<br>- | - | Monocrys.<br>-<br>- | 350 [nm]<br>-<br>100 [fs]<br>1 [kHz] | -<br>- | | 9.00E-20<br>-<br>- | -<br><br>[Kozák2012] |
| Diamond | Z-Scan | 530 [μm]<br>CVD<br>-<br>- | - | Monocrys.<br>-<br>- | 260 [nm]<br>400 [TW/m²]<br>120 [fs]<br>1 [kHz] | -<br>- | | -<br>2.30E-12 | -<br><br>[Almeida2017] |
| Diamond | Z-Scan | 530 [μm]<br>CVD<br>-<br>- | - | Monocrys.<br>-<br>- | 427.5 [nm]<br>400 [TW/m²]<br>120 [fs]<br>1 [kHz] | -<br>- | | 1.70E-19<br>-<br>- | -<br><br>[Almeida2017] |
| Diamond | Other (see Comments) | 700 [μm]<br>CVD<br>-<br>- | - | Polycrys.<br>-<br>- | 351 [nm]<br>-<br>200 [fs] | 515 [nm]<br>-<br>200 [fs] | 2.7E-21 | 1.40E-19<br>-<br>- | Method: Phase Object Pump-Probe<br><br>[Zhang2017] |
| Diamond | Raman Linewidth (TCUPS ‡) | 500 [μm]<br>CVD<br>- | - | Monocrys.<br>-<br><1 [ppm] N$_2$ | 808 [nm]<br>-<br>80 [fs]<br>78 [MHz] | -<br>- | | | Raman Linewidth = 150 [1/m]<br><br>[Lee2010] |
| Diamond | Raman Linewidth (TCUPS ‡) | 250 [μm]<br>Natural<br>- | - | Monocrys.<br>-<br><1 [ppm] N$_2$ | 808 [nm]<br>-<br>80 [fs]<br>78 [MHz] | -<br>- | | | Raman Linewidth = 190 [1/m]<br><br>[Lee2010] |
| Diamond | Raman Linewidth (TCUPS ‡) | 420 [μm]<br>HPHT<br>- | - | Monocrys.<br>-<br>10-100 [ppm] N$_2$ | 808 [nm]<br>-<br>80 [fs]<br>78 [MHz] | -<br>- | | | Raman Linewidth = 190 [1/m]<br><br>[Lee2010] |
| Diamond | SRS Pump-Probe | 6.5 [mm]<br>CVD<br>- | - | Monocrys.<br>-<br>- | 532 [nm]<br>-<br>8 [ns]<br>- | 573 [nm]<br>-<br>8 [ns] | | | g$_{Raman}$ = 4.2E-10 [m/W]<br>Raman shift = 1.33E+5 [1/m]<br><br>[Savitski2013] |
| GaAs | Nonlin. R/T | -<br>-<br>- | - | -<br>-<br>1.42 [eV] | 8700 [nm]<br>-<br>10 [ps]<br>10 [Hz] | 900 [nm]<br>-<br>10 [ps] | | -<br>5.00E-09<br>- | -<br><br>[Fishman2011] |
| (001)-GaAs | Nonlin. R/T | 425 [μm]<br>-<br>- | - | Monocrys.<br>-<br>- | 1305 [nm]<br>1.09 [TW/m²]<br>68.1 [fs]<br>1 [kHz] | 0.5, 1, 2 [THz]<br>-<br>- | | -<br>4.25E-10<br>- | -<br><br>[Tiedje2007] |

| Material | Method | | | | | | | | |
|---|---|---|---|---|---|---|---|---|---|
| GaAs | SPM/XPM | 1 [mm]<br>-<br>- | -<br>-<br>- | -<br>-<br>- | 1064 [nm]<br>-<br>53 [ps]<br>100 [MHz] | -<br>-<br>- | - | -2.83E-10 [esu]<br>-<br>- | -<br>[Garcia2000] |
| GaAs | Z-Scan | 350 [µm]<br>-<br>- | -<br>-<br>- | Monocrys.<br>1.42 [eV]<br>- | 1680 [nm]<br>29 [TW/m²]<br>111 [fs]<br>1 [kHz] | -<br>-<br>- | - | 3.00E-17<br>2.50E-11 | -<br>[Hurlbut2007] |
| GaAs | Other (see Comments) | 700 [µm]<br>-<br>- | -<br>1.33 [1/µm]<br>800 [nm] | Monocrys.<br>-<br>- | 810 [nm]<br>-<br>55 [fs]<br>1 [kHz] | -<br>-<br>- | - | -<br>(110): 2.2E-9<br>(100): 1.8E-9 | Method: Optical-Pump THz-Probe<br>[Kadlec2004] |
| GaN [b] | THG Imaging | 2.5 [µm]<br>-<br>- | -<br>-<br>- | -<br>-<br>- | 1230 [nm]<br>-<br>150 [fs]<br>125 [MHz] | -<br>-<br>- | 1E-20 | -<br>-<br>- | -<br>[Sun2000] |
| GaN | Z-Scan | 10 [µm]<br>MOVPE<br>Sapphire | -<br>-<br>- | -<br>3.39 [eV]<br>1E+23 [1/m³] | 800 [nm]<br>-<br>120 [fs] | -<br>-<br>- | - | 2.80E-18 | -<br>[Almeida2019] |
| GaN | Z-Scan | 10 [µm]<br>MOVPE<br>Sapphire | -<br>-<br>- | -<br>3.39 [eV]<br>1E+23 [1/m³] | 550 [nm]<br>-<br>120 [fs] | -<br>-<br>- | - | 2.90E-11 | -<br>[Almeida2019] |
| GaN | Z-Scan | 500 [µm]<br>HVPE<br>GaN | -<br>-<br>- | Monocrys.<br>-<br>1E+23 [1/m³] | 724 [nm]<br>90 [TW/m²]<br>-<br>- | -<br>-<br>- | - | 2.50E-18<br>9.00E-12 | -<br>[Chen2017] |
| GaN | Z-Scan | 1 [mm]<br>HVPE<br>GaN | -<br>-<br>- | Monocrys.<br>3.39 [eV]<br>- | 760 [nm]<br>200 [TW/m²]<br>190 [fs]<br>20 [Hz] | -<br>-<br>- | - | 1.55E-18 | -<br>[Fang2015] |
| (100)-GaP [a] | Nonlin. R/T | 350 [µm]<br>-<br>- | -<br>610-745; 760-980 [nm]<br>- | -<br>2.26 [eV]<br>- | 650; 800 [nm]<br>-<br>7 [fs] | 650; 800 [nm]<br>-<br>7 [fs] | - | 2E-18; 4E-18<br>3E-11; 8E-11<br>- | -<br>[Grinblat2019] |

| Material | Method | Sample | n / α | Eg / N | Excitation | Excitation 2 | χ(3) [esu] | n2 / β | Reference |
|---|---|---|---|---|---|---|---|---|---|
| (100)-GaP[a] | Z-Scan | 350 [μm] | - | 2.26 [eV] | 800 [nm]<br>14 [TW/m$^2$]<br>7 [fs] | - | - | 1.60E-17<br>8.00E-11 | [Grinblat2019] |
| (100)-GaP[a,b] | Z-Scan | 500 [μm] | 3.1<br>1040 [nm] | 2.27 [eV] | 1040 [nm]<br>61 [fs]<br>52 [MHz] | - | - | 7.00E-18 | [Liu2010] |
| (100)-GaP[a,b] | Z-Scan | 500 [μm] | 2.2 [1/mm]<br>800 [nm] | 2.27 [eV] | 800 [nm]<br>37 [fs]<br>100 [MHz] | - | - | 4.16E-11 | [Liu2010] |
| InN[a,b] | FWM | 680 [nm]<br>PAMBE<br>GaN-on-sapphire<br>1400 [nm] | 2.78<br>1.2 [1/mm]<br>1400 [nm] | 0.82 [eV] | 1400 [nm]<br>35 [TW/m$^2$]<br>100 [fs]<br>1 [kHz] | - | 4.20E-10 [esu] | - | [Naranjo2007] |
| InN[b] | FWM | 1 [μm]<br>PAMBE<br>GaN-on-Si | 2.95<br>870 [1/mm]<br>1500 [nm] | Monocrys.<br>0.75 [eV]<br>1E+25 [1/m$^3$] | 1500 [nm]<br>100 [fs]<br>1 [kHz] | - | 8.00E-10 [esu] | -3.40E-10 | [Naranjo2009] |
| InN | Nonlin. R/T | 750 [nm]<br>Sputtering<br>GaN-on-Sapphire | - | Polycrys.<br>1.74 [eV]<br>1E+26 [1/m$^3$] | 1550 [nm]<br>153 [TW/m$^2$]<br>100 [fs]<br>100 [MHz] | 1550 [nm]<br>15.3 [TW/m$^2$]<br>100 [fs] | - | 1.67E-09 | [Valdueza-Felip2012] |
| InN | Z-Scan | 250 [nm]<br>MBE<br>AlN on c-Sapphire<br>790 [nm] | 2.45<br>8.7 [1/μm]<br>790 [nm] | 0.8 [eV]<br>5.5E+24 [1/m$^3$] | 790 [nm]<br>26 [TW/m$^2$]<br>200 [fs]<br>80 [MHz] | - | - | - | Nonlinear Absorption Cross Section: 4E-21 [m$^2$]<br><br>[Tsai2009] |
| InN | Z-Scan | 1.4 [μm]<br>PAMBE<br>AlN/GaN on R-sapphire | - | Monocrys.<br>0.65 [eV]<br>1.4E25 [1/m$^3$] | 1500 [nm]<br>20 [TW/m$^2$]<br>120 [fs] | - | - | 1.90E-14<br>4.75E-08 | [Ahn2014] |

| Material | Method | | | | | | | | Reference |
|---|---|---|---|---|---|---|---|---|---|
| InP [a,b] | I-Scan | 2 [mm]<br>-<br>- | -<br>-<br>- | -<br>-<br>- | 1064 [nm]<br>6 [TW/m²]<br>10 [ps]<br>- | -<br>-<br>- | - | 2.55E-10<br>- | -<br>[Gonzalez2009] |
| InP:Fe [a] | I-Scan | 2 [mm]<br>-<br>- | -<br>-<br>- | -<br>-<br>- | 1064 [nm]<br>6 [TW/m²]<br>10 [ps]<br>- | -<br>-<br>- | - | 2.55E-10<br>- | -<br>[Gonzalez2009] |
| InP:Zn [a] | I-Scan | 2 [mm]<br>-<br>- | -<br>-<br>- | -<br>-<br>- | 1064 [nm]<br>6 [TW/m²]<br>10 [ps]<br>- | -<br>-<br>- | - | 2.55E-10<br>- | -<br>[Gonzalez2009] |
| (001)-InP:Fe | Nonlin. R/T | 350 [µm]<br>-<br>- | 3.1<br>-<br>1640 [nm] | -<br>-<br>- | 1640 [nm]<br>5 [TW/m²]<br>250 [fs]<br>50 [MHz] | -<br>-<br>- | xxxx: 2.066E-18<br>xxyy: 3.11E-19<br>xyyx: 1.231E-18 | 3.70E-10<br>- | -<br>[Matsusue2011] |
| (001)-InP:Fe | Nonlin. R/T | 365 [µm]<br>-<br>- | -<br>-<br>- | Monocrys.<br>-<br>- | 1305 [nm]<br>1.42 [TW/m²]<br>68.1 [fs]<br>1 [kHz] | 0.5, 1, 2 [THz]<br>-<br>- | - | 7.03E-10<br>- | -<br>[Tiedje2007] |
| (001)-InP:Fe | Nonlin. R/T | 310 [µm]<br>-<br>- | -<br>2 [1/m]<br>1600 [nm] | Monocrys.<br>-<br>- | 1600 [nm]<br>1.1 [TW/m²]<br>160 [fs]<br>- | 1600 [nm]<br>160 [fs]<br>- | - | 3.30E-10<br>- | -<br>[Vignaud2004] |
| (001)-InP:S | Nonlin. R/T | 310 [µm]<br>-<br>- | -<br>490 [1/m]<br>1600 [nm] | Monocrys.<br>-<br>7E+24 [1/m³] | 1600 [nm]<br>1.1 [TW/m²]<br>160 [fs]<br>- | 1600 [nm]<br>160 [fs]<br>- | - | 2.60E-10<br>- | -<br>[Vignaud2004] |
| (001)-InP:Zn | Nonlin. R/T | 310 [µm]<br>-<br>- | -<br>2.83 [1/mm]<br>1600 [nm] | Monocrys.<br>-<br>1.5E+24 [1/m³] | 1600 [nm]<br>1.1 [TW/m²]<br>160 [fs]<br>- | 1600 [nm]<br>-<br>160 [fs] | - | 3.10E-10<br>- | -<br>[Vignaud2004] |

| Material | Method | | | | | | | | Reference |
|---|---|---|---|---|---|---|---|---|---|
| (001)-InP:Fe | Z-Scan | 579 [µm]<br>-<br>- | 3.147<br>-<br>1640 [nm] | -<br>-<br>- | 1640 [nm]<br>-<br>194.7 [fs]<br>47.8 [MHz] | -<br>-<br>- | xxxx: 2.619E-18<br>xxyy: 4.92E-19<br>xyyx: 1.531E-18 | -<br>2.20E-10<br>- | -<br>[Oishi2018b] |
| (001)-InP:Fe | Z-Scan | 581 [µm]<br>-<br>- | 3.147<br>-<br>1640 [nm] | -<br>-<br>- | 1640 [nm]<br>-<br>194.7 [fs]<br>47.8 [MHz] | -<br>-<br>- | 2.65E-18 | -<br>2.30E-10<br>- | -<br>[Oishi2018a] |
| InSb | Nonlin. R/T | 1 [mm]<br>-<br>- | -<br>-<br>10600 [nm] | -<br>0.18 [eV]<br>- | 10600 [nm]<br>-<br>-<br>- | -<br>-<br>- | - | - | Nonlinear Free Carrier Refraction Cross Section: 4.5E-19 [m$^2$]<br><br>[Dubikovskiy2008] |
| InSb | Z-Scan | 540 [µm]<br>-<br>- | -<br>-<br>- | Monocrys.<br>0.18 [eV]<br>9.00E+19 [1/m$^3$] | 9000 [nm]<br>-<br>370 [fs]<br>1 [kHz] | -<br>-<br>- | - | -<br>2.90E-08<br>- | -<br>[Olszak2010] |
| InSn:O | Beam Defl. | 310 [nm]<br>-<br>Float Glass | -<br>-<br>- | -<br>-<br>- | 1242 [nm]<br>-<br>100 [fs]<br>1 [kHz] | 1050 [nm]<br>-<br>- | - | -<br>5.50E-17<br>- | -<br>[Benis2017] |
| InSn:O | THG | 37 [nm]<br>Sputtering<br>Silicon | -<br>-<br>1550 [nm] | -<br>-<br>- | 400 [GW/m$^2$]<br>150 [fs]<br>81 [MHz] | -<br>-<br>- | 3.5E-18 | -<br>-<br>- | -<br>[Capretti2015a] |
| Silicon [a,b] | Z-Scan | 675 [µm]<br>-<br>- | -<br>-<br>- | Monocrys.<br>-<br>1E+21 [1/m$^3$] | 2600 [nm]<br>570 [TW/m$^2$]<br>150 [fs]<br>1 [kHz] | -<br>-<br>- | - | 2.50E-18<br>-<br>2.00E-27 | -<br>[Wang2013] |
| Silicon | Z-Scan | 125 [µm]<br>-<br>- | -<br>-<br>- | Monocrys.<br>1.12 [eV]<br>Intrinsic | 1220 [nm]<br>78 [TW/m$^2$]<br>200 [fs]<br>1 [kHz] | -<br>-<br>- | - | 4.70E-18<br>2.10E-11<br>- | -<br>[Bristow2007b] |
| Silicon | Z-Scan | 480 [µm]<br>-<br>- | -<br>-<br>- | Monocrys.<br>-<br>10 [Ω-cm] | 1540 [nm]<br>7 [TW/m$^2$]<br>130 [fs]<br>76 [MHz] | -<br>-<br>- | - | 4.50E-18<br>7.90E-12<br>- | -<br>[Dinu2003b] |
| Silicon | Z-Scan | 500 [μm]<br>-<br>- | -<br>-<br>- | Monocrys.<br>-<br>Intrinsic | 1550 [nm]<br>-<br>190 [fs]<br>1 [kHz] | -<br>- | - | 5.00E-18<br>1.03E-11 | -<br><br>[Gai2013] |
| Silicon | Z-Scan | 500 [μm]<br>-<br>- | -<br>-<br>- | Monocrys.<br>-<br>20 [Ω-cm] | 1501 [nm]<br>-<br>500 [Hz] | -<br>- | - | 2.20E-18<br>4.80E-12 | -<br><br>[Lin2007] |
| 4H SiC [a] | Z-Scan | 498 [μm]<br>-<br>- | -<br>-<br>- | Monocrys.<br>3.25 [eV]<br>- | 530 [nm]<br>-<br>123 [fs]<br>50 [kHz] | -<br>- | - | 3.19E-19<br>2.00E-12 | -<br><br>[Guo2021] |
| 6H SiC [a] | Z-Scan | 258 [μm]<br>-<br>- | -<br>-<br>- | Monocrys.<br>2.99 [eV]<br>- | 800 [nm]<br>-<br>176 [fs]<br>50 [kHz] | -<br>- | - | 3.88E-19<br>3.95E-13 | -<br><br>[Guo2021] |
| 6H SiC [b] | Z-Scan | 340 [μm]<br>-<br>- | | Monocrys.<br>3.10 [eV]<br>1.00E+21 [1/m$^3$] | 780 [nm]<br>-<br>160 [fs]<br>41 [Hz] | -<br>- | - | 4.75E-19<br>6.40E-13 | -<br><br>[DesAutels2008] |
| 6H SiC [b] | Z-Scan | 220 [μm]<br>-<br>- | | Monocrys.<br>3.10 [eV]<br>2.50E+23 [1/m$^3$] | 780 [nm]<br>-<br>160 [fs]<br>41 [Hz] | -<br>- | - | 4.00E-19<br>5.20E-13 | -<br><br>[DesAutels2008] |
| **Three Photon Absorption** | | | | | | | | | |
| $Al_2Se_{1.5}S_{1.5}$ | Z-Scan | -<br>-<br>- | 2.7<br>-<br>1550 [nm] | -<br>1.74 [eV]<br>- | 1550 [nm]<br>-<br>150 [fs]<br>1 [kHz] | -<br>-<br>- | - | -<br><br>5.50E-26 | -<br><br>[Shabahang2014] |

| Material | Method | | | | Laser | | | | Value | Reference |
|---|---|---|---|---|---|---|---|---|---|---|
| CdS | Nonlin. R/T | - <br> - <br> - | 2.34 <br> - <br> 1064 [nm] | - <br> 2.42 [eV] <br> - | 1064 [nm] <br> - <br> 30 [ps] <br> - | - <br> - <br> - | - | - <br> - | 1.50E-26 | [Benis2020] |
| CdS | Z-Scan | - <br> - <br> - | 2.34 <br> - <br> 1200 [nm] | - <br> 2.42 [eV] <br> - | 1200 [nm] <br> - <br> 150 [fs] <br> 1 [kHz] | - <br> - <br> - | - | - <br> - | 1.10E-26 | [Benis2020] |
| CdSe | Z-Scan | - <br> - <br> - | 2.5 <br> - <br> 1500 [nm] | - <br> 1.9 [eV] <br> - | 1500 [nm] <br> - <br> 150 [fs] <br> 1 [kHz] | - <br> - <br> - | - | - <br> - | 2.40E-25 | [Benis2020] |
| CdTe | Z-Scan | - <br> - <br> - | 2.7 <br> - <br> 1750 [nm] | - <br> 1.44 [eV] <br> - | 1750 [nm] <br> - <br> 150 [fs] <br> 1 [kHz] | - <br> - <br> - | - | - <br> - | 1.20E-24 | [Benis2020] |
| GaAs [a] | Z-Scan | - <br> - <br> - | 3.4 <br> - <br> 2300 [nm] | - <br> 1.42 [eV] <br> - | 2600 [nm] <br> - <br> 15 [ps] <br> 1 [kHz] | - <br> - <br> - | - | - <br> - | 9.00E-25 | [Benis2020] |
| GaAs [a] | Z-Scan | - <br> - <br> - | 3.4 <br> - <br> 2300 [nm] | - <br> 1.42 [eV] <br> - | 2600 [nm] <br> - <br> 150 [fs] <br> 1 [kHz] | - <br> - <br> - | - | - <br> - | 7.50E-25 | [Benis2020] |
| GaAs | Z-Scan | - <br> - <br> - | 3.4 <br> - <br> 2300 [nm] | - <br> 1.42 [eV] <br> - | 2300 [nm] <br> - <br> 100 [fs] <br> 1 [kHz] | - <br> - <br> - | - | - <br> - | 3.50E-25 | [Hurlbut2007] |
| InSb [c] | Z-Scan | - <br> - <br> - | 3.95 <br> - <br> 12000 [nm] | - <br> 0.228 [eV] <br> - | 12000 [nm] <br> - <br> 10 [ps] <br> 1 [kHz] | - <br> - <br> - | - | - <br> - | 2.50E-20 | [Benis2020] |
| 4H SiC | Z-Scan | 498 [µm] <br> - <br> - | - <br> - <br> - | Monocrys. <br> - <br> - | 990 [nm] <br> - <br> 177 [fs] <br> 50 [kHz] | - <br> - <br> - | - | - <br> - | 1.9E-27 | [Guo2021] |

| Material | Method | | | | | | | | Value | Reference |
|---|---|---|---|---|---|---|---|---|---|---|
| 6H SiC | Z-Scan | 258 [µm]<br>-<br>- | -<br>-<br>- | Monocrys.<br>-<br>- | 990 [nm]<br>-<br>177 [fs]<br>50 [kHz] | -<br>-<br>- | -<br>-<br>- | -<br>- | -<br>-<br>4.0E-28 | [Guo2021] |
| ZnO | Nonlin. R/T | -<br>-<br>- | 1.94<br>-<br>1064 [nm] | -<br>3.27 [eV]<br>- | 1064 [nm]<br>-<br>30 [ps]<br>- | -<br>-<br>- | -<br>-<br>- | -<br>- | -<br>2.20E-26 | [Benis2020] |
| ZnO | Z-Scan | -<br>-<br>- | 1.95<br>-<br>900 [nm] | -<br>3.27 [eV]<br>- | 900 [nm]<br>-<br>100 [fs]<br>1 [kHz] | -<br>-<br>- | -<br>-<br>- | -<br>- | -<br>5.40E-27 | [He2005] |
| ZnS [a] | Z-Scan | -<br>-<br>- | 2.3<br>-<br>800 [nm] | -<br>3.54 [eV]<br>- | 960 [nm]<br>-<br>30 [ps]<br>1 [kHz] | -<br>-<br>- | -<br>-<br>- | -<br>- | -<br>1.60E-27 | [Benis2020] |
| ZnS | Z-Scan | -<br>-<br>- | 2.3<br>-<br>800 [nm] | -<br>3.54 [eV]<br>- | 800 [nm]<br>-<br>100 [fs]<br>1 [kHz] | -<br>-<br>- | -<br>-<br>- | -<br>- | -<br>1.70E-27 | [He2005] |
| ZnSe [a] | Nonlin. R/T | -<br>-<br>- | 2.48<br>-<br>1064 [nm] | -<br>2.67 [eV]<br>- | 1064 [nm]<br>-<br>30 [ps]<br>- | -<br>-<br>- | -<br>-<br>- | -<br>- | -<br>1.50E-26 | [Benis2020] |
| ZnSe | Nonlin. R/T | -<br>-<br>- | 2.48<br>-<br>1064 [nm] | -<br>2.67 [eV]<br>- | 1350 [nm]<br>-<br>30 [ps]<br>- | -<br>-<br>- | -<br>-<br>- | -<br>- | -<br>9.10E-27 | [Cirloganu2008] |
| ZnTe | Z-Scan | -<br>-<br>- | 2.8<br>-<br>1200 [nm] | -<br>2.28 [eV]<br>- | 1200 [nm]<br>-<br>150 [fs]<br>1 [kHz] | -<br>-<br>- | -<br>-<br>- | -<br>- | -<br>2.00E-26 | [Benis2020] |

[a] Multiple wavelengths reported, [b] Multiple parameters (e.g., thickness, crystal orientation) reported, [c] Measurement taken at a temperature other than room temperature. *Units as illustrated unless otherwise indicated in table, [†] A negative value of $\alpha_2$ represents saturable absorption, [‡] TCUPS: Transient Coherent Ultrafast Phonon Spectroscopy

## 3.2 Solvents: data table and discussion

*Team: Eric Van Stryland (team leader), Eiji Tokunaga*

As mentioned in Section 3.1 on bulk materials, we include here a separate section dedicated specifically to solvents. In Table 3, we show a list of the $3^{rd}$-order nonlinearities of a variety of liquids, mainly organic solvents at room temperature, taken from representative experimental works in the literature. These are included primarily because our understanding of their NLO responses has advanced and one of them, carbon disulfide ($CS_2$) is often used as a reference. Besides reporting the values of the effective nonlinear refractive indices, $n_{2,eff}$, where listed in the original works, we report the 2PA coefficient, $\alpha_2$, and the 3PA coefficient, $\alpha_3$. However, in most of the publications the 2PA is not listed since the wavelengths used are with photon energies where such absorption is not energetically possible and 3PA is also extremely small or not energetically allowed.

These solvent nonlinearities are important not only because $CS_2$ is used as a reference material, but also because of the many measurements of organic dye nonlinearities in the literature (not included in the current data tables) as well as for measurements of particle suspensions, as they are usually performed for these materials dissolved or dispersed in various solvents. A lack of knowledge of the nonlinearities of the solvents can potentially confuse interpretation of such measurements. We also refer the reader to Section 3.7 on hybrid waveguiding systems that includes, amongst others, hollow fibers filled with solvents.

The primary difficulty in reporting the solvent nonlinearities in Table 3, is that it is now known that there are four separate physical processes leading to the reported NLR, i.e., the bound-electronic nonlinearity, which can be considered instantaneous for any pulse widths reported in this publication, and three nuclear contributions whose temporal characteristics depend on the solvent but are on the order of femtoseconds to picoseconds. These nonlinearities are reorientational, librational, and collisional [McMorrow 1998, Reichert 2014]. For nanosecond and longer pulses, other nonlinearities of electrostriction and thermal origin also become important. We restrict ourselves here to the subnanosecond range. Using single-beam measurements, it is problematic to separate the various nonlinearities. Excite-probe experiments are needed for this. The bound-electronic and collisional nonlinearities can be separated from the rotational and librational nonlinearities by their symmetries given the signals using different relative polarizations of the excitation and probe beams. The nuclear nonlinearities have both rise and fall times and each has an associated response function given, e.g., in [Reichert 2014]. Thus, the response is quite complicated. In rare cases the NLR response is dominated by a single nonlinearity. In fact, this only occurs in the case of extremely short pulses (usually a few femtoseconds) where only the bound-electronic response is activated. Even in 3-dimensionally symmetrical molecules, the nuclear collisional nonlinearity is still present for pulses longer than a few femtoseconds. Thus, the overall NLO response depends on the pulsewidth used for the experiment. This has led to difficulties in reported nonlinearities of multiple materials since $CS_2$, included in this table, has often been used as a reference standard for measurements of other materials. In this regard, we point out [Miguez 2017, Zhao 2018, Reichert 2014] where measurements of $CS_2$ include the nuclear components. The effective nonlinearity can be calculated from the values of the four nonlinear responses along with their temporal responses as in [Zhao 2018, Reichert 2014]. We show results for $CS_2$ in Fig. 1 taken from [Reichert 2014] because of its wide use as a reference material. [Ganeev 2004] shows a dependence consistent with that of [Reichert 2014] for several pulsewidths from 110 fs to 75 ns. The fast and slow nonlinearities referred to in [Miguez 2017] are combinations of the four nonlinearities; thus, attempts to combine information from [Miguez 2017, Zhao 2018, Reichert 2014] which are the only ones in Table 3, reporting the separate nuclear nonlinear responses, into a single table are problematic. Therefore, the reader will have to consult the original papers to get the full details of the NLR [Miguez 2017, Zhao 2018, Reichert 2014]. The two values of $n_2$ reported in Table 3 for [Zhao 2018, Reichert 2014] are the predicted bound-electronic $n_2$ that would be measured using few femtosecond pulses along with an estimate of the effective nonlinear refractive index, $n_{2,eff}$, that would be measured in a single beam experiment such as Z-scan using pulses of >100 ps, i.e. long compared to the rise and fall times of the nuclear nonlinearities but short compared to times where electrostriction and thermal effects become important. These values are estimated from the temporal dependence plots given in those works [ Zhao 2018, Reichert 2014]. These values can be compared to the predictions from [Miguez2017] where pulses from 60 fs to 2 ps were used to measure the nonlinearities and separate slow and fast responses. In table 3 we report the 2 values for the slow and fast components. The sum of these components should give the same long pulse limit as the 2nd value given in the table for [Zhao 2018, Reichert 2014].

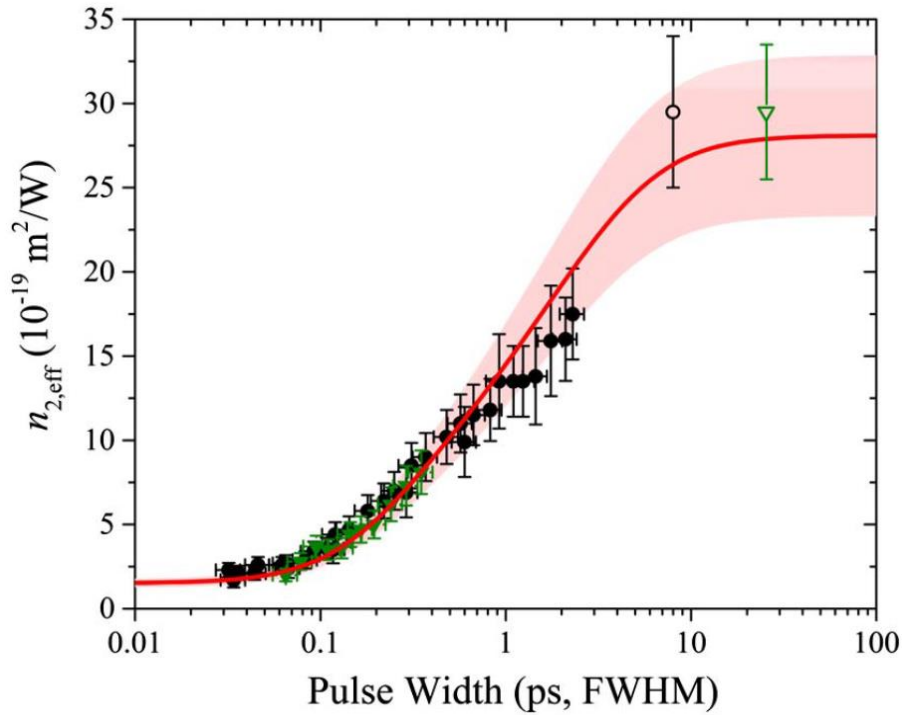

**Fig. 1**. Results of beam deflection (BD) experiments for $CS_2$ yielding the four nonlinearities with their response functions to calculate the predicted $n_{2,eff}$ in a Z-scan experiment (solid red line), along with Z-scan measurements at various pulse widths and 2 wavelengths namely 700nm (black circles) and 1064nm (green triangles). Reprinted with permission from [Reichert 2014] © 2014 Optica Publishing Group.

**Table 3**. NLO measurement results for solvents. The methods used are Z-scan, Beam Deflection (BD), Ellipse Rotation (ER), and Third-Harmonic Generation (THG). All liquid samples reported in this table were held in either 1- or 2-mm internal thickness cuvettes and were at room temperature. The $\chi^{(3)}$ values specified in the table are in fact $\chi^{(3)}_{eff}$ values containing multiple nonlinearity contributions rather than just the bound-electronic contribution. Legend for superscripts: see below the table. The following abbreviations have been used: DMSO = dimethyl sulfoxide; ODCB = o-dichlorobenzene; DCM = dichloromethane; BS = butyl salicylate; ACN = acetonitrile; NB = nitrobenzene; CB = Chlorobenzene; THF = tetrahydrofuran; CH = Cyclohexane; $CCl_4$ = carbon tetrachloride; $CS_2$ = carbon disulfide; $D_2O$ = deuterated water; lin = linear polarization; cir = circular polarization; BE = bound electronic.

| Material | Method | $\chi^{(3)}$ x10$^{-14}$ (esu) | $\chi^{(3)}$ x10$^{-22}$ (m$^2$/W) | $n_{2,eff}$ x10$^{-20}$ (m$^2$/W) | 2:$\alpha_2$(m/W) 3:$\alpha_3$(m$^3$/W$^2$) | $\lambda_{pump}$ (nm) | $\tau_{pump}$ Rep-Rate | $\lambda_{probe}$ (nm) | $\tau_{probe}$ (fs)$^\blacklozenge$ | Ref |
|---|---|---|---|---|---|---|---|---|---|---|
| $CCl_4$ | Z-scan | | 20.8 | 28 | | 1064 | 16ps 10 Hz | | | 1 |
| | THG | | 11.1 | 15 | | 1064 | 16ps 10 Hz | | | |
| Chloroform | Z-scan | | 27.8 | 38 | | 1064 | 16ps 10 Hz | | | 1 |
| | THG | | 9.77 | 13.3 | | 1064 | 16ps 10 Hz | | | |
| DCM | Z-scan | | 26.8 | 37 | | 1064 | 16ps 10 Hz | | | 1 |
| | THG | | 8.56 | 11.8 | | 1064 | 16ps 10 Hz | | | |

| Material | Method | | | | | λ | Pulse | | | Ref |
|---|---|---|---|---|---|---|---|---|---|---|
| CH | Z-scan | | 14.0 | 19 | | 1064 | 16ps 10 Hz | | | 1 |
| | THG | | 9.03 | 12.2 | | 1064 | 16ps 10 Hz | | | |
| n-Hexane | Z-scan | | 11.2 | 17 | | 1064 | 16ps 10 Hz | | | 1 |
| | THG | | 6.17 | 9.35 | | 1064 | 16ps 10 Hz | | | |
| Acetone | Z-scan | | 14.9 | 23 | | 1064 | 16ps 10 Hz | | | 1 |
| | THG | | 6.55 | 10.1 | | 1064 | 16ps 10 Hz | | | |
| Methanol | Z-scan | | 11.1 | 18 | | 1064 | 16ps 10 Hz | | | 1 |
| | THG | | 4.16 | 6.74 | | 1064 | 16ps 10 Hz | | | |
| Ethanol | Z-scan | | 11.5 | 18 | | 1064 | 16ps 10 Hz | | | 1 |
| | THG | | 4.82 | 7.57 | | 1064 | 16ps 10 Hz | | | |
| DMF | Z-scan | | 29.8 | 42 | | 1064 | 16ps 10 Hz | | | 1 |
| | THG | | 7.78 | 11 | | 1064 | 16ps 10 Hz | | | |
| Water | Z-scan | | 8.70 | 14 | | 1064 | 16ps 10 Hz | | | 1 |
| | THG | | 3.91 | 6.28 | | 1064 | 16ps 10 Hz | | | |
| CB | Z-scan | | 95.9 | 117 | | 1064 | 16ps 10 Hz | | | 1 |
| | THG | | 14.8 | 18.1 | | 1064 | 16ps 10 Hz | | | |
| Toluene | Z-scan | | 73.4 | 93 | | 1064 | 16ps 10 Hz | | | 1 |
| | THG | | 13.7 | 17.4 | | 1064 | 16ps 10 Hz | | | |
| $CS_2$ | Z-scan | | 289 | 310 | | 1064 | 16ps 10 Hz | | | 1 |
| | THG | | 32.1 | 34.5 | | 1064 | 16ps 10 Hz | | | |
| Toluene | Z-scan | | | lin 5.7 cir 3.3 | | 800 | 125fs 1 kHz | | | 2 |
| ODCB | Z-scan | | | lin 4.8 cir 2.8 | | 800 | 125fs 1 kHz | | | 2 |
| DMF | Z-scan | | | lin 4.0 cir 2.2 | | 800 | 125fs 1 kHz | | | 2 |
| Acetone | Z-scan | 20.5 | | 43.7 | 2: $6.0 \times 10^{-13}$ | 532 | 30ps 10 Hz | | | 3 |
| | Z-scan | 4.37 | | 9.34 | 3: $7.8 \times 10^{-23}$ | 800 | 100fs 1kHz | | | |
| Chloroform | Z-scan | 27.6 | | 51.9 | 2: $1.25 \times 10^{-12}$ | 532 | 30ps 10 Hz | | | 3 |
| | Z-scan | 4.96 | | 9.37 | 3: $5.2 \times 10^{-23}$ | 800 | 100fs 1kHz | | | |
| DMF | Z-scan | 24.5 | | 47.0 | 2: $1.22 \times 10^{-12}$ | 532 | 30ps 10 Hz | | | 3 |
| | Z-scan | 5.37 | | 10.4 | 3: $9.7 \times 10^{-23}$ | 800 | 100fs 1kHz | | | |
| THF | Z-scan | 20.1 | | 40.1 | 2: $5.70 \times 10^{-13}$ | 532 | 30ps 10 Hz | | | 3 |
| | Z-scan | 6.46 | | 12.9 | 3: $150 \times 10^{-23}$ | 800 | 100fs 1kHz | | | |
| Toluene | Z-scan | 25.4 | | 42.7 | 2: $3.20 \times 10^{-12}$ | 532 | 30ps 10 Hz | | | 3 |
| | Z-scan | 7.95 | | 14 | 3: $20.5 \times 10^{-23}$ | 800 | 100fs 1kHz | | | |
| $CS_2$ | Z-scan | 762 | | 1130 | 2: $2.00 \times 10^{-11}$ | 532 | 30ps 10 Hz | | | 3 |
| | Z-scan | 12.0 | | 17.8 | 3: $74.7 \times 10^{-23}$ | 800 | 100fs 1kHz | | | |
| Ethanol | Z-scan | 5.11 | | 10.9 | 3: $21.0 \times 10^{-23}$ | 800 | 100fs 1kHz | | | 3 |
| $CCl_4$ | Z-scan | 6.31 | | 11.7 | 3: $18.2 \times 10^{-23}$ | 800 | 100fs 1kHz | | | 3 |
| $CS_2$ | Z-scan | 87.1 | 122 | 519 | | 532 | 35ps 10 Hz | | | 4 |
| | Z-scan | 0.668 | 0.932 | 3.97 | | 800 | 40fs 10 Hz | | | |
| NB | Z-scan | 23.3 | 32.6 | 153 | | 532 | 35ps 10 Hz | | | 4 |
| | Z-scan | 0.151 | 0.21 | 0.988 | | 800 | 40fs 10 Hz | | | |
| DB | Z-scan | 18.6 | 26.0 | 122 | | 532 | 35ps 10 Hz | | | 4 |
| | Z-scan | 0.158 | 0.22 | 1.03 | | 800 | 40fs 10 Hz | | | |
| Aniline | Z-scan | 17.8 | 24.9 | 112 | | 532 | 35ps 10 Hz | | | 4 |
| | Z-scan | 0.165 | 0.23 | 1.03 | | 800 | 40fs 10 Hz | | | |

| Toluene | Z-scan | 15.3 | 21.4 | 108 | | 532 | 35ps 10 Hz | | | 4 |
|---|---|---|---|---|---|---|---|---|---|---|
| | Z-scan | 0.143 | 0.20 | 1.01 | | 800 | 40fs 10 Hz | | | |
| Benzene | Z-scan | 12.6 | 17.5 | 88 | | 532 | 35ps 10 Hz | | | 4 |
| | Z-scan | 0.246 | 0.343 | 1.72 | | 800 | 40fs 10 Hz | | | |
| Chloroform | Z-scan | 4.4 | 6.1 | 33 | | 532 | 35ps 10 Hz | | | 4 |
| | Z-scan | 0.11 | 0.153 | 0.827 | | 800 | 40fs 10 Hz | | | |
| DMF | Z-scan | 4.3 | 6.0 | 33 | | 532 | 35ps 10 Hz | | | 4 |
| | Z-scan | 0.101 | 0.141 | 0.781 | | 800 | 40fs 10 Hz | | | |
| DCM | Z-scan | 4.1 | 5.7 | 32 | | 532 | 35ps 10 Hz | | | 4 |
| | Z-scan | 0.100 | 0.14 | 0.781 | | 800 | 40fs 10 Hz | | | |
| DMSO | Z-scan | 3.9 | 5.4 | 28 | | 532 | 35ps 10 Hz | | | 4 |
| | Z-scan | 0.102 | 0.142 | 0.735 | | 800 | 40fs 10 Hz | | | |
| Acetone | Z-scan | 2.6 | 3.6 | 22 | | 532 | 35ps 10 Hz | | | 4 |
| | Z-scan | 0.075 | 0.105 | 0.528 | | 800 | 40fs 10 Hz | | | |
| ACN | Z-scan | 2.3 | 3.2 | 20 | | 532 | 35ps 10 Hz | | | 4 |
| | Z-scan | 0.0474 | 0.0661 | 0.413 | | 800 | 40fs 10 Hz | | | |
| n-Hexane | Z-scan | 2.0 | 2.8 | 17 | | 532 | 35ps 10 Hz | | | 4 |
| | Z-scan | 0.0496 | 0.0692 | 0.413 | | 800 | 40fs 10 Hz | | | |
| THF | Z-scan | 1.4 | 1.9 | 11 | | 532 | 35ps 10 Hz | | | 4 |
| | Z-scan | 0.0721 | 0.101 | 0.574 | | 800 | 40fs 10 Hz | | | |
| $CS_2$ | BD | | | 15.0; 275$^\diamond$ | | 800 | 50fs 1kHz | 650 | 158 | 5 |
| $CS_2$ | Z-scan | | | 15.0; 300$^{\diamond\diamond}$ | | 1064 | 65fs-25ps 1kHz | | | 5 |
| $CS_2$ | Z-scan | | | 15.0; 300$^{\diamond\diamond}$ | | 700 | 32fs-8ps 1kHz | | | 5 |
| $CS_2$ | Z-scan | | | $BEn_2$=30 | 2: $2.0 \times 10^{-11}$ | 420 | ** 1KHz | | | |
| | | | | $BEn_2$=60 | 2: $0.7 \times 10^{-11}$ | 545 | ** 1kHz | | | 5 |
| Toluene | BD | | | 6.0; 49$^\diamond$ | | 800 | 150fs 1kHz | 700 | 150 | 6 |
| NB | BD | | | 6.0; 77$^\diamond$ | | 800 | 150fs 1kHz | 700 | 150 | 6 |
| Benzene | BD | | | 6.0; 44$^\diamond$ | | 800 | 150fs 1kHz | 700 | 150 | 6 |
| p-Xylene | BD | | | 6.2; 52$^\diamond$ | | 800 | 150fs 1kHz | 700 | 150 | 6 |
| Pyridine | BD | | | 6.0; 53$^\diamond$ | | 800 | 150fs 1kHz | 700 | 150 | 6 |
| ODCB | BD | | | 6.0; 51$^\diamond$ | | 800 | 150fs 1kHz | 700 | 150 | 6 |
| DCM | BD | | | 3.0; 15$^\diamond$ | | 800 | 150fs 1kHz | 700 | 150 | 6 |
| Chloroform | BD | | | 4.1; 16$^\diamond$ | | 800 | 150fs 1kHz | 700 | 150 | 6 |
| $CCl_4$ | BD | | | 4.8; 6.8$^\diamond$ | | 800 | 150fs 1kHz | 700 | 150 | 6 |
| Acetone | BD | | | 4.0; 12$^\diamond$ | | 800 | 150fs 1kHz | 700 | 150 | 6 |
| ACN | BD | | | 3.5; 11$^\diamond$ | | 800 | 150fs 1kHz | 700 | 150 | 6 |
| DMF | BD | | | 4.0; 21$^\diamond$ | | 800 | 150fs 1kHz | 700 | 150 | 6 |
| BS | BD | | | 3.8; 26$^\diamond$ | | 800 | 150fs 1kHz | 700 | 150 | 6 |
| THF | BD | | | 3.2; 6.2$^\diamond$ | | 800 | 150fs 1kHz | 700 | 150 | 6 |
| Hexane | BD | | | 3.2; 7.0$^\diamond$ | | 800 | 150fs 1kHz | 700 | 150 | 6 |
| CH | BD | | | 3.5; 5.9$^\diamond$ | | 800 | 150fs 1kHz | 700 | 150 | 6 |
| Methanol | BD | | | 3.0; 4.4$^\diamond$ | | 800 | 150fs 1kHz | 700 | 150 | 6 |
| 1-Octanol | BD | | | 4.0; 4.9$^\diamond$ | | 800 | 150fs 1kHz | 700 | 150 | 6 |
| 1-Butanol | BD | | | 3.3; 4.4$^\diamond$ | | 800 | 150fs 1kHz | 700 | 150 | 6 |
| Ethanol | BD | | | 3.2; 4.2$^\diamond$ | | 800 | 150fs 1kHz | 700 | 150 | 6 |
| $CS_2$ | BD | | | 15.0; 275$^\diamond$ | | 800 | 150fs 1kHz | 700 | 150 | 6 |
| DMSO | BD | | | 4.5; 8.2$^\diamond$ | | 800 | 150fs 1kHz | 700 | 150 | 6 |
| $D_2O$ | BD | | | 2.8; 3.7$^\diamond$ | | 800 | 150fs 1kHz | 700 | 150 | 6 |
| $H_2O$ | BD | | | 2.5; 3.2$^\diamond$ | | 800 | 150fs 1kHz | 700 | 150 | 6 |
| $CS_2$ | ER | | | 23.9; 260$^*$ | | 790 | Fast ~fs Slow ~ps | | | 7 |
| Toluene | ER | | | 11.5; 45.2* | | 790 | Fast ~fs Slow ~ps | | | 7 |
| DMSO | ER | | | 9.15; 5.23* | | 790 | Fast ~fs Slow ~ps | | | 7 |
| Chloroform | ER | | | 5.45; 9.1* | | 790 | Fast ~fs Slow ~ps | | | 7 |
| Acetone | ER | | | 5.66; 9.6* | | 790 | Fast ~fs Slow ~ps | | | 7 |
| Methanol | ER | | | 4.62; 1.44* | | 790 | Fast ~fs Slow ~ps | | | 7 |
| Ethanol | ER | | | 4.43; 1.02* | | 790 | Fast ~fs Slow ~ps | | | 7 |
| $H_2O$ | ER | | | 3.35; 1.06* | | 790 | Fast ~fs Slow ~ps | | | 7 |

[1- Rau 2008] [2- Yan 2012] [3- Bala Murali Krishna 2013] [4- Iliopoulos 2015] [5- Reichert 2014]
[6- Zhao 2018] [7- Miguez 2017]

♦ Repetition Rate for the probe is the same as for the pump.

* The 1st entry is the sum of the fast components, and the 2nd entry, which is the sum of the slow entries, is also the predicted $n_{2,eff}$ for pulses >100ps as described in [Miguez 2017].

◊ The 2 entries are the predicted bound-electronic $n_2$, and the predicted $n_{2,eff}$ for pulses >100ps.

◊◊ The 1st entry is the predicted bound-electronic $n_2$, while the 2nd is from measured Z-scan data for the longest ps pulses. See Fig. 1.

** The peak 2PA is at 420nm, and the peak bound-electronic $n_2$ is at 545 nm. Depending on the wavelength, the minimum pulse width varies from 32 to 165 fs.

There is one final set of data that does not fit on this table, and that is the dispersion of the NLO response. The data in the table only covers wavelengths from 532 nm to 1064 nm so little dispersion information is shown. However, the dispersion of the three nuclear contributions to the NLR should be small [Reichert 2014, Alms 1975, Zahedpour 2015]. On the other hand, the bound-electronic nonlinearity should follow similar dispersion to other materials having UV resonances [Hutchings 1992]. For $CS_2$, the short pulse limit values for $n_{2,eff}$ show essentially no dispersion between 1064 nm and 700 nm (see Fig. 1). However, [Reichert 2014] performed both open and closed aperture Z-scans over a much larger wavelength range from 390 nm to 1550 nm using the shortest pulsewidths available which resulted in signals dominated by bound-electronic $n_2$ and/or 2PA. Furthermore, the values of the nuclear contributions at these pulsewidths could be confidently subtracted resulting in the plot of $n_2$ along with $\alpha_2$ as a function of wavelength as shown in Fig. 2.

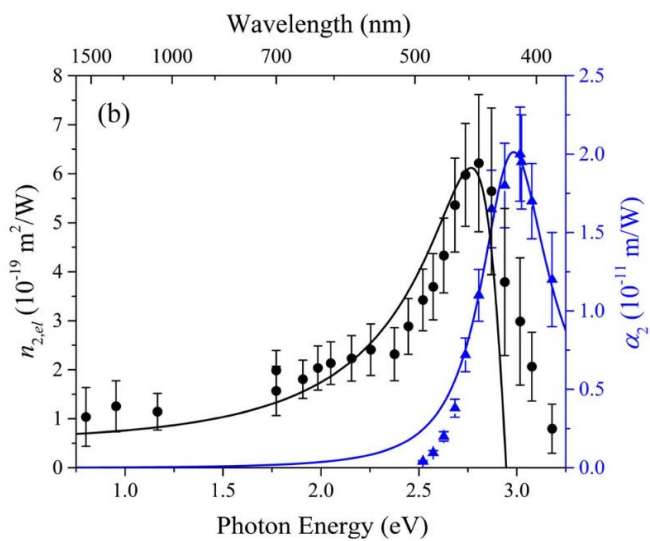

**Fig. 2.** Z-scan measurements of bound-electronic $n_2$ (black circles – labeled as $n_{2,el}$) for femtosecond pulses with non-instantaneous components subtracted, and $\alpha_2$ (blue triangles) for $CS_2$. Curves represent model fits as described in [Reichert 2014]. Reprinted with permission from [Reichert 2014] erratum © 2016 Optica Publishing Group.

Future work on solvents could include measurements of the 2PA spectra along with the dispersion of the bound-electronic $n_2$ to test theory as done in Fig. 2, as well as measurements verifying the lack of dispersion of the nuclear contributions, and the inclusion of more solvents. Additionally, there are few measurements of solvents in spectral regions where there is significant linear loss [Marble2018, Bautista2021]. Determination of the influence of polar versus nonpolar solvents on the second-order hyperpolarizabilities could also be of interest to spur theoretical calculations for some of these molecules. This could also test local field effects as was done for $CS_2$ in [Reichert 2014].

## 3.3 0D-1D-2D materials: data table and discussion

***Team:*** *Cecília Campos, Ksenia Dolgaleva, Daniel Espinosa,* **Anderson Gomes (team leader)**, *Mikko Huttunen, Dragomir Neshev, Lázaro Padilha, Jingshi Yan, Nathalie Vermeulen*

### 3.3.1 Introduction

#### 3.3.1.1 0D-1D-2D materials and their NLO applications

Among the ways to categorize nanomaterials, at least two approaches can be followed [Ngo 2014]. In one case, the bulk material is the starting point, and one employs the general definition that a nanomaterial contains at least one dimension of 100 nm or less. Thus, a thin film whose thickness is 100 nm or less is a nanomaterial, such as a nanofilm or a nanoplate. If two dimensions are under 100 nm, then they are named nanowires, nanofibers or nanorods. Finally, under the above definition, if all three dimensions are smaller than 100 nm, it is a nanoparticle. An alternative way becomes relevant if one looks at the material nanostructure. In this case, the dimensionality of the nanoscale component leads to the definition. When the nanostructure has a length larger than 100 nm in one direction only, it is 1D (wire, fiber or rod, for instance). If no dimension is larger than 100 nm, it is 0D (nanoparticle), and a thin film, with two dimensions larger than 100 nm, is a 2D nanostructure (this holds as well for plates and multilayers). A typical representative diagram of the subcategories 0D-1D-2D for nanostructured materials (NSM), as used in this text, is shown in Fig. 3 (adapted from ref. [Sajanlal 2011]). Besides these general definitions, we note that the dimension shorter than 100 nm might confine the movement of free charge carriers and produce a quantum confinement effect [Fox 2001]. In this case, the optical properties of the material would become size-dependent. 0D, 1D and 2D materials presenting quantum confinement effects are called quantum dots (QDs), quantum wires, and quantum wells (QWs), respectively.

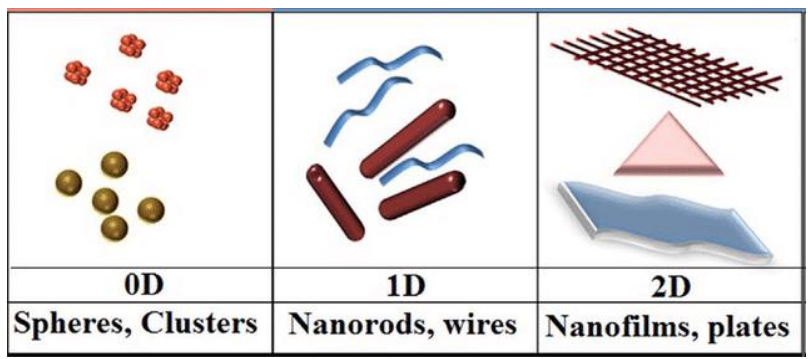

**Fig. 3** - A typical representative diagram of 0D-1D-2D NSM, as used in this text (figure adapted from [Sajanlal 2011]. Reprinted under CC BY license from [Sajanlal 2011].

There is a myriad of NLO and other applications for 0D-1D-2D NSM, which can be revisited in [Liu 2017; Zhang 2017; Xu 2017; Autere 2018a; Jasim 2019; Eggleton 2019; Liu 2019; Wang 2020; Ahmed 2021]. Based upon the already known and well-established fundamental understanding, and upon the NLO applications in bulk materials, researchers have demonstrated that most NLO effects could also occur at the nanoscale or can be engineered in different ways [Zhang 2021] when they take place in nanoscale dimensions. Although the practical applicability of 0D-1D-2D NSM is at times limited by e.g. fabrication quality issues and/or losses, the NLO properties of 0D-1D-2D NSM have already been exploited in LEDs, nanolasers, nanobiosensors, imaging, as well as in nonlinear microscopy, photoacoustics, photocatalysis, energy harvesting, optical limiting, and saturable-absorber-based mode-locked lasing. Nonlinear nanoplasmonics is another field where NLO in 0D-1D-2D materials has been gaining much attention. The field of nanoplasmonics deals with the study of optical phenomena and applications in the nanoscale neighborhood of dielectric-metal interfaces [Stockman2011]. It goes

"beyond" plasmonics – with decreasing interaction dimension – aiming at focusing light below the diffraction limit, determined by the Abbe diffraction formula, which can be roughly approximated by $\lambda/2$, where $\lambda$ is the wavelength of light being used in the light-matter interaction. Physically, it is performed by converting photons into localized charge-density oscillations – so-called surface plasmons– on metallic nanostructures. Further reviews on nanoplasmonics and applications can be found at [Barbillon 2017; Panoiu 2018; Krasavin 2019]. Finally, it should be noted that 0D-1D-2D NSM can also feature unusual NLO properties in the THz domain. Whereas this Section focuses on the nonlinearities of 0D-1D-2D NSM at optical wavelengths, further details on the THz nonlinearities are provided in Section 3.8.

### 3.3.1.2 Background prior to 2000

The history of nanostructured materials has been traced back to the 9[th] century and several other examples of ancient artifacts using nanocomposites emerged throughout the centuries, which of course were only explained after the 19[th] century with the availability of electron microscopes capable of measuring in the nanoworld [Heiligtag 2013]. But the general nanoscience and nanotechnology scientific history and first publication date back to 1857 with Michael Faraday's work [Faraday 1857]. Then came Feynman's lecture (1959), Taniguchi's "nanotechnology" (1974) and Drexler (1981), as reviewed in [Heiligtag 2013]. Thereafter, the broad field of nanoscience and nanotechnology flourished. Nowadays the material subcategories described here have become a reality, and the study of their NLO properties is a continuously growing research field.

#### 3.3.1.2.1 Background for 0D materials

The first studies about the optical properties of so-called semiconductor microspheres date back from the early 1980s after the works from Efros and Efros [Efros 1982] and Brus [Brus 1984]. Those nanomaterials were nano-sized semiconducting spheres embedded in a glass matrix. However, they exhibited poor optical quality, mostly due to the difficulties in controlling the particle size distribution and in eliminating surface trap sites. Later, in the early 1990s, colloidal semiconductor quantum dots were first synthesized, narrowing the size distribution and improving the optical properties after surface treatment with organic ligands [Murray 1993]. Major breakthroughs were achieved after the control over the optical properties in II-VI QDs and the field grew rapidly in the late 1990s and early 2000s. During this period, the scientific interest moved from simple core-only nanospheres to more complex nanostructures, including nanorods, nanoplatelets, tetrapods, and core-shell heterostructures, reaching an unprecedented level of control over the optical and electronic properties of those semiconductors by manipulating their sizes, compositions, and shapes. At this stage, the NLO properties of those nanomaterials came to the spotlight as they emerged as promising candidates for a number of applications including bio-labeling, all-optical signal processing, light detectors, and lighting technologies [Kairdolf 2013; García de Arquer 2021].

#### 3.3.1.2.2 Background for 1D materials

Unidimensional structures of different materials were fabricated before 2000. Single-crystalline nanowires made of semiconductors such as GaN, $SnO_2$, ZnO, and Si, as well as carbon nanotubes were particularly attractive for photonics applications. The confinement of light and the cavity-like resonances in the nanowires are interesting features to realize nanolasers, LEDs, frequency converters, solar cells, photodetectors, and sensors [Pauzauskie 2006]. Some linear optical characterizations of the 1D structures were performed before 2000, but there were no extensive experimental measurements of the optical nonlinearities, although some specific works started exploring NLO responses in unidimensional materials in 2000 [Kishida 2000; Ogasawara 2000]. In 2002, the first measurement of the NLO properties of a single semiconductor nanowire was reported [Johnson 2002]. In the same year, the saturable absorption nonlinearity of single-wall carbon nanotubes (SWCNTs) was demonstrated for the first time [Chen 2002]. Generally speaking, prior to 2002, if the information on the NLO coefficient of a particular 1D material was needed, one usually had to rely on the bulk value.

### *3.3.1.2.3 Background for 2D materials*

The potential of semiconductor QW structures for efficient third-order NLO processes has been explored since the early 1980s, especially GaAs-AlGaAs multiple quantum wells (MQWs). Resonant excitonic effects near the band edge region provide the mechanism for NLA and NLR in these materials [Chemla 1985]. By 2000, the importance of semiconductor QWs had already been established, with various applications in optoelectronics, such as laser diodes, LEDs, solar cells, modulators, and detectors [Fox 1996].

Single-layer 2D materials, as we know today, have had very little or almost no scientific impact before 2000. As pointed out in [Miró 2014], according to classical physics, at any finite temperature they would be thermodynamically unstable due to thermal lattice fluctuations (see [Miró 2014] and references therein). The major experimental breakthrough came only in the post-2000 years, after the 2004 publication by Novoselov *et al.* [Novoselov 2004] who first isolated a single-layer 2D material, namely graphene, through the Scotch tape exfoliation of graphite. After 2004, the field of single-layer 2D materials grew rapidly, with many novel materials beyond graphene [Lv 2015; Zhang 2015a], and among the several review articles that have been published, we highlight those related to NLO in 2D materials published in recent years [Liu 2017; Autere 2018a; You 2018; Eggleton 2019; Yamashita 2019; Zhou 2020; Ahmed 2021; Vermeulen 2022]. Several other excellent review articles on different aspects of 2D materials can be found in the literature and have not been indicated here.

## 3.3.1.3 Considerations for 0D-1D-2D materials when performing NLO measurements

Assessing NLO properties in photonic materials, regardless of their dimensions, requires a great amount of scientific care and proper planning regarding the optical source to be employed, the chosen technique, and whether the technique will adequately address the desired information. From the optical material point of view, preferably all the morphological and optical information should be known in order to perform a proper analysis of the results. Regarding 0D-1D-2D materials, further considerations must be taken into account. For 0D materials, size and shape must be properly characterized, and the same holds for the environment in which the 0D material is embedded, e.g., a solvent, in the case of suspensions, a matrix or a substrate. It is very important to characterize the NLO properties of the solvent, matrix or substrate. This is also valid for 1D and 2D materials. Another aspect to be considered, in the case of 0D-1D-2D materials in suspension, is the scattering, both linear and nonlinear. Therefore, when measuring the linear absorption, the researcher should also insure the knowledge of the extinction coefficient and, in turn, the scattering coefficient of the sample. During the NLO measurement, the influence of nonlinear scattering can contribute to the NLO properties in an undesirable manner. We also point out that for some materials such as graphene, the level of doping strongly influences the NLO response [Jiang 2018] and as such needs to be specified in order to correctly interpret the measurement data. When working with 0D-1D-2D materials in a free-space excitation setup, it is often preferred to use ultrashort excitation pulses (picosecond, femtosecond) to avoid damage to the sample. We point out that the short propagation distance of the excitation beam through nanometer-thick 0D-1D-2D materials typically rules out the need for phase matching in NLO experiments. It is also worth noting that some works on 2D materials report $\chi^{(2)}/\chi^{(3)}$ values either considering sheet susceptibilities or bulk-like susceptibilities, as differentiated in the tables below. The units are different for both cases, as the bulk-like susceptibility is typically assumed to be the sheet susceptibility divided by the monolayer thickness [Autere 2018a]. Finally, we point out that the nonlinearities can also be expressed in terms of conductivities $\sigma^{(2)}/\sigma^{(3)}$ rather than susceptibilities [Cheng 2014]. In addition to these specific considerations for 0D-1D-2D materials, also the best practices described in Section 2 should be taken into account when performing NLO measurements.

#### 3.3.1.4 Description of general table outline

Tables 4A and 4B show a representative list of, respectively, second-order and third-order NLO properties of 0D, 1D and 2D materials taken from the literature since 2000, with the entries arranged in alphabetical order. Based upon the best practices in Section 2, Tables 4A and 4B were put together using inclusion criteria for publications that clearly state the required technical information for an unambiguous and clear identification of the NLO properties being evaluated. In some cases, publications missing only one important parameter could unfortunately not be considered, although they were technically sound. The tables are subdivided into "Material properties," "Measurement details" and "Nonlinear properties." The same set of columns was used for 0D, 1D and 2D materials. Within each column the information is given in the order of the header description, and powers of 10 (e.g., $10^{\pm\alpha}$) are written as E±α for compactness. The NLO technique used is provided in the "Method" column. Some papers relied on non-standard NLO measurement techniques (e.g., method listed as "Other") or have notes associated with their measurement/analysis. These values and information are listed within the "Comments" column. The works included in the tables nominally report data obtained at room temperature, unless denoted otherwise in the "Comments" column. For works that report dependences of the nonlinearity on parameters such as wavelength/energy, polarization, sample size, input fluence, doping level, pulse length/peak power, and concentration, these dependences are denoted by "[1 2 3 4 5 6 7]" respectively, in the "Comments" column. If dispersive values for the NLO parameter were provided, the cited value represents the peak value for the material within the measurement range. The tables are restricted to optical nonlinearities up to the third order, and although higher orders have become of interest [Reyna 2017], they are not covered here.

### 3.3.2 Discussion

#### 3.3.2.1 Advancement since 2000 and remaining challenges

##### 3.3.2.1.1 Advancement and challenges for 0D materials

Since 2000, as the field of 0D semiconductor nanomaterials was growing, stable, monodisperse nanoparticles were becoming available, and the interest in investigating their NLO response has increased. This was mainly due to the initial excitement created by the predictions that the NLO response in 0D materials would be highly enhanced as compared to bulk semiconductors [Brunhes 2000a] since the quantum confinement should enhance the oscillator strength as the nanomaterial size was reduced. Thus, several groups started to experimentally study distinct optical properties regarding QDs, including the second-order susceptibility [Brunhes 2000, Sauvage2001], third-order susceptibility [Valdueza-Felip 2008], NLR [Wang 2019], and NLA [Pu 2006; Wang 2019; Alo 2020]. The first reports on the magnitude of the 2PA cross-section indicated that, for spherical nanomaterials, the magnitude of this process was linearly dependent on the nanoparticle volume [Pu 2006; Padilha 2007; Makarov 2014]. Strategies to better control the NLO response in 0D materials have involved band-structure optimization, obtained with PbS QDs [Padilha 2011], and shape control [Scott 2015]. In the last decade, a fair amount of effort has been put into gaining further control over the electronic and optical properties of these nanostructures by developing sophisticated heterostructures, and by shape control. Strong enhancement of 2PA cross-sections has been reported in CdSe nanoplatelets, with superlinear dependence on the volume [Scott 2015]. On the other hand, sublinear volume dependence has been recently reported for core/shell heterostructures [Alo 2020]. By shape controlling and heterostructuring these nanomaterials, it is expected that one can obtain further control over the nanomaterials' NLO response towards the development of tailored nanostructures for on-demand NLO.

### 3.3.2.1.2    Advancement and challenges for 1D materials

Since the year 2000, NLO characterizations of 1D structures in various arrangements and shapes were reported. We give several examples in the following. While many studies focused on experiments involving many nanowires with a distribution of dimensions, a few involve only single nanowires. SHG and SRS measurements were performed on individual GaP nanowires [Wu 2009; Sanatinia 2014]. SHG was also evaluated in a single GaAs nanoneedle [Chen 2010] or even in areas within a single GaAs nanowire with different crystal phases [Timofeeva 2016]. Often, the NLO technique is applied to a collection of 1D structures, either random or spatially ordered. For example, SHG was measured in an array of GaP nanowires embedded in a polydimethylsiloxane matrix [Fedorov 2020] or grown on a GaP substrate [Sanatinia 2012]. Also, Z-scan experiments were performed in single-walled carbon nanotubes (SWCNTs) deposited on a glass substrate as a thin-film [Seo 2006] or dispersed in a colloidal suspension [Shi 2019]. In the latter case, the authors reported that a semiconductor-SWCNTs colloid exhibits a lower saturation irradiance and a lower 2PA coefficient as compared to a mixture of metallic-SWCNTs and semiconducting-SWCNTs [Shi 2019].

One exciting feature of 1D materials is the optical nonlinearity dependence on the structure dimensions due to resonances or surface effects. For example, in GaP nanowires (with a diameter of 210 nm), the SHG efficiency increases with increasing equivalent thickness (total volume of nanowires per unit area) [Fedorov 2020], while the SRS irradiance decreases with increasing length [Wu 2009]. Furthermore, the SHG irradiance in GaP nanopillars (with diameters from 100 nm to 250 nm) presents a strong dependence on the pillar diameter [Sanatinia 2012]. Moreover, polarization-dependent measurements can be used to distinguish between the second-harmonic light generated in bulk or on the surface [Johnson 2002; Sanatinia 2012; Sanatinia 2014; Sanatinia 2015]. Indeed, by using polarization-dependent measurements, the nonlinear coefficient at the surface of a GaP nanopillar was observed to be approximately 15 times higher than that from the bulk [Sanatinia 2015]. Conversely, the SHG light from a wurtzite (WZ) GaAs single nanoneedle was shown to be primarily originating from the bulk [Chen 2010].

Advances in SHG polarimetry also allow studying the crystal structure of a nonlinear material because the SHG is sensitive to crystallographic symmetry. A per-pixel analysis of SHG images was used to distinguish between GaAs nanowires in the pure WZ, pure zinc blende (ZB), WZ-ZB mixed phases, or with ZB rotational twins [Timofeeva 2016].

Despite the advancements reported on SHG measurements, this material category is far from being thoroughly characterized. There is still room to use other second- and third-order NLO techniques to exploit the large surface-to-volume ratio and strong Mie and Fabry-Perot resonances of the 1D structures.

### 3.3.2.1.3    Advancement and challenges for 2D materials

The study of NLO in quantum-well structures post-2000 emphasizes the effects of intersubband transitions (ISBT) rather than excitonic effects. GaN-AlN and GaN-AlGaN MQW structures have been shown to be particularly suitable for both second-order and third-order processes resonant at the ISBT frequency [Rapaport2003, Nevou2006].

The advancements in single-layer 2D materials, which can be considered a novel field of research with less than two decades of R&D, have been outstanding both from the point of view of scientific understanding and applications. Both refractive and absorptive nonlinearities have been investigated. Particularly, graphene has been shown to exhibit extremely strong third-order nonlinearities and a low saturable absorption threshold [Chen 2015, Miao 2015, Dremetsika 2017, Jiang 2018, Soavi 2018. Thakur 2019], and there are several other 2D materials and particularly Transition Metal

Dichalcogenides (TMDs) such as $MoS_2$ that feature a pronounced second-order NLO response as well (see, e.g., [Woodward 2016, Autere 2018b]). Nevertheless, in view of the large family of 2D materials, further characterization will be required to get the full picture of their NLO behavior, including the contributions from exciton resonances, free carriers, etc., in order to fully exploit their application potential in, e.g., optoelectronics [Sundaram 2018]. The progress made so far has been promoted by advances in the 2D materials synthesis and proper morphological characterization. Also, the tunability of the NLO response using electrical gating [Jiang 2018, Soavi 2018] is an important advantage for the practical use of both the refractive and absorptive nonlinearities in, e.g., wavelength converters and saturable-absorption-based modelocked lasers, respectively [Vermeulen 2022]. Nevertheless, further fabrication improvement is necessary to achieve large-area films on a wide range of substrates, easy placement of contact points, proper etching, and advanced metrology control. The NLO characterization might also provide metrology methods to control parameters of interest for practical NLO applications of 2D materials.

### 3.3.2.2  Recommendations for future works on 0D-1D-2D materials

The future of NLO in 0D-1D-2D materials relies on a more in-depth NLO characterization of the already existing materials and on the development of novel materials. A multidisciplinary approach will also be required to further exploit their optical nonlinearities. Among the novel 0D-1D-2D materials, we highlight nanodiamonds, nanoporous materials, core–shell nanoparticles, perovskite nanostructures, silicene, antimonene, MXenes, 2D metal-organic framework nanosheets, boron nitride nanosheets, and metal-based nanomaterials [Baig 2021]. Metal-based nanomaterials are at the basis of nanoplasmonics, and the field of nonlinear nanoplasmonics is also a flourishing and promising area, opening new avenues for future works [Panoiu 2018; Krasavin 2019]. Further NLO research on 0D materials could benefit, amongst others, biosensing applications [Wang 2020]. Regarding 1D materials, assemblies of 1D nanomaterials have been recently introduced, and their NLO properties need to be further understood [Chen 2019]. 2D materials also have significant potential for future study and applications. For instance, 2D rare-earth based materials are opening new avenues [Chen 2021], since rare-earth materials are quite well studied and have already found a great deal of applications. Also, various methods to enhance the NLO response, including plasmonics, gating, and functionalization [Wei 2019] can be further explored. That said, to enable further progress in the overall understanding of 0D-1D-2D materials, it will be important that future works report in detail both the fabrication aspects, the material properties and the NLO experiments carried out with the materials. Parameters such as the linear absorption loss (which is typically quite high in 0D-1D-2D materials) and the doping level are often overlooked, yet essential to properly evaluate the observed NLO behavior [Vermeulen2022]. In addition, great care is needed to adequately extract and describe the actual NLO coefficients that underpin the measurement data, especially when the dependence of the NLO coefficients on, e.g., wavelength is also being studied. Only such a systematic approach allows building up an extensive and in-depth understanding of the NLO physics of this promising category of materials and the potential they hold for NLO applications.

### 3.3.3 Data table for 0D-1D-2D materials

**Table 4A.** Second-order NLO properties of 0D-1D-2D materials from representative works since 2000. Legend for superscripts: see below the table.

| | | | | | | | |
|---|---|---|---|---|---|---|---|
| **Second-Order Nonlinearities** | | | | | | | |
| | | **Material Properties** | | | **Measurement Details** | **Nonlinear Properties** | |
| Material | Method | Fabrication<br><br>Substrate | Width<br><br>Number of layers | Crystallinity<br><br>Bandgap<br><br>Doping Level | Pump Wavelength<br>Peak Power<br>Peak Irradiance<br>Beam spot size<br>Pulse Width<br>Rep. Rate | $\chi^{(2)}$<br><br>$d_{eff}$ | Reference<br><br>Additional parameters and comments |
| **0D Materials** | | | | | | | |
| InAs/GaAs QD | SHG | MBE<br><br>GaAs | -<br><br>40 | -<br><br>-<br><br>- | 7381 nm<br>-<br>2.00E+12 W/m²<br>-<br>-<br>- | 1.20E-9 m/V<br><br>- | [Brunhes 2000]<br><br>Nonlinearity dependence[1,2] shown in Fig. 2 of reference paper<br><br>InAs QD height/diameter = 0.2 (diameter = 2.3E-8 m)<br><br>$\chi^{(2)}$ :<br>1.2E-9 m/V (sample)<br>2E-7 m/V (1 QD layer)<br><br>Pump polarization:<br>TE (in the layer plane) |
| InAs/GaAs QD | SHG | MBE<br><br>(001)-GaAs | -<br><br>30 | -<br><br>-<br><br>- | 20000 nm<br>-<br>1.00E+12 W/m²<br>1.50E-4 m<br>-<br>- | 2.50E-6 m/V<br><br>- | [Sauvage 2001]<br><br>$\chi^{(2)}$ for 1 QD layer<br><br>1 layer:<br>InAs QD height of 3E-9 m<br>InAs QD base length 1.5E-8 m<br>GaAs barrier thickness of 5E-8 m<br><br>Pump polarization:<br>50% in the layer plane along [110] direction,<br>50% along z-growth axis<br><br>Temperature: 10K |
| **1D Materials** | | | | | | | |
| GaP nanopillars | SHG | Nanosphere lithography / RIE / CAIBE<br><br>GaP (100) | 1.50E-7 m<br><br>- | Monocrystalline<br><br>2.26 eV<br><br>- | 840 nm<br>-<br>1.00E-6 m<br>1.00E-13 s<br>8.20E+7 Hz | -<br><br>- | [Sanatinia 2012]<br><br>$\eta$ = 2.00E-7 %<br><br>Nonlinearity dependence[3] shown in Fig. 2b of reference paper<br><br>The efficiency is for the SHG light generated by the array of nanopillars. The authors did not take into account the light collection efficiency of the microscope objective |
| GaP nanowires | SHG | MBE / G-coating<br><br>Silicon (111) | 1.50E-7 m<br><br>- | Monocrystalline<br><br>-<br><br>- | 1048 nm<br>-<br>5.00E-6 m<br>1.50E-13 s<br>8.00E+7 Hz | -<br><br>- | [Fedorov 2020]<br><br>$\eta$ = 1.00E-2 %<br><br>Nonlinearity dependence[1] shown in Fig. 2c of the reference paper |

| WZ-GaAs nanoneedle | SHG | MOCVD<br><br>Silicon (111) sapphire | 1.50E-6 m<br><br>- | Monocrystalline<br><br>-<br><br>- | 806 nm<br>-<br>-<br>5.00E-7 m<br>1.20E-13 s<br>7.60E+7 Hz | 5.30E-11 m/V<br><br>- | [Chen 2010]<br><br>Diameter:<br>1.5E-6 m (base)<br>2E-9 m to 4E-9 m (tip)<br><br>$\chi^{(2)}$:<br>aca: 5.3E-11 m/V<br>ccc: 1.15E-10 m/V<br>caa: 2.6E-11 m/V |
| WZ-GaAs nanowires | SHG imaging | MBE<br><br>Silicon (111) / $Si_3N_4$ thin film | ~ 1E-7 m<br><br>- | Monocrystalline<br><br>-<br><br>- | 820 nm<br>-<br>2.50E-6 m<br>1.00E-13 s<br>8.00E+7 Hz | -<br><br>$d_{15} = d_{24} = 4.2E-11$ m/V<br>$d_{31} = d_{32} = 2.1E-11$ m/V<br>$d_{33} = 1.15E-10$ m/V | [Timofeeva 2016]<br><br>Nonlinearity dependence[2] shown in Fig. 5 of the reference paper<br><br>While they did not extract the tensor elements' values from the data, the values were used in the theoretical model that fitted well the experimental data<br><br>The deposition substrate was silicon (111), but the nanowire was transferred to the $Si_3N_4$ thin film substrate to perform the SHG measurement |

| 2D Materials | | | | | | | |
|---|---|---|---|---|---|---|---|
| $Al_{0.08}Ga_{0.92}N/$ GaN QW | SHG | MOCVD<br><br>GaN-on-(0001)-c-sapphire | -<br><br>10<br><br>- | Monocrystalline<br>-<br><br>- | 1064 nm<br>-<br>-<br>5.00E-9 s<br>1.40E+1 Hz | $d_{31}$ = 2.2E-12 m/V<br>$d_{33}$ = -4.80E-12 m/V | [Passeri 2004]<br><br>Thickness for 1 layer $Al_{0.08}Ga_{0.92}N$/GaN:<br>1.4E-8 m/2.4E-9 m<br><br>The authors also provided information for $Al_{0.15}Ga_{0.85}N$/GaN QW and $Al_{0.08}Ga_{0.92}N$/GaN with different thickness<br><br>Pump: p and s-polarization |
| GaN-AlN QW | SHG | MBE<br><br>AlN-on-c-sapphire | -<br><br>200<br><br>- | -<br><br>-<br><br>- | 1980 nm<br>2.50E+3 W<br>-<br>1.50E-4 m<br>6.00E-9 s<br>3.00E+1 Hz | 1.14E-10 m/V<br><br>- | [Nevou 2006]<br><br>Thickness for 1 layer GaN-AlN:<br>2.593E-9 m/3E-9 m<br><br>Nonlinearity dependence[1] shown in Fig. 3 of reference paper<br><br>The nonlinear conversion efficiency in MQW sample was 16 times higher compared to that in a reference bulk GaN sample<br><br>Pump: p-polarization |
| Graphene | DFG | CVD<br><br>Quartz | -<br><br>-<br><br>0.3 eV | -<br><br>-<br>Monolayer | 547 nm<br>-<br>-<br>3.00E-4 m<br>1.00E-13 s<br>1.00E+3 Hz | 3.00E-7 m/V<br><br>- | [Constant 2016]<br><br>DFG enhanced through plasmonic excitation<br><br>Probe parameters:<br>λ = 615 nm<br>Pulse duration: 1.00E-13 s<br>Polarization: p-polarization<br><br>Pump: p-polarization |
| Graphene | SHG | CVD<br>Fused silica | - | Monocrystalline<br>-<br>-<br>-0.9 eV | 1035 nm<br>-<br>-<br>2.00E-13 s<br>8.00E+7 Hz | 3.00E-11 m/V<br><br>- | [Zhang 2019]<br><br>Electric-quadrupole SHG<br><br>Pump: p and s-polarization<br><br>Nonlinearity dependence[5] in Fig. 4 of the reference paper |
| $MoS_2$ | SHG | Micromechanical exfoliation<br>Si/SiO$_2$ (285 nm) | ~ 0.65E-9 m<br>Monolayer | -<br>-<br>- | 1560 nm<br>2.70E+3 W<br>-<br>1.50E-13 s<br>5.00E+7 Hz | 5.40E-12 m/V<br><br>- | [Autere 2018b]<br><br>Nonlinearity dependence[2] in Fig. 5 of the reference paper |
| $MoS_2$ | SHG | Mechanical exfoliation<br>Si/SiO$_2$ | -<br>Monolayer | -<br>-<br>- | 870 nm<br>-<br>6.00E-7 m<br>1.40E-13 s<br>8.00E+7 Hz | 8E-20 m$^2$/V<br><br>- | [Malard 2013]<br>Nonlinearity dependence[1] shown in Fig. 3e of reference paper<br>Specified second-order nonlinearity is the sheet susceptibility |
| $MoS_2$ | SHG | Micromechanical cleavage<br>Si/SiO$_2$ | -<br>Monolayer | -<br>-<br>- | 1560 nm<br>8.00E+3 W<br>9.30E-7 m<br>1.50E-13 s<br>5.00E+7 Hz | 2.20E-12 m/V<br><br>- | [Säynätjoki 2017]<br>Nonlinearity dependence[2,3,6] in Fig. 3c, Fig. 4 and Fig. 5 of the reference paper<br><br>The authors also provided information for 2 and 5 layers at 1560 nm<br><br>$\eta_{conv}$ = 6.47E-9 % |
| $MoS_2$ | SHG imaging | CVD<br>SiO$_2$/Si (300 nm) | 7.00E-10 m<br>Monolayer | -<br>-<br>- | 1560 nm<br>-<br>1.00E+15 W/m$^2$<br>1.80E-6 m<br>1.50E-13 s<br>8.90E+5 Hz | 2.90E-11 m/V<br><br>- | [Woodward 2016]<br><br>Nonlinearity dependence[1] shown in Fig. 3(a,b) of the reference paper |
| MoSe$_2$ | SHG | Micromechanical exfoliation<br><br>Si/SiO$_2$ (285 nm) | ~ 0.65E-9 m<br><br>Monolayer | -<br>-<br>- | 1560 nm<br>2.70E+3 W<br>-<br>1.50E-13 s<br>5.00E+7 Hz | 3.70E-11 m/V<br>- | [Autere 2018b]<br><br>Nonlinearity dependence[2] in Fig. 5 of the reference paper |
| WS$_2$ | SHG | Micromechanical exfoliation<br><br>Si/SiO$_2$ (285 nm) | ~ 0.65E-9 m<br><br>Monolayer | -<br>-<br>- | 1560 nm<br>2.70E+3 W<br>-<br>1.50E-13 s<br>5.00E+7 Hz | 1.62E-11 m/V<br>- | [Autere 2018b]<br><br>Nonlinearity dependence[2] in Fig. 5 of the reference paper |
| WS$_2$ | SHG | CVD<br><br>SiO$_2$/Si | 6.50E-10 m<br><br>Monolayer | -<br>-<br>- | 832 nm<br>-<br>1.80E-6 m<br>1.06E-13 s<br>8.80E+7 Hz | -<br>4.50E-9 m/V | [Janisch 2014]<br><br>- |
| WS$_2$–MS | SHG | CVD<br><br>Sapphire | -<br><br>Monolayer | -<br>-<br>- | 804 nm<br>-<br>4.00E-6 m<br>8.00E-15 s<br>8.00E+7 Hz | -<br>- | [Li 2019]<br><br>The authors reported a 20-fold enhancement of the optical SHG from WS$_2$ monolayers by cooperating with SiO$_2$ dielectric microspheres (MSs) |
| WSe$_2$ | SHG | Micromechanical exfoliation<br><br>Si/SiO$_2$ (285 nm) | ~ 0.65E-9 m<br><br>Monolayer | -<br>-<br>- | 1560 nm<br>2.70E+3 W<br>-<br>1.50E-13 s<br>5.00E+7 Hz | 1.65E-11 m/V<br>- | [Autere 2018b]<br><br>Nonlinearity dependence[2] in Fig. 5 of the reference paper |

| WSe$_2$ | SHG | CVT  SiO$_2$/Si (300 nm) | 7.00E-10 m  Monolayer | -  -  - | 816 nm  -  -  1.07E-13 s  8.80E+7 Hz | -  5.00E-9 m/V | [Ribeiro-Soares 2015]  Nonlinearity dependence[2] in Fig. 3 of the reference paper |
| WSe$_2$ | SHG | Micromechanical exfoliation  Fused silica | 1.17E-9 m  Monolayer | -  1.65 eV  - | 1546 nm  -  (0.8 ± 0.1)E-6 m  2.00E-13 s  8.00E+7 Hz | (0.7 ± 0.09)E-19 m²/V  - | [Rosa 2018]  Nonlinearity dependence[2,3] in Fig. 2 and Fig. 4 of the reference paper  The authors also provided information for 4, 5, 6 and 9 layers at 1546 nm  Specified second-order nonlinearity is the sheet susceptibility |

Superscripts indicate the work reports the nonlinearity dependence on [1]wavelength/energy, [2]polarization, [3]sample size (width, diameter, thickness, number of layers), [4]input fluence, [5]doping level, [6]pulse length/peak power, [7]concentration.

**Table 4B.** Third-order NLO properties of 0D-1D-2D materials from representative works since 2000. The $\chi^{(3)}$ values specified in the table might in some cases be $\chi^{(3)}_{eff}$ values containing multiple nonlinearity contributions rather than just the bound-electronic contribution. Legend for superscripts: see below the table.

| Third-Order Nonlinearities | | | | | | | | |
|---|---|---|---|---|---|---|---|---|
| Material Properties | | | | | | Measurement Details | Nonlinear Properties | |
| Material | Method | Fabrication<br><br>Solvent/ Substrate | Width<br><br>Number of layers | Index<br><br>Abs. Coeff.<br><br>Wavelength | Crystallinity<br><br>Bandgap<br><br>Doping Level | Pump<br>Wavelength<br>Peak Power<br>Peak Irradiance<br>Beam spot size<br>Pulse Width<br>Rep. Rate | $\chi^{(3)}$<br><br>$n_{2,eff}$<br><br>$\alpha_2$ | Reference<br><br>Additional parameters and comments |
| 0D Materials | | | | | | | | |
| CdSe QDs | Other (add in comments) | Organic Chemical Route<br><br>Toluene | 6.00E-9 m<br><br>- | -<br><br>-<br><br>- | -<br><br>-<br><br>1.96 eV | 800 nm<br>-<br>-<br>-<br>8.00E-14 s<br>1.00E+3 Hz | -<br><br>-<br><br>2.00E+4 GM | [Alo 2020]<br><br>2PA cross-section in GM<br>(1E-50 cm⁴ . s . photon⁻¹)<br><br>The authors also provided NLO parameters for different diameters (3E-9 m, 4E-9 m and 5E-9 m) and doping levels<br><br>The technique is called MPAPS. The bandgap is defined as the excitonic peak position in Fig. S1 of reference paper |
| CdSe QDs | Other (add in comments) | Organic Chemical Route<br><br>Toluene | 4.80E-9 m | -<br>-<br>- | -<br>-<br>2.04 eV | 800 nm<br>-<br>-<br>1.00E-13 s<br>8.20E+7 Hz | -<br>-<br>1.03E+4 GM | [Pu 2006]<br><br>2PA cross-section in GM<br>(1E-50 cm$^4$ . s . photon$^{-1}$)<br><br>The authors also provided NLO parameters for different diameters (2.4E-9 m to 4.6E-9 m)<br><br>Nonlinearity dependence[3] shown in Fig. 1 of the reference paper<br><br>Two-photon excited photoluminescence |
| CdSe/CdZnS QDs | Other (add in comments) | Organic Chemical Route<br><br>Toluene | 1.24E-8 m | -<br>-<br>- | -<br>-<br>2.03 eV | 800 nm<br>-<br>-<br>8.00E-14 s<br>1.00E+3 Hz | -<br>-<br>2.70E+4 GM | [Alo 2020]<br><br>2PA cross-section in GM<br>(1E-50 cm$^4$ . s . photon$^{-1}$)<br><br>Nuclei diameter: 4.0 nm<br>Shell thickness: 4.2 nm<br><br>The authors also provided NLO parameters for shell thicknesses of 1.2 nm, 2.5 nm, and 3.0 nm<br><br>Nonlinearity dependence[1] shown in Fig. 4 of the reference paper<br><br>The technique is called MPAPS. The bandgap defined as the excitonic peak position is shown in Fig. S1 of the reference paper |
| CdTe QDs | Other (add in comments) | Organic Chemical Route<br><br>Toluene | 5.40E-9 m | -<br>-<br>- | -<br><br>1.8 eV | 840 nm<br>-<br>-<br>1.00E-13 s<br>8.20E+7 Hz | -<br><br>-<br><br>7.96E+3 GM | [Pu 2006]<br><br>2PA cross-section in GM<br>(1E-50 cm$^4$ . s . photon$^{-1}$)<br><br>The authors also provided NLO parameters for different diameters (4.4E-9 m to 5.2E-9 m) and doping levels<br><br>Nonlinearity dependence[3] shown in Fig. 1 of the reference paper<br><br>Two-photon excited photoluminescence |
| GaN-AlN QD | FWM | MBE<br><br>AlN-on-sapphire | -<br><br>20 | -<br>-<br>- | -<br>-<br>- | 1500 nm<br>-<br>-<br>1.00E-13 s<br>1.00E+3 Hz | 1.3E-6 esu<br><br>-<br><br>- | [Valdueza-Felip 2008]<br><br>The authors also provided NLO parameter for 200 layers GaN-AlN QD<br><br>For 1 layer:<br>GaN QD height is 1.1E-9 m<br>AlN barrier thickness is 3E-9 m |
| Graphene QDs | Z-scan | -<br><br>Water | < 5E-9 m<br><br>- | 2.53E+3 1/m<br><br>- | -<br><br>- | 355 nm<br>-<br>2.6E+14 W/m²<br>2.30E+1 m<br>1.00E-11 s<br>1.00E+1 Hz | -<br><br>(5.7 ± 1.2)E-19 m²/W<br><br>-1.40E-11 m/W | [Wang 2019]<br><br>Details regarding fabrication: https://www.strem.com/catalog/v/06-0334/44/nanomaterials_1034343-98-0<br><br>Peak irradiance: 2.5E+13 W/m² to 26E+13 W/m²<br><br>No NLO response is presented at 532 nm and infrared regions. At 355 nm, authors have brought out a clear saturable absorption effect |

| 1D Materials | | | | | | | | |
|---|---|---|---|---|---|---|---|---|
| GaP nanowires | SRS Threshold | Pulsed Laser Vaporization (PLV)<br><br>Silicon | 2.10E-7 m<br>-<br>- | -<br>-<br>- | Monocrystalline<br>-<br>- | 514.5 nm<br>-<br>-<br>4.00E-7 m<br>CW<br>- | -<br>-<br>- | [Wu 2009]<br><br>Authors studied stimulated Raman scattering from GaP NWs as a function of their length<br><br>Raman shift (1/m):<br>3.62E+4 (TO) / 3.98E+4 (LO)<br><br>The quality factor for the nanowire segment was measured as Q ~ 18000 by assuming $g_{Raman}$ = 10 cm/GW (the gain coefficient for GaP bulk material from Rhee B. K., Bron W. E., Kuhl J., 1984, Phys. Rev. B 30, 7358–7361) |
| s-SWCNT | Z-scan | CoMoCAT catalytic CVD process (SigmaAldrich)<br><br>Petroleum ether | 0.6E-9 m to 1.1E-9 m<br><br>- | -<br><br>22.6%<br><br>1064 nm | - | 1064 nm<br>3.20E+12 W/m²<br>-<br>4.00E-9 s<br>1.00E+1 Hz | -<br><br>-<br><br>2.90E-12 m/W | [Shi 2019]<br><br>Saturation irradiance: 2.13E+12 W/m²<br><br>14% non-saturable absorption |
| SWCNT | Z-scan | HIPCO SWNT<br><br>Glass | 1.00E-9 m<br><br>- | -<br><br>8.40E+4 1/m<br><br>532 nm | -<br><br>- | 532 nm<br>1.60E+10 W/m²<br>1.20E-5 m<br>8.00E-9 s<br>1.00E+1 Hz | (-1.4E-15 + i 4.3E-16) m²/V²<br><br>-<br><br>7.10E-7 m/W | [Seo 2006]<br><br>- |
| SWCNT | Z-scan | CoMoCAT catalytic CVD process (SigmaAldrich)<br><br>Petroleum ether | - | -<br><br>17.2%<br><br>1064 nm | -<br><br>- | 1064 nm<br>-<br>3.20E+12 W/m²<br>-<br>4.00E-9 s<br>1.00E+1 Hz | -<br><br>-<br><br>5.20E-12 m/W | [Shi 2019]<br><br>Saturation irradiance: 2.95E+12 W/m²<br><br>12% non-saturable absorption |

| 2D Materials | | | | | | | | |
|---|---|---|---|---|---|---|---|---|
| AlGaN-GaN QW | Nonlinear transm./refl. | MBE<br><br>c-Sapphire | -<br><br>100 | -<br><br>-<br><br>- | -<br><br>-<br><br>- | 1500 nm<br>-<br>7.00E+10 W/m²<br>-<br>1.00E-13 s<br>1.00E+3 Hz | 5.00E-19 m²/V²<br><br>2E-17 m²/W<br><br>4.00E-10 m/W | [Rapaport 2003]<br><br>NLO parameters are for GaN/Al$_{0.65}$Ga$_{0.35}$N/GaN<br><br>The technique used to measure $n_2$ was cross-phase modulation. $\alpha_2$ is due to SA<br><br>The saturation irradiance was measured using many different pump wavelengths (see Fig. 3 of the reference paper)<br><br>Beam radius of 8.5E-5 m for the saturation irradiance measurement<br><br>Pump and probe polarization: p-polarized<br><br>Thickness for 1 layer of GaN/Al$_{0.65}$Ga$_{0.35}$N/GaN<br>1.2E-9 m/1.5E-9 m/7.8E-10 m<br>[(x 3 periods) barrier]<br><br>Information also for GaN/Al$_{0.15}$Ga$_{0.85}$N/GaN |
| Black Phosphorus | THG | Exfoliated<br><br>Glass + AlOx encapsulation | 9.50E-9 m<br><br>20 | -<br><br>7.00E+7 1/m<br><br>520 nm | -<br><br>-<br><br>- | 1560 nm<br>-<br>5.58E+15 W/m²<br>2.00E-6 m<br>1.00E-13 s<br>8.00E+6 Hz | 1.64E-19 m²/V²<br><br>-<br><br>- | [Autere 2017]<br><br>Nonlinearity dependence[3] in Fig. 4(b) of the reference paper |

| | | | | | | | | | |
|---|---|---|---|---|---|---|---|---|---|
| Black Phosphorus | Z-scan | Solvent exfoliation + gradient centrifugation<br><br>NMP | -<br><br>13 to 15 | 3.86E+2 1/m<br><br>800 nm | -<br><br>- | | 800 nm<br>-<br>3.54E+15 W/m²<br>-<br>1.00E-13 s<br>1.00E+3 Hz | -3.01E-14 esu<br><br>-2.07E-20 m²/W<br><br>- | [Xu 2017b]<br><br>Saturation irradiance: 6E+13 W/m²<br><br>Nonlinearity dependence[3] in Table 1 of the reference paper<br><br>Real part of $\chi^{(3)}$ is given<br>28.5% non-saturable absorption |
| Black Phosphorus | Z-scan | Nanoplatelets obtained by grinding bulk material and then dispersed<br><br>NMP/PVA | 30E-9 m to 60E-9 m<br><br>50 to 100 | around 7 %<br><br>500 nm to 2000 nm | -<br><br>- | | 800 nm<br>-<br>1.30E+16 W/m²<br>4.00E-5 m<br>1.00E-13 s<br>1.00E+3 Hz | -<br><br>1.20E-13 m²/W<br><br>4.50E-10 m/W | [Zheng2015a]<br><br>- |
| Electrochemical graphene oxide (GO) | Z-scan | Electrochemical method/ Vacuum filtration<br><br>- | 3.00E-7 m<br><br>Thin film | > 2.0 above 350 nm<br><br>-<br><br>300 nm to 1800 nm | -<br><br>- | 0.88 eV | 800 nm<br>-<br>-<br>1.00E-5 m<br>8.50E-14 s<br>1.00E+4 Hz | -<br><br>3.63E-13 m²/W<br><br>7.00E-11 m/W | [Ren 2016]<br><br>Nonlinearity dependence[4] in Fig.3 and Fig. 4 of the reference paper<br><br>$n_2$ is 3.63E-13 m²/W at 8E+02 J/m²<br>$n_2$ is 2.82E-13 m²/W at 1E+03 J/m²<br>$n_2$ is 1.91E-13 m²/W at 2E+03 J/m²<br>$n_2$ is 0.57E-13 m²/W at 4E+03 J/m²<br><br>Nonlinear absorption coefficient is 7E-11 m/W at 4E+03 J/m² |

|  |  |  |  | | |  |  |  |
|---|---|---|---|---|---|---|---|---|
| GaN-AlN QW | Nonlinear transm./refl. | MBE<br><br>AlN-on-(0001)sapphire | -<br><br>292 | -<br><br>-<br><br>- | -<br><br>-<br><br>- | 1550 nm<br>-<br>1.10E+13 W/m²<br>-<br>1.00E-13 s<br>1.00E+5 Hz | 5.50E-18 m²/V²<br><br>-<br><br>- | [Hamazaki 2004]<br><br>Saturation irradiance: 2.7 E+14 W/m²<br><br>Thickness for 1 layer of GaN-AlN:<br>1.1E-9 m/2.8E-9 m<br><br>Probe parameters:<br>$\lambda$ = 1550 nm<br>Irradiance = 1.1E+12W/m²<br>Pulse width = 1E-13 s<br>Polarization: p-polarized<br><br>Pump polarization:<br>p-polarized |
| GaN-AlN QW | FWM | MBE<br><br>AlN-on-sapphire | -<br><br>100 | -<br><br>-<br><br>- | -<br><br>-<br><br>- | 1500 nm<br>-<br>-<br>-<br>1.00E-13 s<br>1.00E+3 Hz | 2.40E-7 esu<br><br>-<br><br>- | [Valdueza-Felip 2008]<br><br>Thickness for 1 layer of GaN-AlN QW:<br>1.5E-9 m/1.5E-9 m<br><br>Pump polarization:<br>p-polarized |
| Graphene | Z-scan | CVD<br><br>Quartz | -<br><br>Monolayer | -<br>7.1E+7 1/m<br>to<br>10.5E+7 1/m<br><br>435 nm to 1100 nm | -<br><br>-<br><br>0.2 eV | 435 nm to 1100 nm<br>-<br>2.00E+14 W/m²<br>2.5E-5 m to 3.5E-5 m<br>1.50E-13 s<br>1.00E+3 Hz | -<br><br>-<br><br>3.6E-9 m/W to<br>8.6E-9 m/W | [Chen 2015]<br><br>Saturation irradiance:<br>2.3E+13 W/m² to 2.15E+14 W/m²<br><br>Nonlinearity dependence[1] in Fig.3 and in Table S1 of the reference paper<br><br>Authors confirmed via email that $\alpha_0$ equals 7.1E+7 1/m to 10.5E+7 1/m<br>(typo in the exponent in suppl. info) |

| Graphene | Other (add in comments) | CVD<br><br>Glass | Monolayer | -<br>-<br>- | -<br><br>-0.3 eV to<br>-0.2 eV | 1600 nm<br>-<br>5.00E+12 W/m²<br>2.00E-5 m<br>1.80E-13 s<br>8.20E+7 Hz | (-6 - i 9.6)E-16 m²/V²<br>(-1 - i 1.6)E-13 m²/W<br><br>- | [Dremetsika 2017]<br><br>Technique: optically-heterodyne-detected optical Kerr effect<br><br>Probe parameters:<br>λ = 1600 nm<br>Irradiance = 3E+11W/m²<br>Pulse width = 1.8E-13 s<br>Beam radius = 1.5E-05 m<br><br>Specified $\chi^{(3)}$ values are for $\chi_{xyxy}+\chi_{xyyx}$ |
|---|---|---|---|---|---|---|---|---|
| Graphene | FWM | CVD<br><br>Fused silica | Monolayer | -<br>-<br>- | -<br>-<br>0 eV | 1040 nm<br>-<br>-<br>2.00E-13 s<br>8.00E+7 Hz | 3.00E-17 m²/V²<br><br>-<br><br>- | [Jiang 2018]<br><br>Nonlinearity dependence[5] in Fig. 5d of the reference paper<br><br>Probe parameters:<br>λ = 1300 nm<br>Pulse width = 2.00E-13 s |
| Graphene | Z-scan | CVD<br><br>Quartz | 6 to 8 | -<br>14.9%<br>1000 nm to 2500 nm | -<br>-<br>~ 0 eV | 1930 nm<br>5.75E+11 W/m²<br>3.50E-5 m<br>2.80E-12 s<br>3.23E+7 Hz | -<br><br>4.58E-11 m²/W<br><br>- | [Miao 2015]<br><br>Saturation irradiance: 1E+10 W/m²<br><br>Nonlinearity also at 1562 nm specified in the reference paper |
| Graphene | THG | CVD<br><br>Sapphire | Monolayer | -<br>2.3%<br>1250 nm | -<br>-<br>0.25 eV | 3100 nm<br>-<br>2.40E+12 W/m²<br>4.70E-6 m<br>3.00E-13 s<br>8.00E+7 Hz | 8.00E-17 m²/V²<br><br>-<br><br>- | [Soavi 2018]<br><br>Nonlinearity dependence[1,5] in Fig. 3b and Fig. 3d of reference paper<br><br>η: 1E-11 % |

| Graphene | Z-scan | CVD<br><br>Quartz | -<br><br>Monolayer | -<br>-<br>- | -<br>-<br>~ 0 eV | 900 nm<br>-<br>4.59E+13 W/m²<br>-<br>1.00E-13 s<br>8.00E+7 Hz | -<br><br>1.08E-12 m²/W<br><br>- | [Thakur 2019]<br><br>Nonlinearity dependence[1,6] in Fig. 2b and Fig. 5b of the reference paper |
|---|---|---|---|---|---|---|---|---|
| Graphene oxide (GO) | Z-scan | Modified Hummers method/ Self-assembly/ Vacuum filtration process<br><br>$H_2SO_4$ /$H_2O_2$/deionized water/methanol mixture | 1.00E-6 m<br><br>Thin film | -<br><br>-<br><br>- | -<br><br>-<br><br>- | 1560 nm<br>-<br>3.80E+12 W/m²<br>-<br>6.70E-14 s<br>2.00E+7 Hz | -<br><br>4.50E-14 m²/W<br><br>- | [Xu 2017a]<br><br>Nonlinearity dependence[6] in Fig. 4(a) of the reference paper |
| Graphene oxide (GO) | Z-scan | Self-assembly<br><br>$H_2SO_4$ /$H_2O_2$/deionized water/methanol mixture | 8E-7 m to 2E-6 m<br><br>Thin film | -<br><br>-<br><br>- | -<br><br>-<br><br>- | 800 nm<br>-<br>-<br>2.50E-6 m<br>1.00E-13 s<br>- | -<br><br>-<br><br>4.00E-7 m/W | [Zheng 2014]<br><br>$\alpha_2$ is 4.00E-7 m/W at 32 μJ/cm²<br><br>Nonlinearity dependence[4] in Fig. 4(a) of reference paper<br><br>Information also for reduced Graphene Oxide (rGO) |
| Graphene oxide (GO) with gold nanoparticles | Z-scan | Vacuum filtration process<br><br>Water | 1.00E-6 m<br><br>Thin film | -<br>< 20%<br>-<br>800 nm | -<br>-<br>-<br>- | 800 nm<br>-<br>-<br>-<br>- | -<br><br>2.40E-16 m²/W<br><br>1.00E-9 m/W | [Fraser 2015]<br>Nonlinearity dependence[7] in Fig. 4 of the reference paper<br>$n_2$ of reference GO film is 4E-17 m²/W<br>$\alpha_2$ of reference GO film is 1.7E-10 m/W<br>Input fluence is 1.4E+02 J/m² |
| $Mo_{0.5}W_{0.5}S_2$ | Z-scan | CVT<br><br>Glass | 2.20E-5 m<br><br>Multilayer | -<br>6.22E-2 (-)<br>1064 nm | Polycrystalline<br>-<br>- | 1064 nm<br>-<br>8.50E-16 W/m²<br>-<br>2.50E-11 s<br>2.00E+1 Hz | [-(4.18 ± 2.32)E-8<br>+ i(5.73 ± 1.12)E-11] esu<br><br>-(8.73 ± 1.47)E-15 m²/W<br><br>(1.91 ± 0.78)E-10 m/W | [Bikorimana 2016]<br><br>- |
| $MoS_2$ | THG | Micromechanical exfoliation<br><br>Si/SiO₂ (285 nm) | ~ 0.65E-9 m<br><br>Monolayer | -<br>-<br>- | -<br>-<br>- | 1560 nm<br>2.70E+3 W<br>-<br>-<br>1.50E-13 s<br>5.00E+7 Hz | 3.60E-19 m²/V²<br><br>-<br><br>- | [Autere 2018b]<br>Nonlinearity dependence[2] in Fig. 5 of the reference paper |
| $MoS_2$ | THG | Micromechanical cleavage<br><br>Si/SiO₂ | -<br><br>Monolayer | -<br>Close to 0<br>1560 nm | -<br>-<br>- | 1560 nm<br>8.00E+3 W<br>-<br>9.30E-7 m<br>1.50E-13 s<br>5.00E+7 Hz | -<br><br>-<br><br>- | [Säynätjoki 2017]<br>Nonlinearity dependence[2,3,6] in Fig. 4 and Fig. 5 of reference paper<br>η: 4.76E-8 % |
| MoS$_2$ | THG imaging | CVD<br><br>SiO$_2$/Si (300 nm) | 7.00E-10 m<br><br>Monolayer | -<br>-<br>- | -<br>-<br>- | 1560 nm<br>-<br>-<br>1.80E-6 m<br>1.50E-13 s<br>8.90E+5 Hz | 2.40E-19 m²/V²<br><br>-<br><br>- | [Woodward 2016]<br><br>Nonlinearity dependence[1] shown in Fig. 3(a,b) of the reference paper<br><br>The authors provided also the sheet susceptibilities |
| MoS$_2$ | Z-scan | Liquid-phase exfoliation technique<br><br>Cyclohexyl pyrrolidinone | > 8E-9 m<br><br>~ 15 | -<br><br>2.57E+3 1/m<br><br>532 nm | -<br><br>indirect: 1.2 eV/<br>direct: 1.8 eV | 532 nm<br>-<br>-<br>-<br>1.00E-10 s<br>1.00E+4 Hz | [-(14.1 ± 6.5)E-11<br>-i(9.9 ± 3.3)E−12 esu<br><br>-(2.5 ± 1.2)E−16 m²/W<br><br>-(26.2 ± 8.8)E-11 m/W | [Wang 2014]<br><br>Saturation irradiance: (1.13 ± 0.52)E+13 W/m²<br><br>The authors also provided NLO parameters for measurements at 515 nm, 800 nm, 1030 nm and 1064 nm |
| MoS$_2$ | I-scan with microscopic imaging | CVD<br><br>SiO$_2$/Si and quartz (300nm) | -<br><br>1, 4, 6 | -<br>-<br>- | -<br>-<br>- | 1030 nm<br>-<br>-<br>1.73E-5 m<br>3.40E-13 s<br>1.00E+3 Hz | -<br><br>-<br><br>(7.62 ± 0.15)E-8 m/W | [Li 2015]<br><br>Saturation irradiance: (64.5 ± 1.53)E+13 W/m² |
| MoS$_2$ | Z-scan | CVT<br><br>Glass | 2.50E-5 m<br><br>Multilayer | -<br><br>5.03E-1 (-)<br><br>1064 nm | Polycrystalline<br><br>-<br><br>- | 1064 nm<br>5.66E-15 W/m²<br>-<br>2.50E-11 s<br>2.00E+1 Hz | (8.71 ± 1.59)E−10<br>- i(1.50 ± 0.88)E−11 esu<br><br>(1.88 ± 0.48)E−16 m²/W<br><br>-(3.8 ± 0.59)E−11 m/W | [Bikorimana 2016]<br><br><br>α$_2$ due to SA |
| MoSe₂ | THG | Micromechanical exfoliation<br><br>Si/SiO₂ (285 nm) | ~0.65E-9 m<br><br>Monolayer | -<br><br>-<br><br>- | -<br><br>-<br><br>- | 1560 nm<br>2.70E+3 W<br><br>1.50E-13 s<br>5.00E+7 Hz | 2.20E-19 m²/V²<br><br>-<br><br>- | [Autere 2018b]<br><br>Nonlinearity dependence[2] in Fig. 5 of the reference paper |
| MoSe₂ | Z-scan | Liquid-phase exfoliation technique<br><br>Cyclohexyl pyrrolidinone | -<br><br>Mono or triple<br><br>7.93E+2 1/m<br><br>800 nm | | indirect: 1.1 eV/ direct: 1.5 eV | 800 nm<br>-<br>-<br>1.00E-13 s<br>1.00E+3 Hz | -i(1.45 ± 0.34)E−15 esu<br><br>-<br><br>-(2.54 ± 0.60)E-14 m/W | [Wang 2014]<br><br>Saturation irradiance: (590 ± 225)E+13 W/m²<br><br>The authors also provided NLO parameters for measurements at 515 nm, 532 nm, 1030 nm and 1064 nm |
| MoTe₂ | Z-scan | Liquid-phase exfoliation technique<br><br>Cyclohexyl pyrrolidinone | -<br><br>Mono or triple<br><br>1.07E+2 1/m<br><br>1064 nm | | indirect: 1.0 eV/ direct: 1.0 eV | 1064 nm<br>-<br>-<br>1.00E-10 s<br>1.00E+4 Hz | [-(0.92 ± 0.15)E-11 - i(2.27 ± 0.39)E-12] esu<br><br>-(0.160 ± 0.027)E-16 m²/W<br><br>-(2.99 ± 0.52)E-11 m/W | [Wang 2014]<br><br>Saturation irradiance: (0.19 ± 0.04)E+13 W/m²<br><br>The authors also provided NLO parameters for measurements at 515 nm, 532 nm, 800 nm and 1030 nm |
| Reduced GO (rGO) | Z-scan | Self-assembly<br><br>H₂SO₄ /H₂O₂/deionized water/methanol mixture | 8E-7 m to 2E-6 m<br><br>Thin film | -<br><br>-<br><br>- | -<br><br>-<br><br>- | 800 nm<br>-<br><br>2.50E-6 m<br>1.00E-13 s<br>- | -<br><br>-<br><br>-1.00E-13 m²/W<br><br>- | [Zheng 2014]<br><br>Fully rGO with no oxygen-containing group for optical fluence > 50 μJ/cm²<br><br>Nonlinearity dependence[4] in Fig. 4(a) of reference paper<br><br>Information also for Graphene Oxide (GO) |
| WS$_2$ | THG | Micromechanical exfoliation<br>Si/SiO$_2$ (285 nm) | ~ 0.65E-9 m<br>Monolayer | -<br>-<br>- | | 1560 nm<br>2.70E+3 W<br><br>1.50E-13 s<br>5.00E+7 Hz | 2.40E-19 m²/V²<br>-<br>- | [Autere 2018b]<br><br>Nonlinearity dependence[2] in Fig. 5 of the reference paper |
| WS$_2$ | Z-scan | CVT<br>Glass | 2.00E-5 m<br>Multilayer | -<br>2.54E-1 (-)<br>1064 nm | Polycrystalline<br>-<br>- | 1064 nm<br>-<br>2.30E-15 W/m²<br>-<br>2.50E-11 s<br>2.00E+1 Hz | [(2.31 ± 0.21)E−8<br>- i(1.75 ± 0.11)E−11]<br>esu<br><br>(5.83 ± 0.18)E−15 m²/W<br><br>-(5.1 ± 0.26)E−11 m/W | [Bikorimana 2016]<br><br>α$_2$ due to SA |
| WS$_2$ | Z-scan | Vapor phase sulfurization of metal films<br>Quartz | 7.50E-10 m | -<br>7.17E+7 1/m<br>-<br>1040 nm | - | 1040 nm<br>-<br>2.35E+14 W/m²<br>-<br>3.40E-13 s<br>1.00E+2 Hz | (4.82E-9 + i 1.49E-8) esu<br><br>1.28E-14 m²/W<br><br>3.07E-8 m/W | [Dong 2016]<br><br>Nonlinearity dependence[3] in Table 2 and Fig. 4(d) of the reference paper |
| WS$_2$ | Z-scan | Vapor phase sulfurization of metal films<br>Fused quartz | -<br>1 to 3 | -<br>7.17E+7 1/m<br>-<br>1030 nm | - | 1030 nm<br>-<br>3.80E+14 W/m²<br>-<br>3.40E-13 s<br>1.00E+2 Hz | i 4E-8 esu<br><br>-<br><br>1.00E-7 m/W | [Zhang 2015b]<br><br>Nonlinearity dependence[1] in Table 2 of the reference paper<br><br>In reference paper, also values for pump and probe at 800 nm and 515nm |
| WS$_2$ | Z-scan | CVD<br>Sapphire | 7.00E-10 m<br>Monolayer | -<br>8.88E+9 1/m<br>-<br>800 nm | - | 800 nm<br>-<br>-<br>4.00E-7 m<br>1.00E-13 s<br>1.00E+3 Hz | -<br><br>8.10E-13 m²/W<br><br>-3.70E-6 m/W | [Zheng2015b]<br><br>- |
| WSe$_2$ | THG | Micromechanical exfoliation<br><br>Si/SiO$_2$ (285 nm) | ~ 0.65E-9 m<br><br>Monolayer | -<br>-<br>- | -<br>-<br>- | 1560 nm<br>2.70E+3 W<br>-<br>1.50E-13 s<br>5.00E+7 Hz | 1.00E-19 m²/V²<br>-<br>- | [Autere 2018b]<br><br>Nonlinearity dependence[2] in Fig. 5 of the reference paper |
| WSe$_2$ | THG | Micromechanical exfoliation<br><br>Fused silica | 1.17E-9 m<br><br>Monolayer | -<br>-<br>- | -<br>1.65 eV<br>- | 1546 nm<br>-<br>(0.8 ± 0.1)E-6 m<br>2.00E-13 s<br>8.00E+7 Hz | (0.9 ± 0.2)E-28 m³/V²<br>-<br>- | [Rosa 2018]<br><br>Nonlinearity dependence[2,3] in Fig. 2 and Fig. 4 of the reference paper<br><br>The authors also provided information for 4, 5, 6, and 9 layers at 1546 nm<br><br>Specified third-order nonlinearity is the sheet susceptibility |
| WSe$_2$ | Z-scan | CVT<br><br>Glass | 2.20E-5 m<br><br>Multilayer | -<br>1.55E-1 (-)<br>1064 nm | Polycrystalline<br>-<br>- | 1064 nm<br>8.50E-15 W/m²<br>-<br>2.50E-11 s<br>2.00E+1 Hz | -(9.74 ± 1.19)E-7 + i(6.35 ± 1.35)E-12 esu<br><br>-(2.47 ± 1.23)E-13 m²/W<br><br>(1.9 ± 0.57)E-11 m/W | [Bikorimana 2016]<br><br>- |
| WSe$_2$ | Z-scan | Vapor phase sulfurization of metal films<br><br>Quartz | 1.14E-8 m<br><br>- | 1.13E+8 1/m<br>-<br>1040 nm | -<br>-<br>- | 1040 nm<br>-<br>8.11E+13 W/m²<br>-<br>3.40E-13 s<br>1.00E+2 Hz | (-7.2E-10 + i 2.4E-8) esu<br><br>-1.87E-15 m²/W<br><br>4.80E-8 m/W | [Dong 2016]<br><br>Nonlinearity dependence[3] in Table 2 and Fig. 4(d) of the reference paper |

Superscripts indicate the work reports the nonlinearity dependence on [1]wavelength/energy, [2]polarization, [3]sample size (width, diameter, thickness, number of layers), [4]input fluence, [5]doping level, [6]pulse length/peak power, [7]concentration.

## 3.4 Metamaterials: data table and discussion

***Team:*** *Adam Ball, Ksenia Dolgaleva, Daniel Espinosa, Nathaniel Kinsey,* **Mikko Huttunen** *(team leader), Dragomir Neshev, Ray Secondo*

### 3.4.1 Introduction

#### 3.4.1.1 Metamaterials definitions

Efficient NLO interactions are essential for many applications in modern optics. However, they typically require high-irradiance laser sources or long interaction lengths. These requirements cannot be satisfied in the small-footprint optical devices used in many applications. With the growing importance of photonic integrated circuits (PICs) and ultracompact nanostructured optical devices, enabling efficient NLO interactions at the nanoscale is essential. This may be achieved in nanostructured optical devices with engineered NLO responses. Along these lines, metamaterials offer several advantages such as the ability to confine light for enhanced nonlinearities as well as to shape and control scattered light in new ways [Chen2018, Minovich2015, Husu2012, Litchinitser2018, BinHasan2014, Kauranen2012, Smirnova2016, Chang2018, Butet2015, Huang2020, Gigli2019a, Reshef2019b, DeAngelis2020]. It is also worth noting that in ultra-thin metasurfaces the phase-matching condition is largely relaxed. It is expressed in the form of transverse phase matching, together with mode-matching that replaces the conventional longitudinal phase matching for propagating waves in bulk materials. Recent work also demonstrates that stacked metasurface structures can be longitudinally phasematched by utilizing phase-engineered metasurfaces [Stolt2021].

In this Section, we focus on highlighting recent achievements in the broader field of NLO metamaterials and nanostructured materials. Metamaterials are artificial nanostructures comprised of building blocks called meta-atoms, which serve as their structural units. The constituent materials can be metallic [Kauranen2012, Husu2012, BinHasan2014, Litchinitser2018], semiconducting [Vabishchevich2018, Shcherbakov2017] or dielectric [Sain2019, Gigli2019a, Yan2020]. Moreover, one can choose the arrangement geometry of the building blocks in such a way as to change the enhancement mechanism. The structural units can be positioned in a regular or irregular 2D array to form a 2D metamaterial or a metasurface [Kauranen2012, Saad-Bin-Alam2021a]. Alternatively, they can comprise 3D metamaterials through structures such as extruded 2D arrays, multi-layer films [Stolt2021, Suresh2021], or fully 3D nanostructure arrays [Kadic2019]. The latter requires far more sophisticated fabrication approaches and is not discussed in this work. We focus instead on the NLO performance of more practical metasurfaces and planar nanostructures made of metal, semiconductor and dielectric materials. Since metamaterials are manmade artificial materials different from all other material types considered in this article, a short tutorial on metamaterials is provided below.

#### 3.4.1.2 Mechanisms of nonlinearity enhancement

The power of artificial nanostructures resides in their ability to manipulate the flow of light in a manner not accessible in bulk materials by engineering optical confinement, scattering and interference on a subwavelength scale. An artificial nanostructure thus represents a new material comprised of the structural elements (meta-atoms), which has an optical response that can be very different from that of its constituent materials. There are several mechanisms responsible for the enhancement, suppression and/or sign reversal of the NLO effects attainable in nanostructures. Although most of the mechanisms are based on the resonances associated with the meta-atoms, the origin and type of resonance differ depending on the type of metasurface, as discussed below. Moreover, there are some additional mechanisms of enhancement one can explore, which are briefly described as well.

*Metal–dielectric nanostructures* with meta-atoms built of noble metals, such as silver and gold, exhibit surface plasmon polaritons and localized surface plasmon resonances (LSPRs) associated with the meta-atoms (often referred to as antennae) [Chen2018, Gomes2020, Lis2014, Wurtz2008,

Minovich2015, Krasavin2018, Husu2012, Litchinitser2018, BinHasan2014, Deka2017, Kauranen2012, Smirnova2016, Chang2018, Butet2015, Huang2020, Gwo2016]. LSPRs are responsible for the localization, and subsequent increase, of the electric field strength in the vicinity of the meta-atoms, which contributes to the enhancement of the NLO interactions. While the metallic meta-atoms can themselves exhibit nonlinearities due to the strong fields outside of the meta-atom, plasmonic structures are typically employed to enhance the electric field within another nonlinear medium, such as lithium niobate or a nonlinear polymer. This allows for the separation and optimization of the confinement and nonlinear material. Yet, the Q-factors of LSPRs are generally very low ($< 10$) due to high absorption losses associated with metals. High ohmic losses and Joule heating limit the applicability of the noble metal metasurfaces in practical NLO devices unless other mechanisms of enhancement are explored in addition to LSPRs [Reshef2019a, Saad-Bin-Alam2021a, Huttunen2019, Khurgin2013]. For example, in regular metal–dielectric metasurfaces, there exist collective resonances called surface lattice resonances (SLRs) [Michaeli2017], which exhibit much higher Q-factors on the order of $10^3$ [Saad-Bin-Alam2021a]. SLRs arise from the coherent scattering of light where the periodic structure introduces interference to amplify the resonance and its associated Q-factor. In this way, the collective scattering of the entire lattice, as opposed to the enhancement provided by a single meta-atom, contributes to the NLO processes much more efficiently.

*Dielectric metasurfaces*, including those made of semiconductors, have appeared in recent years as a solution to overcome limitations associated with the high losses of metals [Smirnova2016, Gigli2019a, Yan2020, Hopkins2015, Shcherbakov2017, Sain2019, Vabishchevich2018]. Dielectric metasurfaces are typically divided by the values of their refractive indices into high-index ($n > 3.5$), such as Si, GaAs, Ge, mid-index ($2 < n < 3.5$), such as TiO$_2$, Se, SiC, Si$_3$N$_4$, and low-index ($n < 2$), such as silica- and polymer-based metasurfaces, which are not commonly used in NLO due to their poor confinement or weak NLO characteristics [Yan2020]. The primary mechanisms of enhancement of the NLO interactions in dielectric metasurfaces are electric and magnetic Mie resonances associated with their individual building blocks. These resonances result in moderate Q-factors of 10-100, depending upon the index [Rybin2017], and fields localized within the antenna, enabling enhanced effective NLO responses. In this sense, their nonlinearity is derived from the nanoantenna's constituent material, coupling the linear index and nonlinearity. While this coupling generates some restrictions in the case of second-order nonlinearities, for third-order effects, higher index (needed for high-Q) materials also generally exhibit larger nonlinearities [Sheik-Bahae1990]. High-quality-factor phase-gradient dielectric metasurfaces with a Q-factor around 2,500 have recently been demonstrated and hold the potential for NLO applications and on-chip lasing [Lawrence2020]. Moreover, the NLO performance of dielectric metasurfaces can be boosted by exploiting multimode interference, resulting in Fano resonances [Hopkins2015] or non-radiative anapole modes [Yang2018]. Dielectric metasurfaces hold promise for infrared imaging [Camacho-Morales2021], harmonic generation and other areas of application of NLO processes [Yan2020].

*Photonic bound states in the continuum* (BICs) structures create localized eigenstates with an infinite lifetime coexisting with the continuous spectrum of radiative modes [Bykov2020] and complement the more general photonic topological insulating structures [Khanikaev2013]. Localized types of practical BIC nanostructures are a subgroup of dielectric metamaterials where a special type of resonance, having an ultra-high Q-factor on the order of $10^7$ and higher, is employed [Koshelev2019a, Wang2020, Koshelev2019b, Koshelev2020, Bernhard2020, Liu2019, Azzam2021]. They represent localized states with energies embedded in the continuum of the radiating states. The basic idea behind BICs is the lack of coupling between the resonant mode and all the radiation channels of the surrounding space. Different BICs can be categorized based on the radiation suppression mechanisms. In practice, infinitely high Q-factors of BICs are limited by finite sample sizes, material absorption, symmetry breaking and fabrication imperfections. These limitations result in quasi-BICs with large but finite Q-factors, rendering the opportunity for radiation collection, important for practical nonlinear photonic devices. Dynamical nonlinear image tuning via polarization and wavelength tuning in BIC nanostructures has been experimentally demonstrated [Xu2019], opening the door for application of such nanostructures in tunable displays, nonlinear holograms and other areas.

*Epsilon-near-zero* (ENZ) nanostructures [Reshef2019b, Neira2015, Alam2018, Suresh2021, Wen2018, Deng2020, Yang2019] represent a new group of metamaterials designed in such a way that the real part of the effective dielectric permittivity is vanishingly small at a certain wavelength called zero-permittivity wavelength $\lambda_{ENZ}$. In contrast, *near-zero-index* (NZI) metamaterials have the effective refractive index $n_{eff}$ vanishing at a specific wavelength called zero-index wavelength $\lambda_{ZI}$. Such materials are generally achieved through the combination of metal and dielectric constituents or through the use of resonant structures [Kinsey2019]. There are several mechanisms of enhancement of the NLO interactions acting as the basis of ENZ and NZI metamaterials. First, near the ENZ/NZI wavelength, the strong dispersion of the index gives rise to slow-light propagation, which results in temporal compression and enhancement of the electric field [Khurgin2010]. Second, the small magnitude of the permittivity in the ENZ region gives rise to additional electric field-enhancement, through the continuity of the normal component of the displacement field $D_\perp$. Lastly, ENZ and NZI provide reduced phase advance within the bulk and facilitate easier phase matching. However, it should be noted that these properties result primarily from slow light effects and are not present in NZI schemes which maintain a finite group velocity [Khurgin2019, Khurgin2020].

### 3.4.1.3 Historical development

The precursors to the rise and development of the field of metal–dielectric NLO metasurfaces were the earlier demonstrations in the 1980s of the benefit of surface roughness in surface-emitted second- and third-order NLO processes [Chen1981, Shalaev1998] exhibiting several-orders-of-magnitude enhancement in comparison to harmonic generation from a smooth surface. Experiments on hyper-Rayleigh scattering (HRS), reported at the end of the 1990s, have served as the first formal demonstrations of incoherent SHG from spherical gold and silver nanoparticles [Dadap1999, Vance1998]. Noble metal nanoparticles of different dimensions and shapes have been explored in later HRS studies [Hao2002, Nappa2006, Butet2010a, Butet2010b], performed in the 2000s, with an emphasis on the interplay between multipoles [Nappa2006, Butet2010a]. SHG from non-spherical plasmonic objects [Bouhelier2003, Nahata2003, Danckwerts2007, Hanke2012] (before 2012) and more advanced 3D plasmonic nanostructures [Zhang2011, Cai2011, Pu2010, Park2011] (after 2010) have also been investigated.

Structured plasmonic surfaces for efficient NLO interactions were proposed [Wokaun1981, Reinisch1983] and later experimentally demonstrated [Wokaun1981, Coutaz1985] in the 1980s by SHG generation off an extended metal grating into the first order of diffraction. The first experimental demonstration of the NLO effects in a plasmonic array of meta-atoms was SHG in an array of L-shaped gold nanoparticles [Lamprecht1997, Tuovinen2002, Canfield2004] (late 1990s – early 2000s), followed by a split-ring resonator (SRR) metasurface with magnetic resonances [Linden2012, Klein2006, Feth2008] (2006 – 2012). These works laid the foundation to the rapidly evolving field of plasmonic metasurfaces for efficient NLO interactions.

The study of resonant dielectric nanostructures is a relatively new research direction in the field of nanophotonics that started a decade ago [Zhao2009, Schuller2007]. Employing sub-wavelength dielectric nanoparticles with Mie resonances for engineering optical metasurfaces with strong NLO responses resulted in the demonstration of enhanced SHG, THG, FWM and other NLO interactions without paying the cost of high losses associated with metal–dielectric nanostructures [Minovich2015, Smirnova2016, Liu2018a, Liu2018b, Kivshar2018, Grinblat2017a, Grinblat2017b]. Mie-resonant silicon and AlGaAs nanoparticles have received considerable attention for nonlinear frequency generation [Liu2018a, Liu2018b].

Spurred by developments in solid-state physics, work in dielectric metasurfaces has more recently broadened to encompass the physical effects of optical topology, enabling robust unidirectional propagation and control over scattering [Khanikaev2013]. Such effects have been explored in the nonlinear regime to realize devices such as lasers [Bandres2018] and switches [Shalaev2019] for integrated photonics (see on-chip waveguiding materials category in Section 3.6). Among the many physical effects being explored, BIC structures represent a special case for NLO due to their ability to achieve high Q-factors [Koshelev2019a, Wang2020, Koshelev2019b, Koshelev2020, Bernhard2020, Liu2019, Azzam2021] and are being actively explored. Although BICs were mathematically predicted a long time ago and were investigated in other fields of physics in the 1960s –1970s [Fonda1963, Cumpsty1971], the first experimental demonstrations of BIC-based photonic devices were performed only in 2016 [Hsu2016].

ENZ and NZI metamaterials represent other emerging classes of nanostructured materials for efficient NLO interactions. The research in the field started in the 2010s with a plethora of theoretical works [Ciattoni2010a, Ciattoni2010b, Vincenti2011, Ciattoni2012] predicting enhanced NLO interactions in media with vanishing permittivity, which preceded the first experimental demonstrations both in homogeneous thin films and effective ENZ metastructures [Alam2016, Kinsey2015, Luk2015, Capretti2015, Caspani2016]. Following these works, ENZ nanostructures have been proposed, offering tunability and unprecedented strength and tailorability of NLO interactions [Neira2015, Alam2018, Suresh2021, Wen2018].

### 3.4.1.4  Metamaterial-related best practices and considerations when performing NLO measurements

Among the best experimental practices for a proper assessment of the NLO performances of a metamaterial, the following rules-of-thumb can be recommended. If possible, one needs to provide information about the estimated values of the nonlinearities per meta-atom. Since some of the described systems are fragile and can melt under even moderate levels of optical power in a free-space excitation setup, one must follow precautions to avoid optical damage, and to ensure reproducibility of the obtained NLO results. Additionally, as efforts to realize higher Q-factor structures emerge, analysis should also account for the strong dispersion near these resonances. For example, not all of the incident excitation power may be captured by the resonator [Shcherbakov2019] and the effect of pulse chirp near the resonance will begin to play a role in the analysis and extraction of effective NLO parameters. In the end, the caveats and best practices are the same as those listed in Section 2, with enhanced focus on damage, thermal nonlinearities, etc. due to the relatively strong absorption of plasmonic structures.

### 3.4.1.5  Description of general table outline

Tables 5A and 5B show a representative list of, respectively, second-order and third-order NLO properties of metamaterials taken from the literature since 2000, with the entries arranged in alphabetical order. Our aim has been to compile the most relevant advances in NLO metamaterials since 2000. Because numerous studies have been performed during the last two decades, not all of them could be tabulated and/or referenced here. Instead, the studies highlighted here are only representative works, where extra care has been taken in the characterization of the investigated metamaterial following the recommendations in the section above and the best practices in Section 2. Furthermore, our focus is on the studies that report NLO coefficients (e.g., $\chi^{(2)}$, $\chi^{(3)}$, $n_{2(eff)}$, $\alpha_2$, etc.). Many studies exist where the investigated metamaterial has been shown to result in enhancement of NLO signals when compared to some reference material/system. In order to restrict our scope, we chose not to tabulate works providing only enhancement factors, but instead focused on investigations also reporting NLO coefficients and/or conversion efficiencies. An interested reader is referred to Refs. [Chen2018, Minovich2015, Krasavin2018, Husu2012, Litchinitser2018, BinHasan2014, Kauranen2012, Smirnova2016, Chang2018, Butet2015, Huang2020, Gigli2019a, Reshef2019b, DeAngelis2020] for a broader perspective of the topic.

Table 5A and 5B are subdivided into "Material properties," "Measurement details" and "Nonlinear properties." The main material properties that we have tabulated are the material(s), fabrication technique, sample thickness, material crystallinity, substrate and reference to the sub-figure in Figs. 4-5 showing the fabricated device. All reported experiments have nominally been performed at room temperature. Fig. 4 shows SEM pictures of devices with second-order NLO responses, while Fig. 5 shows SEM pictures of devices with third-order NLO responses. The main pump parameters have also been tabulated, along with the enhancement mechanism utilized in the sample. Within each column the information is given in the order of the header description, and the NLO technique used is provided in the "Method" column. In Table 5A, we list the overall conversion efficiency η and/or absolute value of the effective susceptibility as the NLO parameters of key importance. In Table 5B, in addition to the above parameters, the effective $n_2$ and $\alpha_2$ values have also been listed where relevant.

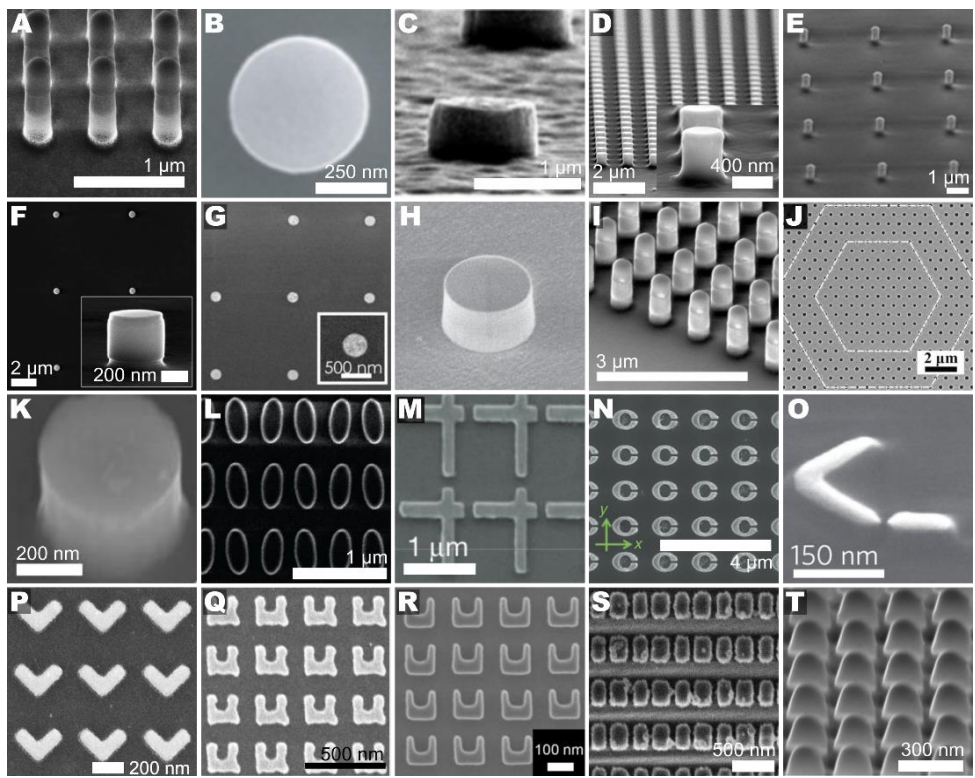

**Fig. 4** SEM images of metamaterial devices with second-order nonlinear properties. The labels of the sub-figures correspond to the labels given in Table 5A. Panels A–D, G, K, L, P, and R–T are reprinted with permission from [Liu2016, Xu2020, Sautter2019, Marino2019a, Camacho-Moralez2016, Cambiasso2017, Anthur2020, Czaplicki2018, Wen2018, Bernhardt2020, Semmlinger2018], ACS. Panels M and O are reprinted with permission from [Lee2014, Celebrano2015], Springer Nature. Panels I and N are reprinted under CC BY license from [Liu2018a, Wolf2015]. Panels E, F, J and Q are reprinted with permission from [Gili2016, Gigli2019b, Wang2020, Klein2007], ©2007,2016,2019,2020 Optica Publishing Group. Panel H is reprinted with permission from [Koshelev2020], AAAS.

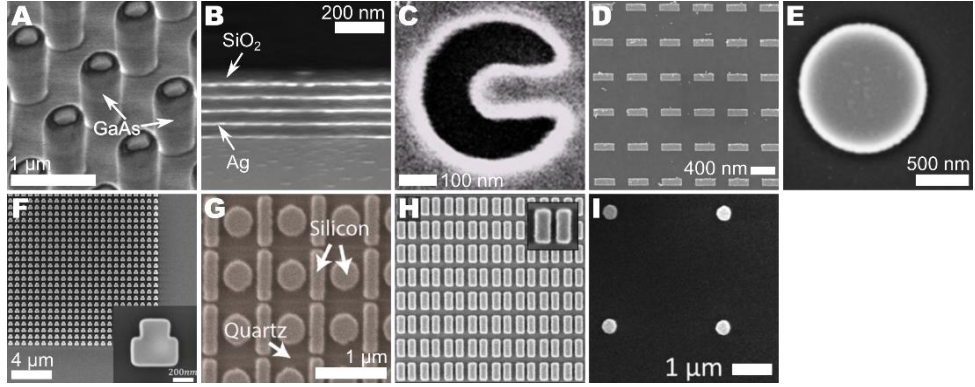

**Fig. 5** SEM images of metamaterial devices with third-order nonlinear properties. The labels of the sub-figures correspond to the labels given in Table 5B. Panels A, B, E, G, H, and I are reprinted with permission from [Zubyuk2019, Suresh2021, Grinblat2017b, Yang2015, Koshelev2019b, Shcherbakov2014], ACS. Panel C is reprinted with permission from [Melentiev2013], ©2013 Optica Publishing Group. Panel D is reprinted with permission from [Lee2014, Celebrano2015], Springer Nature. Panel F is reprinted under CC BY license from [Liu2019].

### 3.4.2 Discussion

#### 3.4.2.1 Advancements and remaining challenges for metamaterials

Because most of the early works were focused on metallic nanoparticles, *de facto* metamaterials of that era, a major advancement in the field of metamaterials can be seen to be the gradual shift from metals to the lower-loss semiconductors and dielectric nanomaterials [Grinblat2017a, Grinblat2017b, Semmlinger2018, Koshelev2019b]. New mechanisms of enhancement, associated with these new kinds of metamaterials, including ENZ, BIC and Fano resonances, have clearly diversified from the early approaches that were almost solely based on material resonances [Czaplicki2018, Koshelev2020]. Apart from the fundamental studies of NLO phenomena, dielectric metasurfaces have enabled the observation of such challenging effects as spontaneous parametric down-conversion (SPDC). Recently, for the first time, the generation of photon pairs via SPDC in lithium niobate quantum optical metasurfaces [Santiago-Cruz2021] and in a single AlGaAs nanocylinder [Marino2019b] has been demonstrated. By engineering the quantum optical metasurface, the authors demonstrated tailoring of the photon-pair spectrum in a controlled way [Santiago-Cruz2021]. These achievements lay the foundation for the application of dielectric metasurfaces in quantum light generation.

Furthermore, a generally progressing trend is visible in the achieved NLO parameters where newer NLO metamaterials seem to (clearly) outperform their earlier counterparts. As an example, many of the recent investigations of dielectric and semiconducting metasurfaces demonstrate conversion efficiencies on the order of ~$10^{-3}$ % for SHG [Lee2014, Gili2016, Sautter2019, Xu2020] and on the order of ~$10^{-4}$ % for THG [Grinblat2017b, Koshelev2019b]. In contrast, conversion efficiencies achieved in earlier works have been around $2\times10^{-9}$ % for SHG and $3\times10^{-10}$ % for THG [Klein2008].

NLO metasurfaces have recently started gaining popularity as efficient sources of THz radiation, as this frequency range is not easily attainable by conventional methods of light generation. There is a growing number of works where THz generation by DFG off a metasurface is reported [Luo2014, Keren-Zur2019, Polyushkin2011, Tal2020, McDonnell2021]. These results compare favorably with the more standard method of THz generation by DFG in bulk crystals. Specifically, the measured sheet NLO susceptibility ~$10^{-16}$ m$^2$ V$^{-1}$ far exceeds that of thin films and bulk non-centrosymmetric materials [Luo2014]. On the other hand, the extraction of the associated NLO coefficients has only been performed in [Luo2014], leaving this a development area for future studies. There is very little knowledge about NLO parameters of metamaterials at THz frequencies, and it is highly advisable for research works to report such values for this frequency window.

On the design side, significant improvements have been made in the development of new analytical [Saad-Bin-Alam2021b] and numerical [Butet2016, Blechman2019, Noor2020] methods for designing NLO metasurfaces optimized for some specific NLO interactions.

Further work is in progress to make NLO metamaterials functional in real-life applications. A partly connected challenge is to develop nanofabrication techniques to the level where fabrication of 3D NLO metamaterials becomes a routine task [Stolt2021]. When these achievements can be combined with recent advances in phase-engineered metasurfaces [Chen2018], one could envisage 3D phase-matched metamaterials that could potentially boost efficiencies up to the levels adequate for practical applications. We note that at that stage, linear absorption/scattering losses would start playing a role [Stolt2021], and actions to mitigate their detrimental effects should be taken.

### 3.4.2.2  Recommendations for future works on metamaterials

It would be extremely beneficial if future publications would in general report more extensive details of the experimental parameters alongside the NLO parameters of the studied metamaterials (see our earlier recommendations for "best practices"). For example, oftentimes publications report only the relative NLO enhancement factors of the metamaterials, making it difficult to estimate the potential importance of the studies. Therefore, we recommend that, in future works, at least the parameters listed in the tables reported here be properly quantified and reported.

In many cases, it would also be important to quantitatively characterize individual meta-atoms acting as the building blocks in the metamaterials (see, for example, [Saad-Bin-Alam2021b]). If such information becomes more abundantly available, designing next-generation NLO metamaterials would become an easier task. Furthermore, such data would improve the transparency of the work because it would facilitate estimating the success of the demonstrated enhancement mechanism and the potential importance of the performed study.

Finally, we recommend the community to continue their effort to estimate the damage thresholds of various metamaterial platforms and to study their possible damage mechanisms. We consider such efforts particularly useful when NLO metamaterials reach the level of maturity where applications start to emerge. At that stage, the relevant parameters should not be limited only to conversion efficiency/susceptibility/$n_2$ values, but can include the measured power of frequency-converted light. In this case, such information becomes critical for an application-oriented researcher to be able to estimate whether the studied metamaterial system could be scaled up in area to be used with high-power pump lasers.

### 3.4.3 Data table for metamaterials

**Table 5A**. Second-order NLO properties of metamaterials from representative works since 2000. Legend for superscripts: see below the table.

| MATERIAL PROPERTIES | | MEASUREMENT DETAILS | | NONLINEAR PROPERTIES | | | | |
|---|---|---|---|---|---|---|---|---|
| Material Substrate Fabrication Thickness Crystallinity | SEM image | Method | Pump wavelength Peak irradiance Beam waist Pulse width Rep. rate | η (%)* | $|\chi^{(2)}|$ (m/V)* | Enhancement mechanism | Additional parameters and Comments | Reference |
| GaAs / AlGaO GaAs MBE/e-beam lithography 0.3 [μm] Monocrystalline / monocrystalline | 4(a) | SHG | 1020 [nm] $3.4 \times 10^7$ [MW/m²] 3 [μm] 0.12 [ps] 80000 [kHz] | $2 \times 10^{-3}$ % / $1.5 \times 10^{-8}$ W⁻¹ | - | Magnetic dipole resonance | - | [Liu2016] |
| (110)-GaAs Fused silica MOCVD/e-beam lithography 0.4 [μm] Monocrystalline | 4(b) | SHG | 1450 [nm] $1.2 \times 10^7$ [MW/m²] 2 [μm] 0.255 [ps] 80000 [kHz] | $3.00 \times 10^{-3}$ | - | - | - | [Xu2020] |
| (111)-GaAs Fused silica MOCVD/e-beam lithography 0.4 [μm] Monocrystalline | 4(c) | SHG | 1556 [nm] $1.0 \times 10^7$ [MW/m²] - 0.1 [ps] 80000 [kHz] | $4.80 \times 10^{-3}$ | - | - | - | [Sautter2019] |
| $Al_{0.18}Ga_{0.82}As$ / AlOx GaAs MBE/e-beam lithography 0.4 / 1.0 [μm] Monocrystalline / amorphous | - | SHG | 1550 [nm] $1.6 \times 10^7$ [MW/m²] - 0.15 [ps] - | $1 \times 10^{-3}$ % / $1.5 \times 10^{-7}$ W⁻¹ | - | - | - | [Ghirardini2017] |
| $Al_{0.18}Ga_{0.82}As$ / AlOx<br>GaAs<br>MBE/e-beam lithography<br>0.4 / 1.0 [$\mu$m]<br>Monocrystalline / amorphous | 4(d) | SHG | 1550 [nm]<br>$5\times10^6$ [MW/m$^2$]<br>25 [$\mu$m]<br>0.16 [ps]<br>1000 [kHz] | $2\times10^{-6}$ W$^{-1}$ | - | - | - | [Marino2019a] |
| $Al_{0.18}Ga_{0.82}As$ / AlOx<br>GaAs<br>MBE/e-beam lithography<br>0.4 / 1.0 [$\mu$m]<br>Monocrystalline / amorphous | 4(e) | SHG | 1554 [nm]<br>$1.6\times10^7$ [MW/m$^2$]<br>-<br>0.15 [ps]<br>- | $1.10\times10^{-3}$ | - | - | - | [Gili2016] |
| $Al_{0.18}Ga_{0.82}As$ / AlOx<br>GaAs<br>MOCVD/e-beam lithography<br>0.35 / 1.0 [$\mu$m]<br>Monocrystalline / amorphous | 4(f) | SHG | 1550 [nm]<br>-<br>2.36 [$\mu$m]<br>0.16 [ps]<br>1000 [kHz] | $6.5\times10^{-6}$ W$^{-1}$ | - | - | - | [Gigli2019b] |
| $Al_{0.2}Ga_{0.8}As$<br>Fused silica<br>MOCVD/e-beam lithography<br>0.3 [$\mu$m]<br>Monocrystalline | 4(g) | SHG | 1556 [nm]<br>$7.0\times10^7$ [MW/m$^2$]<br>1.1 [$\mu$m]<br>0.5 [ps]<br>5000 [kHz] | $8.50\times10^{-3}$ | - | - | - | [Camacho-Morales2016] |
| AlGaAs<br>$SiO_2$/ITO/$SiO_2$<br>e-beam lithography<br>0.635 [$\mu$m]<br>Monocrystalline | 4(h) | SHG | 1570 [nm]<br>-<br>1.8 [$\mu$m]<br>2 [ps]<br>5144 [kHz] | $1.3\times10^{-6}$ W$^{-1}$ | - | Quasi-BIC, Mie resonance | - | [Koshelev2020] |
| GaAs / AlGaO<br>GaAs<br>MBE/e-beam lithography<br>0.45 [μm]<br>Monocrystalline / monocrystalline | 4(i) | SHG | 1570 [nm]<br>-<br>1.5 [μm]<br>0.04 [ps]<br>1 [kHz] | $2.30\times10^{-4}$ | - | - | - | [Liu2018a] |
| GaN<br>Silicon<br>MOCVD, PECVD, e-beam lithography<br>0.2 [μm]<br>Monocrystalline | 4(j) | SHG | 1543.55 [nm]<br>-<br>-<br>CW<br>- | $2.4\times10^{-2}$ W$^{-1}$ | - | Quasi-BIC at SH and cavity resonance at FH | - | [Wang2020] |
| GaP nanoantennas<br>GaP<br>e-beam lithography<br>0.2 [μm]<br>Monocrystalline | 4(k) | SHG | 910 [nm]<br>$2\times10^9$ [MW/m$^2$]<br>-<br>0.18 [ps]<br>100 [kHz] | $2\times10^{-4}$ % / $4\times10^{-9}$ W$^{-1}$ | - | Scattering resonance and surface effect | - | [Cambiasso 2017] |
| GaP nanodimers<br>Sapphire<br>MOCVD on GaAs then transfer-bonded to Sapphire, e-beam lithography<br>0.15 [μm]<br>Monocrystalline | 4(l) | SHG | 1190–1220 [nm]<br>100 [MW/m$^2$]<br>50 [μm]<br>CW<br>- | $2.00\times10^{-5}$ % / $5\times10^{-7}$ W$^{-1}$ | - | Quasi-BIC in arrays with slight asymmetry in nanoparticles | - | [Anthur2020] |
| GaP nanodimers<br>Sapphire<br>MOCVD on GaAs then transfer-bonded to Sapphire, e-beam lithography<br>0.15 [μm]<br>Monocrystalline | 4(l) | SHG | 1190–1220 [nm]<br>$1.0\times10^5$ [MW/m$^2$]<br>50 [μm]<br>0.2 [ps]<br>- | $4.00\times10^{-3}$ % / $1\times10^{-3}$ W$^{-1}$ | - | Quasi-BIC in arrays with slight asymmetry in nanoparticles | - | [Anthur2020] |
| Gold / InGaAs/AlInAs MQW InP MBE 0.1 / 0.03 / 0.656 [μm] Amorphous / monocrystalline / monocrystalline | 4(m) | SHG | 8000 [nm] 150 [MW/m²] 17.5 [μm] 400000 [ps] 250 [kHz] | $2.00\times10^{-4}$ | $55\times10^{-8}$ (yyy) | MQW enhancement | - | [Lee2014] |
| Gold / InGaAs/InAlAs MQW InP e-beam lithography 1 [μm] Amorphous / monocrystalline / monocrystalline | 4(n) | SHG | 10100 [nm] - 50 [μm] 12000 [ps] 1 [kHz] | $2.30\times10^{-3}$ W W$^{-2}$ | - | Resonators on MQWs | - | [Wolf2015] |
| Gold nanoantenna Fused silica Focused ion beam lithography 0.04 [μm] Monocrystalline | 4(o) | SHG imaging | 1560 [nm] $1.69\times10^{7}$ [MW/m²] 0.7 [μm] 0.12 [ps] 80000 [kHz] | $5\times10^{-10}$ W$^{-1}$ | - | Multiresonant antenna | - | [Celebrano2015] |
| Gold nanoparticle arrays Fused silica e-beam lithography 0.02 [μm] Amorphous | 4(p) | SHG | 1150 [nm] $1.2\times10^{5}$ [MW/m²] - 0.2 [ps] 80000 [kHz] | $2.50\times10^{-9}$ | - | Collective lattice effect | - | [Czaplicki2018] |
| Gold nanoparticles Embedded in gelatin matrix e-beam lithography 0.15 [μm] - | - | SHG imaging | 794 [nm] - - 0.18 [ps] 76000 [kHz] | - | $1\times10^{-23}$ (esu) | - | - | [Butet2010b] |
| Gold split-ring resonators Fused silica e- beam lithography 0.025 [μm] Amorphous | 4(q) | SHG | 1500 [nm] 0.0077 [MW/m²] - 0.17 [ps] 81000 [kHz] | $2.00\times10^{-9}$ | - | - | - | [Klein2007] |
| TiN<br>Sapphire<br>Pulsed laser deposition/e-beam lithography<br>0.05 [μm]<br>Monocrystalline | 4(r) | SHG | 1040 [nm]<br>$1.2 \times 10^8$ [MW/m$^2$]<br>2 [μm]<br>0.2 [ps]<br>80000 [kHz] | - | - | LSPR at the fundamental frequency, ENZ at the SH frequency | - | [Wen2018] |
| WS$_2$ / Silicon<br>BK-7<br>PECVD, e-beam lithography<br>- / 0.16 [μm]<br>Monocrystalline | 4(s) | SHG | 832 [nm]<br>-<br>6 [μm]<br>0.08 [ps]<br>80000 [kHz] | - | - | Quasi-BIC | - | [Bernhardt2020] |
| ZnO nanodisks<br>Soda lime glass<br>Sputtering/focused ion beam lithography<br>0.15 [μm]<br>Polycrystalline | 4(t) | SHG | 394 [nm]<br>$1.68 \times 10^7$ [MW/m$^2$]<br>8.5 [μm]<br>0.205 [ps]<br>250 [kHz] | $7.00 \times 10^{-7}$ | $9.6 \times 10^{-13}$ (d-coefficient) | Magnetic dipole resonance | - | [Semmlinger 2018] |

*Units as illustrated unless otherwise indicated in the table

**Table 5B**. Third-order NLO properties of metamaterials from representative works since 2000. Legend for superscripts: see below the table.

| MATERIAL PROPERTIES | | | MEASUREMENT DETAILS | NONLINEAR PROPERTIES | | | | |
|---|---|---|---|---|---|---|---|---|
| Material Substrate Fabrication Thickness Crystallinity | SEM image | Method | Pump wavelength Peak irradiance Beam waist Pulse width Rep. rate | $\eta$ (%)* $n_2$ (m²/W) $\alpha_2$ (m/W) | $\chi^{(3)}$ (m²/V²)* | Enhancement mechanism | Additional parameters and Comments | Reference |
| (100)-GaAs / AlGaO GaAs MBE/e-beam lithography 0.3 [μm] Monocrystalline | 5(a) | I-scan | 830 [nm] $2.5\times10^7$ [MW/m²] 10 [μm] 0.06 [ps] 80000 [kHz] | - - - | - | Higher free-carrier generation rate | Saturation irradiance: $1.6\times10^7$ MW/m² | [Zubyuk2019] |
| Ag nanoantenna / SiO₂ Fused silica e-beam evaporation 0.016 / 0.065 [μm] Amorphous | 5(b) | Z-scan | 410–560 [nm] - - 28 [ps] 0.05 [kHz] | - $1.2\times10^{-12}$ $-1.5\times10^{-5}$ | - | ENZ | - | [Suresh2021] |
| Al split hole resonator / - 40 nm of fused silica e-beam lithography/FIB 0.2 [μm] Amorphous | 5(c) | THG | 1560 [nm] 10 [MW/m²] 2.15 [μm] 0.2 [ps] 70000 [kHz] | $1.0\times10^{-3}$ - - | - - - | Plasmon resonance and lightning-rod effect | - | [Melentiev2013] |
| Au nanoantenna / ITO Fused silica e-beam lithography 0.05 [μm] - | 5(d) | Z-scan | 1240 [nm] $1.5\times10^6$ [MW/m²] - 0.14 [ps] - | - $-3.7\times10^{-13}$ $-2.4\times10^{-7}$ | - | ENZ + optimized field coupling and enhancement by antennae | - | [Alam2018] |
| | | | | | | | | |
| Au nanoparticle<br>Fused silica<br>electrodeposition<br>0.15 [µm]<br>Amorphous | - | Z-scan | 600 [nm]<br>$8.0\times10^8$ [MW/m²]<br>1.5 [µm]<br>0.05 [ps]<br>- | -<br>$-2.4\times10^{-15}$<br>$-9.97\times10^{-8}$ | $(0.28 - 1.48i)\times 10^{-16}$ | ENZ + Au interband transition resonance | $\chi^{(3)}$ for 600 nm at 60° and (0.18 - $1.48i)\times10^{-17}$ for 550 nm at 20° | [Neira2015] |
| Ge nanodisk<br>Borosilicate glass<br>e-beam lithography<br>0.2 [µm]<br>Amorphous | 5(e) | THG | 1650 [nm]<br>0.8 [MW/m²]<br>-<br>0.18 [ps]<br>100 [kHz] | $1.0\times10^{-3}$<br>-<br>- | $2.8\times10^{-16}$ | Higher-order (anapole) modes | - | [Grinblat2017b] |
| Si nanoblock<br>Quartz<br>e-beam lithography<br>0.5 [µm]<br>Monocrystalline | 5(f) | THG | 1587 [nm]<br>-<br>-<br>5 [ps]<br>- | $1.4\times10^{-8}$ W⁻²<br>-<br>- | - | Quasi-BIC in arrays with slight asymmetry in nanoparticles | - | [Liu2019] |
| Si nanoantenna<br>Quartz<br>LPCVD/e-beam lithography<br>0.12 [µm]<br>Polycrystalline | 5(g) | THG | 1350 [nm]<br>0.32 [MW/m²]<br>-<br>0.25 [ps]<br>1 [kHz] | $1.2\times10^{-4}$<br>-<br>- | - | High-Q Fano resonance | - | [Yang2015] |
| Si nanoantenna<br>Fused silica<br>PECVD, e-beam lithography<br>0.538 [µm]<br>Amorphous | 5(h) | THG | 1425 [nm]<br>-<br>-<br>0.2 [ps]<br>80000 [kHz] | $1.0\times10^{-4}$<br>-<br>- | - | Quasi-BIC in arrays with broken symmetry | - | [Koshelev2019] |
| Si nanodisk<br>SOI<br>e-beam lithography<br>0.26 [µm]<br>Monocrystalline | 5(i) | THG | 1260 [nm]<br>0.5 [MW/m²]<br>-<br>0.2 [ps]<br>80000 [kHz] | $0.8\times10^{-5}$<br>-<br>- | - | Magnetic response | - | [Shcherbakov2014] |

*Units as illustrated unless otherwise indicated in the table

## 3.5 Fiber waveguiding materials: data table and discussion

_**Team:** **John Ballato (team leader)**, Peter Dragic_

### 3.5.1 Introduction

#### 3.5.1.1 Background information

Our understanding and use of NLO are very much intertwined with the history and development of fiber optics. This is because optical fibers are unique tools for studying and utilizing NLO phenomena due to their combination of low loss, long lengths, small core sizes and resulting small mode diameters. Indeed, the first observations of nonlinearities [Stolen1972, Ippen1972, Stolen1974, Hill1974] in fibers are contemporaneous with the first low loss fibers themselves [Keck1972a, Keck1972b]. Furthermore, phase matching of NLO processes in optical fibers can be achieved in more than one way [Stolen1974, Stolen1981]. First, in the case of a multimode fiber, different optical modes across a wide span of wavelengths, for example the signal and idler in FWM, may be found which possess the same propagation constant. Alternatively, the design of the fiber can be suitably tailored to control (e.g., "flatten") the dispersion curve of a given mode (such as the fundamental mode) to compensate for bulk chromatic dispersion. Such fibers can take the form of both solid multilayered conventional and microstructured waveguides.

From a practical perspective, a key driver of nonlinear fiber optics was the concurrent development of commercial communications networks. As noted by Stolen, a pioneer in nonlinear fiber optics, _"Fiber nonlinear optics has grown from a novel medium for the study of nonlinear optical effects, through a period where these effects appeared as system impairments, to the present day where optical nonlinearities are an integral part of high-capacity optical systems"_ [Stolen2008]. This sentiment is well-reflected in the history of nonlinearities and communication systems [Smith1972, Chraplyvy1990, Zhang1994, Li2001, Hasegawa2017, Winzer2018, Essiambre2021].

Other applications of modern consequence where fiber nonlinear optics is critical include fiber sensors and high power / high energy fiber lasers. Fiber lasers, of considerable interest for defense, security, remote sensing, and manufacturing uses, generally demand reduced nonlinearities, particularly those that are considered parasitic to power-scaling and beam quality [Richardson2010, Zervas2014, Dawson2008], such as stimulated Brillouin, Raman, and thermal Rayleigh scattering (which manifests as transverse mode instabilities) [Dong2013, Smith2011, Jauregui2020, Zervas2019], as well as thermal lensing [Dong2016]. Fiber-based sensors, on the other hand, including distributed systems [Lu2019] such as those designed for structural health [Barrias2016] and down-hole and geological [Schenato2017] temperature and pressure monitoring, tend to favor higher nonlinearities for enhanced signal-to-noise ratio and therefore higher measurement sensitivity [Bao2021]. 'Linear' systems, namely those based on fiber Bragg grating technology and Rayleigh scattering, have found wide commercial success, as have those based on the Raman scattering nonlinearity. The latter systems work on the principle that the ratio of the anti-Stokes to Stokes Raman scattered signals is a strong function of temperature, and therefore Raman scattering primarily is used for distributed temperature sensing. Both temperature and strain sensing are possible with Brillouin scattering, since the scattering frequency is a strong function of both the fiber thermal and mechanical environments. Unfortunately, Brillouin-based systems have not found wide commercial success due to their complexity and cost [Dragic2018].

#### 3.5.1.2 Considerations for fibers when performing NLO measurements

All this said, care must be taken in assessing nonlinearities in optical fiber. This is because the measured nonlinearity is contributed to by both the materials comprising the fiber and the design of the fiber itself. Being a waveguide, the latter is important because fiber design influences the spatial (modal) and spectral (dispersion) properties of the propagating light. This, in turn, influences the relative contributions of the core and clad materials to the nonlinearities through their respective NLO coefficients. Representative examples of such fiber design influences on, for example, the nonlinear refractive index, $n_2$, can be found in Refs. [Kato1995, Boskovic1996]. To this point, even the $n_2$ value of fused silica, the most canonical and well-studied fiber optic material, is not as precisely known as one would expect given the half-century since the first low loss fibers were reported (see Section 1.1). Another case where considerable care is required when applied to waveguides is Brillouin scattering. Namely, the Brillouin gain coefficient $g_{Brillouin}$ has both material and waveguide influences. Depending on the waveguide design, one may dominate the other, or they may play a more cooperative role, often making it difficult to distinguish between the two. Materially, $g_{Brillouin}$ depends on refractive index, transverse photoelasticity, isothermal compressibility, Brillouin linewidth, and acoustic velocity such that each of these values should be carefully measured and reported for true completeness. More commonly, publications report the Brillouin shift and linewidth / lineshape but these can strongly be impacted by the fiber itself. These include factors such as the number of acoustic modes present and their relative confinement, whether the waveguide is acoustically anti-guiding or guiding, fiber quality (loss) and length, and birefringence, to name just a few. All waveguide effects broaden the spectrum and reduce the observed Brillouin gain relative to the bulk material. In addition to these specific considerations for fibers, also the best practices described in Section 2 should be taken into account when performing NLO measurements.

### 3.5.1.3 Description of general table outline

Table 6 provides a representative list of NLO properties of fiber waveguiding materials taken from the literature since 2000. Our goal has been to compile and highlight advancements in nonlinear materials in fiber form since 2000 with the above-mentioned caveats and the best practices of Section 2 in mind. The publications included in Table 6 were selected based on their representing measurements on fibers across the spectrum of reasonably common materials and respective nonlinearities. The included works nominally report data obtained at room temperature. The Table is subdivided into "Fiber properties," "Measurement details" and "Nonlinear properties." Within each column the information is given in the order of the header description, and the NLO technique used is provided in the "Method" column. The Table compiles representative values for the following fiber material sub-categories: telecommunications-grade silica [Oguama2005, Oguama2005b, Deroh2020, Evert2012], silicate [Tuggle2017, Dragic2014, Lee2005, Cavillon2018], non-silica oxide (e.g., phosphate [Lee2007] and tellurite [Deroh2020]), and non-oxide (e.g., chalcogenides [Deroh2020, Florea2006, Tuniz2006, Fortin2008] and fluoride [Fortin2011, Lambin-Lezzi2013, Deroh2020]) optical fibers. For the more conventional silica and silica-based glass fibers, trends in Brillouin, Raman, and $n_2$ for a wide variety of dopants are provided in Ref. [Cavillon2018a] and references therein. Also noted are NLO properties of representative specialty optical fibers including polymer [Mizuno2010] and crystalline core optical fibers such as those made from sapphire [Harrington2014, Yin2008, Kim2008] and semiconductors [Ren2019, Shen2020]. Since second-order nonlinearities are generally precluded by symmetry in glass and cubic crystalline core fibers, the focus here is on third order $\chi^{(3)}$ wave-mixing as well as nonlinear scattering (e.g., Raman and Brillouin) phenomena. For completeness, second order nonlinearities in poled glass optical fiber also are noted [Canagasabey2009].

Two last points are noteworthy. First, the focus here is on the materials and material nonlinearities in conventional core-cladding configurations and not specifically on nonlinearity contributions from advanced waveguide design (e.g., microstructured or photonic-crystal fibers (PCFs), see, for example [Cordeiro2005, Dudley2009, Hu2019, Nizar2021]) as the range of those enabled influences is virtually endless. Indeed, one can make a very strong case that understanding / quantifying the relevant properties of a material should precede any attempts to design a waveguide from it. Second, nonlinearities can be materially enhanced or reduced (even negated), depending on the application. As such, both are generally treated here by way of discussing trends and ranges. The more fundamental materials science of optical nonlinearities in fibers has been recently reviewed [Ballato2018, Cavillon2018a].

### 3.5.2 Discussion

#### 3.5.2.1 Advancements and remaining challenges for fibers

Not surprisingly, the range of NLO parameter values measured from optical fibers as shown in Table 6 is nearly as expansive as the range of materials from which they are made. Ironically, it is this plenty that precludes an equally abundant coverage of all materials and values. Unless otherwise noted, Table 6 is meant to provide generalized values, ranges, and trends for each fiber material family and associated nonlinearity.

Over the past twenty years, since approximately 2000, the dominant trend one observes is in the range of (strong and weak) NLO materials contained within or integrated into the fibers. Some of this growth is due to the expansion in applications employing fibers, such as power-scaled high energy lasers and sensors, as well as in new fiber processing methods that permit the fiberization of materials previously not possible [Sazio2006, Ballato2018a]. Arguably, this renaissance in fiber materials has driven rapid progress in the range of NLO parameter values. This said, beyond telecommunications grade silica, losses are still relatively high in most glass and crystal core fibers, thus potentially reducing their efficiencies and range of applications. However, markedly higher NLO parameter values offset issues of loss in some cases, such as with the semiconductor core optical fibers since a very large material nonlinearity shortens, often by several orders-of-magnitude, the required fiber lengths.

#### 3.5.2.2 Recommendations for future works on fibers

The aforementioned renaissance in optical fiber materials has opened two particularly intriguing doors to the world of NLO. First, the wide range of new materials has greatly expanded the achievable range of NLO parameter values. When coupled with fiber design (e.g., microstructured and photonic crystal fibers) and post-processing (e.g., tapering), unquestionable benefits to future applications arise [Sylvestre2021]. Second, the availability of new fiber materials has generated new insights and concepts relevant to nonlinear fiber optics. These include, for example, zero nonlinearity values based on carefully balanced fiber compositions. For instance, in the case of Brillouin scattering, positive and negative transverse photoelasticities [Ballato2018, Cavillon2018a] are balanced such that no spontaneous scattering occurs, thereby obviating its stimulated form. Such opportunities are not possible based only on fiber design. Needless to say, there are great possibilities ahead for fiber optics, both in terms of uniquely high and low nonlinearities. The Periodic Table can be construed to be a rich palette from which new materials with novel properties may be derived. Continued characterization sheds light on how to combine the base materials to achieve a desired outcome, be it using a hybrid or designer material approach. That being said, too few literature works go the extra mile to derive the relevant NLO material values from a system-level demonstration. For example, there are numerous examples of supercontinuum generation (SCG) from novel fibers / materials, but a large proportion of these papers lack sufficient detail from which the relevant nonlinearities may be quantified. Many other works, especially those in the area of hybrid fibers (see Section 3.7), refer back to bulk values for the nonlinearities. It is important to remember that many materials, including glasses, undergo very different thermal histories when found in either bulk or fiber form. If a material's structural properties depend on its fabrication history, this may have a significant impact on its NLO behavior. In the case of hybrid materials, secondary effects such as material inter-diffusion during fabrication may have a similar impact.

It is fortunate that many of the works cited herein set out from the beginning to quantify the nonlinearities in fibers made from novel materials, but the difficulty in doing so is evident from Table 6. The vast majority, with some exceptions, focus on one nonlinearity, while none have included three (i.e. $n_2$, Raman, and Brillouin together in the same work). That said, and even where there is a focus on a single nonlinearity, there are still gaps that can be seen in the Table, though not following any particular trend, where information is lacking but desirable. Included in the Table are those parameters that generally are most important to be included in publications reporting NLO measurements in fibers. In other words, we consider the reporting of these parameters to be a "best practice," in addition to the more general best practices described in Section 2. To further complicate things, additional fiber-specific measurement complexities can arise from such characteristics as the number of propagating modes (not just a V-number alone), fiber splicing efficiency, fiber uniformity, attenuation at the operating wavelength(s), mode field diameter, mode index, dispersion, etc. that can have a strong impact on the strength of a nonlinearity and deserve discussion where relevant. For completeness, it is ideal to also include the compositional profile across the core since each chemical constituent influences a given nonlinearity in differing ways; see, for example, Table 1 in [Cavillon2018a], which qualifies changes in Raman, Brillouin, and $n_2$-mediated wave-mixing for a wide variety of glass components often used in silica-based optical fibers. This further epitomizes the need for all relevant data to be provided, through which a reader can at least formulate their own estimates of the NLO coefficients, or at least in the case of hybrid systems, an effective value for the nonlinearity (see Section 3.7). Such data is of critical importance, as it is key to the design of next-generation fibers.

### 3.5.3 Data table for fiber waveguiding materials

**Table 6.** NLO properties of fiber waveguiding materials from representative works since 2000. Legend for superscripts: see below the table.

| Material | Method | Fiber Properties | | Measurement Details | | Nonlinear Properties | | | Reference |
|---|---|---|---|---|---|---|---|---|---|
| | | Length Loss | Core index contrast Cladding material Core Size Effective area | Wavelength Peak power | Rep. rate Pulse width | $n_{2,eff}$ * | $g_{Raman}$ Frequency shift | $g_{Brillouin}$ Frequency shift Linewidth | |
| **Commercial telecommunications fiber** | | | | | | | | | |
| Pure $SiO_2$ | IGA / SPM / SRS Threshold | 20 m $\sim 0$ dB/m | - $SiO_2$ - 50 $\mu m^2$ | 1064 nm $5-25$ W | 100 MHz $50-70$ ps | $1.81 \times 10^{-20}$ $m^2$/W | $0.78 \times 10^{-13}$ m/W | - - - | Oguama 2005b |
| SMF-28 | SpBS Linewidth | 500 m 0.022 dB/m | - $SiO_2$ 8.2 $\mu m$ 101 $\mu m^2$ | 2000 nm - | CW | $2.44 \times 10^{-20}$ $m^2$/W | - | $2.525 \times 10^{-11}$ m/W 8.40 GHz 15 MHz | Deroh 2020 |
| SMF-28 | SpBS Linewidth | 500 m 0.0002 dB/m | - $SiO_2$ 8.2 $\mu m$ 78 $\mu m^2$ | 1550 nm - | CW | - | - | $2.262 \times 10^{-11}$ m/W 10.85 GHz 28 MHz | Deroh 2020 |
| DCF | IGA / SPM / SRS Threshold | 20 m $\sim 0$ dB/m | - $SiO_2$ - 10.61 $\mu m^2$ | 1064 nm $5-25$ W | 100 MHz $50-70$ ps | $2.67 \times 10^{-20}$ $m^2$/W | $1.47 \times 10^{-13}$ m/W | - - - | Oguama 2005b |
| EDFAs | IGA / SPM | 20 m $\sim 0$ dB/m | - $SiO_2$ - $8.9-27.2$ $\mu m^2$ | 1064 nm $5-25$ W | 100 MHz $50-70$ ps | 1.82 to 3.04 $\times$ $10^{-20}$ $m^2$/W | - - | - - - | Oguama 2005 |
| $GeO_2$ doped $SiO_2$ | SpBS Reference Fiber** | 20 m 0.082 dB/m | $\leq 5 \times 10^{-3}$ $SiO_2$ 40 $\mu m$ | 1534 nm - | CW | - | - | $0.53 \times 10^{-11}$ m/W 10.71 GHz 80 MHz | Evert 2012 |
| **Silicate glass fibers** | | | | | | | | | |
| $Y_2O_3$-$Al_2O_3$-$SiO_2$ | FWM / SpBS Reference Fiber | 4 m 0.78, 0.47 dB/m | 47.3, 32.9 $\times 10^{-3}$ $SiO_2$ 11.2, 20.7 $\mu m$ - | 1542 nm 850 W | CW | 1.8 to 2.0 $\times$ $10^{-20}$ $m^2$/W | - | 0.125 to 0.139 $\times 10^{-11}$ m/W $\sim 11.40$ to 12.5 GHz 200 to 500 MHz | Tuggle 2017 |
| $La_2O_3$-$Al_2O_3$-$SiO_2$ | SpBS Reference Fiber | ~ 2 m<br>1.0 dB/m | $100 \times 10^{-3}$<br>$SiO_2$<br>~ 16 µm<br>5.8 µm MFD | 1534 nm<br>- | CW | -<br>- | -<br>- | $0.26 \times 10^{-11}$ m/W<br>11.476 GHz<br>82.0 MHz | Dragic 2014 |
| $Bi_2O_3$-$SiO_2$ | Nonlinear phase shift / SBS Pump-Probe | 1 m<br>0.8 dB/m | $SiO_2$<br><br>3.08 µm² | 1550 nm<br>- | CW | $8.17 \times 10^{-19}$ m²/W | -<br>- | $6.43 \times 10^{-11}$ m/W<br><br>- | Lee 2005 |

**Other oxide glass fibers**

| | | | | | | | | | |
|---|---|---|---|---|---|---|---|---|---|
| $P_2O_5$-based | SBS Threshold | 1.245 m<br>5.47 dB/m | 0.144 NA<br>Phosphate<br>27.7 µm<br>- | 1064 nm<br>≤ 60 W | -<br>1 µs | -<br>- | -<br>- | $2.1 \times 10^{-11}$ m/W<br>27.7 GHz<br>219 MHz | Lee 2007 |
| Oxyfluoride | SpBS Reference Fiber, SpRS, FWM | 800 m<br>0.65 dB/m | $35 \times 10^{-3}$<br>$SiO_2$<br>18.6 µm<br>- | 1540 nm<br>430 W | -<br>- | $3 \times 10^{-20}$ m²/W | $0.48 \times 10^{-13}$ m/W<br>- | $0.56 \times 10^{-11}$ m/W<br>10.75 GHz<br>52 MHz | Cavillon 2018 |

**Heavy metal oxides glass fibers**

| | | | | | | | | | |
|---|---|---|---|---|---|---|---|---|---|
| $GeO_2$-$SiO_2$ | SpBS Linewidth | 3 m<br>0.1 dB/m | -<br>$SiO_2$<br>2 µm<br>5 µm² | 2000 nm<br>- | CW | $4.97 \times 10^{-20}$ m²/W | -<br>- | $7.45 \times 10^{-12}$ m/W<br>6.00 GHz<br>76 MHz | Deroh 2020 |
| $GeO_2$-$SiO_2$ | SpBS Linewidth | 3 m<br>0.2 dB/m | $SiO_2$<br>2 µm<br>3.5 µm² | 1550 nm<br>- | CW | -<br>- | -<br>- | $1.05 \times 10^{-11}$ m/W<br>7.70 GHz<br>98 MHz | Deroh 2020 |
| TZN (TeO₂ based) | SpBS Linewidth | 2 m<br>0.5 dB/m | Tellurite<br>4 µm<br>10 µm² | 2000 nm<br>- | CW | $38 \times 10^{-20}$ m²/W | -<br>- | $13 \times 10^{-11}$ m/W<br>6.17 GHz<br>15 MHz | Deroh 2020 |
| TZN (TeO₂ based) | SpBS Linewidth | 2 m<br>0.5 dB/m | Tellurite<br>4 µm<br>8 µm² | 1550 nm<br>- | CW | | -<br>- | $16 \times 10^{-11}$ m/W<br>7.97 GHz<br>21 MHz | Deroh 2020 |

**Chalcogenide glass fibers**

| | | | | | | | | | |
|---|---|---|---|---|---|---|---|---|---|
| $As_2S_3$ | SBS Threshold | 0.1 m<br>0.57 dB/m | 0.33 NA<br>-<br>4.2 µm<br>- | 1548.4 nm<br>≤ 0.03 W | -<br>- | -<br>- | -<br>- | ~ $3.9 \times 10^{-9}$ m/W<br><br>- | Florea 2006 |
| As$_2$S$_3$ | SpBS Linewidth | 2 m<br>0.2 dB/m | 0.26 NA<br>As-S glass<br>6.1 µm<br>26 µm² | 2000 nm<br>- | CW | -<br>- | -<br>- | 1.17 × 10⁻⁹ m/W<br>6.21 GHz<br>25 MHz | Deroh 2020 |
| As$_2$S$_3$ | SpBS Linewidth | 2 m<br>0.2 dB/m | 0.26 NA<br>As-S glass<br>6.1 µm<br>20 µm² | 1550 nm<br>- | CW | -<br>- | -<br>- | 1.54 × 10⁻⁹ m/W<br>7.96 GHz<br>33 MHz | Deroh 2020 |
| As$_2$Se$_3$ | SBS Threshold | 0.1 m<br>0.9 dB/m | 0.14 NA<br>-<br>6.5 µm | 1548.4 nm<br>≤ 0.133 W | - | -<br>- | -<br>- | ~ 6.75 × 10⁻⁹ m/W<br>-<br>- | Florea 2006 |
| As$_2$Se$_3$ | SPM, XPM | 0.25 m<br>1 dB/m | Single mode,<br>As-Se glass<br>-<br>21 µm² | 1560, 1503, 1470 nm<br>73 W | 100 MHz<br>15 ps | 1.10 to 0.75 × 10⁻¹⁷ m²/W | 2 to 3 × 10⁻¹¹ m/W<br>~ 7 THz | -<br>- | Tuniz 2008 |
| Ge-Sb-S | SpBS, SRS Pump-Probe | 1.5 m<br>5.5 dB/m | -<br>PCF<br>50 µm | 1553 nm<br>≤ 25 W | -<br>10 ns | - | 1.8 × 10⁻¹¹ m/W<br>9.7 THz | 8 × 10⁻¹⁰ m/W<br>8.2 GHz | Fortier 2008 |

**Fluoride glass fibers**

| | | | | | | | | | |
|---|---|---|---|---|---|---|---|---|---|
| Fluoride | SBS Threshold | 29 m<br>0.02 dB/m | 0.23 NA<br>-<br>6.5 µm | 1940 nm<br>≤ 7 W | CW | - | 3.25 to 3.52 × 10⁻¹⁴ m/W<br>17.35 THz | -<br>-<br>- | Fortin 2011 |
| ZBLAN | SBS Pump-Probe | 10.4 m<br>0.25 dB/m | 0.17 NA<br>Fluoride glass<br>- | 1550 nm<br>≤ 5 W | CW | - | -<br>- | 4±3 × 10⁻¹² m/W<br>7.76 GHz<br>< 38 MHz | Lambin-Iezzi 2013 |
| ZBLAN | SpBS Linewidth | 5 m<br>0.25 dB/m | -<br>Fluoride glass<br>9 µm<br>66 µm² | 2000 nm<br>- | CW | 2.93 × 10⁻²⁰ m²/W | - | 5.28 × 10⁻¹² m/W<br>6.00 GHz<br>35 MHz | Deroh 2020 |
| ZBLAN | SpBS Linewidth | 5 m<br>0.125 dB/m | -<br>Fluoride glass<br>9 µm<br>55 µm² | 1550 nm<br>- | CW | - | -<br>- | 4.95 × 10⁻¹² m/W<br>7.75 GHz<br>59 MHz | Deroh 2020 |

**Representative specialty core phase fibers**

| | | | | | | | | | |
|---|---|---|---|---|---|---|---|---|---|
| Perfluoro polymer | SBS Threshold | 100 m<br>0.15 dB/m | 0.185 NA<br>Polymer<br>120 µm<br>- | 1550 nm<br>≤ 1 W | CW | - | -<br>- | 3.09 × 10⁻¹¹ m/W<br>2.83 GHz<br>105 MHz | Mizuno 2010 |
| Sapphire | SCG | 0.001 m *** | - Unclad 60 µm - | 784 nm - | 1 kHz 150 fs | $3 \times 10^{-20}$ $m^2/W$ | - - | - - - | Yin 2008 |
| Sapphire | SCG | 0.035 m *** | - Unclad 115 µm - | 2000 nm $3 - 16$ MW | - 150 fs | $2.8 \times 10^{-20}$ $m^2/W$ | - - | - - - | Kim 2008 |
| Silicon | SPM | 0.008 m 0.2 to 3 dB/cm | 3.9 NA $SiO_2$ Depends on taper | 1540 nm $0.04 - 1.2$ kW | - 700 fs | $3 \times 10^{-18}$ $m^2/W$ | - - | - - - | Ren 2019 |
| Silicon | SPM | 0.01 m 2 dB/cm | 3.9 NA $SiO_2$ Depends on taper | 2400 nm - | - 100 fs | $1 \times 10^{-17}$ $m^2/W$ | - - | - - - | Shen 2020 |
| Poled germanosilicate | SHG | 0.32 m $\sim 0$ dB/m | 0.11 NA $SiO_2$ 6 µm | 1541 nm 200 W | 3 MHz - | $\chi^{(2)}_{eff} = 0.054$ pm/V, 15.2% conversion efficiency | | | Canagasabey 2009 |

* Sometimes the value of $n_{2,eff}$ should be multiplied by 9/8 when comparing to bulk measurements to account for the effects of polarization randomization that occurs for fibers longer than several meters [Buckland1996] (see Section 1.1 on fused silica).

** The Brillouin gain coefficient was determined by comparing the strength of Brillouin scattering to a reference fiber of known $g_{Brillouin}$.

*** Not noted in stated reference but attenuation values of 0.3 dB/m at a wavelength of 2.94 µm have been reported [Harrington2014].

## 3.6 On-chip waveguiding materials: data table and discussion

***Team:*** *Philippe Boucaud, Ksenia Dolgaleva, Daniel Espinosa, Rich Mildren,* **Minhao Pu** *(team leader), Nathalie Vermeulen, Kresten Yvind*

### 3.6.1 Introduction

#### 3.6.1.1 On-chip waveguiding materials and their NLO applications

Integrated photonics deals with miniaturization of bulk-component optical setups including all their key components (light-emitting devices, light-steering optics and detectors) and functionalities on small chips, typically with dimensions $< 1 \times 1$ cm$^2$. The on-chip miniaturization of photonic components allows building compact photonic integrated circuits (PICs) with high complexity, small footprint and potentially low cost when produced in large volumes. Thanks to the strong advancement in manufacturing technology over the past decades, a wide variety of materials can be combined nowadays through heterogeneous integration and bonding, thus covering a very wide spectral range and optimizing the performance of both active and passive devices by selecting the appropriate material platform for each on-chip functionality. Alternatively, monolithic photonic integration where active and passive integrated optical components are built on the same platform, with some variation of the material composition depending on the device functionality, is also possible using III-V semiconductor platforms.

Nonlinear integrated photonics exploits the NLO response of on-chip components. Some examples of on-chip nonlinear photonic structures include straight passive waveguides, micro-ring resonators or other compact structures, such as photonic-crystal waveguides. Fabrication of waveguides with dimensions smaller than the wavelength of operation [Karabchevsky2020] has enabled strong light confinement and, as such, high irradiances beneficial for efficient NLO interactions. Ultra-compact waveguides with large refractive index contrasts between the core and cladding can also exhibit strong waveguide dispersion, enabling dispersion engineering useful for establishing, e.g., phase-matched second-harmonic generation or four-wave mixing. Finally, besides basic waveguide cross-sections consisting of three material layers (guiding layer or core, substrate and cladding), more advanced quantum-well waveguide geometries featuring enhanced NLO interactions have also emerged over the years [Wagner2009a, Hutchings2010, Wagner2011].

In case the waveguide core material lacks inversion symmetry, NLO interactions of the second order can take place; otherwise, the waveguide will only exhibit a third-order NLO response. These two waveguide subcategories offer quite different NLO functionalities: whereas waveguides with a second-order nonlinearity allow, e.g., on-chip frequency doubling, DFG and EO modulation (the latter is not included in the tables), those with a third-order nonlinearity enable Kerr-based supercontinuum generation, Raman amplification, all-optical switching, etc. These functionalities are finding practical use in a wide variety of applications, ranging from on-chip biosensing, spectroscopy and LIDAR to optical datacom, signal processing, and quantum computing.

#### 3.6.1.2 Background prior to 2000

The NLO properties of the majority of materials used for on-chip integration were already studied in bulk form well before 2000, and many NLO experiments in waveguide configurations were reported before the turn of the century. These early studies were primarily concerned with determining effective NLO coefficients of the materials in waveguide arrangements. Furthermore, efforts in optimizing the light confinement and tailoring various phase-matching techniques in waveguides already began before the new millennium. Nevertheless, the field has still experienced tremendous growth after 2000 facilitated by the maturing fabrication technologies for PICs [Ottaviano2016, Roland2020].

### *3.6.1.2.1 Background for on-chip waveguide materials with second-order nonlinearity*

Second-order NLO effects generally require non-centrosymmetric materials. Although it is possible to induce these effects in centrosymmetric materials by means of strain [Cazzanelli2011], by applying all-optical poling techniques [Nitiss2022] or by exploiting surface effects, the strongest second-order nonlinearities are generally found in materials that intrinsically lack inversion symmetry. Before 2000, the following non-centrosymmetric crystalline waveguide materials were used for establishing $\chi^{(2)}$ effects: III-V semiconductors, aluminum gallium arsenide (AlGaAs) and gallium arsenide (GaAs) [Anderson1971, VanDerZiel1974, VanDerZiel1976, Yoo1995, Ramos1996, Fiore1997a, Street1997a, Street1997b, Xu1997, Fiore1998, Fiore1997b, Yoo1996, Bravetti1998, Xu1998]; III-nitrides, aluminum nitride (AlN) and gallium nitride (GaN) [Blanc1995, Zhang1996]; and some II-VI semiconductors such as zinc selenide (ZnSe), zinc telluride (ZnTe), and zinc sulfide (ZnS), grown epitaxially on a GaAs substrate [Angell1994, Wagner1995, Wagner1997, Kuhnelt1998]; as well as some other materials [Sugita1999, Chui1995, Azouz1995]. Most of the III-V semiconductor optical waveguides used to demonstrate $\chi^{(2)}$ effects before the 1990s [Anderson1971, VanDerZiel1974, VanDerZiel1976] as well as all AlN, GaN [Blanc1995, Zhang1996] and II-VI waveguide demonstrations [Angell1994, Wagner1995, Wagner1997, Kuhnelt1998] were in slab waveguides with 1D confinement. As the means and tools for nanofabrication experienced further development, it became possible to routinely fabricate AlGaAs/GaAs channel or ridge waveguides with 2D confinement, exhibiting more efficient NLO interactions [Yoo1995, Ramos1996, Fiore1997a, Street1997a, Street1997b, Xu1997, Fiore1998, Fiore1997b, Yoo1996, Bravetti1998, Xu1998]. Among the specific second-order NLO effects in AlGaAs and GaAs, there were numerous experiments on SHG in the near-IR [VanDerZiel1974, VanDerZiel1976, Yoo1995, Fiore1997a, Street1997a, Street1997b, Xu1997, Fiore1998], mid-IR [Anderson1971], and visible (SHG of blue light) [Ramos1996]. There were also some DFG reports with generation of mid-IR [Fiore1997b, Bravetti1998], as well as near-IR radiation using 770-nm pump light [Yoo1996] and SFG of 780-nm red light with a near-IR pump [Xu1998]. AlN and GaN III-nitride and II-VI slab dielectric waveguides enabled the observation of SHG in the visible [Blanc1995, Zhang1996].

While crystalline birefringence is often used in bulk crystals to phase-match the widely spaced wavelengths in second-order NLO processes, other techniques are needed for materials that do not possess natural birefringence, such as GaAs and AlGaAs. Phase-matching techniques used to enhance the efficiency of second-order NLO effects include dispersion engineering of the slab waveguide structure [Anderson1971], and modal phase matching where the fundamental radiation occupied a lower-order mode and the second harmonic populated a higher-order mode for effective refractive index matching [VanDerZiel1974, Blanc1995, Wagner1995]. Most of the demonstrations of phase-matched $\chi^{(2)}$ effects employed various QPM approaches [VanDerZiel1976, Yoo1995, Fiore1997a, Street1997a, Street1997b, Zu1997, Yoo1996, Xu1998, Angell1994, Wagner1997, Kuhnelt1998, Azouz1995, Sugita1999]. Specifically, many reports explored periodic domain inversion (PDI) where the sign of $\chi^{(2)}$ was periodically modulated by crystalline domain reorientation achieved through wafer bonding and subsequent epitaxial (MOCVD or MBE) regrowth of the structure with the domain orientation following that of the template [Yoo1995, Xu1997, Yoo1996, Xu1998, Angell1994]. This approach featured high propagation losses of 50-120 dB/cm, except for one study reporting much lower propagation losses of 5.5 dB/cm at the 1460-nm fundamental and 25 dB/cm at the 730-nm second-harmonic [Yoo1995]. QPM by PDI through proton exchange followed by thermal annealing in LiTaO$_3$ [Azouz1995] and by periodic electric-field poling of MgO:LiNbO$_3$ (with a very high efficiency of 31%) [Sugita1999] were also demonstrated. QPM by a periodic modulation of the refractive index was achieved via a corrugation formed by a grating etched in one of the waveguide heterostructure interfaces [VanDerZiel1976]. Furthermore, QPM by periodic modulation of the values of $\chi^{(2)}$ (their periodic suppression) was achieved via asymmetric coupled quantum-well reorientation [Fiore1997a], quantum-well intermixing [Street1997a, Street1997b] selective wet oxidation of AlAs in AlGaAs/GaAs structures [Fiore1998, Fiore1997b, Bravetti1998], and maskless focused-ion-beam implantation in ZnTe and ZnSe waveguides [Wagner1997, Kuhnelt1998]. Thanks to the highest modulation contrast of $\chi^{(2)}$ achievable through PDI, this method has been shown to result in the highest conversion efficiencies despite the largest propagation losses. Some specific conversion efficiency figures are 1040 %/W/cm$^2$ for SHG [Xu1997] and 810 %/W/cm$^2$ for DFG [Xu1998] (PDI Al$_{0.6}$Ga$_{0.4}$As core with Al$_{0.7}$Ga$_{0.3}$As claddings, fundamental wavelengths around 1550 nm). Some other promising results include 190 %/W [Fiore1998] (AlGaAs/AlAs with form birefringence by selective wet oxidation), 450 %/W [Azouz1995] (LiTaO$_3$ formed by proton exchange), and 1500 %/W [Sugita1999] (SHG of 426 nm in MgO:LiNbO$_3$ waveguide with a periodically patterned domain structure).

### 3.6.1.2.2   *Background for on-chip waveguide materials with third-order nonlinearity*

Third-order NLO interactions demonstrated in waveguide platforms before 2000 include a wide variety of phenomena such as SPM, XPM, THG, Raman, Brillouin and FWM. The dominant number of experimental demonstrations were performed in the 1990s in III-V semiconductor optical devices based on AlGaAs [Espindola1995, Peschel1999, Millar1999, Le1990, Hamilton1996, Kang1996, Islam1992, Stegeman1994, Kang1995, Kang1998, Villeneuve1995a, Villeneuve1995b Villeneuve1995c, Le1992] and InGaAsP [Nakatsuhara1998, Day1994, DOttavi1995, Donnelly1996, Darwish1996, Tsang1991]. Waveguides of planar configuration (featuring 1D confinement) [Kang1996, Day1994, Kang1998] and channel/rib waveguides (with 2D confinement) [Espindola1995, Stegeman1994, Nakatsuhara1998, Villeneuve1995c, DOttavi1995, Hamilton1996, Donnelly1996, Islam1992, Peschel1999, Le1990, Millar1999, Villeneuve1995b] were considered. The majority of studies were performed in waveguides with 2D confinement, while 1D structures were primarily used for spatial soliton demonstrations [Kang1996, Kang1998]. The NLO phenomena studied in III-V semiconductors included SPM and XPM [Kang1996, Day1994, Hamilton1996, Kang1998, Tsang1991, Villeneuve1995b], FWM [Espindola1995, DOttavi1995, Donnelly1996, Le1990, Darwish1996, Le1992], and other demonstrations of the third-order optical nonlinearity [Islam1992, Peschel1999, Millar1999].

InGaAsP-based works feature optical devices with both resonant [Day1994] and nonresonant [Nakatsuhara1998, DOttavi1995, Donnelly1996, Darwish1996, Tsang1991] nonlinearities in the vicinity of the bandgap. In contrast, AlGaAs-based works primarily concentrated on operation at wavelengths below half of the bandgap energy, typically the telecom C-band wavelengths, where it is possible to eliminate 2PA by manipulating the material compositions of the waveguide heterostructures [Kang1996, Villeneuve1995c, Stegeman1994, Peschel1999, Kang1998, Villeneuve1995b]. InGaAsP- and AlGaAs-based quantum-well devices have been shown to exhibit strong nonlinearities due to band-filling and excitonic effects [Day1994, Donnelly1996, Darwish1996, Tsang1991, Islam1992].

Most of the studies on the third-order nonlinearities in III-V semiconductor waveguides before 2000 were focusing on measuring nonlinear coefficients such as $n_{2(eff)}$ and $\alpha_2$. The values of $n_{2(eff)}$ in AlGaAs of various material compositions at various wavelengths have been reported to fall within the range between $2.1 \times 10^{-18}$ and $3.3 \times 10^{-17}$m$^2$/W, with the most cited values near half-the-bandgap of $(1.3 - 1.5) \times 10^{-17}$ m$^2$/W in the telecom C-band [Espindola1995, Stegeman1994, Hamilton1996, Islam1992, Aitchison1997]. The 2PA coefficients $\alpha_2$ of AlGaAs below half-the-bandgap have been reported to span between $2.6 \times 10^{-13}$ and $6 \times 10^{-13}$ m/W [Espindola1995, Islam1992, Aitchison1997]. The values of $n_{2(eff)}$ (in the range between $6 \times 10^{-17}$ and $9.5 \times 10^{-16}$ m$^2$/W) and $\alpha_2$ ($3 \times 10^{-11}$ to $6 \times 10^{-10}$ m/W), typically measured in the vicinity of the bandgap, have been reported in InGaAsP waveguides [Donnelly1996, Darwish1996, Tsang1991]. The values of $\chi^{(3)}$ in AlGaAs [Le1990] and InGaAsP [Donnelly1996] waveguides have also been measured. A few studies report the conversion efficiencies for FWM processes in AlGaAs [Le1990, Le1992] and InGaAsP [Donnelly1996, Darwish1996] waveguides.

Organic polymers in waveguide configurations (not included in the data table) represent another large group of materials that has been extensively studied before 2000 [Prasad1987, Grabler1997, Rossi1991, Huang1999, Okawa1991, Chon1994, Muto1992, Hosoda1992, Driessen1998, Bartuch1997,

Malouin1996, Malouin1998, Murata1998, Marques1991, Asobe1995, Lee1993, Konig1999]. Many studies report the effective values of $\chi^{(3)}$ of a variety of polymer-based waveguides [Huang1999, Okawa1991, Chon1994, Rossi1991, Hosoda1992, Lee1993]. The typical value ranges for the nonlinear coefficients $n_2$ of these waveguide materials are between $2.7 \times 10^{-18}$ and $1.7 \times 10^{-16}$ m²/W, while $\alpha_2$ ranges between $8 \times 10^{-13}$ and $5.8 \times 10^{-11}$ m/W [Chon1994, Driessen1998, Bartuch1997, Malouin1998, Murata1998, Marques1991, Asobe1995, Malouin1996, Grabler1997]. Thanks to the wide transparency windows of the majority of the polymer waveguide platforms, most of these measurements have been performed in the visible wavelength range.

There were a few reports demonstrating third-order NLO interactions in waveguides made of other materials, namely: $Si_3N_4$ [Bertolotti1999], QPM KTP waveguides [Sundheimer1993], gelatin-gold nanoparticle composites [Bloemer1990], $SiO_2$-$TiO_2$ [Toruellas1991], and chalcogenide glasses (ChGs) [Cerqua-Richardson1998]. Notably, there was also a series of works on planar iron-doped and titanium-indiffused $LiTaO_3$ and $LiNbO_3$ waveguides where phase conjugation by anisotropic FWM and two-wave mixing has been demonstrated [Kip1994, Popov1992, Normandin1979, Kip1992, Fujimura1999, Kip1995].

The dominant driver behind studying the third-order phenomena in waveguide configurations before 2000 was the potential of developing all-optical signal processing devices for optical communication networks and optical information processing. As the fastest response time associated with most of the nonresonant third-order processes is at the sub-femtosecond time scale, high-speed switching at moderate optical powers is possible. Many proof-of-principle all-optical-switching devices such as nonlinear directional couplers and asymmetric Mach-Zender interferometers [Kang1995, Stegeman1994, Villeneuve1995a] have been demonstrated before 2000. Optical bistable devices in various waveguide platforms have also been realized [Nakatsuhara1998, Huang1999]. Yet, despite this interest in datacom and information processing applications, the NLO properties of Si waveguides remained largely unexplored before 2000.

### 3.6.1.3  Considerations for on-chip waveguides when performing NLO measurements

Determining the irradiance inside the on-chip waveguide is not straightforward and can represent the main source of error in NLO measurements. Apart from the determination of the optical beam characteristics such as the pulse width and the peak power, a careful evaluation of the coupling efficiency and the effective mode area is crucial for the irradiance determination. If the waveguide's cross-sectional dimensions change along the propagation direction, as in the case of devices with tapers, the subsequent change in the irradiance and phase matching conditions must also be accounted for.

The propagation loss must be properly evaluated. It includes the linear absorption, which depends only on the constituent materials, the scattering off the waveguide's walls and the field leakage. The determining factors in the loss characteristics are the waveguide's fabrication process, its geometry, material composition, and the quality of the substrate. Although most of the selected waveguide materials do not exhibit 2PA at the wavelength of interest in their bulk states, the large surface area of the waveguides may give rise to linear absorption from impurities and defects generating free carriers and/or excited impurity or defect states that subsequently absorb. This 2-step process mimics 2PA making it difficult to experimentally separate its effects from direct 2PA. However, this is a fluence dependent rather than an irradiance dependent process, and thus its effects are reduced in short-pulse measurements [Lin2007a, VanStryland1985, Christodoulides2010]. Using CW or long pulse high-repetition-rate sources helps to determine this contribution.

An effective NLO coefficient measured in heterostructure waveguides usually results from an irradiance-weighted average of the NLO coefficients of different layers. The contribution of the guiding layer is dominant in most cases. However, the precise effective NLO coefficient values depend on the confinement factor [Grant1996]. For this reason, the waveguide's NLO coefficients might deviate from the corresponding bulk values.

As for the measurements of the efficiency of the NLO process, the approach used to perform the calculation must be explicitly given. For example, one should state whether the power used in the calculation is the average or peak power, external or internal (coupled in), at the input or output of the device.

In multi-beam experiments, the group velocity mismatch (GVM) and phase mismatch, which depend on the material and waveguide dispersions, should also be properly quantified because for several NLO processes, such as FWM, these are some of the most crucial factors in determining the efficiency of the process. GVM is responsible for separating the probe pulse from the pump pulse in time as they propagate along the waveguide, while the phase mismatch between different wavelength components involved in the NLO process determines the coherence length of the process [Tishchenko2022]. The beauty of on-chip waveguides is that some of them enable dispersion engineering with the possibility of zero GVD at the wavelengths of interest to improve the efficiency of a NLO process [Pu2018, Dolgaleva2015, Meier2007]. Estimating the efficiency of the NLO process in a waveguide from the theoretical models that account for the dispersion, the irradiances, and NLO coefficients, and comparing it to the measured value is also a good practice [Espinosa2021a, Pu2018, Foster2007].

On-chip waveguides used in NLO are often quite short (< 1cm). For waveguide materials that are not highly nonlinear, one should test/account for possible nonlinearities in the rest of the characterization system by performing reference measurements. This is especially true for tapered fiber-coupled devices where a reference measurement without the device can be performed with the input power corrected for the insertion loss.

Besides the special considerations outlined here, also the general best practices of Section 2 should be taken into account when performing NLO measurements with on-chip waveguides.

### 3.6.1.4  Description of general table outline

Tables 7A and 7B show a representative list of, respectively, second-order and third-order NLO properties of on-chip waveguiding materials taken from the literature since 2000. The selection of the papers included in the Tables has been based on the best practices in Section 2 and the considerations outlined above. Table 7B has the entries grouped into five subcategories to facilitate the search for a specific waveguide composition: III-V semiconductors; silicon and silicon carbide, nitride, and oxide; chalcogenide glasses; diamond; tantalum oxide and titanium oxide. Within each subcategory, the entries are ordered alphabetically. Tables 7A and 7B are subdivided into "Material properties," "Measurement details" and "Nonlinear properties". Within each column the information is given in the order of the header description, and powers of 10 (e.g., $10^{\pm\alpha}$) are written as E$\pm\alpha$ for compactness. "Material properties" include the waveguide dimensions (length and cross-section), fabrication methods, propagation loss, refractive index, and a reference to the sub-figure in Figs. 6-7 showing the fabricated device. Fig. 6 shows SEM pictures of devices with second-order NLO responses, while Fig. 7 shows SEM pictures of devices with third-order NLO responses. The peak power values in the Tables are nominally incoupled powers as specified in the papers. The table entries cover various material platforms and compositions, as well as various waveguide configurations and phase-matching techniques. In some instances, the waveguide structure was rather complex, and it was difficult to fit its detailed description in the Tables. In these cases, simplified descriptions were provided with references to the more detailed descriptions in the corresponding papers. The NLO technique used in each of the papers is provided in the "Method" column. Lastly, some papers specify the dependence of the NLO parameters or conversion efficiency η on wavelength, waveguide dimension, temperature, etc. or have notes associated with their measurement/analysis. This information is listed within the "Comments" column. The works included in the Tables nominally report data obtained at room temperature, unless specified otherwise in the "Comments" column. If dispersive values for the NLO parameter were provided, the cited value represents the peak value for the material within the stated measurement range.

### 3.6.2 Discussion

The data in Tables 7A and 7B indicate a clear trend towards the development of high-confinement waveguides because of the benefit of enhanced NLO interactions for both the second- and third-order NLO processes. Novel and improved fabrication procedures have been developed to ensure high-quality materials and low surface roughness. Several new waveguide material platforms have been entering the scene, bringing about excellent NLO performance. These platforms will further stimulate the development of practical NLO integrated devices.

#### 3.6.2.1 Advancement since 2000 and remaining challenges

Over the past twenty years, there have been significant advancements in the methods of micro- and nanofabrication. Apart from the major progress in photo-lithography and electron-beam lithography, novel nanofabrication approaches and techniques have appeared, and also photonic foundries offering multi-project wafer runs at a reduced cost per chip have facilitated the field's growth. In the early 2000s, the NLO community turned its attention to silicon-on-insulator (SOI) photonics as a promising nonlinear platform because of its potentially lower cost and high compatibility with CMOS technology. Various SOI components for parametric amplification, wavelength conversion, and Raman lasing have been demonstrated in the telecom wavelength range.

The success of silicon photonics, inspired by CMOS silicon electronics, relies on the tight light confinement enabled by the combination of a high-index material (Si) surrounded by a low-index oxide cladding. Not surprisingly, this semiconductor-on-oxide approach has also been applied to other material systems, such as AlGaAs, resulting in NLO devices with improved performances. Another enabling factor in the new platform development is significant progress in bonding and substrate removal technologies, opening the route to various types of heterogeneous integration. Nonlinear photonics with suspended semiconductor membranes has consequently emerged. Below we provide more detailed insight into the progress since 2000, the remaining challenges and the applications of on-chip waveguides with second- and third-order NLO effects.

##### 3.6.2.1.1 Advancement and challenges for on-chip waveguiding materials with second-order nonlinearity

There were many improvements in lowering the propagation losses and increasing conversion efficiencies of the second-order NLO interactions in the existing AlGaAs waveguide technologies. Specifically, selectively oxidized AlGaAs/Al$_x$O$_y$ multilayer stacks with 2D-confined ridge waveguides with much lower propagation losses of 1.5 - 5 dB/cm featuring much higher conversion efficiencies ranging between 2.7 and 5 %/W were demonstrated [Savanier2011b (SHG), Savanier2011a (SFG), Ozanam2014]. Moreover, 2D-confined AlGaAs waveguides with symmetric Bragg reflectors in their top and bottom claddings have been realized [Han2009, Bijlani 2008, Han2010, Abolghasem2009], exhibiting SHG and SFG conversion efficiencies around 9 %/W [Han2009, Bijlani2008] and DFG conversion efficiencies of 5.2 ×10$^{-4}$ %/W [Han2010]. Fabrication of QPM AlGaAs waveguides by PDI [Yu2005, Ota2009, Yu2007] and by a periodic suppression of $\chi^{(2)}$ (periodic domain disorientation) achieved through quantum-well intermixing [Wagner2011, Hutchings2010, Wagner2009b] have seen significant improvements. PDI waveguides with propagation losses as low as 5 dB/cm and SHG conversion efficiencies over 40 %/W [Yu2007] have been demonstrated.

Another advancement in the established AlGaAs waveguide technologies resides in the capability to fabricate deeply etched waveguides with sub-micrometer dimensions. An example of such a waveguide has been demonstrated in [Duchesne2011] where 300-1000 nm-wide ultracompact AlGaAs waveguides with tight light confinement were used for SHG with modal phase matching. The normalized conversion efficiency was moderate, 1.4 × 10$^5$ %/(W m$^2$), due to the high propagation losses of 18 dB/cm characteristic to such devices with exposed guiding-layer sidewalls.

In addition to the existing AlGaAs waveguide technologies, there appeared new waveguide platforms where an AlGaAs or GaAs guiding layer is surrounded by an insulator such as $SiO_2$ [Stanton2020, Chang2019, Chang2018] or air [Morais 2017, Roland2020]. The propagation loss of 1.5 - 2 dB/cm in AlGaAs on $SiO_2$ is similar to that of the conventional AlGaAs strip-loaded waveguides, while conversion efficiencies as high as 40 %/W [Stanton2020] and 250 %/W [Chang2018] in the CW regime have been reported. Simple type-II phase matching between a fundamental transverse-electric (TE) mode and a high-order second-harmonic transverse-magnetic (TM) mode was implemented, combined with directional QPM in [Morais2017]. Further advancements in the reduction of the propagation losses will lead to even more efficient AlGaAs devices for second-order NLO interactions. However, there exists another factor limiting the conversion efficiencies of longer waveguide structures. The narrow phase-matching bandwidth combined with an extreme sensitivity of the phase-matched wavelength to the waveguide dimensions (a precision on the order of a few nm is required) makes phase matching over a long distance technologically difficult [Stanton2020].

There are several reports on other waveguide platforms used for phase-matched second-order NLO interactions, among which are AlN on $SiO_2$ [Bruch2018], GaP on $SiO_2$ [Wilson2020], GaN on $SiO_2$ [Xiong2011], AlGaN [Gromovyi2017], and periodically polled $LiNbO_3$ [Wang2018b]. A remarkably high SHG conversion efficiency of 17,000 %/W for the fundamental wavelength around 1550 nm has been achieved in a 2D-confined AlN ring resonator [Bruch2018]. The 2D confinement represents the most significant advancement in AlN- and GaN-based waveguides since only structures with 1D confinement (slab dielectric waveguides) were accessible before 2000. It was not until recently that the fabrication of 2D-confined channel waveguides based on these platforms became possible. AlN and GaN represent wide-bandgap semiconductors exhibiting large hardness and posing challenges in both epitaxial growth and defect-free nanofabrication. Further mitigation of the threading-dislocation density by optimizing the epitaxial growth and substrate choice [Awan2018] is expected to result in a significant improvement and more widespread use of such platforms in nonlinear photonics.

The observation of the second-order NLO interactions in waveguides is not limited to SHG or SFG. Spontaneous parametric down-conversion (SPDC) generating correlated photon pairs for applications in quantum technologies has been demonstrated in AlGaAs waveguides [Guo2017, Sarrafi2013]. OPO based on second-order processes has been demonstrated in GaAs/AlGaAs waveguides [Savanier2013], and, more recently, in AlN ring resonators [Bruch2019]. Supercontinuum generation (SCG) [Lu2019, Zheng2021, Okawachi2020] and frequency comb generation [Zhang2019] based on second-order processes have also been reported.

To illustrate the wide variety of components discussed here, Fig. 6 shows SEM images of some integrated waveguide devices used for second-order NLO interactions (see Table 7A).

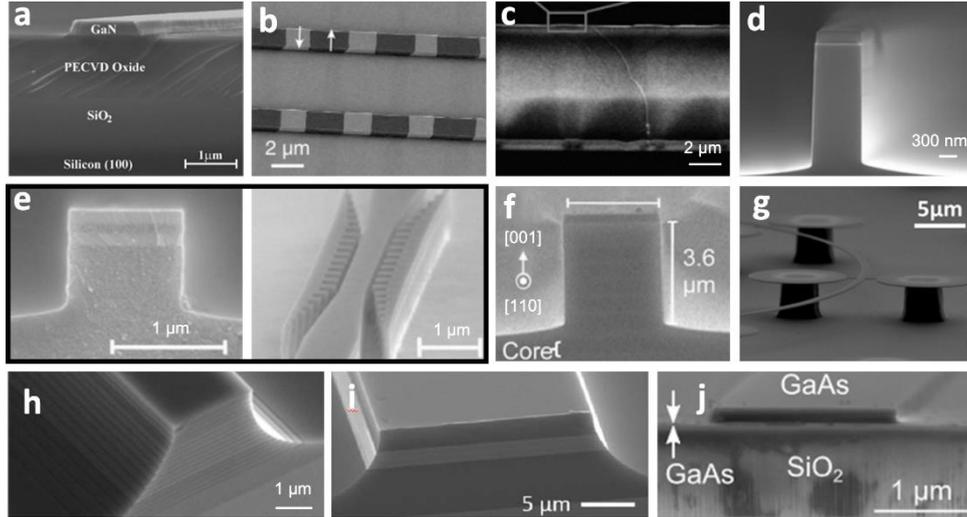

**Fig. 6**. SEM images of integrated photonic devices for second-order nonlinear interactions. The labels of the sub-figures correspond to the labels given in Table 7A. Panels a, b, c, d, e, f, g, h, i and j are reprinted with permission from [Xiong2011, Wang2018b, Yu2005, Duchesne2011, Scaccabarozzi2006, Han2009, Morais2017, Savanier2011b, Ozanam2014, Stanton2020], ©2005, 2006, 2009, 2011, 2014, 2017, 2018, 2020 Optica Publishing Group.

### 3.6.2.1.2 *Advancement and challenges for on-chip waveguiding materials with third-order nonlinearity*

There have been significant improvements in mitigating the propagation losses, optimizing the waveguide geometries, and achieving higher conversion efficiencies in the existing third-order nonlinear optical waveguide platforms, such as AlGaAs. Furthermore, there is a plethora of new works featuring novel waveguide materials that appeared after 2000 or waveguide materials that already existed but were not yet considered for NLO experiments until after 2000. These include various silicon-based platforms (SOI [Claps2002, Tsang2002, Liu2004, Rong2005, Zlatanovic2010, Turner-Foster2010, Liu2011, Kuyken2011, Liu2012, VanLaer2015, Zhang2020], $Si_3N_4$ [Levy2009, Tien2010, Kruckel2015b, Wang2018a], silicon-rich nitride [Wang2015, Kruckel2015a, Lacava2017, Ooi2017], SiC [Zheng2019b, Xing2019], amorphous silicon [Wang2012a, Wang2012b, Lacava2016, Girouard2020]), various chalcogenide glasses (ChGs) [Madden2007, Lamont2008, Gai2010], $TiO_2$ [Evans2013, Guan2018, Hammani2018], Hydex glass [Ferrera2008, Ferrera2009, Duchesne2009], diamond [Hausmann2014, Latawiec2015, Latawiec2018], and additional III-V semiconductor platforms not previously explored for nonlinear photonic devices (GaP [Wilson2020], InGaAs/AlAsSb [Tsuchida2007, Cong2008, Lim2010, Feng2013]). The overall trend indicates that higher effective nonlinearities can be obtained in high-index-contrast waveguides, often with a low-index dielectric like silica or sapphire below the waveguide core. High-index-contrast silicon-based waveguides typically feature sub-micron dimensions [Wang2015, Lacava2017, Ooi2017, Turner-Foster2010, Kuyken2011, Liu2012, VanLaer2015]. Furthermore, ultracompact AlGaAs waveguides with sub-micron dimensions, appearing after 2000, have also shown superior efficiencies of NLO interactions in comparison to their more conventional low-index-contrast counterparts [Apiratikul2014, Dolgaleva2015, Espinosa2021a]. Moreover, the SOI approach has been extended to other semiconductor platforms, enabling an increase in their refractive index contrasts and a decrease in the waveguide dimensions. Among those are AlGaAs-on-insulator (AlGaAs-OI) [Stassen2019, Pu2016, Pu2018, Kaminski2019, Hu2018, Ottaviano2016, Zheng2019a], GaP-OI [Wilson2020], and InGaP-OI [Colman2010, Dave2015a, Dave2015b]. Thanks to the strong light confinement, it became possible to push the effective nonlinear coefficient $\gamma_{eff}$ to the level of 720 $m^{-1}W^{-1}$ in NLO waveguides based on these material platforms [Stassen2019]. Further enhancement of the NLO interactions has been achieved through resonant effects in microring resonators [Absil2000, Li2021, Ramelow2019, Shi2021, Fu2020, Zheng2019a,

Zheng2019b, Ottaviano2016] and photonic-crystal structures [Jandieri2021, Monat2009 Liu2007, Zhu2006, Monat2010, Martin2017, Corcoran2009].

The effective nonlinearity is only one of the critical elements that determine the overall device performance. Another crucial parameter is propagation loss: the linear loss limits the effective nonlinear interaction length, while the nonlinear loss limits the maximum pump power that can be used for the specific NLO process. Significant efforts have been put into lowering both types of losses. As the linear loss is dominated by the scattering loss associated with the waveguide surface roughness and the refractive index contrast between the core and cladding, various approaches have been proposed to smoothen the waveguide surfaces (see, for example, [Awan2018, Liao2017]) and to passivate the waveguide surface for scattering reduction [Apiratikul2014, Wathen2014]. The nonlinear loss induced by multiphoton absorption is determined by the bandgap of the nonlinear material. The latter can be mitigated by using an appropriate material platform that lacks 2PA in the wavelength range of interest, which stimulated an effort in developing wide-bandgap NLO material platforms such as SiN, Hydex, $Ta_2O_5$ [Wu2015], SiC, AlN, and GaN. Some of the aforementioned platforms also exhibit modest index contrast compared to that associated with SOI and similar platforms. Although the effective nonlinearity that can be achieved in such low-index-contrast platforms is relatively low, their propagation losses are also extremely low. Ultra-low loss waveguides (with propagation losses < 0.5 dB/cm) can be easily realized, allowing for much longer propagation lengths and ultra-high-Q microring resonators, thereby magnifying the efficiency of the NLO interactions. In contrast, the propagation losses in most high-index-contrast waveguides are much higher (> 1 dB/cm) and do not permit long-length devices. Nevertheless, superior waveguide compactness and dispersion-management capability make them efficient NLO material platforms. Mitigating propagation losses, and improving waveguide designs, fabrication procedures, and NLO efficiencies continue to be the challenges on the path towards more practical integrated NLO devices.

Among the third-order NLO processes investigated after 2000, there are SPM, XPM, FWM, Raman effects, and Brillouin processes. If the works published before 2000 were focusing on determining the effective NLO coefficients of various waveguide structures, the trend after 2000 was partially shifted towards achieving higher NLO efficiencies with lower incident powers. The underlying idea is to push the proof-of-concept demonstrations towards practical implementations. In many emerging platforms, including AlGaAs-OI [Pu2016, Zheng2019a], SiN [Ramelow2019, Wu2021, Xue2016], Hydex [Reimer2014], SiC [Shi2021], and AlN [Jung2013, Jung2016], ultra-high-Q resonators have been realized to demonstrate FWM-based Kerr frequency comb generation, which has a wide variety of applications in spectroscopy, metrology, telecommunication, and quantum information processing. Furthermore, ultra-broadband (larger than half an octave) FWM bandwidths have been demonstrated [Pu2018, Moille2021], and supercontinuum generation (SCG) spanning more than an octave [Porcel2017, Lu2019, Zhao2015, Liu2016, Yu2019] has also been achieved in different waveguide platforms. All-optical signal processing operations and optical logic gates have been demonstrated [Pelusi2008, Willner2014, Koos2009, Yan2012, Jandier2018, Ma2016, Eggleton2012], targeting applications in optical communication networks and optical computing. An emerging application of on-chip nonlinear photonic devices that appeared after 2000 is to generate correlated photon pairs for use in quantum optics [Ma2018, Shi2019, Kultavewuti2016, Kultavewuti2019].

To illustrate the variety of structures discussed above, Fig. 7 displays SEM images of some integrated waveguide devices used for third-order NLO interactions (see Table 7B).

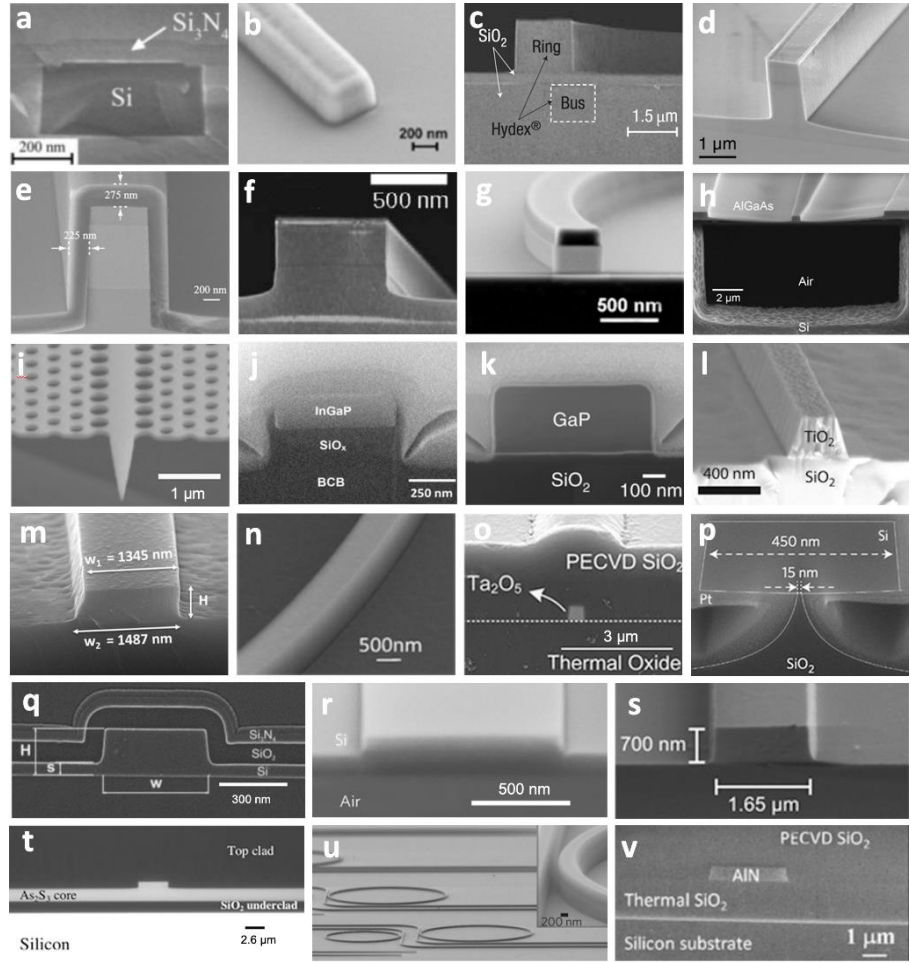

**Fig. 7**. SEM images of integrated photonic devices for third-order NLO interactions. The labels of the sub-figures correspond to the labels given in Table 7B. Panels a, b, d, e, h, i, l, n, o, q, s, t, and v are reprinted with permission from [Liu2011, Wang2012b, Wathen2014, Apiratikul2014, Chiles2019, Dave2015a, Evans2013, Zheng2019b, Wu2015, Gajda2012, Kruckel2015a, Madden2007, Jung2013], ©2007, 2011, 2012, 2013, 2014, 2015, 2019 Optica Publishing Group. Panels c, j, k, p, r, and u are reprinted with permission from [Ferrera2008, Colman2010, Wilson2020, VanLaer2015, Kittlaus2016, Hausmann2014], Springer Nature. Panel f is reprinted with permission from [Espinosa2021a], Elsevier. Panels g and m are reprinted under CC BY license from [Stassen2019] and [Hammani2018].

### 3.6.2.2  Recommendations for future works on on-chip waveguiding materials

Most papers noted in Tables 7A and 7B present the measurements of one physical effect, e.g., SPM-induced spectral broadening, Raman scattering or harmonic generation. We encourage researchers to take into consideration the category-specific recommendations outlined above as well as the best practices in Section 2. Following these tips will help eliminate possible errors encountered in the process of determining the values of NLO coefficients and conversion efficiencies.

We recommend that the authors of papers reporting on the NLO performance of waveguides specify at least those parameters that are included in Tables 7A and 7B. Furthermore, the completeness of such works could benefit from adding additional information. For example, it is instructive to separate the absorption and scattering losses, and to add the thermo-optic coefficients in resonator structures, whenever it is possible. Self-heating is an important consideration in resonator-based applications, while these parameters are rarely provided in waveguide characterization papers.

Another point of concern is nonlinear losses. The parameter $\alpha_2$ in the Tables is most often attributed to the "instantaneous" 2PA that would have been retrieved from low-duty-cycle pulsed-source experiments. The highly nonlinear semiconductor waveguides with high refractive index contrast present a special challenge due to the high field amplitudes at the surfaces where surface states may be present [Espinosa2021, Viswanath2001]. The same applies to amorphous/polycrystalline materials where surface effects may dominate the losses [Wronski1981, Grillanda2015, Girouard2020]. In all cases, a detailed description of the fabrication and surface passivation (if performed) should be provided. These surface effects can result in significant linear absorption creating free carriers and/or excited impurity or defect states which may have long lifetimes. These carriers/excited states can subsequently absorb resulting in an effective 2PA process not easily distinguished from direct 2PA [Espinosa2021b, Grillanda2015]. Such processes may be best characterized using long pulses with high repetition rates and/or CW sources, where the average power makes these nonlinear processes dominant [Grillanda2015]. In addition, if energetically allowed, direct 2PA can also lead to free-carrier generation which results in a higher-order nonlinearity [VanStryland1985, Christodoulides2010].

Although many of the NLO PICs aim in the end to use integrated lasers which have a limited output power, it would be good if the authors report damage thresholds of the waveguides. Special attention should also be paid to quantifying the actual incoupled powers inside the waveguides rather than the incident powers.

Most of the NLO measurements performed in waveguide structures yield an effective NLO coefficient, which represents an irradiance-weighted average over various waveguide constituents [Grant1996]. We recommend performing an "inverse" calculation to estimate nonlinear material parameters from the measured values. By comparing this result to the known bulk-material nonlinear coefficients, one can validate the model and determine a possible need for improvement.

### 3.6.3 Data table for on-chip waveguiding materials

**Table 7A.** Second-order NLO properties of on-chip waveguiding materials from representative works since 2000. Legend for superscripts: see below the table.

# Second-Order Nonlinearities

| Material Properties | | | | | Measurement Details | | Nonlinear Properties | |
|---|---|---|---|---|---|---|---|---|
| Core<br><br>Cladding<br><br>Substrate | Method<br><br>Phase matching | Length[1]<br>Cross-section[2]<br>Width<br>Height[3]<br>SEM image[4] | Growth/Deposition<br><br>Lithography<br><br>Other fabrication | Refractive Index<br><br>Propagation Loss<br><br>Wavelength | **Pump**<br>*Wavelength*<br>*Peak Power*<br>*Pulse Width*<br>*Polarization*<br>*Rep. Rate* | **Probe/Signal**<br>*Wavelength*<br>*Peak Power*<br>*Pulse Width*<br>*Polarization* | $\chi^{(2)}$ [m/V]<br>d [m/V]<br>$\eta$<br>Normalized $\eta$<br>Bandwidth | Reference<br><br>Additional parameters and comments |
| $Al_{0.19}Ga_{0.81}As$<br><br>Air (suspended)<br><br>GaAs [001] | SHG<br><br>Birefringent phase matching | 1.46 [mm]<br>Figs. 1 and 2<br>0.9 [µm]<br>0.123 [µm]<br>Fig. 6(g) | MBE<br><br>e-beam, photolithography<br><br>- | -<br><br>4.5, 64 (SHG) [cm⁻¹]<br><br>1593 [nm] | 1596 [nm]<br>8.0E-4 [W]<br>CW<br>TE<br>- | -<br>-<br>-<br>- | -<br>-<br>1.2 [%/W]<br>-<br>- | [Morais2017]<br><br>Periodically bent (snake-like) waveguide.<br><br>Conversion efficiency dependence on pump wavelength shown in Fig. 4(b) of the reference paper. |
| $Al_{0.19}Ga_{0.81}As$<br><br>Air (suspended)<br><br>GaAs [001] | SHG<br><br>Birefringent phase matching | 1 [mm]<br>Figs. 1, and 2<br>0.9 [µm]<br>0.123 [µm]<br>- | MBE<br><br>e-beam, photolithography<br><br>- | -<br><br>3.7, 38 (SHG) [cm⁻¹]<br><br>1593 [nm] | 1593 [nm]<br>8.0E-4 [W]<br>CW<br>TE<br>- | -<br>-<br>-<br>- | -<br>-<br>2.7 [%/W]<br>-<br>- | [Morais2017]<br><br>Straight waveguide.<br><br>Conversion efficiency dependence on pump wavelength shown in Fig. 4(b) of the reference paper. |
| $Al_{0.2}Ga_{0.8}As$<br><br>$Al_{0.7}Ga_{0.3}As$<br><br>GaAs [001] | SHG<br><br>Modal phase matching | 1.26 [mm]<br>Fig. 1<br>0.65 [µm]<br>0.5 [µm]<br>Fig. 6(d) | -<br><br>e-beam<br><br>- | -<br><br>18.3 [dB/cm]<br><br>1582 [nm] | 1582.6 [nm]<br>1.6E-4 [W]<br>CW<br>TE and TM<br>- | -<br>-<br>-<br>- | -<br>-<br>1.4E+5 [%/(W.m²)]<br>- | [Duchesne2011] |

| Material | Type | Dimensions | Growth | Loss | Pump | Signal | Efficiency | Reference |
|---|---|---|---|---|---|---|---|---|
| Al$_{0.4}$Ga$_{0.6}$As<br><br>Al$_{0.6}$Ga$_{0.4}$As and Al$_{0.2}$Ga$_{0.8}$As<br><br>GaAs | SHG<br><br>Bragg reflection waveguide | 2.12 [mm]<br>Fig. 1<br>3 [µm]<br>4 [µm]<br>- | MOCVD<br><br>photolithography<br><br>- | -<br><br>5.9 [cm⁻¹]<br><br>1555 [nm] | 1600 [nm]<br>-<br>2.0E-12 [s](FWHM)<br>TE<br>7.6E+7 [Hz] | -<br>-<br>-<br>- | -<br>8.6 [%/W][a]<br>2.1E+6 [%/(W.m²)]<br>5.3E+11 [Hz]<br>(FWHM) | [Bijlani2008]<br><br>The upper and lower reflectors are made of 8 and 10 periods, respectively. Each period: 278 nm of Al$_{0.6}$Ga$_{0.4}$As and 118 nm of Al$_{0.2}$Ga$_{0.8}$As. |
| Al$_{0.5}$Ga$_{0.5}$As/Al$_x$O$_y$ multilayer<br><br>Al$_x$O$_y$<br><br>- | SHG<br><br>Birefringent phase matching | 0.6 [mm]<br>Fig. 1<br>0.8-1 [µm]<br><br>Fig. 6(e) | MBE<br><br>e-beam<br><br>- | -<br><br>23 [dB/cm]<br><br>1550 [nm] | 1570 [nm]<br>2.3E-5 [W]<br>CW<br>TE<br>- | -<br>-<br>-<br>- | -<br>-<br>5 [%/W][a]<br>-<br>- | [Scaccabarozzi2006]<br><br>Conversion efficiency dependence on pump wavelength shown in Fig. 2(b) of the reference paper.<br><br>The reported pump power is taken at the waveguide's output.<br><br>The Al$_x$O$_y$ layers are obtained by thermal oxidation of the Al$_{0.93}$Ga$_{0.07}$As layers. |
| Al$_{0.61}$Ga$_{0.39}$As<br><br>Al$_{0.70}$Ga$_{0.30}$As and Al$_{0.25}$Ga$_{0.75}$As<br><br>GaAs [100] | DFG<br><br>Bragg reflection waveguide | 1.5 [mm]<br><br>4.4 [µm]<br>3.6 [µm]<br>- | MOCVD<br><br>photolithography<br><br>- | -<br><br>2 (TE), 2.2 (TM) [cm⁻¹]<br><br>1550 [nm] | 775 [nm]<br>2.0E-12 [s]<br>TM<br>7.6E+7 [Hz] | 1545.9 [nm]<br>2.9E-3 [W]<br>CW<br>TE | -<br>1.2E-4 [%/W][b]<br>5.4E+1 [%/(W.m²)]<br>>5E+12 [Hz] | [Han2010]<br><br>The symmetric top and bottom Bragg reflectors are made of 6 periods of Al$_{0.70}$Ga$_{0.30}$As/ Al$_{0.25}$Ga$_{0.75}$As. |
| Al$_{0.61}$Ga$_{0.39}$As<br><br>Al$_{0.70}$Ga$_{0.30}$As and Al$_{0.25}$Ga$_{0.75}$As<br><br>GaAs [100] | DFG<br><br>Bragg reflection waveguide | 1.5 [mm]<br><br>4.4 [µm]<br>3.6 [µm]<br>- | MOCVD<br><br>photolithography<br><br>- | -<br><br>2 [cm⁻¹]<br><br>1550 [nm] | 775 [nm]<br>2.0E-12 [s]<br>TE<br>7.6E+7 [Hz] | 1545.9 [nm]<br>2.9E-3 [W]<br>CW<br>TE | -<br>5.2E-4 [%/W][b]<br>2.3E+2 [%/(W.m²)]<br>>5E+12 [Hz] | [Han2010]<br><br>The symmetric top and bottom Bragg reflectors are made of 6 periods of Al$_{0.70}$Ga$_{0.30}$As/ Al$_{0.25}$Ga$_{0.75}$As. |
| Al$_{0.61}$Ga$_{0.39}$As<br><br>Al$_{0.70}$Ga$_{0.30}$As and Al$_{0.25}$Ga$_{0.75}$As<br><br>GaAs [100] | DFG<br><br>Bragg reflection waveguide | 1.5 [mm]<br><br>4.4 [µm]<br>3.6 [µm]<br>- | MOCVD<br><br>photolithography<br><br>- | -<br><br>2 (TE), 2.2 (TM) [cm⁻¹]<br><br>1550 [nm] | 775 [nm]<br>6.3E-2 [W]<br>CW<br>TM<br>- | 1545.9 [nm]<br>2.9E-3 [W]<br>CW<br>TE | -<br>1.7E-4 [%/W][b]<br>7.6E+1 [%/(W.m²)]<br>>5E+12 [Hz] | [Han2010]<br><br>The symmetric top and bottom Bragg reflectors are made of 6 periods of Al$_{0.70}$Ga$_{0.30}$As/ Al$_{0.25}$Ga$_{0.75}$As. |

| Material / Substrate | Type | Length | Fabrication | Loss / Period | Pump | Signal | Efficiency | Reference / Notes |
|---|---|---|---|---|---|---|---|---|
| $Al_{0.61}Ga_{0.39}As$<br>$Al_{0.70}Ga_{0.30}As$ and $Al_{0.25}Ga_{0.75}As$<br>GaAs [100] | DFG<br><br>Bragg reflection waveguide | 1.5 [mm]<br>-<br>4.4 [µm]<br>3.6 [µm]<br>- | MOCVD<br><br>photolithography<br>- | -<br>2 [cm$^{-1}$]<br>1550 [nm] | 775 [nm]<br>6.3E-2 [W]<br>CW<br>TE | 1545.9 [nm]<br>2.9E-3 [W]<br>CW<br>TE | -<br>1.3E-3 [%/W][b]<br>5.8E+2 [%/(W.m²)]<br>>5E+12 [Hz] | [Han2010]<br>The symmetric top and bottom Bragg reflectors are made of 6 periods of $Al_{0.70}Ga_{0.30}As$/ $Al_{0.25}Ga_{0.75}As$. |
| $Al_{0.61}Ga_{0.39}As$<br>$Al_{0.70}Ga_{0.30}As$ and $Al_{0.25}Ga_{0.75}As$<br>GaAs [100] | SFG<br><br>Bragg reflection waveguide | 2.2 [mm]<br>Fig. 1(b)<br>4.4 [µm]<br>3.6 [µm]<br>Fig. 6(f) | -<br><br>-<br>- | -<br>2 (TE), 2.2 (TM) [cm$^{-1}$]<br>1550 [nm] | 1555 [nm]<br>4.2E-3 [W]<br>CW<br>TM | 1552-1556 [nm]<br>3.5E-4 [W]<br>CW<br>TE | 4.0E-11<br>2.9E-2 [%]<br>3.0E+6 [%/(W.m²)][c]<br>>7.5E+12 [Hz]<br>(FWHM) | [Han2009]<br><br>The effective mode areas:<br>Pump: 6.7E-12 m²<br>Probe: 6.6E-12 m²<br><br>The symmetric top and bottom Bragg reflectors are made of 6 periods of $Al_{0.70}Ga_{0.30}As$/ $Al_{0.25}Ga_{0.75}As$. |
| $Al_{0.67}Ga_{0.33}As$<br>$Al_{0.70}Ga_{0.30}As$<br>Ge-on-GaAs | SHG<br><br>Orientation patterned | 8 [mm]<br>Fig. 1<br>7 [µm]<br>1.1 [µm]<br>- | MBE<br><br>e-beam<br>- | -<br>5.5, 25-45 (SHG) [dB/cm]<br>1550 [nm] | 1550 [nm]<br>3.9E-3 [W]<br>CW | -<br>-<br>- | -<br><br>4.3E+1 [%/W] | [Yu2007]<br><br>Conversion efficiency dependence on sample length shown in Fig. 4(b) of the reference paper.<br><br>Quasi-phase matching periods from 4.7E-6 to 4.9E-6 m. |
| $Al_{0.67}Ga_{0.33}As$<br>$Al_{0.7}Ga_{0.3}As$<br>GaAs [001] | SHG<br><br>Periodic domain disordering | 5 [mm]<br>Fig. 1<br>6 [µm]<br>1.1 [µm]<br>Fig. 6(c) | MBE<br><br>photolithography<br>- | -<br>6-7 [dB/cm]<br>1550 [nm] | 1559.1 [nm]<br>1.9E-3 [W]<br>CW<br>TE | -<br>-<br>- | -<br>2.3E+1 [%/W]<br>- | [Yu2005] |

| Material | Type | Dimensions | Growth | Loss / PM | Pump | Signal | Efficiency | Reference / Notes |
|---|---|---|---|---|---|---|---|---|
| AlGaAs/AlO$_x$ multilayer | SHG | 1.5 [mm] | MBE | Fig. 1 | 4420 [nm] | - | - | [Ozanam2014] |
| | | Fig. 1(a) | | | 1.6E-2 [W] | - | - | |
| GaAs | Birefringent phase | 15.3 [µm] | e-beam | 1 [cm$^{-1}$] | 3.0E-7 [s] | - | - | The multilayer is formed by $Al_{0.19}Ga_{0.81}As$, GaAs and AlO$_x$. |
| | | - | | | TE | | | |
| GaAs | matching | Fig. 6(i) | - | 4500 [nm] | - | - | 4.4E+3 [%/(W.m²)][d] | The AlO$_x$ layers are obtained by oxidation of the $Al_{0.98}Ga_{0.02}As$ layers. |
| AlGaAs/AlO$_x$ multilayer | DFG | 0.5 [mm] | MBE | - | 773.2 [nm] | 1559 [nm] | - | [Savanier2011a] |
| | | - | | | - | | - | |
| GaAs | Birefringent phase | 4 [µm] | - | 1.6 [cm$^{-1}$] | CW | CW | 9.7E+4 [%/(W.m²)][e] | The multilayer is formed by $Al_{0.2}Ga_{0.8}As$/AlO$_x$ and $Al_{0.25}Ga_{0.75}As$/AlO$_x$. |
| | | - | | | TE | TE | | |
| GaAs | matching | - | - | 1580 [nm] | - | | | The AlO$_x$ layers are obtained by selective oxidation of the $Al_{0.98}Ga_{0.02}As$ layers. |
| AlGaAs/AlO$_x$ multilayer | SFG | 0.5 [mm] | MBE | - | 1543 [nm] | 1550.4 [nm] | - | [Savanier2011a] |
| | | - | | | 1.4E-3 [W] | 1.9E-4 [W] | - | |
| GaAs | Birefringent phase | 4 [µm] | - | 1.6 [cm$^{-1}$] | CW | CW | 2.7 [%/W] | The multilayer is formed by $Al_{0.2}Ga_{0.8}As$/AlO$_x$ and $Al_{0.25}Ga_{0.75}As$/AlO$_x$. |
| | | - | | | TE | TE | 1.1E+7 [%/(W.m²)] | |
| GaAs | matching | - | - | 1580 [nm] | - | | 7.2 [nm] (at -3 dB) | The AlO$_x$ layers are obtained by selective oxidation of the $Al_{0.98}Ga_{0.02}As$ layers. |
| AlGaAs/AlO$_x$ multilayer | SHG | 0.5 [mm] | MBE | Fig. 1 | 1550 [nm] | - | - | [Savanier2011b] |
| | | Fig. 4 | | | 1.6E-2 [W] | - | - | |
| GaAs | Birefringent phase | 4 [µm] | - | 1.13 [cm$^{-1}$] | CW | - | 2.8 [%/W] | The multilayer is formed by $Al_{0.2}Ga_{0.8}As$/AlO$_x$ and $Al_{0.25}Ga_{0.75}As$/AlO$_x$. |
| | | - | | | TE | - | 1.1E+7 [%/(W.m²)] | |
| GaAs | matching | Fig. 6(h) | - | 1550 [nm] | - | | 2.9 [nm] (at -3 dB) | The AlO$_x$ layers are obtained by post-etching lateral oxidation of the $Al_{0.98}Ga_{0.02}As$ layers. |
| AlN<br><br>Sapphire and SiO₂<br><br>Sapphire c-plane | SHG<br><br>Temperature tuning | 0.188 [mm]<br>1.2-1.3 [µm]<br>1.1 [µm]<br>- | MOCVD<br>e-beam<br>- | -<br>-<br>- | 1590 [nm]<br>2.6E-4 [W]<br>CW<br>TM<br>- | -<br>-<br>-<br>- | -<br>1.7E+4 [%/W][a] | [Bruch2018]<br><br>Conversion efficiency dependence on temperature shown in Fig. 3(b) of the reference paper. |
| AlN<br><br>Sapphire and SiO₂<br><br>Sapphire c-plane | SHG<br><br>Temperature tuning | 0.188 [mm]<br>1.2-1.3 [µm]<br>1.1 [µm]<br>- | MOCVD<br>e-beam<br>- | -<br>-<br>- | 1556 [nm]<br>2.6E-4 [W]<br>CW<br>TM<br>- | -<br>-<br>-<br>- | 6.2E-12<br>1.5E+4 [%/W][a] | [Bruch2018]<br><br>Conversion efficiency dependence on temperature shown in Fig. 3(b) of the reference paper. |
| GaAs<br><br>Air and SiO₂<br><br>Silicon | SHG<br><br>Birefringent phase matching | 2.9 [mm]<br>Fig. 4(a)<br>-<br>0.158 [µm]<br>Fig. 6(j) | MBE<br>e-beam<br>wafer bonding | -<br>1.5 [dB/cm]<br>1968 [nm] | 1968 [nm]<br>-<br>CW<br>TE<br>- | -<br>-<br>-<br>- | -<br>1.8E-10<br>3.95E+3 [%/W][a]<br>1.48E+11 [Hz] | [Stanton2020]<br><br>Conversion efficiency dependence on pump wavelength and temperature shown in Fig. 7 of the reference paper.<br><br>The linear loss at the SH wavelength is one order of magnitude larger than that at the FF wavelength. |
| GaAs<br><br>SiO₂<br><br>Silicon | SHG<br><br>Modal phase matching | 1.4 [mm]<br>Fig. 1(a)<br>1.53 [µm]<br>0.15 [µm]<br>- | MOCVD<br>DUV<br>wafer bonding | -<br>2 [dB/cm]<br>2000 [nm] | 2009.8 [nm]<br>2.5E-3 [W]<br>CW<br>TE<br>- | -<br>-<br>-<br>- | -<br>2.5E+2 [%/W]<br>1.3E+8 [%/(W.m²)][d]<br>0.93 [nm] (at -3 dB) | [Chang2018]<br><br>Conversion efficiency dependence on pump wavelength shown in Fig. 4(c) of the reference paper. |
| GaAs<br><br>SiO₂<br><br>Silicon | SHG<br><br>Ring resonator phase matching | -<br>Fig. 1(a)<br>1.3 [µm]<br>0.15 [µm]<br>- | MOCVD<br>DUV<br>wafer bonding | -<br>2 [dB/cm]<br>2000 [nm] | 2000 [nm]<br>6.1E-5 [W]<br>CW<br>TE<br>- | -<br>-<br>-<br>- | -<br>4 [%], 6.5E+4 [%/W] | [Chang2019]<br><br>The efficiency reported is for a ring resonator device. |

| Material | Process | Dimensions | Fabrication | Loss | Operating (pump) | Operating (signal/idler) | Efficiency | Reference |
|---|---|---|---|---|---|---|---|---|
| GaAs/Al$_{0.85}$Ga$_{0.15}$As superlattice<br><br>Al$_{0.56}$Ga$_{0.44}$As, Al$_{0.60}$Ga$_{0.40}$As and Al$_{0.85}$Ga$_{0.15}$As<br><br>GaAs | DFG<br><br>Periodic domain disordering | 1 [mm]<br>Fig. 1<br>3 [μm]<br>0.6 [μm]<br>- | MBE<br><br>-<br>- | -<br><br>1.8 (TE), 3.7 (TM) [cm$^{-1}$]<br><br>1550 [nm] | 792.9 [nm]<br>3.1E-2 [W]<br>CW<br>TE | 1535-1555 [nm]<br>5.2E-2 [W]<br>CW<br>TE and TM | -<br>-<br>1.2E+2 [%/(W.m$^2$)][e] | [Wagner2011]<br>14:14 GaAs:Al$_{0.85}$Ga$_{0.15}$As monolayers superlattice. |
| GaAs/Al$_{0.85}$Ga$_{0.15}$As superlattice<br><br>Al$_{0.56}$Ga$_{0.44}$As, Al$_{0.60}$Ga$_{0.40}$As and Al$_{0.85}$Ga$_{0.15}$As<br><br>GaAs | DFG<br><br>Periodic domain disordering | 1 [mm]<br>Fig. 1<br>3 [μm]<br>0.6 [μm]<br>- | MBE<br><br>-<br>- | -<br><br>1.8 (TE), 3.7 (TM) [cm$^{-1}$]<br><br>1550 [nm] | 791.7 [nm]<br>2.4E-2 [W]<br>CW<br>TM<br>- | 1535-1555 [nm]<br>4.9E-2 [W]<br>CW<br>TE | -<br>-<br>7.2E+2 [%/(W.m$^2$)][e]<br>>12.5E+12 [Hz] | [Wagner2011]<br>14:14 GaAs:Al$_{0.85}$Ga$_{0.15}$As monolayers superlattice.<br><br>The conversion bandwidth is the maximum separation between the signal and idler. |
| GaN<br><br>Al$_{0.65}$Ga$_{0.35}$N<br><br>Sapphire | SHG<br><br>Modal phase matching | 4 [mm]<br>Fig. 3<br><br>1.2 [μm]<br>- | MBE, MOVPE<br><br>-<br>- | -<br><br>1 [dB/cm]<br><br>1260 [nm] | 1263 [nm]<br>9.0E+1 [W]<br>4.0E-9 [s]<br>TM<br>1.0E+3 [Hz] | -<br>-<br>- | -<br>2 [%]<br>1.5E+3 [%/(W.m$^2$)] | [Gromovyi2017] |
| GaN<br><br>SiO$_2$<br><br>Silicon | SHG<br><br>- | -<br>Figs. 2(b), 4(b)-(c)<br>0.86 [μm]<br>0.4 [μm]<br>Fig. 6(a) | MOCVD, PECVD<br><br>e-beam<br><br>wafer bonding | 2.3<br><br>-<br><br>1550 [nm] | 1550 [nm]<br>1.2E-1 [W]<br>CW<br>-<br>- | -<br>-<br>-<br>- | 1.6E-11<br><br>3.2E-3 [%] | [Xiong2011] |
| LiNbO$_3$<br><br>Air and SiO$_2$<br><br>LN-on-insulator | SHG<br><br>Periodically poled | 4 [mm]<br>Fig. 1(a)-(b)<br>1.4 [μm]<br>0.6 [μm]<br>Fig. 6(b) | -<br><br>photolithography, e-beam | -<br><br>2.5 [dB/cm]<br><br>1550 [nm] | 1550 [nm]<br>2E-3–2.2E-1 [W]<br>CW<br>TE<br>- | -<br>-<br>-<br>- | -<br>-<br>2.6E+7 [%/(W.m$^2$)][d]<br>7 [nm] (at -3 dB) | [Wang2018b]<br><br>Conversion efficiency dependence on pump wavelength shown in Fig. 3 of the reference paper. |

1: The length of ring resonators was calculated from the ring radius.

2: Figure of the reference paper showing the cross-section geometry.

3: For heterostructure waveguides, the reported height is the guiding layer thickness.

4: Selected SEM images presented in this work for each geometry and material platform.

a: Conversion efficiency formula: $\eta = P_{SH}/P_{FF}^2$  [SH – Second Harmonic, FF – Fundamental Frequency]

b: Conversion efficiency formula: $\eta = P_{DF}/(P_P P_S)$  [DF – Difference frequency, P – Pump, S – Signal]

c: Conversion efficiency formula: $\eta = P_{SF}\lambda_{SF}/(P_s\lambda_s)$ [SF – Sum frequency, P – Pump, S – Signal]

d: Normalized conversion efficiency formula: $\eta_N = P_{SH}/(P_{FF}L)^2$  [SH – Second Harmonic, FF – Fundamental Frequency, L - Length]

e: Normalized conversion efficiency formula: $\eta_N = P_{DF}/(P_P P_S L^2)$   [DF – Difference frequency, P – Pump, S – Signal, L - Length]

**Table 7B.** Third-order NLO properties of on-chip waveguiding materials from representative works since 2000. Legend for superscripts: see below the table.

| Third-Order Nonlinearities | | | | | | | |
|---|---|---|---|---|---|---|---|
| **Material Properties** | | | | **Measurement Details** | | **Nonlinear Properties** | |
| Core<br><br>Cladding    Method<br><br>Substrate | Length[1]<br>Cross section[2]<br>Width<br>Height[3]<br>SEM Image[4] | Crystallinity<br><br>Growth/Deposition<br><br>Lithography<br><br>Other fab | Refractive Index<br>Propagation Loss<br>Wavelength<br>Bandgap<br>GVD<br>Dispersion curve[5] | **Pump**<br>*Wavelength*<br>*Peak Power*<br>*Pulse Width*<br>*Polarization*<br>*Rep. Rate*<br>*Effective Area* | **Probe/Signal**<br>*Wavelength*<br>*Peak Power*<br>*Pulse Width*<br>*Polarization*<br>*Rep. Rate*<br>*Effective Area* | $n_2$<br>$\alpha_2$<br>$\gamma_{eff}$<br>$\eta$<br>Bandwidth | Reference<br><br>Additional parameters and comments |
| **III-V semiconductors** | | | | | | | |
| AlN<br><br>SiO$_2$    FWM<br><br>Silicon | 0.376 [mm]<br>Fig. 1(c)<br>3.5 [μm]<br>0.65 [μm]<br>Fig. 7(v) | -<br>Sputtering<br>-<br><br>- | -<br>-<br>1555 [nm]<br>-<br>-<br>Fig. 1(a) | -<br>5.0E-1 [W]<br>CW<br>TE<br>-<br>1.3E-12 [m$^2$] | -<br>-<br>-<br>-<br>-<br>- | 2.3E-19 [m$^2$/W]<br>-<br>-<br>-<br>- | [Jung2013] |
| Al$_{0.12}$Ga$_{0.88}$As<br><br>SiO$_2$    FWM<br><br>Sapphire | 0.81 [mm]<br>-<br>0.44 [μm]<br>0.3 [μm]<br>- | -<br>MOVPE<br>E-beam<br><br>Wafer bonding,<br>substrate removal | -<br>1.2 [dB/cm]<br>1550 [nm]<br>-<br>-<br>- | 1566 [nm]<br>3.8E-4 [W]<br>CW<br>TE<br>-<br>- | 1564 [nm]<br>-<br>CW<br>-<br>-<br>- | -<br>-<br>-<br>-19.8 [dB][b]<br>- | [Zheng2019a]<br><br>The device is a racetrack resonator, with 17-μm-radius curved waveguide and 700-μm-long straight waveguide parts. |

| | | | | | | | | |
|---|---|---|---|---|---|---|---|---|
| Al$_{0.18}$Ga$_{0.82}$As<br><br>Al$_{0.65}$Ga$_{0.35}$As and Al$_{0.35}$Ga$_{0.65}$As<br><br>GaAs [100] | FWM | 1 [mm]<br>Fig. 1<br>0.9 [µm]<br>0.8 [µm]<br>- | Monocrystalline<br><br>-<br><br>E-beam<br><br>- | -<br>2 [dB/cm]<br>1520 [nm]<br>-<br>1.0E-24 [$s^2/m$]<br>Fig. 5 | 1470-1540 [nm]<br>-<br>3.0E-12 [s](FWHM)<br>TE<br>7.7E+7 [Hz]<br>1.4E-12 [$m^2$] | 1540-1560 [nm]<br>2.8E-2 [W]<br>CW<br>TE<br>-<br>- | -<br>5.1E+1 [$m^{-1}W^{-1}$]<br>15.8 [%][c]<br>3.2E+12 [Hz]<br>(at -20dB) | [Espinosa2021a]<br><br>The length of the 900-nm-wide part is 1 mm, but the total length is 5.26 mm, which includes the 2-µm-wide couplers.<br><br>Conversion efficiency dependence on pump-probe detuning shown in Fig. 9 of the reference paper. |
| Al$_{0.18}$Ga$_{0.82}$As<br><br>Al$_{0.65}$Ga$_{0.35}$As<br><br>GaAs [100] | FWM | 2 [mm]<br>Fig. 1<br>0.6 [µm]<br>1.4 [µm]<br>- | Monocrystalline<br><br>-<br><br>E-beam<br><br>- | -<br>40 [dB/cm]<br>1520 [nm]<br>-<br>-7.5E-25 [$s^2/m$]<br>Fig. 5 | 1470-1540 [nm]<br>-<br>3.0E-12 [s](FWHM)<br>TE<br>7.7E+7 [Hz]<br>3.2E-13 [$m^2$] | 1540-1560 [nm]<br>2.8E-2 [W]<br>CW<br>TE<br>-<br>- | -<br>1.8E+2 [$m^{-1}W^{-1}$]<br>31.6 [%][c]<br>3.8E+12 [Hz]<br>(at -20dB) | [Espinosa2021a]<br><br>The length of the 600-nm-wide part is 2 mm, but the total length is 5.33 mm, which includes the 2-µm-wide couplers.<br><br>Conversion efficiency dependence on pump-probe detuning shown in Fig. 9 of the reference paper. |
| Al$_{0.18}$Ga$_{0.82}$As<br><br>Al$_{0.65}$Ga$_{0.35}$As<br><br>GaAs [100] | FWM | 1 [mm]<br>Fig. 1<br>0.8 [µm]<br>0.7 [µm]<br>Fig. 7(f) | Monocrystalline<br><br>-<br><br>E-beam<br><br>- | -<br>7 [dB/cm]<br>1520 [nm]<br>-<br>1.5E-24 [$s^2/m$]<br>Fig. 5 | 1470-1540 [nm]<br>-<br>3.0E-12 [s](FWHM)<br>TE<br>7.7E+7 [Hz]<br>7.3E-13 [$m^2$] | 1540-1560 [nm]<br>2.8E-2 [W]<br>CW<br>TE<br>-<br>- | -<br>9.0E+1 [$m^{-1}W^{-1}$]<br>12.6 [%][c] | [Espinosa2021a]<br><br>The length of the 800-nm-wide part is 1 mm, but the total length is 5.87 mm, which includes the 2-µm-wide couplers.<br><br>Conversion efficiency dependence on pump-probe detuning is shown in Fig. 9 of the reference paper. |
| Al$_{0.18}$Ga$_{0.82}$As<br><br>Al$_{0.7}$Ga$_{0.3}$As<br><br>GaAs | FWM | 25 [mm]<br>Fig. 1<br>1.2 [µm]<br>1.7 [µm]<br>- | Monocrystalline<br><br>MBE<br><br>Photolithography<br><br>- | -<br>1.9 [dB/cm]<br>1550 [nm]<br>1.64 [eV]<br>9.0E-25 [$s^2/m$]<br>- | 1550.5 [nm]<br>8.0E-1 [W]<br>CW<br>TE<br><br>7.3E-13 [$m^2$] | 1551.9 [nm]<br><br>CW<br>TE<br>-<br>- | 1.1E-17 [$m^2/W$]<br>5.0E-13 [m/W]<br>10 [%][a]<br>- | [Apiratikul2014]<br><br>Conversion efficiency dependence on pump power is shown in Fig. 4 of the reference paper. |

| | | | | | | | | | |
|---|---|---|---|---|---|---|---|---|---|
| $Al_{0.18}Ga_{0.82}As$<br>$Al_{0.7}Ga_{0.3}As$<br>GaAs | FWM | 25 [mm]<br>Fig. 1<br>1.35 [µm]<br>1.7 [µm]<br>- | Monocrystalline<br>MBE<br>Photolithography<br>Reflown resist | -<br>0.56 [dB/cm]<br>1550<br>1.64 [eV]<br>$9.0E{-}25$ [s$^2$/cm]<br>Fig. 6 | 1550.5 [nm]<br>$8.0E{-}1$ [W]<br>CW<br>TE<br>-<br>$8.2E{-}13$ [m$^2$] | 1551.9 [nm]<br>-<br>CW<br>TE<br>-<br>- | $7.0E{-}18$ [m$^2$/W]<br>$3.0E{-}13$ [m/W]<br>- | 22.9 [%][a] | [Apiratikul2014]<br>Photoresist reflow was used to achieve smoother sidewalls. |
| $Al_{0.18}Ga_{0.82}As$<br>$Al_{0.7}Ga_{0.3}As$<br>GaAs | FWM | 5 [mm]<br>Fig. 1<br>0.69 [µm]<br>1.7 [µm]<br>- | Monocrystalline<br>MBE<br>Photolithography<br>- | -<br>1550 [nm]<br>1.64 [eV]<br>$2.2E{-}25$ [s$^2$/m]<br>Fig. 6 | 1550.5 [nm]<br>$8.0E{-}1$ [W]<br>CW<br>TE<br>-<br>$4.4E{-}13$ [m$^2$] | 1551.9 [nm]<br>-<br>CW<br>TE<br>-<br>- | -<br>-<br>- | 0.16 [%][a]<br>$8.0E{+}12$ [Hz]<br>(at -3dB) | [Apiratikul2014]<br>Uncoated. |
| $Al_{0.18}Ga_{0.82}As$<br>$Al_{0.7}Ga_{0.3}As$<br>GaAs | FWM | 5 [mm]<br>Fig. 1<br>0.69 [µm]<br>1.7 [µm]<br>Fig. 7(e) | Monocrystalline<br>MBE<br>Photolithography<br>- | -<br>1550 [nm]<br>1.64 [eV]<br>$4.4E{-}25$ [s$^2$/m]<br>Fig. 6 | 1550.5 [nm]<br>$8.0E{-}1$ [W]<br>CW<br>TE<br>-<br>$4.7E{-}13$ [m$^2$] | 1551.9 [nm]<br>-<br>CW<br>TE<br>-<br>- | -<br>-<br>- | 0.25 [%][a]<br>$5.5E{+}12$ [Hz]<br>(at -3dB) | [Apiratikul2014]<br>Coated with SiN$_x$. |
| $Al_{0.20}Ga_{0.80}As$<br>$Al_{0.50}Ga_{0.50}As$ and $Al_{0.24}Ga_{0.76}As$<br>GaAs | Nonlinear transm./refl. | 10 [mm]<br>Fig. 1<br>2 [µm]<br>1.2 [µm]<br>- | Monocrystalline<br>MOCVD<br>E-beam, photolithography<br>- | -<br>3.2 [dB/cm]<br>1550 [nm]<br>-<br>$1.2E{-}24$ [s$^2$/m]<br>- | 1550 [nm]<br>$1.5E{+}2$ [W]<br>$2.0E{-}12$ [s](FWHM)<br>-<br>$3.6E{+}7$ [Hz]<br>$4.4E{-}12$ [m$^2$] | -<br>-<br>-<br>-<br>-<br>- | -<br>-<br>-<br>- | - | [Dolgaleva2011]<br>3PA coefficient = $8E{-}26$ m$^3$/W$^2$ at 1550 nm. |
| $Al_{0.21}Ga_{0.79}As$<br>$SiO_2$ and HSQ<br>SOI | FWM | 9 [mm]<br>Fig. 1<br>0.63 [µm]<br>0.29 [µm]<br>- | -<br>Epitaxy, PECVD<br>E-beam<br>Wafer bonding | -<br>1.5 [dB/cm]<br>1549.5 [nm]<br>1.69 [eV]<br>250 [ps/(nm.km)]<br>Fig. 3 | 1549.5 [nm]<br>$1.1E{-}1$ [W]<br>CW<br>TE<br>-<br>- | -<br>-<br>-<br>-<br>-<br>- | -<br>- | -12 [dB][d]<br>55 [nm]<br>(at -3 dB) | [Ros2017]<br>Conversion efficiency dependence on length, pump power and signal-pump spacing shown in Fig. 5 of the reference paper. |

| Material | Method | Dimensions | Structure / Fabrication | Loss & Wavelength | Pump | Signal | Efficiency / Bandwidth | Notes |
|---|---|---|---|---|---|---|---|---|
| $Al_{0.23}Ga_{0.77}As$<br>$Al_{0.7}Ga_{0.3}As$ and $SiN_x$<br>GaAs | FWM | 10.8 [mm]<br>Fig. 1<br>0.2-1 [µm]<br>2.1 [µm]<br>- | Monocrystalline<br>-<br>Projection lithography<br>- | -<br>5.1 [dB/cm]<br>1550 [nm]<br>-<br>- | 1538-1563 [nm]<br>1.2E-1 [W]<br>CW<br>TM<br>-<br>6.2E-13 [m²] | 1565 [nm]<br>3.2E-2 [W]<br>CW<br>TM<br>-<br>- | -<br>-<br>-<br>0.2 [%][e]<br>1.0E12 [Hz]<br>(at -3dB) | [Mahmood2014]<br><br>Conversion efficiency dependence on pump-probe detuning shown in Fig. 2 of the reference paper. |
| $Al_{0.26}Ga_{0.74}As$ PhC<br><br>Air<br><br>- | SRS Pump Probe | 1 [mm]<br>-<br>-<br>-<br>- | Monocrystalline<br>-<br>-<br>-<br>- | -<br>9.1 [cm⁻¹]<br>1606 [nm]<br>-<br>- | 1550 [nm]<br>4.4 [W]<br>5.0E-12 [s]<br>TE<br>2.0E+7 [Hz] | 1605.8 [nm]<br>2.8E-4 [W]<br>CW<br>TE<br>- | -<br>-<br>-<br>-<br>- | [Oda2008]<br><br>$g_{Raman}$: 5.1E-11 m/W.<br>Raman shift: 2.85E+4 m⁻¹.<br>Linewidth: 3.6E+2 m⁻¹.<br><br>The lattice constant, thickness, and hole diameter are 452, 260, and 120 nm, respectively. |
| $Al_{0.32}Ga_{0.68}As$<br><br>Air<br><br>Silicon | SPM/XPM spectral broadening | 4 [mm]<br>Fig. 1(a)-(b)<br>0.48 [µm]<br>0.54 [µm]<br>Fig. 7(h) | Monocrystalline<br>Epitaxy<br>E-beam<br>Wafer bonding | -<br>0.45 [dB/cm]<br>2400 [nm]<br>1.82 [eV]<br>20 [ps/(nm.km)]<br>Figs. 4(c) and 5(c) | 1560 [nm]<br>-<br>6.1E-14 [s]<br>TE<br>1.6E+8 [Hz]<br>2.2E-13 [m²] | -<br>-<br>-<br>-<br>- | -<br>-<br>-<br>1200 [nm]<br>(at -20dB) | [Chiles2019]<br><br>The reported bandwidth is for the supercontinuum generation (from 1100 nm to 2300 nm).<br><br>Supercontinuum generation bandwidth dependence on pulse energy shown in Fig. 4(a) of the reference paper. |
| $Al_{0.32}Ga_{0.68}As$<br><br>Air<br><br>Silicon | SPM/XPM spectral broadening | 2.3 [mm]<br>Fig. 1(a)-(b)<br>2.15 [µm]<br>0.54 [µm]<br>Fig. 7(h) | Monocrystalline<br>Epitaxy<br>E-beam<br>Wafer bonding | -<br>0.45 [dB/cm]<br>2400 [nm]<br>1.82 [eV]<br>20 [ps/(nm.km)]<br>Figs. 4(c) and 5(c) | 3060 [nm]<br>-<br>8.5E-14 [s]<br>TE<br>1.0E+8 [Hz]<br>1.2E-12 [m²] | -<br>-<br>-<br>-<br>- | -<br>-<br>-<br>4200 [nm]<br>(at -20dB) | [Chiles2019]<br><br>The reported bandwidth is for the supercontinuum generation (from 2300 nm to 6500 nm).<br><br>Supercontinuum generation bandwidth dependence on pulse energy shown in Fig. 5(a) of the reference paper. |

| Material | Method | Length / Figs. | Crystal | Loss / λ / Energy | λ / Power / Mode | λ / Mode | Nonlinear parameters | Reference / Notes |
|---|---|---|---|---|---|---|---|---|
| AlGaAs<br>Al$_{0.7}$Ga$_{0.3}$As<br>GaAs | FWM | 10 [mm]<br>Figs. 1 and 2<br>-<br>-<br>Fig. 7(d) | Monocrystalline<br>-<br>-<br>- | -<br>1.5 [dB/cm]<br>1550 [nm]<br>1.6 [eV]<br>1.1E-24 [s$^2$/m] | 1552.45 [nm]<br>1.0E-1 [W]<br>CW<br>TE<br>- | 1551.9 [nm]<br>-<br>CW<br>TE<br>- | 2.3E-17 [m$^2$/W]<br>3.3E-11 [m/W]<br>-<br>1.6 [%][a]<br>- | [Wathen2014]<br><br>The 2PA coefficient was measured at 1550 nm by the nonlinear transmittance technique with:<br>Pulse width: 1.26E-12 s<br>Rep. rate: 1E+7 Hz<br>Peak irradiance: <3.5E+13 W/m$^2$ |
| AlGaAs<br>Al$_{0.7}$Ga$_{0.3}$As<br>GaAs | FWM | 10 [mm]<br>Figs. 1 and 2<br>-<br>-<br>Fig. 7(d) | Monocrystalline<br>-<br>-<br>- | -<br>0.74 [dB/cm]<br>1550 [nm]<br>1.66 [eV]<br>- | 1552.45 [nm]<br>1.0E-1 [W]<br>CW<br>TE<br>- | 1551.9 [nm]<br>-<br>CW<br>TE<br>- | 1.5E-17 [m$^2$/W]<br>-<br>-<br>0.6 [%][a]<br>- | [Wathen2014]<br><br>3PA coefficient = 8.3E-26 m$^3$/W$^2$ at 1550 nm, measured by nonlinear transmittance with:<br>Pulse width: 1.26E-12 s<br>Rep. rate: 1E+7 Hz<br>Peak irradiance <8E+13 W/m$^2$ |
| AlGaAs<br>Al$_{0.7}$Ga$_{0.3}$As<br>GaAs | FWM | 25 [mm]<br>Figs. 1 and 2<br>-<br>-<br>Fig. 7(d) | Monocrystalline<br>-<br>-<br>- | -<br>0.56 [dB/cm]<br>1550 [nm]<br>1.77 [eV]<br>- | 1552.45 [nm]<br>1.0E-1 [W]<br>CW<br>TE<br>- | 1551.9 [nm]<br>-<br>CW<br>TE<br>- | 9.0E-18 [m$^2$/W]<br>-<br>-<br>22.9 [%][a]<br>1.6E+12 [Hz]<br>(at -3dB) | [Wathen2014] |
| AlGaAs<br>Al$_{0.7}$Ga$_{0.3}$As<br>GaAs | FWM | 5 [mm]<br>Figs. 1 and 2<br>-<br>-<br>Fig. 7(d) | Monocrystalline<br>-<br>-<br>- | -<br>3 [dB/cm]<br>1550 [nm]<br>1.79 [eV]<br>4.5E-25 [s$^2$/m] | 1552.45 [nm]<br>1.0E-1 [W]<br>CW<br>TE<br>- | 1551.9 [nm]<br>-<br>CW<br>TE<br>- | 8.5E-18 [m$^2$/W]<br>-<br>-<br>0.4 [%][a]<br>5.5E+12 [Hz]<br>(at -3dB) | [Wathen2014] |
| AlGaAs<br><br>SiO$_2$ and HSQ<br><br>SOI | FWM | 3 [mm]<br>Fig. 1(c)<br>0.465 [μm]<br>0.29 [μm]<br>Fig. 7(g) | -<br>MOVPE, PECVD<br>E-beam<br>Wafer bonding, substrate removal | -<br>2 [dB/cm]<br>1545 [nm]<br>-<br>- | 1543.6 [nm]<br>-<br>CW<br>TE<br>- | 1538.2 [nm]<br>-<br>CW<br>TE<br>- | -<br>7.2E+2 [m$^{-1}$W$^{-1}$]<br>-16 [dB]<br>130 [nm] (at -3dB) | [Stassen2019]<br><br>The nonlinear coefficient was measured on a straight waveguide device, while the FWM efficiency was measured on a microring. The conversion bandwidth is the same for the waveguide and microring. |
| AlGaAs<br><br>SiO$_2$ and HSQ<br><br>SOI | FWM | 3 [mm]<br>Fig. 1(b)<br>0.64 [μm]<br>0.28 [μm]<br>- | -<br>MOVPE, PECVD<br>E-beam<br>Wafer bonding, substrate removal | 3.3<br>1.3 [dB/cm]<br><br>46 [ps/(nm.km)] | 1550 [nm]<br>4.0E-1 [W]<br>CW<br>TE<br>- | 1549 [nm]<br>1.0E-4 [W]<br>CW<br>TE<br>- | -<br>6.3E+2 [m$^{-1}$W$^{-1}$]<br>-4.2 [dB]<br>750 [nm] (at -3dB) | [Pu2018] |
| AlGaAs<br><br>SiO$_2$ and HSQ<br><br>SOI | FWM | 9 [mm]<br>-<br>-<br>- | -<br>MOVPE, PECVD<br>E-beam<br>Wafer bonding, substrate removal | -<br>8 [dB/cm]<br>-<br>164 [ps/(nm.km)]<br>- | 1550 [nm]<br>1.6E-1 [W]<br>CW<br>TE<br>- | 1545 [nm]<br>-<br>CW<br>TE<br>- | -<br>3.5E+2 [m$^{-1}$W$^{-1}$]<br>-15 [dB] | [Kaminski2019]<br><br>Conversion efficiency dependence on signal wavelength shown in Fig. 4 of the reference paper.<br><br>The authors also reported a conversion efficiency of -23 dB for a 3-mm-long waveguide. |
| AlGaAs<br><br>SiO$_2$ and HSQ<br><br>SOI | SPM/XPM spectral broadening | 5 [mm]<br>Fig. 1(b)<br>0.6 [μm]<br>0.28 [μm]<br>- | -<br>Epitaxy, PECVD<br>E-beam<br>Wafer bonding, substrate removal | -<br>1.5 [dB/cm]<br>1542 [nm]<br><br>Fig. S1 | 1542 [nm]<br>5.6 [W]<br>1.5E-12 [s](FWHM)<br>TE<br>1.0E+10 [Hz]<br>- | -<br>-<br>-<br>-<br>-<br>- | -<br>-<br>66 [%]$^{f}$<br>44 [nm]<br>(at -20 dB) | [Hu2018]<br><br>The generated comb wavelength ranges from 1520 nm to 1539.8 nm, and from 1546.1 nm to 1570 nm, as informed in the supplementary material. |

| Material | Method | Dimensions | Growth | Loss / Wavelength | Parameters | Probe | Nonlinearities | Reference |
|---|---|---|---|---|---|---|---|---|
| GaAs/Al$_{0.85}$Ga$_{0.15}$As superlattice<br><br>Al$_{0.56}$Ga$_{0.44}$As and Al$_{0.60}$Ga$_{0.40}$As<br><br>GaAs | SPM/XPM spectral broadening | 5.7 [mm]<br>-<br>3 [µm]<br>1 [µm]<br>- | Monocrystalline<br>MBE<br>-<br>- | -<br>0.65 [cm$^{-1}$]<br>1550 [nm]<br>-<br>- | 1545 [nm]<br>4.0E+2 [W]<br>2E-12 [s](FWHM)<br>TE, TM<br>7.6E+7 [Hz]<br>4.5E-12 [m$^2$] | -<br>-<br>-<br>-<br>- | 7.5E-18 [m$^2$/W]<br>-<br>-<br>- | [Wagner2009b]<br><br>Nonlinearities dependence on wavelength shown in Fig. 3 of the reference paper. |
| GaAs/AlAs superlattice<br><br>Al$_{0.56}$Ga$_{0.44}$As and Al$_{0.60}$Ga$_{0.40}$As<br><br>GaAs | SPM/XPM spectral broadening | 12 [mm]<br>Fig. 1<br>3 [µm]<br>0.8 [µm]<br>- | Monocrystalline<br>MBE<br>Photolithography<br>- | -<br>0.25 [cm$^{-1}$]<br>1500 - 1600 [nm]<br>-<br>1.0E-24 [s$^2$/m]<br>- | 1545 [nm]<br>3.0E+2 [W]<br>2E-12 [s](FWHM)<br>TE<br>7.6E+7 [Hz]<br>9.7 - 11.2E-12 [m$^2$] | -<br>-<br>-<br>-<br>- | 3.2E-17 [m$^2$/W]<br>2.0E-11 [m/W]<br>-<br>- | [Wagner2007]<br><br>Nonlinearities dependence on wavelength shown in Figs. 2, 5, 8 and 10 of the reference paper. |
| GaAs/AlAs superlattice<br><br>Al$_{0.56}$Ga$_{0.44}$As and Al$_{0.60}$Ga$_{0.40}$As<br><br>GaAs | SPM/XPM spectral broadening | 12 [mm]<br>Fig. 1<br>3 [µm]<br>0.8 [µm]<br>- | Monocrystalline<br>MBE<br>Photolithography<br>- | -<br>0.7 [cm$^{-1}$]<br>1500 - 1600 [nm]<br>-<br>- | 1545 [nm]<br>3.0E+2 [W]<br>2E-12 [s](FWHM)<br>TM<br>7.6E+7 [Hz]<br>16 - 24E-12 [m$^2$] | -<br>-<br>-<br>-<br>- | 1.7E-17 [m$^2$/W]<br>1.0E-11 [m/W]<br>-<br>- | [Wagner2007]<br><br>Nonlinearities dependence on wavelength shown in Figs. 2, 5, 8 and 10 of the reference paper. |
| In$_{0.53}$Ga$_{0.47}$As/ AlAs$_{0.56}$Sb$_{0.44}$ CDQW<br><br>-<br><br>InP | SPM/XPM spectral broadening | 0.25 [mm]<br>Fig. 1<br>3 [µm]<br>-<br>- | -<br>-<br>-<br>-<br>- | -<br>-<br>-<br>-<br>- | 1560 [nm]<br>-<br>2.0E-12 [s]<br>TM<br>1.0E+10 [Hz] | 1360 [nm]<br>CW<br>TE<br>-<br>- | -<br>-<br>-<br>- | [Cong2008]<br><br>XPM efficiency = 2.0E11 [rad/J].<br><br>See the reference paper for quantum well layers thicknesses. |
| InGaAs/AlAs/AlAsSb CDQW<br><br>InAlAs and InP<br><br>InP | SPM/XPM spectral broadening | 0.24 [mm]<br>-<br>-<br>-<br>- | -<br>-<br>-<br>- | -<br>278 [cm$^{-1}$]<br>1600 [nm]<br>-<br>-<br>Fig. 2 | 1550 [nm]<br>-<br>7.0E-13 [s]<br>TM<br>8.0E+7 [Hz]<br>- | 1539.67 [nm]<br>-<br>CW<br>TE<br>- | -<br>-<br>-<br>- | [Lim2010]<br><br>XPM efficiency = 5.2E10 [rad/J].<br><br>See the reference paper for quantum well layers thicknesses.<br><br>$I_{\text{sat}}$ = 19 J/m$^2$. |

| Material | Effect | Length | Method | Parameters | Pump/Signal 1 | Pump/Signal 2 | $n_2$ | Reference / Notes |
|---|---|---|---|---|---|---|---|---|
| InGaAs/AlAs/AlAsSb CDQW<br><br>InAlAs<br><br>- | SPM/XPM spectral broadening | 0.24 [mm]<br>2 [µm]<br>-<br>- | -<br>-<br>-<br>- | -<br>-<br>-<br>- | 1550.1 [nm]<br>-<br>2.3E-12 [s]<br>TM<br>1.0E+10 [Hz]<br>- | 1559.9 [nm]<br>-<br>CW<br>TE<br>-<br>- | -<br>-<br>-<br>- | [Tsuchida2007]<br><br>XPM efficiency = 4.9E11 [rad/J].<br><br>See the reference paper for quantum well layers thicknesses. |
| InGaAs/AlAsSb CDQW<br><br>AlGaAsSb and InP<br><br>InP | SPM/XPM spectral broadening | 0.25 [mm]<br>Fig. 1(b)<br>2 [µm]<br>- | -<br><br><br>- | -<br>400 [cm$^{-1}$]<br>1560 [nm]<br>-<br>-<br>Figs. 2(d) and 2(e) | 1560 [nm]<br>-<br>2.0E-12 [s]<br>TM<br>1.0E+10 [Hz]<br>- | 1541 [nm]<br>-<br>CW<br>TE<br>-<br>- | -<br>-<br>-<br>- | [Cong2009]<br>XPM efficiency = 3.3E11 [rad/J].<br>See the reference paper for quantum well layers thicknesses.<br><br>XPM efficiency dependence on doping density shown in Fig. 2(c) of the reference paper. |
| InGaAs/AlAsSb CDQW<br><br>InP<br><br>InP | SPM/XPM spectral broadening | 1 [mm]<br>-<br>1.3 [µm]<br>- | -<br>MBE<br>E-beam<br>- | -<br>-<br>-<br>- | 1560 [nm]<br>-<br>2.4E-12 [s]<br>TM<br>1.0E+10 [Hz]<br>- | 1545 [nm]<br>-<br>CW<br>TE<br>-<br>- | -<br>-<br>-<br>- | [Feng2013]<br><br>XPM efficiency = 9.3E11 [rad/J].<br><br>The experimental setup and the waveguide fabrication details given in Feng J. et al, Optics Express 20, B279-B287 (2012). |
| GaP<br><br>SiO$_2$<br><br>Silicon | FWM | 0.314 [mm]<br>Fig. 2(d)<br>0.5 [µm]<br>0.3 [µm]<br>Fig. 7(k) | Monocrystalline<br><br>MOCVD<br><br><br>Wafer bonding | 3.1<br>1.2 [dB/cm]<br>1560 [nm]<br>2.1 [eV]<br><br>Fig. 2(e) | 1560 [nm]<br>-<br>CW<br>-<br>-<br>1.5E-13 [m$^2$] | -<br>-<br>-<br>-<br>-<br>- | 1.1E-17 [m$^2$/W]<br>-<br>2.4E+2 [m$^{-1}$W$^{-1}$]<br>-<br>- | [Wilson2020] |
| InGaP<br>SiO$_2$<br>- | SPM/XPM spectral broadening | 1.3 [mm]<br>Fig. 1(a)<br>-<br>Fig. 7(j) | -<br>-<br>-<br>- | 3.13<br>10 [dB/cm]<br>1544 [nm]<br>1.9 [eV]<br>-1.1E-21 [s$^2$/m]<br>Fig. 1(b) | 1551 [nm]<br>-<br>3.2E-12 [s](FWHM)<br>2.2E+7 [Hz]<br>- | -<br>-<br>-<br>- | -<br>-<br>9.2E+2 [m$^{-1}$W$^{-1}$]<br>-<br>- | [Colman2010]<br><br>The GVD in the table is for 1555 nm. |
| InGaP<br>SiO$_2$<br>Silicon | FWM | 2 [mm]<br>Fig. 1<br>0.63 [μm]<br>0.25 [μm]<br>Fig. 7(i) | -<br>MOCVD<br>E-beam<br>Wafer bonding | -<br>12 [dB/cm]<br>1540 [nm]<br>1.9 [eV]<br>-5.0E-25 [s$^2$/m]<br>Fig. 2 | 1552.4 [nm]<br>3.8E-2 [W]<br>CW<br>TE<br>-<br>2.4E-13 [m$^2$] | 1551.1 [nm]<br>1.5E-3 [W]<br>CW<br>TE<br>-<br>2.4E-13 [m$^2$] | -<br>4.8E+2 [m$^{-1}$W$^{-1}$]<br>0.08 [%][a]<br>- | [Dave2015a]<br><br>3PA coefficient = 2.5E-26 m$^3$/W$^2$. |
| InGaP<br>SiO$_2$<br>Silicon | SPM/XPM spectral broadening | 2 [mm]<br>Figs. 1 and 3<br>0.7 [μm]<br>0.25 [μm]<br>- | -<br>MOCVD<br>E-beam<br>Wafer bonding | -<br>12 [dB/cm]<br>1550 [nm]<br>-<br>-6.0E-25 [s$^2$/m]<br>Fig. 1 | 1550 [nm]<br>1.1E+1 [W]<br>1.7E-13 [s](FWHM)<br>TE<br>8.2E+7 [Hz]<br>2.1E-13 [m$^2$] | -<br>-<br>-<br>-<br>- | -<br>-<br>-<br>1.6E+14 [Hz]<br>(at -30 dB) | [Dave2015b]<br><br>Supercontinuum generation bandwidth dependence on waveguide width shown in Fig. 2 of the reference paper. |
| In$_{0.63}$Ga$_{0.37}$As$_{0.8}$P$_{0.2}$<br>InP<br>InP | FWM | 8 [mm]<br>Fig. 1<br>1.7 [μm]<br>0.9 [μm]<br>- | Monocrystalline<br>MOCVD<br>E-beam<br>- | 3.4<br>3 [dB/cm]<br>1568 [nm]<br>-<br>2.2E-23 [s$^2$/m]<br>- | 1568 [nm]<br>9.0 [W]<br>3.0E-12 [s]<br>TM<br>7.6E+7 [Hz]<br>1.1E-12 [m$^2$] | 1551 [nm]<br>5.0E-1 [W]<br>CW<br>TM<br>-<br>- | 1.0E-17 [m$^2$/W]<br>-<br>0.001 [%][a]<br>5.5E+12 [Hz] | [Saeidi2018]<br><br>Conversion bandwidth calculated as the maximum signal-to-idler wavelength separation. |
| In$_{0.63}$Ga$_{0.37}$As$_{0.8}$P$_{0.2}$<br><br>InP<br><br>InP | Nonlinear transm./refl. | 8 [mm]<br>Fig. 1<br>1.7 [μm]<br>0.9 [μm]<br>- | Monocrystalline<br><br>MOCVD<br><br>E-beam<br><br>- | 3.4<br>3 [dB/cm]<br>1568 [nm]<br>-<br>2.2E-23 [s$^2$/m]<br>- | 1568 [nm]<br>4.8E+1 [W]<br>3.0E-12 [s]<br>TM<br>7.6E+7 [Hz]<br>1.1E-12 [m$^2$] | -<br>-<br>-<br>-<br>- | -<br>1.9E-10 [m/W]<br>-<br>- | [Saeidi2018] |
| InGaAsP QW<br><br>InP<br><br>InP | FWM | 10 [mm]<br>-<br>-<br>-<br>- | Monocrystalline<br><br>-<br><br>-<br><br>- | -<br>-<br>0.83 [eV]<br>-<br>- | 1548 [nm]<br>2.5E-2 [W]<br>CW<br>-<br>- | 10-nm shift<br>-<br>CW<br>-<br>- | -<br>-<br>-<br>0.16 [%]<br>- | [Thoen2000]<br><br>10-nm-thick InGaAsP quantum wells with 15-nm InP barriers centered in InGaAsP guiding layer. |
| **Silicon and silicon carbide, nitride, and oxide** | | | | | | | | |
| a-Si<br><br>SiO$_2$<br><br>Silicon | FWM | 18.3 [mm]<br>-<br>0.48 [μm]<br>0.225 [μm]<br>- | 3.39<br><br>PECVD<br><br>E-beam<br><br>- | 6 [dB/cm]<br>1550 [nm]<br>-<br>- | 1602 [nm]<br>2.3E-1 [W]<br>CW<br>TE<br>-<br>- | 1-nm shift<br>-<br>CW<br>-<br>- | 1.5E-17 [m$^2$/W]<br>-<br>3.39 [%]$^a$<br>- | [Girouard2020] |
| a-Si<br><br>SiO$_2$<br><br>Silicon | FWM | 1 [mm]<br>Fig. 1<br>0.48 [μm]<br>0.22 [μm]<br>- | -<br><br>PECVD<br><br>E-beam<br><br>- | 4.7 [dB/cm]<br>1550 [nm]<br>-<br>-4.8E-25 [s$^2$/m]<br>- | 1550 [nm]<br>7.0E-2 [W]<br>CW<br>TE<br>-<br>- | 1550.5 [nm]<br>-<br>CW<br>TE<br>-<br>- | -<br>8.0E+2 [m$^{-1}$W$^{-1}$]<br>0.25 [%]$^a$<br>6.0E+12 [Hz]<br>(at -3dB) | [Lacava2016] |
| a-Si<br>SiO$_2$<br>Silicon | FWM | 8 [mm]<br>Fig. 1<br>0.5 [μm]<br>0.205 [μm]<br>- | -<br>-<br>PECVD<br>E-beam<br>- | -<br>7.2 [dB/cm]<br>1550 [nm]<br>-<br>-1.6E-25 [s$^2$/m]<br>- | 1540 [nm]<br>3.0E-4 [W]<br>CW<br>TE<br>-<br>- | -<br>-<br>CW<br>TE<br>-<br>- | 7.4E-17 [m$^2$/W]<br>-<br>0.1 [%][a]<br>2.0E+13 [Hz]<br>(at -3dB) | [Wang2012a]<br>Conversion efficiency dependence on converted wavelength shown in Fig. 2 of the reference paper. |
| a-Si<br>SiO$_2$<br>Silicon | FWM | -<br>0.5 [μm]<br>0.205 [μm]<br>- | -<br>PECVD<br>E-beam<br>- | -<br>7.2 [dB/cm]<br>1550 [nm]<br>-<br>- | 1560 [nm]<br>1.3E-2 [s]<br>2.8E-12 [s]<br>TE<br>1.0E+10 [Hz] | 1540 [nm]<br>-<br>2.8E-12 [s]<br>-<br>1.0E+10 [Hz]<br>- | 7.4E-17 [m$^2$/W]<br>-<br>5 [%][a]<br>2.0E+13 [Hz]<br>(at -3dB) | [Wang2012a] |
| a-Si<br>SiO$_2$<br>Silicon | FWM | 6 [mm]<br>Fig. 1<br>0.5 [μm]<br>0.198 [μm]<br>Fig. 7(b) | -<br>PECVD<br>E-beam<br>- | -<br>3.5 [dB/cm]<br>1550 [nm]<br>-<br>-1.3E-26 [s$^2$/m]<br>Fig. 1 | 1550 [nm]<br>6.3E-2 [W]<br>1.9E-12 [s]<br>TE<br>1.0E+10 [Hz]<br>- | 1560 [nm]<br>8.0E-4 [W]<br>2.1E-12 [s]<br>TE<br>1.0E+10 [Hz]<br>- | -<br>-<br>3.0E+3 [m$^{-1}$W$^{-1}$]<br>5 [%][a]<br>- | [Wang2012b] |
| Si and PIN<br>SiO$_2$ and Si$_3$N$_4$<br>Silicon | FWM | 40 [mm]<br>-<br>0.5 [μm]<br>0.22 [μm]<br>Fig. 7(q) | -<br>-<br>Photolithography<br>- | -<br>2 [dB/cm]<br>1552.5 [nm]<br>-<br>-1.3E-24 [s$^2$/m]<br>- | 1552.5 [nm]<br>4.0E-1 [W]<br>CW<br>TE<br>-<br>- | -<br>-<br>CW<br>TE<br>-<br>- | 9.0E-12 [m/W]<br>2.0E+2 [m$^{-1}$W$^{-1}$]<br>79.4 [%][a]<br>- | [Gajda2012] |

| Material | Method | Dimensions | Crystalline / Fabrication | Loss | Parameters | | Nonlinear coeff. | Reference / Notes |
|---|---|---|---|---|---|---|---|---|
| Si<br>-<br>- | Nonlinear transm./refl. | 24 [mm]<br>-<br>-<br>-<br>- | Monocrystalline<br>-<br>-<br>- | -<br>0.46 [cm$^{-1}$]<br>1560 [nm]<br>-<br>- | 1560 [nm]<br>4.0E+2 [W]<br>9.0E-13 [s]<br>-<br>2.5E+7 [Hz]<br>8.1E-12 [m$^2$] | -<br>-<br>-<br>-<br>- | -<br>4.4E-12 [m/W]<br>-<br>- | [Claps2003] |
| Si<br>Air (suspended)<br>Silicon | SBS Pump Probe | 2.9 [mm]<br>Fig. 1(a)-(g)<br>1 [μm]<br>0.08 [μm]<br>Fig. 7(r) | Monocrystalline<br>-<br>E-beam<br>- | -<br>0.18 [dB/cm]<br>1550 [nm]<br>-<br>- | 1550 [nm]<br>6.2E-2 [W]<br>CW<br>TE<br>- | -<br>-<br>-<br>-<br>- | -<br>-<br>-<br>- | [Kittlaus2016]<br><br>Forward Brillouin amplification.<br><br>$\gamma_{Brillouin}$ = 1152 m$^{-1}$W$^{-1}$<br>Brillouin shift = 4.35 GHz<br>Linewidth = 7 MHz<br>Free-carrier lifetime = 2.2E-9 s |
| Si<br>Air and SiO$_2$<br>Silicon | SBS Pump Probe | 2.7 [mm]<br>Fig. 1(c)<br>0.45 [μm]<br>0.23 [μm]<br>Fig. 7(p) | Monocrystalline<br>-<br>Photolithography<br>- | -<br>2.6 [dB/cm]<br>1550 [nm]<br>-<br>- | 1550 [nm]<br>2.5E-2 [W]<br>CW<br>TE<br>-<br>- | -<br>-<br>-<br>-<br>- | -<br>-<br>-<br>- | [VanLaer2015]<br><br>Forward Brillouin amplification.<br><br>$\gamma_{Brillouin}$ = 3218 m$^{-1}$W$^{-1}$<br>Brillouin shift = 9.2 GHz<br>Linewidth = 30 MHz<br>Free-carrier lifetime = 5.7E-9 s |
| Si<br>Air and SiO$_2$<br>Silicon | SBS Pump Probe | 20 [mm]<br>Fig. 1(c)<br>0.45 [μm]<br>0.23 [μm]<br>Fig. 7(p) | Monocrystalline<br>-<br>Photolithography<br>- | -<br>2.6 [dB/cm]<br>1550 [nm]<br>-<br>- | 1550 [nm]<br>1.2E-2 [W]<br>CW<br>TE<br>-<br>- | -<br>-<br>-<br>-<br>- | -<br>-<br>-<br>- | [VanLaer2015]<br><br>Backward Brillouin amplification.<br><br>$\gamma_{Brillouin}$ = 359 m$^{-1}$W$^{-1}$<br>Brillouin shift = 13.66 GHz<br>Linewidth = 15 MHz<br>Free-carrier lifetime = 5.7E-9 s |

| Material | Type | Geometry | Fabrication | Loss | Pump 1 | Pump 2 | Efficiency | Notes |
|---|---|---|---|---|---|---|---|---|
| Si<br><br>$SiO_2$ and air<br><br>Silicon | SRS Pump Probe | 48 [mm]<br>Fig. 1<br>1.52 [µm]<br>1.45 [µm]<br>- | Monocrystalline<br>-<br>Photolithography<br>- | -<br>0.22 [dB/cm]<br>1550 [nm]<br>-<br>- | 1545 [nm]<br>4.7E-1 [W]<br>1.7E-8 [s](FWHM)<br>TE<br>1.0E+4 [Hz]<br>1.6E-12 [m$^2$] | 1680 [nm]<br>2.0E-3 [W]<br>CW<br>TM<br>-<br>1.4E-12 [m$^2$] | -<br>-<br>-<br>- | [Liu2004]<br><br>$g_{Raman}$ = 1.05E-10 m/W<br>Raman shift = 5.2E+4 m$^{-1}$<br>Pump peak irradiance = 3E11 W/m$^2$<br>Free-carrier lifetime = 2.5E-8 s |
| Si<br><br>$SiO_2$ and air<br><br>Silicon | SRS Pump Probe | 48 [mm]<br>Fig. 1(a)<br>1.5 [µm]<br>1.55 [µm]<br>- | Monocrystalline<br>-<br>Photolithography<br>- | -<br>0.35 [dB/cm]<br>1550 [nm]<br>-<br>- | 1536 [nm]<br>2.0 [W]<br>1.3E-7 [s]<br>-<br>1.0E+4 [Hz]<br>1.6E-12 [m$^2$] | -<br>-<br>-<br>-<br>-<br>- | -<br>-<br>-<br>10 [%]<br>- | [Rong2005]<br><br>$g_{Raman}$ = 7.5E-11 m/W<br>Raman shift = 5.2E+4 m$^{-1}$<br>Free-carrier lifetime = 1E-8 s.<br><br>The conversion efficiency is the slope efficiency of average Raman lasing output vs average input power. |
| Si<br><br>$SiO_2$ and air<br><br>Silicon | SpRS cross-section | 24 [mm]<br>Fig. 2<br>5 [µm]<br>2.5 [µm]<br>- | Monocrystalline<br>-<br>-<br>- | -<br>2.8 [dB/cm]<br>1542 [nm]<br>-<br>- | 1427 [nm]<br>1.0 [W]<br>CW<br>TE<br>-<br>- | -<br>-<br>-<br>-<br>-<br>- | -<br>-<br>-<br>-<br>- | [Claps2002]<br><br>$g_{Raman}$ = 7.6E-10 m/W<br>Raman shift = 5.2E+4 m$^{-1}$ |
| Si<br><br>$SiO_2$ and air<br><br>Silicon | Raman Laser Threshold | 2.8 [mm]<br>Fig. 1(f)<br>2 [µm]<br>0.22 [µm]<br>- | Monocrystalline<br>-<br>Photolithography<br>- | -<br>0.51 [dB/cm]<br>1325 [nm]<br>-<br>- | 1240 [nm]<br>1.0E-3 [W]<br>CW<br>TE<br>-<br>3.7E-13 [m$^2$] | -<br>-<br>-<br>-<br>-<br>- | -<br>-<br>-<br>-<br>- | [Zhang2020]<br><br>$g_{Raman}$ = 3.66E-10 m/W<br>Raman shift = 5.2E+4 m$^{-1}$<br><br>Gain dependence on pump wavelength shown in Fig. 6(b) of the reference paper. |

| Material | Type | Dimensions | Fabrication | Loss / Dispersion | Pump | Signal | Nonlinear params | Reference |
|---|---|---|---|---|---|---|---|---|
| Si<br>SiO$_2$ and Si$_3$N$_4$<br>Silicon | Nonlinear transm./refl. | 4 [mm]<br>Fig. 1(a)-(c)<br>0.6 [µm]<br>0.22 [µm]<br>Fig. 7(a) | Monocrystalline<br>-<br>Photolithography<br>-<br>- | -<br>4.5 - 7.0 [dB/cm]<br>1800-2300 [nm]<br>-<br>- | 2200 [nm]<br>6.0 [W]<br>2.0E-12 [s](FWHM)<br>TE<br>7.6E+7 [Hz]<br>- | -<br>-<br>-<br>-<br>- | 3.0E-13 [m/W]<br>-<br>-<br>-<br>- | [Liu2011]<br><br>Nonlinearity dependence on wavelength shown in Table 1 of the reference paper. |
| Si<br>SiO$_2$<br>Silicon | FWM | 20 [mm]<br>Fig. 1(b)<br>0.9 [µm]<br>0.22 [µm]<br>- | -<br>-<br>-<br>- | -<br>2.6 [dB/cm]<br>1810-2410 [nm]<br>-5.0E-25 [s$^2$/m]<br>Fig. S4(b) | 1946 [nm]<br>3.7E+1 [W]<br>2.0E-12 [s]<br>TE<br>7.6E+7 [Hz] | 2440 [nm]<br>3.5E-5 [W]<br>CW<br>TE<br>- | -<br>2.8E+2 [m$^{-1}$W$^{-1}$]<br>8912.5 [%][e]<br>6.2E+13 [Hz] | [Liu2012]<br><br>The conversion efficiency refers to the parametric gain.<br><br>Parametric gain dependence on wavelength shown in Fig. 2 of the reference paper. |
| Si<br>SiO$_2$<br>Silicon | FWM | 20 [mm]<br>-<br>0.9 [µm]<br>0.22 [µm]<br>- | -<br>-<br>-<br>- | -<br>2.8 [dB/cm]<br>2173 [nm]<br>-6.0E-25 [s$^2$/m]<br>- | 2173 [nm]<br>1.4E+1 [W]<br>2.0E-12 [s]<br>TE<br>7.6E+7 [Hz]<br>- | 2209-2498 [nm]<br>6.0E-5 [W]<br>CW<br>TE<br>- | -<br>1.5E+2 [m$^{-1}$W$^{-1}$]<br>1000000 [%]<br>580 [nm] | [Kuyken2011]<br><br>The conversion efficiency refers to the parametric gain.<br><br>Parametric gain dependence on wavelength shown in Fig. 3 of the reference paper. |
| Si<br>SiO$_2$<br>Silicon | FWM | 3.8 [mm]<br>Fig. S1<br>1.06 [µm]<br>0.25 [µm]<br>- | -<br>-<br>E-beam<br>- | -<br>2.8 [dB/cm]<br>2025 [nm]<br>-<br>- | 2025 [nm]<br>1.8E-1 [W]<br>1.0E-9 [s]<br>TE<br>1.0E+6 [Hz]<br>3.5E-13 [m$^2$] | 1912-1994 [nm]<br>-<br>CW<br>TE<br>- | 1.1E-17 [m$^2$/W]<br><br>9.7E+1 [m$^{-1}$W$^{-1}$]<br>0.5 [%][a]<br>292 [nm] (at -3dB) | [Zlatanovic2010]<br><br>Conversion efficiency dependence on pump-signal detuning shown in Fig. 2(b) of the reference paper. |

| Material | Effect | | Fabrication | | Pump | Signal | Nonlinearity | Reference |
|---|---|---|---|---|---|---|---|---|
| Si<br><br>SiO$_2$<br><br>Silicon | FWM | 15 [mm]<br>Fig. 1<br>0.55 [µm]<br>0.3 [µm]<br>- | -<br>-<br>E-beam<br>- | -<br>-<br>1554 [nm]<br>-<br>- | 1554 [nm]<br>1.1E-1 [W]<br>CW<br>TE<br>- | 1555-2078 [nm]<br>2.5E-4 [W]<br>CW<br>TE<br>- | -<br>-<br>1.58 [%][a]<br>9.7E+13 [Hz]<br>(at -3dB) | [Turner-Foster2010]<br><br>Conversion efficiency dependence on converted wavelength shown in Fig. 2 of the reference paper. |
| Si<br><br>SiO$_2$<br><br>Silicon | Nonlinear transm./refl. | 17 [mm]<br>Fig. 1 | Monocrystalline<br>-<br>-<br>- | -<br>0.1 [dB/cm]<br>1540 [nm]<br>-<br>1.2E-24 [s$^2$/m]<br>Fig. 5 | 1540 [nm]<br>2.0E+1 [W]<br>5.4E-11 [s](FWHM)<br>-<br>-<br>6.2E-12 [m$^2$] | -<br>-<br>-<br>-<br>- | -<br>4.5E-12 [m/W]<br>-<br>-<br>- | [Tsang2002] |
| 4H-SiC<br><br>SiO$_2$<br><br>Silicon | FWM | 3 [mm]<br><br>0.51 [µm]<br>0.5 [µm]<br>Fig. 7(n) | -<br>-<br>E-beam<br>Wafer bonding | 2.4<br>-<br>1565 [nm]<br>>2.4 [eV]<br>325 [ps/(nm.km)]<br>Fig. 5(a) | 1565 [nm]<br>1.3E-1 [W]<br>CW<br>TE<br>-<br>- | -<br>1.0E-3 [W]<br>-<br>TE<br>-<br>- | 6.0E-19 [m$^2$/W]<br>7.4 [m$^{-1}$W$^{-1}$]<br>0.00032 [%][a]<br>1.6E+13 [Hz]<br>(at -3dB) | [Zheng2019b]<br><br>Nonlinearity dependence on waveguide width shown in Fig. 4 of the reference paper. |
| a-SiC<br><br>SiO$_2$<br><br>Silicon | SPM/XPM spectral broadening | 12 [mm]<br>Fig. 2(a)<br>1.2 [µm]<br>0.35 [µm]<br>- | -<br>PECVD<br>E-beam<br>- | 2.45<br>3 [dB/cm]<br>1550 [nm]<br>2.3 [eV]<br>-400 [ps/(nm.km)]<br>- | 1550 [nm]<br>1.3E+1 [W]<br>-<br>-<br>4.2E-13 [m$^2$] | -<br>-<br>-<br>- | 4.8E-18 [m$^2$/W]<br>4.0E+1 [m$^{-1}$W$^{-1}$]<br>-<br>- | [Xing2019] |

| Material | Effect | Geometry | Fabrication | Loss/Dispersion | Pump | Signal | Nonlinearity | Reference |
|---|---|---|---|---|---|---|---|---|
| $Si_3N_4$<br>$SiO_2$<br>Silicon | FWM | 1000 [mm]<br>Fig.1(a)<br>2.8 [µm]<br>0.1 [µm]<br>- | -<br>-<br>-<br>- | -<br>0.06 [dB/cm]<br>1550 [nm]<br>-<br>-5.0E-25 [s²/m] | 1563 [nm]<br>2.1 [W]<br>CW<br>TE<br>-<br>- | 1562 [nm]<br>7.7E-2 [W]<br>CW<br>TE<br>-<br>- | -<br>2.9E-1 [m⁻¹W⁻¹]<br>0.245 [%]a<br>6.2E+11 [Hz]<br>(at -3dB) | [Kruckel2015b] |
| $Si_3N_4$<br>$SiO_2$<br>Silicon | FWM | 22 [mm]<br>Fig. 2(a)<br>2 [µm]<br>0.72 [µm]<br>- | -<br>PECVD<br>Photolithography | -<br>0.58 [dB/cm]<br>1550 [nm]<br>3.27 [eV]<br>- | 1550 [nm]<br>1.6E-1 [W]<br>CW<br>TE<br>-<br>1.1E-12 [m²] | 1550.8 [nm]<br>1.1E-1 [W]<br>CW<br>TE<br>- | 6.9E-19 [m²/W]<br>-<br>2.6 [m⁻¹W⁻¹]<br>0.00012 [%]e<br>- | [Wang2018a] |
| $Si_3N_4$<br>$SiO_2$<br>Silicon | FWM | 61 [mm]<br>Fig. 2(a)<br>1.45 [µm]<br>0.725 [µm]<br>- | -<br>LPCVD, PECVD<br>E-beam | -<br>0.5 [dB/cm]<br>1550 [nm]<br>-<br>6.4E-27 [s²/m]<br>Fig. 2(a) | 1550 [nm]<br>2.4E+1 [W]<br>1.0E-11 [s]<br>TE<br>1.0E+9 [Hz]<br>- | 1470-1630 [nm]<br>-<br>CW<br>TE<br>- | 2.5E-19 [m²/W]<br>-<br>229 [%]a<br>150<br>(at -3.6dB) | [Levy2009]<br><br>The reported conversion efficiency is the parametric OPO gain achieved in microresonators.<br><br>Signal gain dependence on wavelength shown in Fig. 2(b) of the reference paper. |
| $Si_3N_4$<br>$SiO_2$<br>Silicon | SPM/XPM spectral broadening | 6000 [mm]<br>Fig. 1(a)<br>2.8 [µm]<br>0.08 [µm]<br>- | -<br>LPCVD | -<br>-<br>1550<br>-<br>- | 1549.9 [nm]<br>4.0E-1 [W]<br>CW<br>TE<br>-<br>- | 1550.3 [nm]<br>4.0E-1 [W]<br>CW<br>TE<br>- | 9.0E-20 [m²/W]<br>-<br>6.0E-2 [m⁻¹W⁻¹]<br>-<br>- | [Tien2010]<br><br>Nonlinearity dependence on core thickness shown in Fig. 2 of the reference paper. |

| Materials | Effect | Dimensions | Fabrication | Properties | Pump | Signal/Idler | Nonlinearity | Reference |
|---|---|---|---|---|---|---|---|---|
| $Si_7N_3$<br>$SiO_2$<br>Silicon | FWM | 7 [mm]<br>-<br>0.55 [μm]<br>0.3 [μm]<br>- | -<br>PECVD<br>E-beam<br>- | 3.1<br>4.5 [dB/cm]<br>1550 [nm]<br>2 [eV]<br>$-2.5E{-}25$ [$s^2/m$]<br>Fig. 2(b) | 1555 [nm]<br>2.3E-2 [W]<br>CW<br>TE<br>-<br>- | -<br>2.3E-4 [W]<br>CW<br>TE<br>-<br>- | -<br>5.0E+2 [$m^{-1}W^{-1}$]<br>0.323 [%]<br>2.2E+13 [Hz]<br>(at -3dB) | [Ooi2017]<br>Conversion efficiency dependence on pump wavelength shown in Fig. 3(b) of the reference paper. |
| $Si_7N_3$<br>$SiO_2$<br>Silicon | SPM/XPM spectral broadening | 7 [mm]<br>0.6 [μm]<br>0.3 [μm]<br>- | -<br>PECVD<br>E-beam<br>- | 3.1<br>10 [dB/cm]<br>1550 [nm]<br>2.05 [eV]<br>$-2.4E{-}25$ [$s^2/m$]<br>Fig. 2(c) | 1550 [nm]<br>1.4E+2 [W]<br>1.8E-12 [s]<br>TE<br>2.0E+7 [Hz]<br>2.1E-13 [$m^2$] | -<br>-<br>-<br>-<br>-<br>- | 2.8E-17 [$m^2/W$]<br>-<br>5.5E+2 [$m^{-1}W^{-1}$]<br>-<br>7.7E+13 [Hz]<br>(at -30 dB) | [Wang2015] |
| $Si_xN_y$<br>$SiO_2$<br>Silicon | FWM | -<br>-<br>0.7 [μm]<br>0.22 [μm]<br>- | -<br>PECVD<br>E-beam<br>- | 2.49<br>1.5 [dB/cm]<br>1550 [nm]<br>-<br>-<br>- | 1550.1 [nm]<br>3.2E-1 [W]<br>CW<br>TE<br>-<br>4.0E-13 [$m^2$] | 1550 [nm]<br>3.2E-2 [W]<br>CW<br>TE<br>-<br>- | 1.6E-18 [$m^2/W$]<br>8.0E-12 [m/W]<br>1.6E+1 [$m^{-1}W^{-1}$]<br>0.0032 [%][a]<br>- | [Lacava2017]<br>Conversion efficiency dependence on waveguide width shown in Fig. 6 of the reference paper.<br><br>The reference paper also reports the nonlinearities for a standard $Si_3N_4$ waveguide, and for a $Si_xN_y$ waveguide with a different silicon content. |
| $Si_xN_y$<br>$SiO_2$<br>Silicon | FWM | 9.4 [mm]<br>Fig. 1(b)<br>1.65 [μm]<br>0.7 [μm]<br>Fig. 7(s) | PECVD<br>E-beam<br>- | 2.1<br>1.2 [dB/cm]<br>1570 [nm]<br>2.3 [eV]<br>$-2.0E{-}26$ [$s^2/m$]<br>Fig. 3(b) | 1563 [nm]<br>1.0 [W]<br>CW<br>TE<br>-<br>9.0E-13 [$m^2$] | 1562 [nm]<br>7.9E-2 [W]<br>CW<br>TE<br>-<br>- | 1.4E-18 [$m^2/W$]<br>-<br>6.1 [$m^{-1}W^{-1}$]<br>0.016 [%][a]<br>2.0E+12 [Hz]<br>(at -3dB) | [Kruckel2015a] |
| Hydex<br><br>$SiO_2$ and air<br><br>$SiO_2$ | FWM | 0.3 [mm]<br>Fig. 1<br>1.5 [µm]<br>1.45 [µm]<br>Fig. 7(c) | Amorphous<br><br>CVD<br><br>Photolithography<br><br>- | 1.7<br>0.06 [dB/cm]<br>1550 [nm]<br>-<br>- | 1553.38 [nm]<br>5.0E-3 [W]<br>CW<br>TM<br>-<br>2.0E-12 [$m^2$] | 1558.02 [nm]<br>5.5E-4 [W]<br>CW<br>TM<br>-<br>- | 1.2E-19 [$m^2$/W]<br>-<br>2.3E-1 [$m^{-1}W^{-1}$]<br>0.0013 [%]<br>- | [Ferrera2008]<br><br>Ring radius around 48 µm, Q-factor of 65000, FSR 575 GHz. |
| Hydex<br><br>$SiO_2$ and air<br><br>$SiO_2$ | FWM | 0.85 [mm]<br>-<br>-<br>- | Amorphous<br><br>CVD<br><br>Photolithography<br><br>- | 1.7<br>0.06 [dB/cm]<br>1550 [nm]<br>-1.0E-26 [$s^2$/m]<br>Fig. 3 | 1551 [nm]<br>8.8E-3 [W]<br>CW<br>TE<br>-<br>- | 1553 [nm]<br>1.3E-3 [W]<br>CW<br>TE<br>-<br>- | -<br>-<br>0.25 [%]<br>- | [Ferrera2009]<br><br>Ring radius 135 µm, FSR 200 GHz. |
| Hydex<br><br>$SiO_2$<br><br>$SiO_2$ | SPM/XPM spectral broadening | 450 [mm]<br>Fig. 1<br>1.5 [µm]<br>1.45 [µm]<br>- | Amorphous<br><br>CVD<br><br>Photolithography<br><br>- | 1.7<br>0.06 [dB/cm]<br>1550 [nm]<br>-<br>2.0E-25 [$s^2$/m]<br>- | 1560 [nm]<br>3.9E+1 [W]<br>1.7E-12 [s](FWHM)<br>-<br>1.7E+7 [Hz]<br>2.0E-12 [$m^2$] | -<br>-<br>-<br>-<br>-<br>- | 1.1E-19 [$m^2$/W]<br>-<br>2.2E-1 [$m^{-1}W^{-1}$]<br>- | [Duchesne2009]<br><br>Peak irradiance <25E+13 $W/m^2$. |
| **Chalcogenide glasses** | | | | | | | | |

| Material / Cladding / Substrate | Application | Dimensions | Fabrication | n₂ / Loss / Dispersion | Pump | Signal | Nonlinear parameters | Reference / Notes |
|---|---|---|---|---|---|---|---|---|
| As$_2$S$_3$<br>SiO$_2$<br>Silicon | SPM/XPM spectral broadening | 60 [mm]<br>Fig. 1<br>2 [μm]<br>0.87 [μm]<br>- | Amorphous<br>Thermal evaporation<br>-<br>- | 2.38<br>0.6 [dB/cm]<br>1550 [nm]<br>-<br>3.7E-26 [s$^2$/m]<br>Fig. 2 | 1550 [nm]<br>6.8E+1 [W]<br>6.1E-13 [s](FWHM)<br>TM<br>1.0E+7 [Hz]<br>1.2E-12 [m$^2$] | -<br>-<br>-<br>-<br>- | 6.2E-15 [m/W]<br>1.0E+1 [m$^{-1}$W$^{-1}$]<br>-<br>750 [nm]<br>(at -30 dB) | [Lamont2008] |
| As$_2$S$_3$<br>SiO$_2$<br>Silicon | FWM | 70 [mm]<br>Fig. 1(a)<br>2 [μm]<br>0.85 [μm]<br>- | Amorphous<br>Thermal evaporation<br>Photolithography<br>- | 4 [dB/cm]<br>1550 [nm]<br>-<br>28 [ps/nm km]<br>- | 1547 [nm]<br>1.35E-1 [W]<br>CW<br>-<br>-<br>1E-12 [m$^2$] | 1554-1564 [nm]<br>-<br>CW<br>-<br>- | -<br>10 [m$^{-1}$W$^{-1}$]<br>-37.8 [dB]<br>- | [Pelusi2010]<br><br>Broadband all-optical wavelength conversion of high-speed Differential Phase-Shift Keyed and On-Off Keyed signals with bit rates of 40-160 Gb/s was demonstrated. |
| Ge$_{11.5}$As$_{24}$Se$_{64.5}$<br>SiO$_2$ and polymer<br>Silicon | FWM | 18 [mm]<br>Fig. 3(a)<br>0.63 [μm]<br>0.5 [μm]<br>- | Amorphous<br>Thermal evaporation<br>E-beam<br>- | 2.66<br>2.6 [dB/cm]<br>1550 [nm]<br>-<br>- | 1550.8 [nm]<br>135E-3 [W]<br>CW<br>TM<br>-<br>0.27E-12 [m$^2$] | 1552.6 [nm]<br>2.8E-3 [W]<br>CW<br>TM<br>- | -<br>9.3E-14 [m/W]<br>1.36E+2 [m$^{-1}$W$^{-1}$]<br>-<br>- | [Gai2010]<br><br>$n_2$ = 8.6E-14 cm$^2$/W quoted from other paper<br><br>Supercontinuum generation in the range between 1200 and 2400 nm was demonstrated. |
| **Diamond** | | | | | | | | |
| Diamond<br>SiO$_2$ and air<br>Silicon | Raman Laser Threshold | 0.188 [mm]<br>Fig. 2(a)<br>0.3 [μm]<br>0.3 [μm]<br>- | Monocrystalline<br>-<br>E-beam<br>- | -<br>-<br>5.5 [eV]<br>-<br>- | 750 [nm]<br>2.0E-2 [W]<br>CW<br>TE<br>-<br>- | -<br>-<br>-<br>-<br>- | -<br>-<br>-<br>1.7 [%] | [Latawiec2018]<br><br>$g_{Raman}$ = 3.2E-11 m/W.<br>Raman shift = 1.33E+5 m$^{-1}$.<br><br>The conversion efficiency is the external Raman lasing slope efficiency. |

| Material | Method | Length / Fig. | Fabrication | Loss / λ / Bandgap | Pump | Signal | Nonlinearity | Reference |
|---|---|---|---|---|---|---|---|---|
| Diamond<br>SiO$_2$<br>Silicon | FWM | 0.125 [mm]<br>Fig. 1(a)<br>0.875 [µm]<br>0.85 [µm]<br>Fig. 7(u) | Monocrystalline<br>HPHT<br>E-beam<br>- | -<br>0.34 [dB/cm]<br>1545.1 [nm]<br>5.5 [eV]<br>Figs. 1(b) and 5 | 1600 [nm]<br>7.8E-2 [W]<br>CW<br>TE<br>-<br>5.0E-13 [m$^2$] | -<br>-<br>-<br>-<br>-<br>- | 8.2E-20 [m$^2$/W]<br>-<br>5 [%][g]<br>- | [Hausmann2014]<br><br>OPO based on FWM. |
| Diamond<br>SiO$_2$<br>Silicon | Raman Laser Threshold | 0.6 [mm]<br>Fig. 1(b)<br>0.8 [µm]<br>0.7 [µm]<br>- | Monocrystalline<br>CVD<br>E-beam<br>- | -<br>-<br>-<br>5.5 [eV]<br>- | 1575 [nm]<br>8.5E-2 [W]<br>CW<br>TE<br>-<br>- | -<br>-<br>-<br>-<br>-<br>- | -<br>-<br>-<br>0.43 [%]<br>- | [Latawiec2015]<br><br>$g_{Raman}$ = 2.5E-11 m/W.<br>Raman shift = 1.33E+5 m$^{-1}$.<br><br>The conversion efficiency is the external Raman lasing slope efficiency. |
| **Tantalum oxide, titanium oxide** | | | | | | | | |
| Ta$_2$O$_5$<br>SiO$_2$<br>Silicon | FWM | 12.6 [mm]<br>Fig. 1<br>0.7 [µm]<br>0.4 [µm]<br>Fig. 7(o) | -<br>Sputtering, PECVD<br>E-beam<br>- | 2.1<br>1.5 [dB/cm]<br>1550 [nm]<br>-<br>-1400 [ps/(nm.km)]<br>Fig. 5(a) | 1555.465 [nm]<br>3.5E-2 [W]<br>CW<br>TE<br>-<br>7.7E-13 [m$^2$] | 1556.08 [nm]<br>4.0E-3 [W]<br>CW<br>TE<br>-<br>- | 1.0E-18 [m$^2$/W]<br>-<br>5.2 [m$^{-1}$W$^{-1}$]<br>0.001 [%][e]<br>- | [Wu2015] |
| TiO$_2$<br>SiO$_2$<br>Silicon | FWM | 11 [mm]<br>Fig. 2<br>1.15 [µm]<br>0.38 [µm]<br>- | -<br>Sputtering<br>E-beam<br>- | 2.31<br>5.4 [dB/cm]<br>1550 [nm]<br>3.4 [eV]<br>-50 [ps/(nm.km)]<br>Fig. 7 | 1550.1 [nm]<br>6.0E-1 [W]<br>CW<br>TE<br>-<br>4.3E-13 [m$^2$] | 1551.3 [nm]<br>1.9E-2 [W]<br>CW<br>TE<br>-<br>- | 3.6E-18 [m$^2$/W]<br>-<br>3.4E+1 [m$^{-1}$W$^{-1}$]<br>0.024 [%][a]<br>- | [Guan2018] |
| TiO$_2$<br>SiO$_2$<br>Silicon | SPM/XPM spectral broadening | 22 [mm]<br>Fig. 1<br>1.345 [µm]<br>0.45 [µm]<br>Fig. 7(m) | -<br>Sputtering<br>Photolithography<br>- | 2.35<br>5.5 [dB/cm]<br>1640 [nm]<br>3.4 [eV]<br>20 [ps/(nm.km)]<br>Fig. 4(a) | 1640 [nm]<br>1.3E+3 [W]<br>9.0E-14 [s]<br>TE<br>8.0E+7 [Hz]<br>5.4E-13 [m$^2$] | -<br>-<br>-<br>-<br>-<br>- | -<br>-<br>-<br>-<br>860 [nm]<br>(at -20dB) | [Hammani2018] |

| | | | | | | | | | |
|---|---|---|---|---|---|---|---|---|---|
| TiO$_2$<br><br>SiO$_2$<br><br>Silicon | SPM/XPM spectral broadening | 9 [mm]<br>Fig. 1<br>0.9 [μm]<br>0.25 [μm]<br>Fig. 7(l) | -<br><br>Sputtering<br><br>E-beam<br><br>- | 2.4<br>8 [dB/cm]<br>1560 [nm]<br>3.1-3.3 [eV]<br>1.5E-24 [s$^2$/m] | 1565 [nm]<br>2.9E+4 [W]<br>1.7E-13 [s]<br>TE<br>8.0E+7 [Hz]<br>4.3E-13 [m$^2$] | -<br>-<br>-<br>-<br>-<br>- | 1.6E-19 [m$^2$/W]<br>-<br>1.5 [m$^{-1}$W$^{-1}$]<br>-<br>- | [Evans2013]<br><br>Broadening factor 3.8 at -15dB. |
| TiO$_2$<br><br>SiO$_2$<br><br>Silicon | SPM/XPM spectral broadening | 6 [mm]<br>Fig. 1<br>0.9 [μm]<br>0.25 [μm]<br>Fig. 7(l) | -<br><br>Sputtering<br><br>E-beam<br><br>- | 2.4<br>8 [dB/cm]<br>794 [nm]<br>3.1-3.3 [eV]<br>1.5E-24 [s$^2$/m] | 794 [nm]<br>2.9E+4 [W]<br>8.5E-14 [s]<br>TM<br>1.1E+7 [Hz]<br>1.6E-13 [m$^2$] | -<br>-<br>-<br>-<br>-<br>- | 1.6E-18 [m$^2$/W]<br>7.0E-12 [m/W]<br>7.9E+1 [m$^{-1}$W$^{-1}$]<br>-<br>- | [Evans2013]<br><br>Broadening factor 3.8 at -15dB. |

1: The length of ring resonators was calculated from the ring radius.

2: Figure of the reference paper with the cross-section image or drawing.

3: For heterostructure waveguides, the reported height is the guiding layer thickness.

4: Selected SEM images presented in this work for each geometry and material platform.

5: Figure of the reference paper showing a dispersion curve.

a: Conversion efficiency formula: $\eta = P_{\text{I,output}}/P_{\text{S,output}}$ [I – Idler, and S – Signal for this and the following formulas]

b: Conversion efficiency formula: $\eta = 10\log(P_{\text{I,output}}/P_{\text{S,input}})$

c: Conversion efficiency formula: $\eta = P_{\text{I,output,peak}}/P_{\text{S,input}}$

d: Conversion efficiency formula: $\eta = 10\log(P_{\text{I,output}}/P_{\text{S,output}})$

e: Conversion efficiency formula: $\eta = P_{\text{I,output}}/P_{\text{S,input}}$

f: Conversion efficiency formula: $\eta = \sum P_{\text{all comb lines}}/P_{\text{pump}}$

g: Conversion efficiency formula: $\eta = \sum P_{\text{side bands}}/P_{\text{pump}}$

## 3.7 Hybrid waveguiding systems: data table and discussion

***Team:*** *John Ballato, Peter Dragic, Minhao Pu,* **Nathalie Vermeulen (team leader)**, *Kresten Yvind*

### 3.7.1 Introduction

#### 3.7.1.1 Hybrid waveguiding systems and their NLO applications

Optical fibers and on-chip waveguides are often employed for NLO applications as the strong light confinement enabled by these waveguiding structures benefits the efficiency of NLO phenomena. At the same time, NLO researchers have been exploring whether fibers and on-chip waveguides, typically made of glasses or semiconductor materials, can be combined with very different materials, such as low-dimensional materials, metals, organic solids, liquids, gasses, etc. The goal of adding other materials to the waveguiding structures is to modify, and in most cases to enhance, the efficiency of a given NLO process. The resulting hybrid waveguiding systems find applications in the same domains as their 'bare' counterparts (see Sections 3.5 – 3.6) but can also acquire other NLO functionalities, depending on the properties of the added material. For example, whereas 'bare' fibers have always been the preferred medium when targeting low-loss NLO applications, hybrid fibers enhanced with low-dimensional materials such as carbon nanotubes or graphene can also fulfill absorptive functionalities such as fiber-based saturable absorption in laser cavities with pulsed operation [Liu2020, Teng2020]. As such, the development of hybrid fibers and hybrid on-chip waveguides has allowed combining the best of both worlds, i.e. the intrinsic strengths of the bare waveguiding structures plus the special NLO properties of the added materials. Further details on the importance of NLO hybrid fibers and NLO hybrid on-chip waveguides can be found in several recent review papers (see, for example, [Li2018, Debord2019, Guo2019, Liu2020, Teng2020, Tuniz2021, Steglich2021, Vermeulen2022]).

#### 3.7.1.2 Background prior to 2000

##### 3.7.1.2.1 *Background for hybrid fibers*

Research on NLO hybrid fibers started already in the 1970s, with a focus on hollow fiber capillaries filled with NLO solids [Stevenson1974, Babai1977]. Around the same period, light guidance in fiber capillaries with liquid cores was demonstrated by several groups [Stone1972, Payne1972, Ogilvie1972]. In contrast, laser transmission in fiber capillaries containing gasses became a subject of study from the 1990s onward [Olshanii1993]. Not only can hollow fiber structures be employed for realizing NLO hybrid fibers, but also solid-core fibers with their cross-sectional geometry modified by, e.g., side polishing or tapering can be used for this purpose. Because of their modified cross-section, the added NLO material can be brought in close proximity to the light-guiding fiber core, such that it can interact with the evanescent tails of the fiber mode. The first demonstrations of this concept date back to the 1980s [Bergh1980, Lamouroux1983, Lacroix1986]. Both the hybrid fibers based on hollow capillaries and those relying on side-polished or tapered solid-core fibers were successfully employed in various NLO experiments before the turn of the century (see, e.g., [Stevenson1974, Kanbara1992, Nesterova1996, Lee1998]). Nevertheless, the research area of NLO hybrid fibers has experienced the strongest growth after 2000, thanks to the emergence of new special fiber structures as well as new exotic materials that can enhance the fibers' NLO response (see further on).

##### 3.7.1.2.2 *Background for hybrid on-chip waveguides*

Similarly, as for hybrid fibers, the first hybrid on-chip waveguides developed for NLO applications comprised NLO organics, and this approach allowed demonstrating the EO effect on a silicon chip already in the 1990s [Faderl 1995]. Also here, the evanescent tails of the waveguide mode allowed 'sensing' the presence of the NLO organics deposited on top, even when working with a standard strip waveguide geometry [Faderl 1995]. Due to their planar structure, photonic chips in fact provided a very suitable platform for the deposition of NLO solids on top, whereas NLO liquids and gasses were more easily combined with hollow fiber structures. Nevertheless, the most important breakthroughs in the development of hybrid on-chip waveguides would be triggered by major advancements in both waveguide technology and material science shortly after 2000 (see further on).

### 3.7.1.3 Considerations for hybrid waveguiding systems when performing NLO measurements

When characterizing hybrid fibers and hybrid on-chip waveguides, both the considerations for the bare fibers/waveguides (see Sections 3.5.1.2 and 3.6.1.3) and those for the added materials need to be taken into account. For NLO measurements requiring phase matching, it is important to account for the phase matching conditions of the entire hybrid system. Also interfacial aspects such as (lack of) adhesion of the added materials, uniformity and/or diffusion are important to verify prior to the NLO experiments. From the measured NLO response, the effective nonlinearity for the hybrid system as a whole can be extracted. If, however, one wants to go one step further and assess the contributions from each constituent to the overall NLO response, careful analysis is required: the cross-sectional distribution of the modal power in the hybrid waveguiding structure needs to be evaluated, and this information subsequently fed into a weighted contributions model to extract the individual nonlinear properties of the fiber/waveguide structure and the added material (see, e.g., [Vermeulen2016a] and Appendix of [Vermeulen2016b]). Here, it is also important to keep in mind that the different NLO contributions do not always add up. For example, when studying nonlinear refraction in a silicon waveguide covered with undoped graphene while using excitation wavelengths in the near-IR telecom domain, the silicon will have a positive nonlinearity whereas the graphene top layer will exhibit a negative nonlinearity [Vermeulen2016a, CastelloLurbe2020]. Therefore, in the weighted contributions analysis, it is crucial to implement the correct nonlinearity signs for each of the constituting media. In case the signs are not known upfront, this information can be obtained, for example, using a phase-sensitive NLO technique such as SPM-based spectral broadening and by comparing the NLO response of the hybrid waveguiding structure to that of the bare waveguide/fiber [Vermeulen2016a]. This allows evaluating whether there are nonlinearities of opposite sign present in the hybrid system.

### 3.7.1.4 Description of general table outline

Tables 8A and 8B show a representative list of NLO properties of, respectively, hybrid fibers and hybrid on-chip waveguides taken from the literature since 2000, with the entries arranged in alphabetical order. The selection of works included in the Tables has been based on the general best practices in Section 2 and the considerations outlined above. Taking into account that an extremely large number of material combinations are possible in hybrid systems, this selection has been limited to just a few representative works for different combinations of bare fibers/waveguides with other media, and therefore does not provide an extensive overview. The selection also contains a few papers with organic solids, liquids and gasses, although in this article the focus is rather on inorganic solids, as mentioned in the general introduction. The included works nominally report data obtained at room temperature. Tables 8A and 8B are subdivided into "Material properties," "Measurement details" and "Nonlinear properties." Within each column the information is given in the order of the header description, and powers of 10 (e.g., $10^{\pm\alpha}$) are written as E±α for compactness. "Material properties" include the characteristics of both constituents of the hybrid system, as well as references to the sub-figures in Fig. 8 showing SEM images of the fabricated structures. The peak power values in the Tables are nominally incoupled powers as specified in the papers. The NLO technique used in each of the papers is provided in the "Method" column. As most works report on third-order NLO effects, the focus in the Tables is on third-order NLO parameters such as $|\mathrm{Re}(\gamma_{\mathrm{eff}})|$ and the effective saturation irradiance $I_{\mathrm{sat,eff}}$ together with the saturable loss, extracted from the hybrid system as a whole. That said, also a limited number of papers reporting on second-order NLO parameters are included in the Tables, with their parameter values specified in a separate "Comments" column. Also included in the "Comments" column are individual nonlinearity contributions that could be separately determined per constituent material, when stated as such. Lastly, some papers specify the dependence of the NLO parameters on wavelength, waveguide/fiber dimension, doping level expressed as Fermi level (eV) or carrier concentration ($m^{-2}$), etc., or have notes associated with their measurement/analysis such as the formulas used to calculate conversion efficiencies η. This information is listed within the "Comments" column.

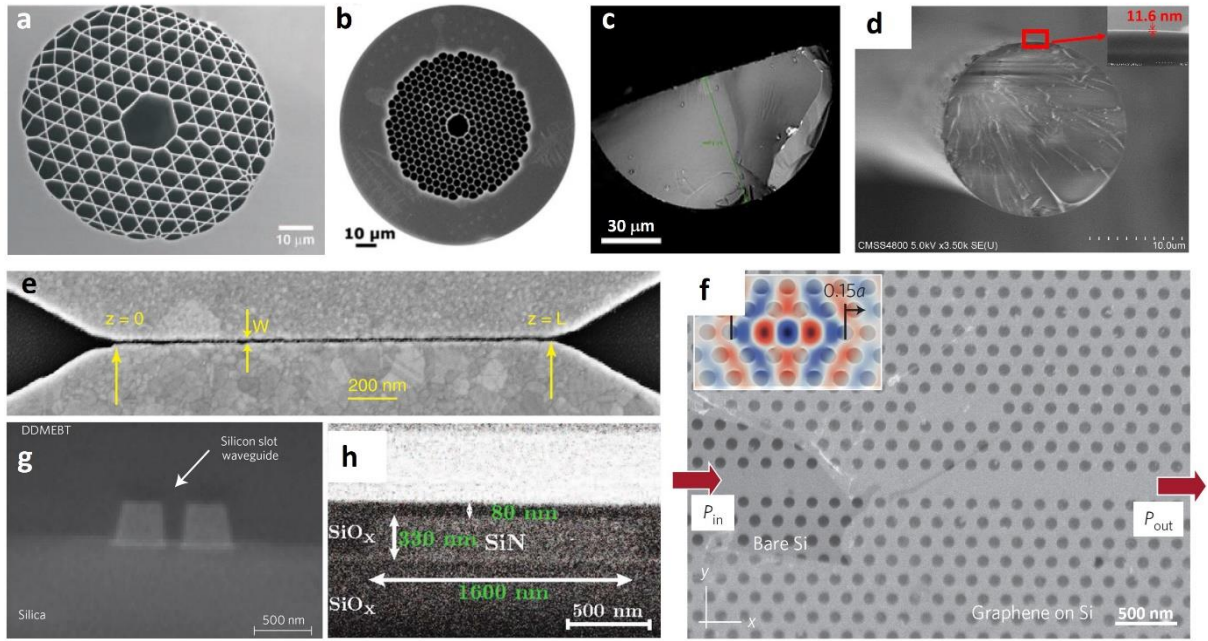

**Fig. 8**. SEM images of NLO hybrid fibers and hybrid on-chip waveguides. The labels of the sub-figures correspond to the labels given in Tables 8A and 8B. Panels a and e are reprinted with permission from [Benabid2002, Nielsen2017], AAAS. Panels b, f, and g are reprinted with permission from [Yang2020, Gu2012, Koos2009], Springer Nature. Panels c and h are reprinted with permission from [An2020, Alexander2017], ACS. Panel d is reprinted with permission from [Wang2018], CLP Publishing.

### 3.7.2 Discussion

In view of the relatively recent emergence of low-dimensional materials, their combination with on-chip waveguides and fibers is quite a new development with most progress being made over the last 5 to 10 years. Within this short period of time, a wide variety of 2D materials has been explored for this purpose, as can be seen in Tables 8A and 8B. At the same time, the Tables also show that organic solids, liquids and gasses still are very interesting NLO media for constructing hybrid waveguiding systems.

#### 3.7.2.1 Advancement since 2000 and remaining challenges

##### 3.7.2.1.1 Advancement and challenges for hybrid fibers

Table 8A for NLO hybrid fibers shows two distinct novelties as compared to the works before 2000, the first novelty being the emergence of $SiO_2$ hollow-core photonic-crystal fibers (PCFs) [Russell2014]. In contrast to hollow fiber capillaries which intrinsically are leaky waveguides even when filled with, e.g., gasses, hollow-core PCFs efficiently trap light within their hollow core by exploiting the physics of photonic bandgaps [Russell2014]. As such, various NLO effects have been demonstrated in hollow-core PCFs filled with liquids and gasses [Baghwat2008, Russell2014], exhibiting strong NLO responses as shown by the examples listed in the Table [Benabid2002, Vieweg2010, Renninger2016, Yang2020]. That said, combining ultra-low loss and truly robust single-mode propagation in a hollow-core PCF still represents a challenge and requires special design approaches [Amrani2021]. It should also be noted that, because of the intrinsic simplicity of fiber capillaries as compared to PCFs, several research groups continued using capillaries after 2000 to realize hybrid fibers for different NLO functionalities [Schneebeli2013, Chemnitz2017].

A second important novelty in NLO hybrid fiber development after 2000 has been the rise of low-dimensional materials [Liu2020]. Both 1D carbon nanotubes and 2D materials such as graphene, graphene oxide, black phosphorus and several types of Transition Metal Dichalcogenides (TMDs) were deposited on side-polished or tapered fibers (see, for example, [Xu2013, Park2015, Lee2015, Martinez2017, Steinberg2018, Wang2018, Chen2019, An2020]). The low-dimensional materials either enhanced the already present NLO response of the bare fibers [Xu2013, An2020] and sometimes even made them tunable by means of electrical gating [An2020], or they enriched the fiber components with additional NLO functionalities such as frequency doubling [Chen2019] and saturable absorption [Park2015, Lee2015, Martinez2017, Steinberg2018, Wang2018]. The latter is particularly useful for the development of pulsed fiber lasers [Liu2020]. Still, transferring low-dimensional materials onto side-polished or tapered fiber surfaces can be quite challenging. Efforts for the direct growth of, e.g., graphene on dielectrics such as glass are currently ongoing [Khan2018] and could circumvent the need for complex material transfer procedures while improving standardization and upscaling of the hybrid fiber fabrication. To assess the quality of fabricated hybrid fibers, careful characterization is required [Vermeulen2022]. Note that particularly for graphene-based devices it is important to know the doping level of the graphene layer as the latter determines whether or not a strong NLO response can be expected from the 2D sheet [CastelloLurbe2020]. Finally, as shown in Table 8A, the strong nonlinear refractive response of fibers enriched with low-dimensional materials is often accompanied by high linear losses [Xu2013, An2020]. At the same time, taking into account that NLO research with low-dimensional materials still is at a relatively early stage, it is remarkable how much progress has been made over the past few years on the implementation of these materials in NLO hybrid fibers.

### 3.7.2.1.2   Advancement and challenges for hybrid on-chip waveguides

A first novelty seen in Table 8B for NLO hybrid on-chip waveguides is that NLO organics are nowadays combined with advanced waveguiding structures such as plasmonic or slot waveguides [Nielsen2017, Koos2009]. This advancement has been enabled by the major progress in photonic chip fabrication technology over the past decades. Both plasmonic and slot waveguides intensify the electromagnetic field in the NLO polymer: the former by focusing the light down to the nanoscale inside a metallic structure close to the polymer [Nielsen2017, Tuniz2021], the latter by having its waveguide mode centered around the polymer located within the slot [Koos2009, Steglich2021]. These two approaches give rise to a strong enhancement of the electromagnetic field strength yielding very high NLO efficiencies, albeit with significant propagation losses in the waveguides. Today's research on hybrid plasmonic or slot waveguides mainly focuses on how to maintain such a strong field enhancement while keeping the losses low [Tuniz2021, Steglich2021].

Hybrid on-chip waveguide development has also greatly benefitted from the emergence of NLO 2D materials [Li2018]. Similarly to the case of hybrid fibers, the combination of on-chip waveguides with graphene, graphene oxide, and TMDs either strengthened the already present NLO response of the waveguides [Gu2012, Vermeulen2016a, Alexander2017, Vermeulen2018, Yang2018, Zhang2020, Qu2020] and sometimes made the response electrically tunable [Alexander2017], or they introduced additional NLO functionalities such as saturable absorption [Demongodin2019], frequency doubling [Chen2017] and difference frequency generation [Yao2018]. Most on-chip waveguides used in these experiments were Si or $Si_3N_4$ waveguides with a standard strip geometry. Transferring 2D materials onto such waveguides embedded within a planar photonic chip poses fewer practical challenges as compared to the transfer onto fiber surfaces. Nevertheless, direct synthesis of the 2D materials on the waveguides would be beneficial from the point of view of standardization and upscaling [Kahn2018]. Just like their fiber-based counterparts, hybrid on-chip waveguides require careful material characterization of their constituents [Vermeulen2022], and for graphene-based devices this also includes evaluating the doping level of the 2D layer (see Section above). Finally, as shown in Table 8B, the strong nonlinear refractive response of on-chip waveguides covered with 2D materials often comes with high losses (see, for example, [Vermeulen2016a, Alexander2017, Vermeulen2018, Yang2018, Zhang2020]). That said, the obtained results are promising, and further progress can be expected in the coming years as the fabrication of photonic chips and 2D materials will continue to improve and the fundamental understanding of the NLO physics of 2D materials will be further extended.

### 3.7.2.2 Recommendations for future works on hybrid waveguiding systems

Our general recommendation for future works on NLO hybrid waveguides is to report at least those parameters that are included in Tables 8A and 8B, as these provide essential information on the hybrid systems. So, including these parameters in future publications can be considered as a best practice in addition to the more general best practices described in Section 2.

To maximize the impact of future works, it is of particular importance to extract the actual nonlinearity, e.g., $\gamma_{eff}$, of the hybrid system (rather than just specifying the relative NLO enhancement enabled by the added material) and to study the dependence of the nonlinearity on, e.g., wavelength. Ideally, one could also go the extra mile of determining the individual nonlinearities (magnitude and sign) of the fiber/waveguide medium and the added medium along the considerations outlined above. By then comparing the retrieved NLO coefficients with measurements of the same materials in a 'stand-alone' configuration, it is possible to evaluate if combining them in a hybrid waveguiding system leaves their individual NLO properties unchanged or not.

We would also like to point out the need for a detailed description of the material properties (including the doping level when relevant) and of the linear optical properties of all constituents in the hybrid system. This is not always done in the existing literature, which makes it difficult to properly interpret and reproduce the reported results. For example, in addition to specifying the linear optical loss for the complete hybrid system, one should also quantify the initial optical loss of the bare fiber/waveguide. In a similar way, the material to be added needs to be optically characterized beforehand. To allow for reproducibility, future works should also provide a detailed description of the fabrication procedure for realizing the hybrid system and specify its geometrical outline, both in the cross-sectional plane and along the optical axis. The latter comprises specifying the length over which the added material is present as well as the remaining length of bare fiber/waveguide. This way, the reader can get the full picture of the system under study, which is essential in view of the inherent complexity of hybrid waveguiding structures. As such, the research community will be able to efficiently build upon state-of-the-art results so that the 'best of both worlds' idea behind hybrid waveguiding systems can be exploited to its full potential.

### 3.7.3 Data table for hybrid waveguiding systems

**Table 8A.** NLO properties of hybrid fibers from representative works since 2000. Legend for superscripts: see below the table.

## Hybrid fibers

| Material Properties | | | Measurement Details | | | Nonlinear Properties | | | | |
|---|---|---|---|---|---|---|---|---|---|---|
| **Constituent 1** *Fiber type* | **Constituent 2** *Added material Thickness Fabrication* | **Hybrid fiber** *Length Loss / wavelength[1] Effective area SEM image[2]* | **Method** | **Pump** *Wavelength Peak power Pulse width Rep. rate* | **Probe** *Wavelength Peak power Pulse width* | $\|Re(\gamma_{eff})\|$ [m$^{-1}$ W$^{-1}$] | $I_{sat,eff}$ [W/m$^2$] | Sat. loss [%] | Additional parameters and comments | Reference |
| SiO$_2$ hollow-core fiber Capillary | B-carotene - - | 500 [mm] - - - | IRS Pump-Probe | 1550 [nm] - 3 [ps] 50 [MHz] | - - 8 [ps] | - | - | - | g$_{Raman}$: 6.00E-13 [m/W] Raman shift: 1.16E+05 [m$^{-1}$] 5 μm core diameter | Schneebeli2013 |
| SiO$_2$ hollow-core fiber Capillary | CCl$_4$ - - | 1 [m] - - - | IRS Pump-Probe | 1550 [nm] - 3 [ps] 50 [MHz] | - - 3 [ps] | - | - | - | g$_{Raman}$: 1.50E-12 [m/W] Ramanshift: 4.59E+04 [m$^{-1}$] 32 mW average pump power; 10 μm core diameter | Schneebeli2013 |
| SiO$_2$ hollow-core fiber Capillary | CS$_2$ - - | 450 [mm] - - - | IRS Pump-Probe | 1550 [nm] - 3 [ps] 50 [MHz] | - - 8 [ps] | - | - | - | g$_{Raman}$: 3.00E-11 [m/W] Raman shift: 6.6E+04 [m$^{-1}$] 1.5 mW average pump power; 2 μm core diameter | Schneebeli2013 |
| SiO$_2$ hollow-core fiber Capillary | CS$_2$<br>-<br>- | 140 [mm]<br>-<br>-<br>- | Other | 1950 [nm]<br>-<br>460 [fs]<br>5.6 [MHz] | -<br>-<br>- | 2.80E-01 | - | - | Supercontinuum generation; 14 nJ pump pulse energy; 4.7 µm core diameter | Chemnitz2017 |
| SiO$_2$ hollow-core fiber PCF | Air<br>-<br>- | 1.61 [m]<br>-<br>-<br>- | Two-color SBS Pump-Probe | 1535 [nm]<br>48 [mW]<br>CW | -<br>-<br>- | 1.60E-04 | - | - | $\gamma_{Brillouin}$: 9.00E-04 [W$^{-1}$m$^{-1}$]<br>Brillouin shift: 3.50E+07 [Hz]<br>Linewidth: 4.00E+06 [Hz]<br>Forward Brillouin scattering; 1 atm air pressure; 5.65 µm core diameter | Renninger2016 |
| SiO$_2$ hollow-core fiber PCF | CCl$_4$<br>-<br>- | 190 [mm]<br>-<br>- | Other | 1030 [nm]-<br>210 [fs]<br>44 [MHz] | -<br>-<br>- | 3.70E-01 | - | - | Supercontinuum generation; 330 mW average pump power; 2.5 µm core diameter | Vieweg2010 |
| SiO$_2$ hollow-core fiber PCF | CO$_2$<br>-<br>- | 50 [m]<br>-<br>80 [µm$^2$]<br>Fig. 8(b) | Two-color SBS Pump-Probe | 1550 [nm]<br>6 [mW]<br>CW<br>- | -<br>-<br>- | - | - | - | $\gamma_{Brillouin}$: 1.68 [W$^{-1}$m$^{-1}$]<br>Brillouin shift: 3.20E+08 [Hz]<br>Linewidth: 3.65E+06 [Hz]<br>Backward Brillouin scattering; 41 bar CO$_2$ pressure; 10 µm core diameter | Yang2020 |
| SiO$_2$ hollow-core fiber PCF | H$_2$<br>-<br>- | 320 [mm]<br>-<br>-<br>Fig. 8(a) | SRS | 532 [nm]<br>-<br>6 [ns]<br>20 [Hz] | -<br>-<br>- | - | - | - | 30% pump-to-Stokes conversion efficiency for 4.5 uJ pump pulse energy; 17 bar H$_2$ pressure; 15 µm core diameter | Benabid2002 |
| SiO$_2$ solid-core fiber Side polished | Black phosphorus<br>10 [nm]<br>Mechanical exfoliation | 100 [mm]<br>73.6 [%]<br>/1566 [nm]<br>-<br>- | Nonlinear transm./refl. | 1566 [nm]<br>-<br>1.2 [ps]<br>14.2 [MHz] | -<br>-<br>- | - | 1.25E+11 | 3.31 | - | Park2015 |
| SiO₂ solid-core fiber Side polished | Graphene Monolayer CVD | 500 [μm] 47 [%] /1561 [nm] 40 [μm²] Fig. 8(c) | FWM | 1561 [nm] 300 [W] 265 [fs] 37.8 [MHz] | 1480-1610 [nm] 10 [mW] CW | 1.14 | - | - | Graphene doping: variable (values here for 0.2 eV) $|\chi^{(3)}|$: 8E-17 [m²/V²] Nonlinearity dependence on graphene doping level shown in Fig. 2 of reference paper | An2020 |
| SiO₂ solid-core fiber Side polished | Graphene 2 layers CVD | 5 [mm] 8.7 [%] /1550 [nm] - - | Nonlinear transm./refl. | 1609 [nm] - 423 [fs] 30.9 MHz | - | - | 2.57E+12 | 1.06 | Graphene doping: variable (values here for -1.2V gate voltage) | Lee2015 |
| SiO₂ solid-core fiber Side polished | Graphene oxide 100 [nm] Modified Hummers method | 10 [mm] 42 [%] /1550 [nm] - - | Nonlinear transm./refl. | 1550 [nm] 19.4 [kW] 150 [fs] 89 [MHz] | - - - | - | - | 22 | Saturation fluence: 7E-2 [J/m²] | Steinberg2018 |
| SiO₂ solid-core fiber Tapered | Carbon nanotubes - High-pressure CO process | 100 [mm] 68 [%] /1550 [nm] 50 [μm²] - | FWM | 1550 [nm] 1.6 [W] CW | 1552 [nm] 0.1 [ns] - | 1.82 | - | - | Semiconducting nanotubes; 1ppm in PTFEMA Conversion efficiency: 0.2 [%] Conversion efficiency formula: idler_out/signal_in Conversion bandwidth: 9 [nm] | Xu2013 |
| SiO₂ solid-core fiber Tapered | Carbon nanotubes - High-pressure CO process | 4 [mm] 34 [%] /1550 [nm] - - | Nonlinear transm./refl. | 1550 [nm] - 600 [fs] 25 [MHz] | - - - | - | - | 8.5 | Semiconducting nanotubes; 50% in PTFEMA; variable fiber diameter (values here for 3 μm) Average saturation power: 4.50E-4 [W] Nonlinearity dependence on diameter of tapered fiber shown in Table II of reference paper | Martinez2017 |
| SiO₂ solid-core fiber Tapered | MoTe₂ 11.6 [nm] Magnetron sputtering deposition | 10 [mm] 44.6 [%] /1572.4 [nm] Fig. 8(d) | Nonlinear transm./refl. | 1572.4 [nm] - 642 [fs] 50.12 [MHz] | - - - | - | 9.60E+10 | 25.5 | - | Wang2018 |
| SiO₂ solid-core fiber Tapered | MoTe₂ 11.6 [nm] Magnetron sputtering deposition | 10 [mm] 43.3 [%] /1915.5 [nm] - Fig. 8(d) | Nonlinear transm./refl. | 1915.5 [nm] - 1.25 [ps] 18.72 [MHz] | - - - | - | 1.23E+11 | 22.1 | - | Wang2018 |

| SiO₂ solid-core fiber Tapered | WS₂ Monolayer CVD | 60 [μm] <25 [%] /1530-1590 [nm] - - | SHG | 1550 [nm] - 10 [ns] 1 [MHz] | - - - | - | - | - | 20-fold enhancement of SHG signal as compared to bare fiber for 50 mW average pump power | Chen2019 |
|---|---|---|---|---|---|---|---|---|---|---|

[1]: Wavelength at which the specified loss has been measured

[2]: Selected SEM images presented in this work

**Table 8B**. NLO properties of hybrid on-chip waveguides from representative works since 2000. Legend for superscripts: see below the table.

## Hybrid on-chip waveguides

| Material Properties | | | Measurement Details | | | Nonlinear Properties | | | |
|---|---|---|---|---|---|---|---|---|---|
| **Constituent 1**<br>*Waveguide (wg)*<br>*Lithography* | **Constituent 2**<br>*Added material*<br>*Thickness*<br>*Fabrication* | **Hybrid wg**<br>*Length*<br>*Loss / wavelength[1]*<br>*Effective area*<br>*SEM image[2]* | **Method** | **Pump**<br>*Wavelength*<br>*Peak power*<br>*Pulse width*<br>*Rep. rate*<br>*Polarization* | **Probe**<br>*Wavelength*<br>*Peak power*<br>*Pulse width*<br>*Polarization* | $\|\mathrm{Re}(\gamma_{eff})\|$<br>$[\mathrm{m}^{-1}\ \mathrm{W}^{-1}]$ | $\eta$ [%] | Additional parameters and comments | Reference |
| Hydex wg<br>Photolith. | Graphene oxide<br>2 layers<br>Solution-based process | 15 [mm]<br>0.2 [dB/mm] /1550 [nm]<br>-<br>- | FWM | 1550 [nm]<br>160 [mW]<br>CW<br>-<br>TE mode | 1551 [nm]<br>160 [mW]<br>CW<br>TE mode | 9.00E-01 | 2.00E-03 | Conversion efficiency formula: idler$_{out}$/signal$_{out}$<br>Conversion bandwidth: 2.50E12 [Hz]<br>Nonlinearity dependence on wavelength detuning and length shown in Fig. 4 of reference paper | Yang2018 |
| Plasmonic (gold) wg<br>E-beam lith. | Polymer MEH-PPV<br>-<br>Spin-coating | 2 [μm]<br>-<br>-<br>Fig. 8(e) | FWM | 1480 [nm]<br>30 [W]<br>1.04 [ps]<br>10 [kHz]<br>TE mode | 1450 [nm]<br>-<br>1.11 [ps]<br>TE mode | 3.09E+04 | 4.60 | Im($\gamma_{eff}$): 7.00E2 [m$^{-1}$ W$^{-1}$]<br>Conversion efficiency formula: idler$_{out}$/signal$_{in}$ | Nielsen2017 |
| Si photonic-crystal wg<br>Photolith. | Graphene<br>Monolayer<br>CVD | 1.5826 [μm]<br>(cavity length)<br>-<br>-<br>Fig. 8(f) | FWM | 1562.36 [nm]<br>0.6 [mW]<br>CW<br>-<br>TE mode | 1562.09 [nm]<br>600 [μW]<br>CW<br>TE mode | - | 1.00E-01 | Graphene doping: 5.00E16 [m$^{-2}$]<br>Free-carrier lifetime: 2.00E-10 [s]<br>$\|n_{2,eff}\|$: 4.80E-17 [m²/W] | Gu2012 |

| Si slot wg Photolith. | Organic molecules DDMEBT - Molecular beam deposition | 4 [mm] 1.6 [dB/mm] /1550 [nm] - Fig. 8(g) | FWM | 1548 [nm] 375 [mW] 3 [ps] 42.7 [GHz] TE mode | 1556 [nm] - 3.00 [ps] TE mode | 1.04E+02 | 6.00E-02 | Conversion efficiency formula: $idler_{out}/signal_{out}$ | Koos2009 |
|---|---|---|---|---|---|---|---|---|---|
| Si wg Photolith. | Graphene Monolayer CVD | 400 [µm] 132 [dB/mm] /1550 [nm] - - | SPM/XPM spectral broadening | 1553 [nm] 1.68 [W] 3 [ps] 80 [MHz] TE mode | - - - - | 1.40E+03 | - | Graphene doping: -0.2 [eV] $n_{2,eff}$ of graphene only: -1E-13 [m$^2$/W] | Vermeulen2016a |
| Si wg E-beam lith. | MoS$_2$ 3 [nm] CVD | 1.6 [mm] 5.5 [dB/mm] /1560 [nm] - - | FWM | 1550 [nm] - CW - TM mode | 1550.4 [nm] - CW TM mode | - | 6.00E-02 | Conversion efficiency formula: $idler_{out}/signal_{out}$ Nonlinearity dependence on wavelength detuning shown in Fig. 6 of reference paper | Zhang2020 |
| Si$_3$N$_4$ wg Photolith. | Graphene oxide Monolayer Solution-based process | 20 [mm] 0.61 [dB/mm] /1550 [nm] - - | FWM | 1549 [nm] 63 [mW] CW - TE mode | 1550 [nm] 63 [mW] CW TE mode | 1.31E+01 | 1.40E-04 | Conversion efficiency formula: $idler_{out}/signal_{in}$ Conversion bandwidth: 1.25E12 [Hz] Nonlinearity dependence on graphene-oxide layer number shown in Fig. 8 of reference paper | Qu2020 |
| Si$_3$N$_4$ wg Photolith. | Graphene Monolayer CVD | 100 [µm] 50 [dB/mm] /1550.18 [nm] - Fig. 8(h) | FWM | 1550.18 [nm] 11.2 [mW] CW - TM mode | 1550.7 [nm] - CW TM mode | 6.40E+03 | 1.00E-03 | Graphene doping: variable (values here for -0.35 eV, i.e. -0.5 V gate voltage) Conversion efficiency formula: $idler_{out}/signal_{out}$ Nonlinearity dependence on graphene doping level and wavelength detuning shown in Fig. 4 of reference paper | Alexander2017 |
| Si₃N₄ wg Photolith. | Graphene Monolayer - | 3.2 [mm] 12.6 [dB/mm] /1540 [nm] 0.81 [μm²] - | Nonlinear transm./refl. | 1547 [nm] - 200 [fs] 20 [MHz] TE mode | - - - | - | - | Graphene doping: -0.3 [eV] Free-carrier lifetime: 1.50E-13 [s] Saturation density: 1.30E+16 [m⁻²] Non-saturable loss: 7.5 [dB/mm] | Demongodin2019 |
| Si₃N₄/SiO₂ wg Photolith. | Graphene Monolayer CVD | 1.1 [mm] 20 [dB/mm] /1563 [nm] - - | SPM/XPM spectral broadening | 1563 [nm] 2.7 [W] 3 [ps] 80 [MHz] TE mode | - | Proportional to free-carrier refraction coefficient | - | Graphene doping: 6.50E16 [m⁻²] Free-carrier lifetime: 1.00E-12 [s] Free-carrier refraction coefficient: 1.00E-5 [-] | Vermeulen2018 |
| a-Si wg E-beam lith. | MoSe₂ Monolayer Mechanical exfoliation | 22 [μm] - - - | SHG | 1550 [nm] - 82 [fs] 80 [MHz] TE mode | - - - | - | - | 5-fold enhancement of SHG signal as compared to free-space excitation of MoSe₂ Pump irradiance: 150 [TW/m²] | Chen2017 |
| Si₃N₄ wg Photolith. | Graphene Monolayer CVD | 80 [μm] 100 [dB/mm] /1550 [nm] - - | DFG | 1531.9 [nm] 200 [W] 2.2 [ps] 39.1 [MHz] TM mode | 1593.2 [nm] 1.6 [W] CW TM mode | - | 4.00E-3 [W⁻¹] | Plasmon-enhanced DFG Graphene doping: variable (values here for 0.05 eV) χ⁽²⁾: 1.20E-6 [esu] Conversion efficiency formula: idler_out / (signal * pump) Nonlinearity dependence on graphene doping shown in Fig. 4 of reference paper | Yao2018 |

[1]: Wavelength at which the specified loss has been measured

[2]: Selected SEM images presented in this work

## 3.8 THz NLO: data table and discussion

***Team:*** *Dmitry Turchinovich*

### 3.8.1 Introduction

#### 3.8.1.1 THz NLO mechanisms

As already mentioned in Sections 3.1.1 and 3.3.1, some materials exhibit pronounced nonlinear responses not only at optical wavelengths but also in the THz domain. These THz nonlinearities can originate from various physical processes. The free carrier response typically constitutes the strongest, and most broadband, contribution to the dielectric function of materials at THz frequencies [Dressel and Grüner 2002; Huber2001; Jepsen2011; Ulbricht2011]. Consequently, the strongest NLO response at THz frequencies is also usually dominated by free-carrier effects, and hence is characteristic of conductive materials, or materials that become conductive under intense THz excitation (see, for example, [Blanchard2011; Cheng2020; Deinert2021; Fan2013; Giorgianni2016; Grady2013; Hafez2018, Hafez2020; Hirori2011; Hoffmann2009b; Hoffmann and Turchinovich 2010; Hohenleutner2015; Hwang2013; Jadidi2016; Junginger2012; König-Otto2017; Kovalev2020, Kovalev2021; Lee2020; Liu2012; Matsunaga2013; Mayer2015; Mics2015; Schubert2014; Sharma2010; Shimano2012; Turchinovich2012]). Here, the THz electromagnetic field couples to the free carriers via optical conductivity mechanisms, leading to absorption of a part of the electromagnetic energy by the electronic system of the material. This energy transfer from the driving THz field to the free electrons typically results in heating of the electron population of the material, or quasi-coherent ponderomotive acceleration of carriers within the band structure, both leading to the concomitant temporal modification of the THz optical conductivity of the material, and hence to its nonlinear response to the driving THz field.

THz nonlinearities may also arise from THz-driven phase transitions (e.g., [Liu2012]), or from direct, quasi-resonant excitation of THz intersubband transitions in semiconductor nanostructures such as quantum wells or superlattices (e.g., [Houver2019; Kuehn2011; Raab2019, Raab2020]). IR-active optical phonons with resonant frequencies in the THz range also contribute to the THz nonlinearity in crystals [Dekorsy2003; Lu2021; Mayer and Keilmann 1986a].

The THz nonlinearities resulting from the direct THz excitation of electrons or IR-active phonon modes in crystals are dissipative in nature, as the driving THz field is physically absorbed by the material in the first steps of the light-matter interaction. The materials exhibiting the strongest nonlinearities at THz frequencies are, typically, doped semiconductors, superconductors, and doped quantum materials such as graphene, topological insulators, and 3D Dirac semimetals (see, for example, [Blanchard2011; Cheng2020; Deinert2021; Fan2013; Giorgianni2016; Grady2013; Hafez2018, Hafez2020; Hirori2011; Hoffmann2009b; Hoffmann and Turchinovich 2010; Hohenleutner2015; Hwang2013; Jadidi2016; Junginger2012; König-Otto2017; Kovalev2020, Kovalev2021; Lee2020; Matsunaga2013; Mayer2015; Mics2015; Schubert2014; Sharma2010; Shimano2012; Turchinovich2012]). Doped graphene was shown to have by far the strongest electronic nonlinearity at THz frequencies [Hafez2018]. Its nonlinear coefficients at THz frequencies surpass those of all other known materials by many orders of magnitude, and this possibly holds for other spectral ranges as well [Hafez2018, Hafez2020].

The non-dissipative THz nonlinearities such as the THz Kerr effect [Hoffmann2009a; Sajadi2015] or the THz-driven quantum-confined Stark effect (QCSE) [Hoffmann2010], also reported in the literature, are usually weaker in strength than the dissipative electronic nonlinearities. These non-dissipative nonlinear effects are typically observed in a THz pump – optical probe arrangement, where the strong THz field modifies the conditions of light-matter interactions for the optical-frequency probe in the material. For THz Kerr measurements, optically transparent solids and liquids are typically used as samples, whereas for the observation of THz QCSE, quantum nanostructures featuring resonant absorption at the optical probe wavelength are used. A somewhat related type of THz NLO experiment is the THz-driven side-band generation on a CW optical carrier wave, which propagates through a THz-nonlinear material or device [Dhillon2007; Zaks2012]. These experiments are usually performed in a quasi-CW mode, and the THz sideband generation is registered in the spectrum of the carrier probe signal at optical frequencies.

### 3.8.1.2 Brief history of THz NLO research

THz NLO is a relatively young discipline. The key reason for this is the relative difficulty of generating strong fields at THz frequencies, as compared to visible and IR optical signals. In the early works on THz NLO from the 1980-1990s gas lasers and THz free-electron lasers (FELs) were used as sources, and the experiments were performed in a quasi-CW mode (see, for example, [Bewley1993; Van Dantzig and Planken 1999; Dekorsy2003; Ganichev and Prettl 2006; Heyman1994; Markelz1994a; Markelz1994b; Mayer and Keilmann 1986a, Mayer and Keilmann 1986b; Pellemans and Planken 1998; Winnerl2000]). Almost all these experiments were performed on semiconductors, both bulk and nanostructured.

Since the 1990s THz spectroscopy is dominated by the THz time-domain spectroscopy (THz-TDS) method (see, for example, [Grischkowsky1990; Jepsen2011; Tonouchi2007; Ulbricht2011]). In THz-TDS, single-cycle THz pulses are generated from femtosecond laser pulses in photoconductive antennas or via optical rectification in $\chi^{(2)}$-nonlinear crystals. The detection of such THz transients in THz-TDS is enabled via photoconductive detection, or via free-space electrooptic sampling (FEOS) in $\chi^{(2)}$-nonlinear crystals [Gallot and Grischkowsky 1999; Planken2001; Zhang and Turchinovich 2021]. Both of these techniques involve a femtosecond laser pulse as a time gate, providing for the field-resolved detection of THz electromagnetic transients with sub-cycle temporal resolution. Importantly, FEOS also permits the calibrated detection of THz fields, yielding the temporal evolution of the instantaneous electric field strength in absolute units within the detected THz field transient [Gallot and Grischkowsky 1999; Planken2001; Zhang and Turchinovich 2021]. If certain conditions are observed, the recorded FEOS signals can be rigorously reconstructed back to the electric field evolution in the propagating THz signal, and even to the initial polarization or magnetization dynamics in the THz emitter (see, for example, [Hafez2018; Zhang2020; Zhang and Turchinovich 2021]). We note that standard table-top THz generation via photoconductive mechanisms or via optical rectification yields THz fields not exceeding a few kV/cm in strength, thus only providing for spectroscopy in the linear regime.

In 2002 highly efficient strong-field THz generation via optimized optical rectification by tilted pulse front pumping (TPFP) of $\chi^{(2)}$-nonlinear crystals was proposed by Hebling et al. [Hebling2002], which later led to the demonstration of THz pulses with peak field strengths reaching 250 kV/cm by Yeh et al. [Yeh2007]. This demonstration paved the way to modern nonlinear THz spectroscopy using table-top sources. Since then, the TPFP of lithium niobate (LN) crystals by mJ-level femtosecond laser pulses dominates the table-top strong-field THz generation, typically yielding single-cycle THz pulses with the spectrum covering the range $0 - 3$ THz, and with (sub-)MV/cm peak electric fields. Amplified Ti:Sapphire femtosecond lasers, delivering mJ-level femtosecond pulses with 800 nm central wavelength remain the most popular pumping sources for strong-field THz generation via the TPFP method in LN, however other types of femtosecond lasers, e.g., Yb-based [Hoffmann2007], can also be used. Other modern methods of table-top strong-field THz generation include air-plasma THz generation [Tani2012], typically yielding ultrabroadband single-cycle THz pulses with the spectrum covering the range of $0 - 20$ THz or even broader, and optical rectification in organic nonlinear crystals [Shalaby and Hauri 2015]. We refer the reader to the following reviews and article collections on modern methods of strong-field THz generation and nonlinear THz spectroscopy [Elsaesser2019; Hafez2016; Hoffmann and Fülöp 2011; Hwang2015; Kampfrath2013; Leitenstorfer2014].

Furthermore, latest-generation relativistic accelerator-based sources, such as those at the TELBE facility at Helmholtz-Zentrum Dresden-Rossendorf [Green2016], are also actively used in modern THz NLO research. They deliver quasi-monochromatic multi-cycle THz pulses with peak fields currently reaching 100s kV/cm (and with the potential to reach MV/cm levels), and are precisely synchronized to a table-top femtosecond laser [Kovalev2017], thus enabling calibrated FEOS of the THz fields in the experiment [Hafez2018; Kovalev2021].

### 3.8.1.3 Considerations for THz NLO research

Since the strongest THz nonlinearities are based on the physical absorption of the driving THz field by the electrons or polar phonons in the materials, such THz nonlinear effects are therefore dissipative. Furthermore, almost all reported THz nonlinearities are based on field effects. Therefore, the THz field strength is the key parameter characterizing the pump signal in THz NLO, and not the integrated THz pulse energy, power or irradiance, as is common in "traditional" NLO.

Interestingly, almost all reported strong THz nonlinearities are generally non-perturbative, i.e., they do not follow a clear power law scaling over the entire range of applied THz peak fields, and often demonstrate saturation behavior. However, in certain cases it is possible to observe a clear perturbative, power law scaling over a considerable range of pumping THz field strengths, usually around the lower end of pumping strengths. In this case it is possible to use this small-signal nonlinearity regime and establish the effective nonlinear coefficients of the material similar to "traditional" NLO [Cheng2020; Deinert2021; Hafez2018].

Owing to a generally dissipative nature of THz NLO processes, "classical" phase matching between the pump wave and the generated NLO signal is not the dominating concept dictating the conversion efficiency in THz NLO. The conversion efficiency here is rather dependent on the efficiency of energy transfer from the pump signal to the material itself (pump absorption), reabsorption of the generated THz NLO signal within the material, and the outcoupling efficiency of the generated NLO signal.

Finally, and in stark contrast to "traditional" NLO, most of the experiments on THz NLO are performed using strong-field single-cycle THz pulses, usually produced via TPFP in LN. Such single-cycle pulses are per default ultra-broadband, and contain many octaves of frequencies. This often leads to the observation of frequency-dependent effective nonlinearities, sometimes even of different signs (see, e.g., [Turchinovich2012]), co-existing within the ultrabroadband bandwidth of a single-cycle THz pulse. However, observation of THz nonlinearities in a multi-cycle, quasi-monochromatic THz pumping regime is also possible. Such THz pump fields are either generated at large-scale facilities such as FELs (see, e.g., [Jadidi2016; König-Otto2017; Zaks2012]) and TELBE (see, e.g., [Deinert2021; Hafez2018; Kovalev2020, Kovalev2021]), or are produced via the monochromatization of a strong-field LN-generated single-cycle THz pulse using bandwidth filtering [Cheng2020].

### 3.8.1.4 Description of general table outline

Table 9 shows a representative list of THz nonlinearities taken from the literature since 2000. As THz NLO is quite different from "traditional" NLO, Table 9 has been organized in a different way as compared to the previously presented tables and contains the following columns:

- Material type and relevant material information such as doping level (specified either as doping concentration $N_c$ or Fermi energy $E_F$) and type (electrons or holes)
- Additional material information if available
- Material dimensionality (dim.) which can be 0D/1D/2D/3D
- Parameters of THz pump signal used in the measurement: source type, signal type namely single-cycle / few cycle / quasimonochromatic, central frequency (for quasimonochromatic / few cycle pulses) or frequency range (for single-cycle THz pulses), and peak THz field strength
- Physical mechanism of THz nonlinearity
- Nonlinearity type: dissipative (diss.)/non-dissipative (non-diss.)
- Type of observed nonlinear effect, such as nonlinear absorption, high-harmonics generation (HHG) etc.
- Measured nonlinear coefficients. Note that the coefficients $\chi^{(n)}$ are reported for the pumping field range corresponding to the perturbative nonlinearity regime.

Within each column the information is given in the order of the header description, and powers of 10 (e.g., $10^{\pm\alpha}$) are written as E±α for compactness.

In Table 9 we only listed papers presenting nonlinear coefficients of various materials, measured and quantified in a fashion compatible with "traditional" NLO ($\chi^{(n)}$ coefficients, conversion efficiency, saturable absorption parameters, nonlinear refractive index or refractive index modulation etc). We thus did not include papers demonstrating the nonlinear effect, but not quantifying its strength, nor papers that only report a relative transmission change of the THz field through the material.

Furthermore, we only listed papers where the THz pump field was characterized using the calibrated FEOS [Gallot and Grischkowsky 1999; Planken2001; Zhang and Turchinovich 2021] method at the sample position, with one exception of an FEL signal where the THz field at the sample position could be precisely calibrated from the power measurement [König-Otto 2017]. We thus excluded papers reporting only the integrated power/energy/irradiance as the characteristic of the pump THz signal, since it is the THz field that is key for the THz nonlinearity. In addition, we excluded the works reporting on THz Z-scans, especially those using single-cycle ultrabroadband THz pulses. The focal spot for such an ultrabroadband pulse is per definition strongly frequency-dependent, and the measurement of a precise spot shape is highly nontrivial since the THz cameras usually have a rather limited, and strongly frequency-dependent sensitivity. As a result, the precise quantitative determination of NLO coefficients of materials using THz Z-scan is a highly challenging task, prone to too many experimental uncertainties. Finally, we remark that the works included in Table 9 nominally report data obtained at room temperature, unless denoted otherwise.

### 3.8.2  Discussion

Most, and also the strongest nonlinearities presented in Table 9 are dissipative and are based on nonlinear conduction of free electrons in strong THz fields. These nonlinearities are observed directly in the THz field interacting with the material. The non-dissipative nonlinearities such as the THz Kerr effect or THz-driven QCSE, are rather observed in THz pump – optical probe measurements, where the THz field modifies the propagation conditions for the optical-frequency probe in the material, such as its polarization state (Kerr effect) or resonant absorption coefficient (QCSE).

#### 3.8.2.1  Advancement and remaining challenges in THz NLO research

The major advancement in THz NLO since the demonstration of table-top generation of strong-field THz pulses via TPFP of LN around 2002-2007 is the ability to implement a nonlinear THz-TDS scheme. Here one is able to "look inside an optical cycle" during the NLO interaction, i.e., to time-resolve the nonlinear propagation of a THz field transient with sub-cycle resolution. Such a type of measurement, giving an unprecedented direct look into the initiation and development of an optical nonlinearity at the level of the light field, is presently unattainable at other frequency ranges featuring much faster oscillating optical fields, that are therefore more problematic to directly sample in the time domain.

The future progress in THz NLO will be most likely driven by the wider availability of strong-field THz sources in combination with highly sensitive field- and time-resolved THz signal detection, the development of novel nonlinear spectroscopy techniques, and the availability of novel nonlinear materials. The following factors will contribute: (i) broader availability of strong-field single-cycle and multi-cycle THz field sources, both table-top and at large-scale facilities; (ii) improved sensitivity in the THz field detection, in particular by increasing the repetition rate of the experiment using high-pulse-energy (multi-mJ level), high-repetition-rate (100 kHz or higher) femtosecond lasers; (iii) development of more advanced data acquisition and analysis protocols; (iv) further development of novel nonlinear THz spectroscopy techniques, e.g., multi-dimensional THz spectroscopy (see, e.g., [Elsaesser2019; Grechko2018; Junginger2012; Kuehn2011; Woerner2013]), and (v) broader availability of novel nonlinear materials such as quantum materials with higher potential for nonlinear ultrafast electron conduction.

### 3.8.2.2  Recommendations for future works on THz NLO research

In the future works on THz NLO, we do strongly recommend to precisely characterize the THz field at the position of the sample via calibrated FEOS and to provide this calibrated field transient in the publication. Once all the THz signals in the experiment – pump and nonlinear product(s) – are calibrated, it becomes rather straightforward to quantify the parameters of the NLO interaction, such as the effective nonlinear coefficient, in the fashion used in "traditional" NLO.

For example, a broader use of quasi-monochromatic strong-field pumping, either at large-scale facilities such as TELBE (e.g., [Deinert2021; Hafez2018; Kovalev2021]), or via table-top quasi-monochromatic strong-field generation (e.g., [Cheng2020; Lee2020]), combined with fully calibrated FEOS [Gallot and Grischkowsky 1999; Planken2001; Zhang and Turchinovich 2021] detection, should lead to more rigorously calibrated experiments on discrete THz HHG in various materials (see [Deinert2021; Hafez2018; Kovalev2021]). Such measurements allow in particular to extract the THz nonlinear susceptibility coefficients $\chi^{(n)}$ by measuring the conversion efficiency from the pump field to the n-th harmonic, over a wide range of pumping field strengths and frequencies [Hafez2018; Hafez2020; Cheng2020].

Finally, we strongly recommend providing comprehensive information on the nonlinear material used, on the details of the experimental setups, and on the protocols of data processing. This should allow for the reproducibility of the THz NLO results by the broader community.

### 3.8.3  Data table for THz NLO

**Table 9**: THz nonlinearities from representative works since 2000. Legend for superscripts: see below the table.

| Material type; Fabrication; Key material properties; | Dim. | THz source; Single-cycle / quasimonochromatic; THz pump frequency range (for single cycle) or central pump frequency (for quasimonochromatic); THz peak field strength range | Physical mechanism of THz-induced NLO effect | Diss./ Non-diss. | Observed nonlinear effects | Measured THz nonlinear coefficients | Reference |
|---|---|---|---|---|---|---|---|
| Al$_2$O$_3$ | 3D | TPFP of LN; single-cycle; 0 - 3 THz; max. 2.1 MV/cm | THz-induced Kerr effect in bulk material probed by an 800 nm probe pulse | Non-diss. | Transient birefringence at optical probe wavelength of 800 nm, measured via optical polarization evolution | n$_2$ (1 THz) = 0.7E-16 cm$^2$/W; max. Δn (800 nm) = 0.6E-6 | [Sajadi 2015] |
| Benzene (liquid) | 3D | TPFP of LN; single-cycle; 0 - 3 THz; 30 - 150 kV/cm | THz-induced Kerr effect in liquid, probed by an 800 nm probe pulse | Non-diss. | Transient birefringence at optical probe wavelength of 800 nm, measured via optical polarization evolution | χ$^{(3)}$ = 0.22E-20 m$^2$/V$^2$; n$_2$ = 56E-16 cm$^2$/W; K = 0.26E-14 m/V$^2$ (Kerr coefficient*) | [Hoffmann 2009a] |

| | | | | | | |
|---|---|---|---|---|---|---|
| CCl$_4$ (liquid) | 3D | TPFP of LN; single-cycle; 0 - 3 THz; 30 - 150 kV/cm | THz-induced Kerr effect in liquid, probed by an 800 nm probe pulse | Non-diss. | Transient birefringence at optical probe wavelength of 800 nm, measured via optical polarization evolution | $\chi^{(3)}$ = 0.10E-20 m$^2$/V$^2$; n$_2$ = 27E-16 cm$^2$/W; K = 0.12E-14 m/V$^2$ (Kerr coefficient*) | [Hoffmann 2009a] |
| CHCl$_3$ (liquid) | 3D | TPFP of LN; single-cycle; 0 - 3 THz; 30 - 150 kV/cm | THz-induced Kerr effect in liquid, probed by an 800 nm probe pulse | Non-diss. | Transient birefringence at optical probe wavelength of 800 nm, measured via optical polarization evolution | $\chi^{(3)}$ = 0.04E-20 m$^2$/V$^2$; n$_2$ = 10E-16 cm$^2$/W; K = 0.045E-14 m/V$^2$ (Kerr coefficient*) | [Hoffmann 2009a] |
| CH$_2$I$_2$ (liquid) | 3D | TPFP of LN; single-cycle; 0 - 3 THz; 30 - 150 kV/cm | THz-induced Kerr effect in liquid, probed by an 800 nm probe pulse | Non-diss. | Transient birefringence at optical probe wavelength of 800 nm, measured via optical polarization evolution | $\chi^{(3)}$ = 0.70E-20 m$^2$/V$^2$; n$_2$ = 140E-16 cm$^2$/W; K = 0.75E-14 m/V$^2$ (Kerr coefficient*) | [Hoffmann 2009a] |
| CS$_2$ (liquid) | 3D | TPFP of LN; single-cycle; 0 - 3 THz; 30 - 150 kV/cm | THz-induced Kerr effect in liquid, probed by an 800 nm probe pulse | Non-diss. | Transient birefringence at optical probe wavelength of 800 nm, measured via optical polarization evolution | $\chi^{(3)}$ = 2.08E-20 m$^2$/V$^2$; n$_2$ = 440E-16 cm$^2$/W; K = 2.4E-14 m/V$^2$ (Kerr coefficient*) | [Hoffmann 2009a] |
| Diamond; polycrystalline | 3D | TPFP of LN; single-cycle; 0 - 3 THz; max. 2.1 MV/cm | THz-induced Kerr effect in bulk material probed by an 800 nm probe pulse | Non-diss. | Transient birefringence at optical probe wavelength of 800 nm, measured via optical polarization evolution | n$_2$ (1 THz) = 3.0E-16 cm$^2$/W; max. Δn (800 nm) = 1.03E-6 | [Sajadi 2015] |
| GaAs; doped, $N_c = 8E15$ cm$^{-3}$ (electrons), thickness d = 0.4 mm; | 3D | TPFP of LN; single-cycle; 0.2 - 2.5 THz; 9 - 292 kV/cm | intervalley transfer and increase of effective mass of THz-driven free electrons | Diss. | Frequency-dependent nonlinear conductivity and saturable absorption | Frequency-dependent index change $\Delta n$ = -0.13 - 0.08; Reduction of power absorption coefficient by ca. 50% across the whole measurement spectrum | [Turchinovich 2012] |
| GaAs; doped, $N_c = 8E15$ cm$^{-3}$ (electrons), thickness d = 0.4 mm; | 3D | TPFP of LN; single-cycle; 0.2 - 2.5 THz; 9 - 292 kV/cm | intervalley transfer and increase of effective mass of THz-driven free electrons | Diss. | Frequency-integrated saturable absorption; nonlinear pulse group delay | Saturation fluence = 8.2 $\mu$J/cm$^2$; Max. nonlinear pulse compression $\Delta T/T \sim$ 0.1; Max. group index change $\Delta n_g \sim$ 0.1 | [Hoffmann and Turchinovich 2010] |
| GaAs in external electric field; intrinsic, photoexcited at 400 nm to create conductivity; | 3D | Unspecified source; single-cycle, monochromatized using bandpass filter; quasimonochromatic; 0.6 THz; 50 kV/cm; | THz E-FISH in static applied symmetry-breaking electric bias field up to $E_b$ = 15 kV/cm | Diss. | SHG and THG | Maximum effective $\chi^{(2)} = \chi^{(3)} * E_b = 1.7E-7$ m/V, corresponding to natural $\chi^{(3)} = 1E-14 - 1E-13$ m$^2$/V$^2$ ; Max power conversion efficiency $\eta_{SHG} \sim$ 5E-5; $\eta_{THG} \sim$ 1E-5, at optical pump fluence $F_p$ = 4 $\mu$J/cm$^2$ | [Lee 2020] |
| GaP;<br>doped, $N_c$ = 1E16 cm$^{-3}$ (electrons),<br>thickness d = 0.3 mm; | 3D | TPFP of LN;<br>single-cycle;<br>0.2 - 2.5 THz;<br>9 - 292 kV/cm | intervalley transfer and increase of effective mass of THz-driven free electrons | Diss. | Frequency-integrated saturable absorption; nonlinear pulse group delay | Saturation fluence = 20.9 µJ/cm$^2$;<br>Max. nonlinear pulse compression ΔT/T ~ 5E-2;<br>Max. group index change Δn$_g$ ~ 0.05 | [Hoffmann and Turchinovich 2010] |
| Ge;<br>doped, $N_c$ = 1E14 cm$^{-3}$ (electrons),<br>thickness d = 6 mm; | 3D | TPFP of LN;<br>single-cycle;<br>0.2 - 2.5 THz;<br>9 - 292 kV/cm | intervalley transfer and increase of effective mass of THz-driven free electrons | Diss. | Frequency-integrated saturable absorption; nonlinear pulse group delay | Saturation fluence = 3.1 µJ/cm$^2$;<br>Max. nonlinear pulse compression ΔT/T ~ 5E-2;<br>Max. group index change Δn$_g$ ~ 2E-3 | [Hoffmann and Turchinovich 2010] |
| Graphene;<br>produced by thermal decomposition of SiC on the C-face of 4H-SiC;<br>n-doped (inhomogeneously within the sample) | 2D | FEL (FELBE);<br>quasimonochromatic;<br>19 THz;<br>max. 25 kV/cm | FWM in Landau-quantized graphene (under 4.5 T magnetic field and 10 K cryogenic temperature) | Diss. | Degenerate FWM and transient grating generation | χ$^{(3)}$ ~ 9.2E-20 m$^3$/V$^2$ (bulk susceptibility)<br>Temperature = 10 K | [König-Otto 2017] |
| Graphene;<br>CVD-grown;<br>doped, $E_F$ = 0.2 eV (holes); | 2D | air-plasma source;<br>single-cycle;<br>0 - 15 THz;<br>100 - 300 kV/cm | THz-driven impact ionization of carriers | Diss. | Modulation of optical density at 800nm probe wavelength | Optical density change ΔOD ~ 0.01 - 0.1 at 800 nm | [Tani 2012] |
| Graphene;<br>CVD-grown;<br>doped, $N_c$ = 2.1E12 cm$^{-2}$ (holes), $E_F$ = 0.17 eV | 2D | TELBE;<br>quasimonochromatic;<br>0.3 THz, 0.37 THz, 0.68 THz;<br>0-100 kV/cm | thermodynamic response of free electrons | Diss. | HHG (odd): 3, 5, 7 | $\chi^{(3)}\sim$ 1E-9 m$^2$/V$^2$;<br>$\chi^{(5)}\sim$ 1E-22 m$^4$/V$^4$;<br>$\chi^{(7)}\sim$ 1E-38 m$^6$/V$^6$;<br>Max. field conversion efficiencies:<br>THG: η = 2E-3;<br>5HG: η = 2.5E-4;<br>7HG: η = 8E-5 | [Hafez 2018] |
| Graphene;<br>CVD-grown;<br>doped, $N_c$ = 6E11 cm$^{-2}$ (holes), $E_F$ = 0.07 eV | 2D | TPFP of LN;<br>single-cycle;<br>0.3 - 2 THz;<br>2 - 120 kV/cm | thermodynamic response of free electrons | Diss. | Frequency-dependent nonlinear conductivity and saturable absorption | Power loss = 12.5%, non-saturable power loss = 12.5% | [Mics 2015] |
| Graphene | 2D | TPFP of LN,<br>monochromatization using bandpass filter;<br>quasimonochromatic;<br>0.8 THz;<br>12-31 kV/cm | thermodynamic response of free electrons | Diss. | THG | $\chi^{(3)}\sim$ 1E-9 m$^2$/V$^2$; | [Cheng 2020] |
| InGaAs/GaAs QD-based SESAM for 1040 nm;<br>MBE-grown, 80 QD layers;<br>intrinsic; | 0D | TPFP of LN;<br>single-cycle;<br>0.2 - 3.0 THz;<br>10 - 100 kV/cm | THz-driven QCSE in QDs | Non-diss. | Modulation of absorption at 1040 nm probe wavelength | Absorption change Δα = 3% at 1040 nm | [Hoffmann 2010] |
| LDPE (low-density polyethylene) | 3D | TPFP of LN; single-cycle; 0 - 3 THz; max. 2.1 MV/cm | THz-induced Kerr effect in bulk material probed by an 800 nm probe pulse | Non-diss. | Transient birefringence at optical probe wavelength of 800 nm, measured via optical polarization evolution | $n_2$ (1 THz) = 2.0E-16 $cm^2$/W; max. $\Delta n$ (800 nm) = 3.15E-6 | [Sajadi 2015] |
| $LiNbO_3$ | 3D | TPFP of LN; few-cycle; 0.35 THz and 1.1 THz; max. 10 kV/cm | THz DFG at 0.76 THz | Diss. | DFG | $\chi^{(2)} > 1.5E-6$ m/V | [Lu 2021] |
| Metamaterial based on GaAs; metallic (Au/Cr) split-ring resonators deposited on n-doped ($N_c$ = 1E16 $cm^{-3}$) and semi-insulating GaAs. | 3D | TPFP of LN; single-cycle; 0.2 - 1.2 THz; 24 - 400 kV/cm | THz-driven intervalley scattering and impact ionization in GaAs | Diss. | Modulation of THz relative permittivity $\varepsilon$ | Relative permittivity change at the metamaterial resonant frequency. Doped GaAs: $\varepsilon = -2.5 - 1$ at f = 0.82 THz, THz field range 24 - 400 kV/cm; Semi-insulating GaAs: $\varepsilon = -12.5 - 1$ at f = 0.85 THz, THz field range 100 - 400 kV/cm | [Fan 2013] |
| Metamaterial based on golden grating deposited on graphene; CVD-grown graphene; doped, $E_F \sim 0.1$ eV (holes) | 2D | TELBE; quasimonochromatic; 0.7 THz; 5 - 70 kV/cm | thermodynamic response of free electrons; plasmonic THz field concentration in a grating | Diss. | THG | $\chi^{(3)} \sim$ 3E-8 $m^2$/$V^2$; Field conversion efficiency $\eta \sim$ 1E-2 | [Deinert 2021] |

| MgO | 3D | TPFP of LN; single-cycle; 0 - 3 THz; max. 2.1 MV/cm | THz-induced Kerr effect in bulk material probed by an 800 nm probe pulse | Non-diss. | Transient birefringence at optical probe wavelength of 800 nm, measured via optical polarization evolution | $n_2$ (1 THz) = 0.5E-16 cm$^2$/W; max. $\Delta n$ (800 nm) = 0.7E-6 | [Sajadi 2015] |
|---|---|---|---|---|---|---|---|
| Semiconducting SWCNT; produced by CoMoCAT-process; nominally intrinsic | 1D | TPFP of LN; single-cycle; 0.2 - 3 THz; 50 - 420 kV/cm | THz-driven QCSE and interband transitions in CNTs | Non-diss. | Modulation of absorption at 1.2-1.25 eV probe energy | Absorption change $\Delta\alpha$ = 4% at 1.2-1.25 eV | [Shimano 2012] |
| Si; nominally intrinsic | 3D | TPFP of LN; single-cycle; 0 - 3 THz; max. 2.1 MV/cm | THz-induced Kerr effect in bulk material probed by an 800 nm probe pulse | Non-diss. | Transient birefringence at optical probe wavelength of 800 nm, measured via optical polarization evolution | $n_2$ (1 THz) = 56E-16 cm$^2$/W; $\Delta n$ (800 nm) = 65E-6 | [Sajadi 2015] |
| SiN | 3D | TPFP of LN; single-cycle; 0 - 3 THz; max. 2.1 MV/cm | THz-induced Kerr effect in bulk material probed by an 800 nm probe pulse | Non-diss. | Transient birefringence at optical probe wavelength of 800 nm, measured via optical polarization evolution | $n_2$ (1 THz) = 0.08E-16 cm$^2$/W; max. $\Delta n$ (800 nm) = 0.05E-6 | [Sajadi 2015] |
| TPX (polymethylpentene) | 3D | TPFP of LN; single-cycle; 0 - 3 THz; max. 2.1 MV/cm | THz-induced Kerr effect in bulk material probed by an 800 nm probe pulse | Non-diss. | Transient birefringence at optical probe wavelength of 800 nm, measured via optical polarization evolution | $n_2$ (1 THz) = 0.3E-16 cm$^2$/W; max. $\Delta n$ (800 nm) = 0.36E-6 | [Sajadi 2015] |

* Kerr coefficient is defined in the paper as $K = \Delta n / \lambda E^2$

# 4 Conclusion

We have identified general best practices for performing and reporting NLO measurements regardless of the NLO technique used, and have also highlighted several technique-specific best practices. Furthermore, we have introduced a set of tables with representative NLO data from the literature since 2000 for bulk materials, solvents, 0D-1D-2D materials, metamaterials, fiber waveguiding materials, on-chip waveguiding materials, hybrid waveguiding systems, and materials suitable for THz NLO. The data were selected based on the identified best practices and on special considerations for the different material types. For each of the material categories, we have also discussed the background prior to 2000, highlighted the recent advancements and remaining challenges, and concluded with recommendations for future NLO studies.

As shown in the discussions, the field of NLO has gained considerable momentum over the past two decades thanks to major breakthroughs in material science and technology. This has given rise to an enormous growth in NLO publications. However, many of them were not included in the tables presented here as they provided too limited information to comply with the best practices. The publications that brought most value to the tables are those that provide one or several NLO coefficients – and possibly also conversion efficiencies – for one or several materials, wavelengths, pulse durations, etc., measured and reported along the best practices. The dependence of NLO coefficients on wavelength, pulse duration, etc. is very insightful information not only from a fundamental science perspective (as it allows distinguishing NLO processes while ruling out, e.g., thermal effects) but also from an application point of view. To assess the practical applicability of a NLO material, it is key that papers also clearly specify the material properties and fabrication details, and provide information on both the nonlinear and linear optical characteristics, such as the linear loss. This will allow evaluating the suitability of the material for specific NLO applications along well-defined figures-of-merit.

We encourage the NLO community to take all these aspects into account and implement the presented best practices in future works. In fact, there is still much to be discovered in NLO research, and in the coming years we intend to update the data tables by considering additional NLO processes and by adding materials that are currently not included or yet to be investigated. Hence, for those future investigations we want to stimulate the use of the listed best practices to allow a more adequate comparison, interpretation and practical implementation of the published parameters and as such further the fundamental understanding of NLO as well as its exploitation in real-life applications.

# 5 Description of author contributions

In the order of the author list:

Nathalie Vermeulen: co-coordinated the data table initiative together with Eric Van Stryland; contributed to the data collection and text of hybrid waveguiding systems (as team leader), on-chip waveguiding materials, and 0D-1D-2D materials; compiled the data table for hybrid waveguiding systems; contributed to the general introduction and conclusion; wrote the best practices Sections 2.2.2.1, 2.2.3, and the introductory part of Section 3; compiled the manuscript and contributed to the final editing of the manuscript

Daniel Espinosa: contributed to the data collection and text of bulk materials, 0D-1D-2D materials, metamaterials, and on-chip waveguiding materials; compiled the data table for on-chip waveguiding materials

Adam Ball: contributed to the data collection and text of bulk materials and metamaterials

John Ballato: contributed to the data collection and text of fiber waveguiding materials (as team leader) and hybrid waveguiding systems; compiled the data table for fiber waveguiding materials; contributed to the final editing of the manuscript

Philippe Boucaud: contributed to the data collection and text of bulk materials and on-chip waveguiding materials

Georges Boudebs: contributed to the data collection and text of bulk materials

Cecília Campos: contributed to the text of 0D-1D-2D materials; compiled the data table for 0D-1D-2D materials

Peter Dragic: contributed to the data collection and text of fiber waveguiding materials and hybrid waveguiding systems

Anderson Gomes: contributed to the data collection and text of 0D-1D-2D materials (as team leader), and bulk materials; contributed to the best practices Section 2.2.2.4

Mikko Huttunen: contributed to the data collection and text of metamaterials (as team leader) and 0D-1D-2D materials; compiled the data table for metamaterials; wrote the best practices Sections 2.2.1.1, 2.2.1.2, 2.2.2.6

Nathaniel Kinsey: contributed to the data collection and text of bulk materials (as team leader) and metamaterials; compiled the data table for bulk materials; contributed to the best practices section 2.2.2.4

Rich Mildren: contributed to the data collection and text of bulk materials and on-chip waveguiding materials; wrote the best practices Section 2.2.2.2

Dragomir Neshev: contributed to the data collection and text of 0D-1D-2D materials and metamaterials

Lázaro Padilha: contributed to the data collection and text of 0D-1D-2D materials

Minhao Pu: contributed to the data collection and text of on-chip waveguiding materials (as team leader) and hybrid waveguiding systems

Ray Secondo: contributed to the data collection and text of bulk materials and metamaterials

Eiji Tokunaga: contributed to the data collection and text of bulk materials and solvents; wrote the best practices Section 2.2.2.5

Dmitry Turchinovich: collected the data and wrote the text for THz NLO (as team leader); compiled the data table for THz NLO

Jingshi Yan: contributed to the data collection and text of 0D-1D-2D materials

Kresten Yvind: contributed to the data collection and text of on-chip waveguiding materials and hybrid waveguiding systems

Ksenia Dolgaleva: contributed to the data collection and text of bulk materials and 0D-1D-2D materials; contributed to the data collection and wrote most of the text of metamaterials and on-chip waveguiding materials

Eric Van Stryland: co-coordinated the data table initiative together with Nathalie Vermeulen; collected the data and wrote the text for fused silica and solvents (as team leader); compiled the data tables for fused silica and solvents; contributed to the data collection and text of bulk materials; contributed to the general introduction, conclusion and the best practices Section 2.2.2.4; wrote the best practices Section 2.2.2.3; contributed to the final editing of the manuscript

# Acknowledgements

NV acknowledges the financial support from Fonds Wetenschappelijk Onderzoek (FWO) under Grants G005420N and G0F6218N (EOS-convention 30467715).
JB acknowledges support from the J. E. Sirrine Foundation.
PB acknowledges the French National Research Agency (Agence Nationale de la Recherche, ANR) - OPOINt project (ANR-19-CE24-0015).
GB acknowledges the support from the University of Angers and from the NNN-TELECOM Program, region des Pays de la Loire, contract n° 2015 09036.
PD acknowledges the U.S. Department of Defense Directed Energy Joint Transition Office (DE JTO) (N00014-17-1-2546) and the Air Force Office of Scientific Research (FA9550-16-1-0383).
ASLG and CLAVC acknowledge support from Brazilian INCT of Photonics (CNPq, CAPES, FACEPE), and Air Force Office of Scientific Research (AFOSR) under grant FA9550-20-1-0381.
MJH acknowledges the support of the Flagship of Photonics Research and Innovation (PREIN) funded by the Academy of Finland (Grant No. 320165).
NK, AB, RS acknowledge support from Air Force Office of Scientific Research (FA9550-18-1-0151) and the National Science Foundation (1808928).
RPM acknowledges funding from AFOSR FA2386-21-1-4030 and Australian Research Council LP200301594.
DN and JY acknowledge the support by the Australian Research Council through the Centres of Excellence program (CE20010001) and NATO SPS program (OPTIMIST).

MP and KY acknowledge the financial support from Danish National Research Foundation through the Research Centre of Excellence, Silicon Photonics for Optical Communications (SPOC) (ref. DNRF123). MP also acknowledges the financial support from the European Union's Horizon 2020 research and innovation programme (grant agreement N° 853522 REFOCUS).

DT acknowledges the financial support from European Union's Horizon 2020 research and innovation programme (grant agreement N° 964735 EXTREME-IR), and from Deutsche Forschungsgemeinschaft (DFG) within the project 468501411–SPP2314 INTEGRATECH under the framework of the priority programme SPP2314 – INTEREST.

KD and DE are grateful to Roberto Morandotti and Sisira Suresh for suggesting relevant references for the tables. KD and DE acknowledge the financial support from Canada Research Chairs program and Natural Science and Engineering Council's Discovery program RGPIN-2020-03989.